\documentclass[bibyear]{aa} 

\usepackage{natbib,twoopt}
\usepackage{ulem}
\usepackage{soul}
\usepackage{ragged2e}
\usepackage[breaklinks=true, colorlinks=true, allcolors=blue]{hyperref} 
\makeatletter
\newcommandtwoopt{\citeads}[3][][]{\href{http://adsabs.harvard.edu/abs/#3}%
{\def\hyper@linkstart##1##2{}%
\let\hyper@linkend\@empty\citealp[#1][#2]{#3}}}
\newcommandtwoopt{\citepads}[3][][]{\href{http://adsabs.harvard.edu/abs/#3}%
{\def\hyper@linkstart##1##2{}%
\let\hyper@linkend\@empty\citep[#1][#2]{#3}}}
\newcommandtwoopt{\citetads}[3][][]{\href{http://adsabs.harvard.edu/abs/#3}%
{\def\hyper@linkstart##1##2{}%
\let\hyper@linkend\@empty\citet[#1][#2]{#3}}}
\newcommandtwoopt{\citeyearads}[3][][]%
{\href{http://adsabs.harvard.edu/abs/#3}
{\def\hyper@linkstart##1##2{}%
\let\hyper@linkend\@empty\citeyear[#1][#2]{#3}}}
\makeatother

\usepackage{graphicx}
\usepackage{txfonts}
\usepackage{comment}
\usepackage{url}
\usepackage{threeparttablex}
\usepackage{makecell}
\usepackage{colortbl}
\newcommand{\angstrom}{\mbox{\normalfont\AA}}
\usepackage{longtable}
\usepackage{verbatim}
\usepackage{graphics,epsfig,amssymb,color,times}
\usepackage{lscape}
\usepackage{pdflscape}
\usepackage{arydshln}
\usepackage[table,xcdraw]{xcolor}
\usepackage{subcaption}
\usepackage{rotating}
\usepackage{float}
\usepackage{soul}
\usepackage{orcidlink}

\titlerunning{SUBWAYS. III.}
\authorrunning{V. E. Gianolli et al.}
\begin{document}

   \title{Supermassive Black Hole Winds in X-rays -- SUBWAYS. III. A population study on ultra-fast outflows}

   \author{V. E. Gianolli$^{1,2}$\thanks{E-mail: vittoria.gianolli@univ-grenoble-alpes.fr}\orcidlink{0000-0002-9719-8740}, S. Bianchi$^{2}$\orcidlink{0000-0002-4622-4240}, P-O Petrucci$^{1}$\orcidlink{0000-0001-6061-3480}, M. Brusa$^{3,4}$\orcidlink{0000-0002-5059-6848}, G. Chartas$^{6}$\orcidlink{0000-0003-1697-6596}, G. Lanzuisi$^{4}$\orcidlink{0000-0001-9094-0984}, G. A. Matzeu$^{3,4,5}$\orcidlink{0000-0003-1994-5322}, M. Parra$^{1,2}$\orcidlink{0009-0003-8610-853X}, F. Ursini$^{2}$\orcidlink{0000-0001-9442-7897}, E. Behar$^{7}$ \orcidlink{0000-0001-9735-4873}, M. Bischetti$^{8}$\orcidlink{0000-0002-4314-021X}, A. Comastri$^{4}$\orcidlink{0000-0003-3451-9970}, E. Costantini$^{9}$\orcidlink{0000-0001-8470-749X}, G. Cresci$^{10}$, M. Dadina$^{4}$\orcidlink{0000-0002-7858-7564}, B. De Marco$^{11}$\orcidlink{0000-0003-2743-6632}, A. De Rosa$^{21}$\orcidlink{0000-0001-5668-6863}, F. Fiore$^{12}$\orcidlink{0000-0001-5742-5980}, M. Gaspari$^{13}$\orcidlink{0000-0003-2754-9258}, R. Gilli$^{4}$\orcidlink{0000-0001-8121-6177}, M. Giustini$^{14}$\orcidlink{0000-0002-1329-658X}, M. Guainazzi$^{15}$, A. R. King$^{16,17}$\orcidlink{0000-0002-2315-8228}, S. Kraemer$^{18}$, G. Kriss$^{19}$\orcidlink{0000-0002-2180-8266}, Y. Krongold$^{20}$\orcidlink{0000-0001-6291-5239}, F. La Franca$^{2}$\orcidlink{0000-0002-1239-2721}, A. L. Longinotti$^{20}$\orcidlink{0000-0001-8825-3624}, A. Luminari$^{21,22}$\orcidlink{0000-0002-1035-8618}, R. Maiolino$^{34,35}$\orcidlink{0000-0002-4985-3819}, A. Marconi$^{23}$\orcidlink{0000-0002-9889-4238}, S. Mathur$^{24,25}$, G. Matt$^{2}$\orcidlink{0000-0002-2152-0916}, M. Mehdipour$^{26}$\orcidlink{0000-0002-4992-4664}, A. Merloni$^{33}$\orcidlink{0000-0002-0761-0130} , R. Middei$^{27}$\orcidlink{0000-0001-9815-9092}, G. Miniutti$^{14}$\orcidlink{0000-0003-0707-4531}, E. Nardini$^{10}$\orcidlink{0000-0001-9226-8992}, F. Panessa$^{21}$\orcidlink{0000-0003-0543-3617}, M. Perna$^{14}$\orcidlink{0000-0002-0362-5941}, E. Piconcelli$^{22}$\orcidlink{0000-0001-9095-2782}, G. Ponti$^{28,33}$\orcidlink{0000-0003-0293-3608}, F. Ricci$^{2}$\orcidlink{0000-0001-5742-5980}, R. Serafinelli$^{22}$\orcidlink{0000-0003-1200-5071}, F. Tombesi$^{29,22,30,31,32}$\orcidlink{0000-0002-6562-8654}, C. Vignali$^{3,4}$\orcidlink{0000-0002-8853-9611}, L. Zappacosta$^{22}$\orcidlink{0000-0002-4205-6884}
    }

   \institute{Affiliations are shown at the end of the paper}

   \date{Received XXX; accepted YYY}

 
  \abstract
   {The detection of blueshifted absorption lines likely associated with ionized iron K-shell transitions in the X-ray spectra of many active galactic nuclei (AGNs) suggests the presence of a highly ionized gas outflowing with mildly relativistic velocities (0.03c - 0.6c) named ultra-fast outflow (UFO). Within the SUBWAYS project, we characterized these winds starting from a sample of 22 radio-quiet quasars at an intermediate redshift (0.1 $\leq$ z $\leq$ 0.4) and compared the results with similar studies in the literature on samples of local Seyfert galaxies (i.e., 42 radio-quiet AGNs observed with {\it XMM-Newton} at z $\leq$ 0.1) and high redshift radio-quiet quasars (i.e., 14 AGNs observed with {\it XMM-Newton} and {\it Chandra} at z $\geq$ 1.4). The scope of our work is a statistical study of UFO parameters and incidence considering the key physical properties of the sources, such as supermassive black hole (SMBH) mass, bolometric luminosity, accretion rates, and spectral energy distribution (SED) with the aim of gaining new insights into the UFO launching mechanisms. 
   We find indications that highly luminous AGNs with a steeper X-ray/UV ratio, $\alpha_{\mathrm{ox}}$, are more likely to host UFOs. The presence of UFOs is not significantly related to any other AGN property in our sample.  
   These findings suggest that the UFO phenomenon may be transient.
   Focusing on AGNs with UFOs, other important findings from this work include: (1) faster UFOs have larger ionization parameters and column densities; (2) X-ray radiation plays a more crucial role in driving highly ionized winds compared to UV; (3) the correlation between outflow velocity and luminosity is significantly flatter than what is expected for radiatively driven winds; (4) more massive black holes experience higher wind mass losses, suppressing the accretion of matter onto the black hole; (5) the UFO launching radius is positively correlated with the Eddington ratio. 
   Furthermore, our analysis suggests the involvement of multiple launching mechanisms, including radiation pressure and magneto-hydrodynamic processes, rather than pointing to a single, universally applicable mechanism.}


   \maketitle
%
\section{Introduction}
\label{sec:intro}

It is well established that active galactic nuclei (AGNs) are powered by supermassive black holes (SMBHs), which reside in the gravitational center of galaxies and actively accrete matter. Many observational correlations have set the basis to the co-evolution paradigms of AGNs and galaxies, suggesting that their formation and evolution are connected \citep[see][for a review]{kormendy13}. However, the underlying mechanisms that drive this co-evolution are still debated. Recent studies have suggested that highly ionized gas outflows may play an important role in regulating the intricate interplay between AGNs and their host galaxies \citep{king15, gaspari17,harrison18}. Therefore, studies of AGN outflows across different scales are essential for advancing our understanding of these phenomena. In particular, various types of ionized outflows have been identified in AGNs, including broad absorption line (BAL) outflows \citep[e.g.,][]{arav01,xu20}; warm absorber (WA) outflows \citep[e.g.,][]{halpern84,mathur97,crenshaw12,tombesi13,laha21}; transient obscuring winds \citep[e.g.,][]{markowitz14,kaastra14}; and ultra-fast outflows \citep[UFOs; e.g.,][]{chartas02,pounds03,cappi06,tombesi10,gofford13}.
Among these, UFOs seem to be capable of injecting substantial amounts of momentum and energy into the interstellar medium (ISM) of the host galaxy, and thus, they are one of the main candidates as prime agents of feedback \citep[e.g.,][for reviews]{king03,king05,tombesi15,gaspari20,laha21}, along with relativistic jets. As a consequence, ejection of material from the inner regions up to the host galaxy scale can proceed in the forms of ionized and molecular winds \citep[e.g.,][]{sturm11,kakkad17} or powerful radio jets \citep[e.g.,][]{whittle92,mukherjee18}.
The primary detection of UFOs occurs through the analysis of X-ray spectra, where they manifest as absorption troughs often associated with blueshifted transitions of highly ionized elements, such as \ion{Fe}{xxv} He$\alpha$, and \ion{Fe}{xxvi} Ly$\alpha$. Mildly relativistic velocities ($\sim$0.03\footnote{This lower limit to the outflow velocity was introduced by \citet{tombesi10}, and it is typically adopted in the literature.} up to 0.6c\footnote{The highest observed outflow velocity is detected by \citet{chartas21} in APM 08279+5255 with v$_\mathrm{out} / {\mathrm{c}}$ = 0.59 $\pm$ 0.03.}) are their main characteristic, together with column densities N$_{\mathrm{H}}$ in the range 10$^{22} - 10^{24}$ cm$^{-2}$ and ionization parameters $\log(\xi / \mathrm{erg\,cm\, s^{-1}}) \simeq 4 - 5.6 $ \citep[see e.g.,][]{chartas02,chartas03,reeves03,braito07,cappi09,tombesi10,giustini11,gofford13, tombesi14,matzeu17,reeves18a, braito18}. Recent observations have revealed the existence of lower-ionization counterparts to highly ionized UFOs in the ultraviolet and soft X-ray bands, highlighting the complex structure of these outflows that should be taken into account by theory and models \citep{longinotti15,kriss18,venturi18,serafinelli19,chartas21,krongold21,vietri22,mehdipour22}. These studies hold the potential to shed new light on the origin and driving mechanisms of UFOs, which are not fully understood.

Due to the observed physical properties, these Fe K absorbers are thought to be launched by radiative \citep{elvis00,kp03,proga04,everett04,sim08,sim10,sim12,schurch09,higginbottom14} and/or magneto-hydrodynamic \citep[MHD;][]{proga00,everett05,kazanas12,fukumura10,fukumura14,fukumura15,sadowski17} processes. 
In the first case, when SMBHs are undergoing substantial accretion, the emitted radiation, which interacts with and applies pressure to the surrounding material, may form a highly ionized outflow. These radiation-driven outflows may be accelerated by the radiation pressure of the continuum or spectral lines \citep[line driven, e.g.,][]{murray95,proga00,proga04,giustini19}. The effectiveness of the latter mechanism largely depends on the ionization state of the gas \citep[i.e., being most powerful at low/moderate ionization states, $\log(\xi / \mathrm{erg\,cm\, s^{-1}})\sim2$, e.g.,][]{arav94}. Nonetheless, \citet{dannen19} demonstrate that with a typical AGN spectral energy distribution (SED), line driving is operative up to $\log(\xi / \mathrm{erg\,cm\, s^{-1}})\sim3$, potentially explaining the acceleration of moderately ionized UFOs.
Highly ionized winds can also be ejected by intense magnetic fields from different regions of the accretion disk, leading to a stratification characterized by an increase in column density, ionization, and velocity closer to the SMBH. The outflow velocity is then directly proportional to the rotational velocity of the disk at each radius, reaching up to relativistic values \citep{fukumura10, fukumura14}. In addition, these magnetic processes can amplify the acceleration of outflows produced by other mechanisms, such as the radiation pressure \citep{everett05,cao14}.
A third acceleration mechanism can also be taken into account that considers the pressure gradient of X-ray-heated gas as the driving force behind the so-called thermal winds \citep{begelman83,dorodnitsyn11,dorodnitsyn12}. However, these winds are expected to exhibit significantly lower velocities (i.e., with a maximum value of about 1000 km s$^{-1}$), as they originate at larger distances from the ionizing source, and thus, they are unlikely to be classified as UFOs.

The influence of UFOs is thought to be able to affect different galactic scales. Depending on the amount of expelled mass, these winds are expected to provide changes to the disk accretion rate, thus regulating the growth of the central BH. 
In particular, the removal of accreting material may affect the optical/UV and, consequently, the X-ray luminosity.
Hence, a connection between the emitted luminosity in both bands and the outflow properties (e.g., velocity, outflow rate, etc.) is likely to be present. 
Potentially, these winds may affect the overall structure of the galaxy \citep[e.g.,][]{marasco20,bertola20,tozzi21}. By removing large amounts of gas and dust from the central regions of the galaxy, they would be able to quench star formation \citep{hopkins10,king15,kraemer18,laha21,salome23} as well as cooling flows \citep[e.g.,][]{gaspari12,mizumoto19}. For this reason, comprehensive studies of AGN outflows employing both detailed case studies and large-scale statistical surveys are crucial. By exploring the relationships between different types of outflows, their origins, and their driving mechanisms, it is possible to understand the complex interplay between SMBHs, the AGN environment, and the formation and evolution of galaxies.

In the first two papers of the SUpermassive Black hole Winds in the x-rAYS (SUBWAYS) series, \citet{matzeu22} report the results of their X-ray spectroscopy study, while \citet{mehdipour22} analyze the ionized outflows in the UV band using the HST/COS instrument \citep{green12}. In particular, \citet{matzeu22} find that the fraction of UFO detections in the SUBWAYS sample (i.e., at moderate redshift; see Sect. \ref{S23sample}) aligns with the findings in the local Universe. Additionally, on the basis of the observed relation between the outflow velocity and the bolometric luminosity, \citet{matzeu22} suggest that radiation pressure is likely the primary launching mechanism of these winds (for more, see Sect.~\ref{ufo_lum_corr} of the present paper). 
From \citet{mehdipour22}, it appears that the properties of the UV outflows detected in the SUBWAYS sample are similar to those seen in local Seyfert-1 galaxies. Interestingly, sources with detected X-ray UFOs do not often exhibit UV absorption counterparts, likely due to the highly ionized nature of the gas, but they consistently display lower-velocity UV outflows \citep[with few exceptions, e.g.,][]{kriss18,mehdipour22}.

The primary objective of this paper is to assess possible relations and differences, or lack thereof, between AGNs hosting UFOs and sources without. By doing so, we aim at gaining further insights into the physical processes occurring near the SMBH that may be responsible for launching UFOs.
The paper is organized as follows: Sect.~\ref{global_par} presents the three samples studied here and the AGN and UFO parameters retrieved from the literature. In Sect.~\ref{derived_par}, we describe the AGN and UFO properties derived during our study. In Sect.~\ref{par_distr}, we evaluate the statistical properties of each sample, and we present all the parameters' distributions. In Sect.~\ref{par_corr}, we describe the most significant results of the extended correlation analysis we performed. In Sect.~\ref{conclusions} is a summary of the results.
Throughout this paper, the following cosmological constants are assumed: H$_{\rm 0}$ = 70 km s$^{-1}$ Mpc$^{-1}$, $\Omega _{\rm \Lambda 0}$ = 0.73, and $\Omega _{\rm M}$ = 0.27.\footnote{We note that the luminosities reported in \citet{chartas21} were derived using different cosmological constants from those adopted here. Consequently, we have corrected them, and although the difference is minimal (resulting in a median increase of $\sim$1\%  after correction), the corrected luminosity will be used in the following analysis.} Errors are quoted at the 90\% confidence level unless otherwise stated.

\section{Sample selection and global properties} \label{global_par}

In order to characterize possible correlations between the AGN and the outflow properties, we provide in this paper a statistical study of three different AGN samples. The main data set is the SUBWAYS sample (``S23 sample'' hereafter), presented by \citet{matzeu22}, which covers an intermediate range of redshift (0.1 $\leq$ z $\leq$ 0.4) and luminosity (6 $\times 10^{44}$ $\leq$ L$_{\mathrm{bol}}$/erg s$^{-1}$ $\leq$ 2$\times 10^{46}$). With the purpose of extending the ranges of both parameters to low and high values, we include as ``comparison samples'' the data sets analyzed by \citet{tombesi10} and \citet{chartas21} (``T10'' and ``C21 sample'', respectively).
We acknowledge the presence of two further systematic UFO studies in the literature: \citet{igo20} and \citet{gofford13}. 
Both samples share the redshift range (and most of the sources) covered by the T10 sample, thus presenting very significant overlaps. On one hand, \citet{igo20} use a completely different methodology with respect to all the other studies, adopting the variability detection method defined by \citet{parker17,parker18}. On the other hand, \citet{gofford13} perform a more standard spectroscopic analysis, but including also radio-loud AGNs and using {\it Suzaku} data. 
Therefore, we have chosen not to consider these additional works in our analysis. However, for the sake of completeness, we will mention their results in the following Sections when appropriate.

\subsection{SUBWAYS (S23) sample}
\label{S23sample}
The SUBWAYS sample is composed by AGNs in the 3XMM-DR7 catalog \citep[XMM-{\it Newton} EPIC Serendipitous Source catalog,][]{rosen16} matched to the SDSS-DR14 catalog \citep[Sloan Digital Sky Survey Quasar Catalog,][]{paris18}, or to the Palomar-Green Bright QSO catalog \citep[PG QSO;][]{schmidt83}.
The adopted selection criteria consider intermediate redshifts, ranging from z = 0.1 to z = 0.4, and bolometric luminosities in the range 10$^{44.5-46}$ erg s$^{-1}$. This roughly translates into a count rate of at least $\sim$0.12 cts/s in the XMM-{\it Newton} EPIC-pn spectra in a single XMM-{\it Newton} orbit, to ensure proper continuum characterization up to 10 keV and detection of faint absorption features. Moreover, Narrow Line Seyfert 1 and AGN in clusters or radio-loud systems were excluded, and thus the sample focuses on isolated radio-quiet AGNs with L$_{bol} \geq 10^{44.5}$ erg $\, $s$^{-1}$.
As a result, the S23 sample counts 22 radio-quiet X-ray AGNs with a total of 81 observations.  

In order to search for \ion{Fe}{xxv} He$\alpha$ and \ion{Fe}{xxvi} Ly$\alpha$ absorption lines, after performing a fit of the broad-band spectrum of each source in the 0.3–10 keV band, a systematic narrow-band (i.e., 5–10 keV) analysis of the {\it XMM-Newton} EPIC-pn observations was performed by \citet{matzeu22}. Afterward, they carried out extensive Monte Carlo (MC) simulations to evaluate the statistical significance of the lines and then obtained a detailed physical modeling of the absorbers with the \texttt{XSTAR} photo-ionization code v2.54a \citep{kallman01,kallman04}, assuming an illuminating SED with a photon index $\Gamma$ fixed to 2 and turbulence velocity in the range $\sigma_{\mathrm{turb}} \sim$ 1000-10000 km s$^{-1}$. 

Based on the characterization of the sample achieved by \citet{matzeu22}, we divided it in two sub-groups: a first sub-sample in which the detection of blueshifted Fe K absorption lines have been made with a Monte Carlo derived confidence level higher than 95\%, and with outflows velocities larger than 0.03c (in the following it will be called ``UFOs sub-sample''); and a second sub-sample whose sources do not show absorption lines in the iron band (``no-UFOs sub-sample''). In particular, the UFOs sub-sample is composed of 7 sources, that is $\sim$32\% of the entire sample. 
It must be noted that due to the transient nature of these outflows \citep[e.g.,][]{tombesi10,pounds12,king15,igo20}, the same source may present observations with a detected UFO and observations without. Hence, we included in the UFO sub-sample all the sources with at least one absorption line detection among the different observations performed.
For example, PG 1114+445 and PG 0804+761 show iron absorption lines in the K band with outflow velocities larger than 0.03c in only two (out of eleven) and one (out of nine) observations, respectively.
It is important to note, however, that the lack of detections could simply be attributed to a combination of observational issues (e.g., insufficient S/N) and/or wind duty cycle (see Sect.~\ref{ufo_vs_noufo}). This should be taken into account when comparing  the incidence of UFOs in each sample (see Table \ref{tab:9}, Fig.~\ref{fig:6}, Sects. \ref{global_par} and \ref{ufo_vs_noufo}).

\subsection{Tombesi et al. (T10) sample}\label{t10_sample}

The sample studied by \citet{tombesi10} includes 35 type 1 and 7 type 2 radio-quiet AGNs (for a total of 101 observations) with $z\leq 0.1$, drawn from the RXTE All-Sky Slew Survey Catalog \citep[XSS;][]{revnivtsev04} and cross-correlated with the XMM-{\it Newton} Accepted Targets Catalog, considering pointed observations available at the date of October 2008. The spectra of the XMM-{\it Newton} EPIC-pn observations must have a net exposure time exceeding 10 ks and an intrinsic equivalent hydrogen column density, N$_{\rm H}$, lower than 10$^{24}$ cm$^{-2}$, to ensure the direct observation of the nuclear continuum in the Fe K band (4-10 keV energy band).

The absorption features have been modeled using \texttt{XSTAR} v2.2 \citep{kallman01}, specifically developing tables computed assuming an illuminating SED with a spectral photon index $\Gamma$=2 and the turbulence velocity ranges between 1000 and 5000 km s$^{-1}$ \citep{tombesi11}. As for the S23 sample, we divided the T10 sample into UFO and no-UFO sub-samples. In particular, 15 AGNs are hosting UFOs (i.e., $\sim$36\% of the total sample). 
We note here that Fe K absorption lines exhibiting N$_{\mathrm{H}}$ and $\xi$ values consistent with typical UFO sources, but with outflowing velocities lower than the UFO threshold (i.e., 0.03c), have been detected in four sources. 
While these outflows share velocities in the range of standard warm absorbers, they instead exhibit column densities and ionization parameters closer to what observed in UFOs.
In any case, these AGN, reported in Table \ref{tab:1} under the "Fe-K sub-sample" label, will be considered as no-UFO sources during the statistical analysis. Indeed, the inclusion of these sources in the UFO sub-sample does not significantly alter the results, apart from marginal differences that will be commented in the corresponding Sections. Similar objects are not present in the other two samples.

\subsection{Chartas et al. (C21) sample}\label{c21_pres}
In relation to the AGN-galaxy co-evolution paradigm that proposes the outflow of highly ionized gas as one of the main feedback mechanisms, it is crucial to consider sources near the peak of the AGN and star formation activity. Therefore, we took into account the quasar sample studied by \citet{chartas21}, in the 1.41–3.91 redshift range.
The authors focus on the gravitationally lensed narrow absorption line (NAL) quasars with blueshifted \ion{C}{iv} troughs present in the Sloan Digital Sky Survey (SDSS) surveys. In addition, they added 7 z > 1 quasars with already reported UFOs and SDSS J1029 +2623, a lensed quasar at z = 2.197.

The spectral fits were performed considering the energy range between 0.3 keV and 11 keV. It must be noted that to assess the physical properties of the UFOs, C21 used the analytic version of \texttt{XSTAR}, \texttt{warmabs}, instead of employing table models as in T10 and S23. In particular, they produced ad hoc new population files with appropriate $\Gamma$ for each observation and they allowed the turbulent velocity to vary ($3000 < \sigma_{\mathrm{turb}}$ / km s$^{-2} < 36000$). 
In order to compare the outflow properties of high-z sources with those of the T10 and S23 samples, consistent procedures are crucial. We thus refitted the C21 quasar sample using the same tables adopted by \citet{matzeu22}. 
We observe that, while v$_{\mathrm{out}}$ lay within the errors of the \texttt{warmabs} model values, the other fit parameters (i.e., photon index, ionization parameters, and column densities) are significantly larger than those measured by \citet{chartas21}. This shows the dependence of these parameters on the model used to fit the data. As said before, with the aim of consistency, we will use the new values, obtained by using a similar procedure to that in the SUBWAYS and T10 sample, keeping in mind that these values are SED dependent. 

We here note that the quasar PID 352 lies within the {\it Chandra} Deep Field South and it was observed with XMM-{\it Newton} during 2001–2002 and 2008–2010 for a total of 33 exposures. Given the complexity of the spectra stacking procedure, we have chosen not to re-analyze this source. Consequently, PID 352 will not be taken into account in our study.
A putative UFO is reported in SDSS J0904+1512 with a significance of only $\sim$90\%. As in both S23 and T10 the threshold to detect significant absorption lines is set to 95\%, this quasar will be included in the no-UFO sub-sample. Consequently, the high-z sample will consist of 13 AGNs of which 12 are hosting UFOs. 
The extremely high UFO incidence in the C21 sample is due to a clear selection bias, since the sources were targeted a priori for their larger probability of hosting UFOs. \\ 

\subsection{Global properties}

\begin{figure}
\centering
    \includegraphics[width=.5\textwidth]{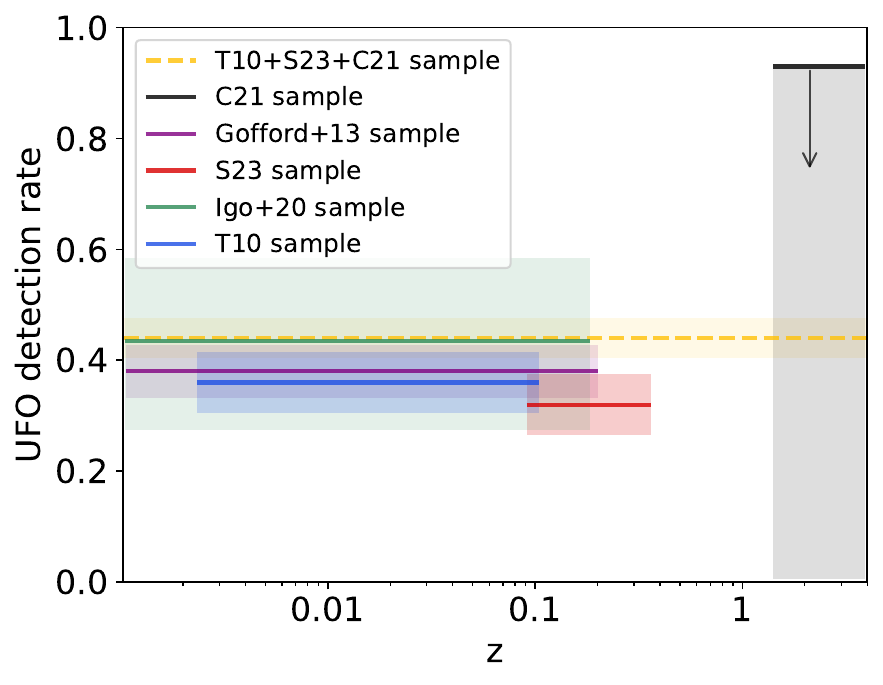}
    \caption[]{{Percentage of UFO detections at $>$95\% confidence level. The red symbolize S23, T10 is in blue, and C21 is shown with black. The upper limit to the detection fraction of AGNs hosting UFOs at high redshift is represented by the C21 sample due to its selection bias toward AGNs hosting UFOs. The hatched dark yellow line represent the total UFOs fraction in the combined sample (34/77 = 44\%). We report in green and purple the fractions obtained by \citet{igo20} and \citet{gofford13}. The colored bands indicate 68\% confidence intervals, calculated adopting the Bayesian approach described in \citet{cameron11}. In the case of \citet{igo20}, the lower limit of the shaded area represents the detection rate for sources with strong evidence of UFOs, whereas the upper limit encompasses the total rate (strong + weak evidence).}
    }
    \label{fig:6}
\end{figure}

In Table \ref{tab:9} a summary of the different sample properties (in terms of redshift, total number of sources and number of sources in the UFO and no-UFO sub-samples) is reported. The comparison between the percentage of UFO detections in the three samples versus redshift is shown in Fig.~\ref{fig:6}. 
We consider the C21 UFOs fraction as an upper limit due to its strong selection bias (see Sect.~\ref{c21_pres}).
In Fig.~\ref{fig:6}, we also add the fraction of UFOs obtained by \citet{igo20} and \citet{gofford13}. 
In particular, the former group identified $\sim$28\%-59\% (i.e., considering both AGNs with strong and weak evidence of UFOs, respectively, 13/58 and 21/58 AGNs) of sources with signatures of UFOs, while the latter observed that 38\% (i.e., 17/45 sources) of radio-quiet AGNs hosted UFOs.

\begin{table}
\centering
\renewcommand{\arraystretch}{1.5}
\setlength{\tabcolsep}{2.5pt}
\caption{Summary of the different sample properties. }
\label{tab:9}
\begin{tabular}{cccccc}
Sample & z & Total & no-UFOs & UFOs & Detection \\ \hline
 &  &  &  &  & [\%] \\ \hline
T10 & $\leq$ 0.1 & 42 & 27 (15) & 15 (7) & 36$^{+6}_{-5}$ \\
S23 & 0.1 - 0.4 & 22 & 15 (14) & 7 (5) & 32$^{+3}_{-7}$ \\
C21 & 1.4 - 3.9 & 13 & 1 & 12 (6) & < 93 \\ 
T10+S23+C21 & 0.02 - 3.9 & 77 & 43 & 34 & 44$^{+4}_{-3}$ \\ \hline
\end{tabular}\par
\begin{FlushLeft}
\textbf{Notes.} The second column reports the redshift ranges. The last four columns show the total number of AGNs, the number of sources in the no-UFO and UFO sub-samples and the UFO detection fraction. For each no-UFO and UFO sub-sample, we also report the number of sources, if any, that belong to the ``unabsorbed sample'' (see Sect.~\ref{x_uv_slope}) between parentheses.
\end{FlushLeft}
\end{table}

For each sample we performed a literature search to collect the properties that characterize the sources. In particular, we consider the following parameters: 
\begin{itemize}
    \item redshift z;
    \item SMBH mass M$_{\mathrm{BH}}$;
    \item full width at half maximum (FWHM) of the broad H$\beta$ emission line;
    \item 2-10 keV X-ray intrinsic luminosity L$_{\mathrm{x}}$;
    \item X-ray power-law photon index $\Gamma$;
\end{itemize}

The BH masses have been estimated through single-epoch spectroscopy (i.e., relying on H$\alpha$, H$\beta$, MgII, or CIV emission lines), stellar velocity dispersion and reverberation mapping\footnote{23/77 BH masses have been estimated through reverberation mapping, 7/77 adopting the stellar velocity dispersion and 47/77 with single-epoch spectroscopy. For additional information, see references listed in Table \ref{tab:1}.}.
In the corresponding papers, the 2–10 keV intrinsic luminosity and the photon indices have been obtained by modeling the XMM-{\it Newton} data, with the exception of the high-z sample, where they have been obtained both from XMM-{\it Newton} and {\it Chandra} data. The broad H$\beta$ FWHM values are listed only for type 1 and intermediate (type 1.2 and 1.5) AGN.
For each source, we report the values and references in Table \ref{tab:1}. Most of these parameters have been collected as originally tabulated in the T10, S23, and C21 works and we refer also to these papers for the appropriate references. In the following section, these properties will be adopted to derive other important physical parameters of the sources.

In addition, for each AGN with a detected UFO, we include the following observed parameters that characterize the outflow: 
\begin{itemize}
    \item column density N$_{\mathrm{H}}$;
    \item ionization parameter $\xi$;
    \item outflow velocity v$_{out}$.
\end{itemize} These parameters are listed in Table \ref{tab:2}. 
As presented in \citet{luminari20}, special relativistic effects impact on the measured column density of the winds. To compensate for these effects, the observed N$_{\mathrm{H}}$ should be multiplied by a factor that, for a radially outflowing wind, can be written as $\Psi= (1+\beta)/(1-\beta)$, where $\beta = v_{out}/c$ (see Luminari et al., in prep). We report in Table \ref{tab:2} each AGN correction factor. In our analysis we always adopt the corrected column density values.
As previously mentioned, the UFO observed properties are necessarily model-dependent. A comprehensive analysis with self-consistent photo-ionization models based on the observed SEDs of each source, and the subsequent derivation of the outflow parameters is beyond the scope of this paper, but it will be presented in a future work.

\section{Derived parameters}\label{derived_par}

In addition to the AGN and UFO global parameters retrieved from the literature and presented in Sect.~\ref{global_par}, we derived the bolometric luminosity L$_{\mathrm{bol}}$, the ionizing luminosity L$_{\mathrm{ion}}$, the Eddington ratio $\mathrm{\lambda}_{\mathrm{Edd}}$, the optical to X-ray spectral slope $\alpha_{\mathrm{ox}}$ and the location and energetics of the winds. To estimate the corresponding uncertainties, we adopted the \texttt{python} package \textit{uncertainties}, which calculate them from the uncertainties of the involved parameters in accordance to the error propagation theory.

\subsection{Bolometric, ionizing luminosity, and Eddington ratio}\label{bol_lum}

In the S23 and C21 samples, the bolometric luminosities are derived by considering L$_x$ as a proxy and applying an X-ray bolometric correction factor based on the empirical relations computed by \citet{duras20}, using their Eq.~3 (with 15.33 $\pm$ 0.06, 11.48 $\pm$ 0.01, and 16.20 $\pm$ 0.16 as best-fit parameters from their Table 1). 
In T10, the bolometric luminosities were instead estimated by applying a fixed bolometric correction of 10 to the ionizing luminosity. We thus re-estimated L$_{\mathrm{bol}}$ following the same methodology, as in S23 and C21\footnote{\citet{duras20} reported a spread of 0.37 dex for the relation used to obtain the bolometric corrections, which is not taken into account in the derived L$_{\mathrm{bol}}$.}. From L$_{\mathrm{bol}}$ we derived the corresponding Eddington ratio and ionizing luminosity. The former is defined as $\lambda _{\mathrm{Edd}} = \mathrm{L}_{\mathrm{bol}}/\mathrm{L}_{\mathrm{Edd}}$, where $\mathrm{L}_{\mathrm{Edd}} \equiv 4 \pi \mathrm{G} \mathrm{M}_{\mathrm{BH}} \mathrm{m}_{\mathrm{p}} c / \sigma_{\mathrm{T}}$ $\simeq 1.26 \times 10^{38} (\mathrm{M}_{\mathrm{BH}}/\mathrm{M}_{\odot})$ erg s$^{-1}$ is the Eddington luminosity. For the latter, we adopt L$_{\mathrm{ion}}$ = 1/2 L$_{\mathrm{bol}}$\footnote{Where L$_{\mathrm{ion}}$ is the ionizing luminosity between 1 and 1000 Ryd (1 Ryd = 13.6 eV).}, as appropriate for a standard AGN SED \citep[e.g.,][]{panda22}. 
A detailed SED modelling of each source is beyond the scope of our statistical analysis and will be presented for the SUBWAYS sample in a future paper.
All the estimated values are reported in Table \ref{tab:1}.

\subsection{X-ray/UV ratio ($\alpha _{ox}$)}\label{x_uv_slope}

The X-ray/UV ratio, that is, the relationship between the X-ray and optical/UV luminosity of AGN, is usually described in terms of a hypothetical power-law slope between 2500 $\angstrom$ and 2 keV rest-frame frequencies \citep[e.g.,][]{tananbaum79,vagnetti10}:

\begin{equation}
\label{eq:1}
    \alpha _{ox} = \frac{\log(L_{2 \, keV}/L_{2500 \, \angstrom})}{\log(\nu_{2 \, keV}/\nu_{2500 \angstrom})} = 0.3838 \, \log \left (\frac{L_{2 \, keV}}{L_{2500 \angstrom}}\right )
\end{equation}

In order to calculate the $\alpha_{\mathrm{ox}}$ index in our samples, the X-ray and UV monochromatic luminosities must be determined. For the X-ray measurements of the S23 and T10 samples, we derived the X-ray 2-10 keV energy band fluxes from the luminosity values reported in the corresponding papers and we evaluated the specific luminosity at 2 keV (rest-frame) using the observed photon index. For the C21 sample, we directly derived the L$_{\mathrm{2 keV}}$ from the data.
Meanwhile, to evaluate the rest-frame monochromatic UV luminosity, we considered the fluxes obtained by the set of filters on board XMM-{\it Newton} Optical Monitor (OM). The UV filters, UVW1, UVM2, and UVW2, have central wavelength $2675 \angstrom$, $2205 \angstrom$, $1894 \angstrom$\footnote{\url{https://www.mssl.ucl.ac.uk/www_astro/XMM-OM-SUSS/SourcePropertiesFilters.shtml}}, respectively. 

The following procedure has been adopted:
\begin{itemize}
    \item if the available filters cover the rest-frame $2500 \angstrom$ wavelength, then the luminosity is calculated as a linear interpolation of the two nearest filter fluxes.
    
    \item if the available filters do not extend to the $2500 \angstrom$ wavelength, the $L_{2500 \angstrom}$ is calculated through a power-law extrapolation of the nearest filter flux assuming a standard UV spectral shape for type 1 AGN \citep[i.e., $f_{\mathrm{\nu}} \propto \nu^{\alpha}$, where $\alpha_{\mathrm{\nu}} = -0.5$,][]{richards06}, following for example \citet{vagnetti10,martocchia17,chiaraluce18,serafinelli21}.
\end{itemize}

In PG 1416-129 (S23 sample) and Mrk 205 (T10 sample), the OM UV filters are not available. Thus, we considered the {\it Swift}'s Ultraviolet/Optical Telescope (UVOT) data closer in time to the studied XMM-{\it Newton} observation (i.e, Obs ID 00049481002 and Obs ID 00091003002, respectively). In this case, UVW1, UVM2, and UVW2 filters central wavelengths are: $2600 \angstrom$, $2246 \angstrom$, $1928 \angstrom$\footnote{\url{https://www.swift.ac.uk/analysis/uvot/filters.php}}, respectively. We then applied the same procedure as for the XMM/OM data. In addition, neither OM nor UVOT filters are available for NGC 2110 (T10 sample). However, the neutral column density reported for this source is 2.21 $\pm$ 0.11 $\times 10^{22}$ cm$^{-2}$ \citep{laha20}, so it would be in any case removed from the unabsorbed sample, as explained below.
We then corrected the derived UV fluxes for extinction, estimating the galactic extinction for each source from \citet{schlegel98}\footnote{\url{https://irsa.ipac.caltech.edu/applications/DUST/}} and following \citet{lusso16} method.

The presence of gas and dust along the line of sight can affect both the UV and the X-ray intrinsic luminosities and thus, it cannot be neglected. 
In the three samples, some obscured AGNs are indeed present (e.g., see the cumulative distributions of the sources neutral absorber N$_{\mathrm{H}}$ in Fig.~\ref{fig:14} panel a). In order to define a sub-sample of AGNs that are not affected by absorption and reddening in UV or X-rays, we simulated the effect of the equivalent hydrogen column density on the $\alpha_{\mathrm{ox}}$. To do so, we estimated the E(B-V) values by assuming the Galactic E(B-V)/N$_{\mathrm{H}}$ ratio, 1.7 $\times$ 10$^{-22}$ mag cm$^2$ \citep{bohlin78}. Then we computed the corresponding decrease of UV and X-ray flux, using the \texttt{redden*powerlw*phabs} model in \textsc{xspec} v.12.11.1 \citep{arnaud96} considering an intrinsic $\alpha_{\mathrm{ox}}$ of -1.5 ($\alpha_{\mathrm{ox}}^{\mathrm{int}}$), and varying the neutral absorber column densities between 10$^{20}$ up to 5 $\times$ 10$^{22}$ cm$^{-2}$ and the reddening accordingly.
We then calculated the X-ray/UV ratio as affected by reddening and absorption ($\alpha_{\mathrm{ox}}^{\mathrm{red}}$) and the corresponding expected deviation from the initial intrinsic value, $\alpha_{\mathrm{ox}}^{\mathrm{red}} - \alpha_{ox}^{\mathrm{int}}$. The latter is plotted in Fig.~\ref{fig:14}, panel b, with respect to the neutral absorber column density. 
Taking ($\alpha_{\mathrm{ox}}^{\mathrm{red}} - \alpha_{ox}^{\mathrm{int}}$) $<$ 0.1 as the acceptable threshold, we could then derive a maximum neutral absorber column density above which the observed $\alpha_{\mathrm{ox}}$ cannot be considered as the intrinsic one, that is, N$_{\mathrm{H}} = 5 \times 10^{20}$ cm$^{-2}$. We verified that the same procedure applied with different intrinsic $\alpha_{\mathrm{ox}}$, in the -1.8 to -1.2 range, leads to a very similar N$_{\mathrm{H}}$ threshold.
In Table \ref{tab:1}, we present the $\alpha_{\mathrm{ox}}$ indices that were obtained only for the 46 sources\footnote{ Since the intrinsic N$_{\mathrm{H}}$ threshold closely matches Galactic column density levels, we also investigated the N$_{\mathrm{H}}^{\rm Gal}$ distribution across the three samples to verify that the $\alpha_{\mathrm{ox}}$ values are not potentially affected. We found four AGNs (WISE~J053756-0245 and IRAS 5078+1626 from the no-UFO sub-sample, Ark~120 and MG~J0414+0534 from the UFO sub-sample) with intrinsic N$_{\mathrm{H}}$ below the adopted cut-off, but with N$_{\mathrm{H}}^{\rm Gal}$ exceeding it. We thus excluded these sources.} (i.e., 21/42 sources for the T10, 19/22 sources for the S23 and 6/13 sources for the C21 sample\footnote{In the C21 sample, the real unaffected sub-sample may be smaller than what reported. Indeed, at high redshift, no precise constrains can be achieved on the low neutral N$_{\mathrm{H}}$ in the soft X-ray band.}) with N$_{\mathrm{H}} < 5 \times 10^{20}$ cm$^{-2}$. We note that much larger N$_{\mathrm{H}}$ for the neutral absorber (and therefore unacceptable deviations for $\alpha_{\mathrm{ox}}$) would be needed to significantly enlarge this sub-sample (see Fig.~\ref{fig:14}). 
For our subsequent analysis, when accounting for the $\alpha_{\mathrm{ox}}$ and $L_{2500 \angstrom}$ values, we will solely consider these 46 AGNs (referred as ``unabsorbed sample'' in the following)\footnote{ We note that the adopted cut-off in the intrinsic neutral N$_{\rm H}$ does not affect the other AGN and UFO parameters, but only $L_{2500 \angstrom}$, $\alpha_{\mathrm{ox}}$, and $\Delta \alpha_{ox}$.}.
As a result of our procedure, the $\alpha_{\mathrm{ox}}$ distribution of the unabsorbed sample (see Fig.~\ref{fig:4}) covers the approximate range between -1.8 and -1.2 (as expected e.g., \citealt{lusso10}).

\begin{table*}
\centering
\renewcommand{\arraystretch}{1.5}
\caption{Comparison between each sample (i.e., considering both UFO and no-UFO sub-groups).}
\label{tab:4}
\begin{tabular}{lllllllllllll}
\multicolumn{1}{c}{Sample} & log(M$_{\textrm{BH}}$) & log(L$_{x}$) & log(L$_{bol}$) &  $\lambda_{\rm Edd}$ & $\Gamma$ & FWHM H$\beta$ & $\alpha_{\mathrm{ox}}$ & $\Delta \alpha_{ox}$ \\ 
\cline{1-12} 
S23 vs T10 & -5.47 & -7.10 & -7.10 & \multicolumn{1}{c}{{\bf x}} & \multicolumn{1}{c}{{\bf x}} & \multicolumn{1}{c}{{\bf x}} & -2.13 & -2.17 \\
S23 vs C21 & -4.13 & \multicolumn{1}{c}{{\bf x}} & \multicolumn{1}{c}{{\bf x}} & -1.54 & \multicolumn{1}{c}{{\bf x}} & \multicolumn{1}{c}{{\bf x}} & -1.70 & \multicolumn{1}{c}{{\bf x}} \\ 
T10 vs C21 & -10.39 & -7.78 & -7.78 & \multicolumn{1}{c}{{\bf x}} & -1.70 & \multicolumn{1}{c}{{\bf x}} & -1.85 & \multicolumn{1}{c}{{\bf x}} \\ 
T10+S23 vs C21 & -8.48 & -4.49 & -4.49 & -1.42 & -1.40 & \multicolumn{1}{c}{{\bf x}} & -1.82 & \multicolumn{1}{c}{{\bf x}} \\
\hline
\end{tabular}\par
\begin{FlushLeft}
\textbf{Notes.} We report the logarithm of the \textit{NHP} for a KS test with respect to the parameters used in this paper. The lower the $\log\mathrm{NHP}$ values, the more statistically different the compared samples become. Meanwhile, we mark ``\textbf{x}'' when the difference between two samples is below the adopted significance threshold ($\log\mathrm{NHP}> -1.30$, i.e., the compared samples are statistically the same).
\end{FlushLeft}
\end{table*}

A strong anticorrelation between $\alpha_{\mathrm{ox}}$ and the monochromatic luminosity at $2500 \angstrom$ has been identified in many studies \citep[e.g.,][]{zamorani85,wilkes94,vignali03,strateva05,steffen06,just07,vagnetti13,lusso16}, corresponding to a non linear relation between the UV and X-ray luminosity. Hence, more luminous objects are  weaker in X-rays relatively to UV. We observe the same correlation when considering the T10+S23+C21 combined sample (see Table \ref{tab:8}). 
In order to identify sources that may diverge from this standard population, we calculated the difference between the observed $\alpha_{\mathrm{ox}}$ and that expected from the best fit $\alpha_{\mathrm{ox}}$-$L_{2500 \angstrom}$ relation used in Eq.~13 of \citet{vagnetti13}: 
\begin{equation}
    \Delta \alpha_{ox}= \alpha_{ox} -\alpha_{ox}^{fit}(L_{2500 \angstrom}).
\end{equation}
The derived $\Delta \alpha_{ox}$ are reported in Table \ref{tab:1}. This parameter is usually adopted as an X-ray weakness proxy (e.g., \citealt{nardini19,zappacosta20,pu20}; see Sect.~\ref{comparison_samples}).

\subsection{UFO global properties}
\label{sec:out_global}

By combining the UFO and AGN global properties, the location and energetics of the winds can be derived. 
There are two possible estimates for the distance between the wind and the illuminating central source. The first can be obtained from the definition of the ionization parameter, $\displaystyle\xi \equiv \frac{L_{ion}}{n_H r^2}$ \citep{tarter69} where $n_H$ is the hydrogen number density of the absorbing gas and $r$ its distance from the ionizing source. By requiring the size of the absorber to not exceed its distance to the BH, $N_H \simeq n_H(r) \Delta r < n_H(r)r$ \citep[where $n_H(r)$ is the number density of the gas at a certain radius; e.g.,][]{behar03,crenshaw12}, we then derived the following expression:

\begin{equation}
    r_{1} \equiv \frac{L_{ion}}{\xi N_H} \, .
\end{equation}

Another estimate of the radial distance of the absorbing gas producing the UFO can be inferred by comparing the observed outflow velocity along the line of sight to the escape velocity (i.e, v$_{\mathrm{esc}}= \displaystyle\sqrt{\frac{2GM_{\mathrm{BH}}}{r}}$ for a Keplerian disk). The radius at which this happens is equal to (in the Newtonian limit):
\begin{equation}
    r_{2} \equiv \frac{2 G M_{BH}}{v_{out}^2} = r_s \left(\frac{c}{v_{out}}\right)^2, 
\end{equation}
where $r_{\mathrm{s}}$ = 2GM$_{\mathrm{BH}}$/c$^2$ is the Schwarzschild radius. 
This represents the radius at which a disk wind can be accelerated, as its outflow velocity must overcome v$_{\mathrm{esc}}$ for the wind to be successfully launched.
Since r$_{\mathrm{2}}$ is always smaller than or consistent with (within the errors) r$_{\mathrm{1}}$ (see Fig. \ref{fig:radii}), our analysis will focus solely on r$_{\mathrm{1}}$, which will be referred to as r${_\mathrm{wind}}$ from now on. 

From the estimates of r$_{\mathrm{wind}}$, the energetics of the wind can be derived. The mass outflow rate is computed using the following expression derived by \citet{krongold07}:
\begin{equation}
    \dot{M}_{wind} \equiv f(\delta,\varphi) \, \pi \, \mu  \, m_p \, N_H \, v_{out} \, r_{wind},
\end{equation}
where $f(\delta,\varphi)$ is a geometric factor of the order of unity which depends on the angles $\delta$ and $\varphi$ between the line of sight and the wind direction with the accretion disk plane respectively \citep[for details see][]{krongold07}. We adopt $f(\delta,\varphi)\sim 1.5 $, appropriate for a vertical disk wind ($\varphi \simeq \pi$/2) and an average optical type 1 line-of-sight angle of $\delta \simeq$ 30°. 
Meanwhile, we use $\displaystyle \mu \equiv \frac{n_H}{n_e} \simeq 1.2$ for fully ionized gas and solar abundances.

Finally, by considering the velocity of the outflow as constant, any acceleration is thus neglected, the mechanical power can be derived as:
\begin{equation}
    \dot{E}_k^{wind} \equiv \frac{1}{2} \, \dot{M}_{wind} \, v_{out}^2
\end{equation}
and the outflow momentum rate:
\begin{equation}
    \dot{P}_{wind} \equiv \dot{M}_{wind} \, v_{out}.
\end{equation}
We report the derived values of these parameters for each AGN in Table \ref{tab:2}.

\subsection{Selection and bias effects}\label{biases}

In our work, we adopt UFO and AGN properties as derived in other papers, notably \citet{tombesi10},\citet{chartas21}, and \citet{matzeu22}. In Sections \ref{global_par} and \ref{derived_par} we tried to maximize the consistency between each sample, yet we are aware that the sample selection, as well as the accuracy and reliability of some parameters could affect the outcomes of our analysis. For instance, the samples analyzed here, by construction, contain among the brightest AGNs from relatively deep pointed observations across various sky regions. Sample incompleteness might also arise from AGN obscuration effects. The inability to detect UFOs in highly obscured sources results from an observational bias that prevents conclusions about the overall incidence of UFOs in these AGNs from being drawn. Moreover, incompleteness could be due to orientation effects (e.g., outflows not intersecting our line of sight).
Additionally, we must note that the flux detection threshold unavoidably limits the selection at high-z of observable AGNs to even smaller numbers (see Sect.~\ref{ufo_vs_noufo}). In this respect, the C21 sample is clearly subject to a significant selection bias by containing almost purely AGNs hosting UFOs. Nonetheless, the inclusion of high-z samples is crucial for our and future population studies.

As pointed out in Sect.~\ref{S23sample}, when assessing the incidence of UFO detections across different samples, one must consider the transient and variable nature of these winds and their duty cycle (see also Sections \ref{ufo_vs_noufo} and \ref{windradius}). On the other hand, all three studies considered here (T10, C21, and S23) systematically searched for blueshifted Fe K absorption lines and evaluated their statistical significance through Monte Carlo simulations, effectively mitigating potential publication biases.
Furthermore, the observed UFO properties (N$_{\mathrm{H}}$ and $\xi$ against v$_{\mathrm{out}}$) are highly model-dependent and, to mitigate this effect, we re-obtain the C21 values with the same model as for the other samples. A self-consistent analysis with ad-hoc photo-ionization models for each source is devoted to a future paper.


\section{Parameter distributions}\label{par_distr}

The first part of our analysis consists in assessing the statistical properties of each sample. In particular, we made use of the two-sample KS test \citep{hodges58} to determine whether the distributions of the parameters of the three data sets exhibit significant differences. In this paper, we consider a probability of 0.05 (roughly corresponding to $\sim$2$\sigma$ for a Gaussian distribution) as a statistically significant threshold for the null hypothesis probability (NHP; i.e., $\log\mathrm{NHP}<-1.30$).

\begin{figure*}
\centering
    \includegraphics[width=\textwidth]{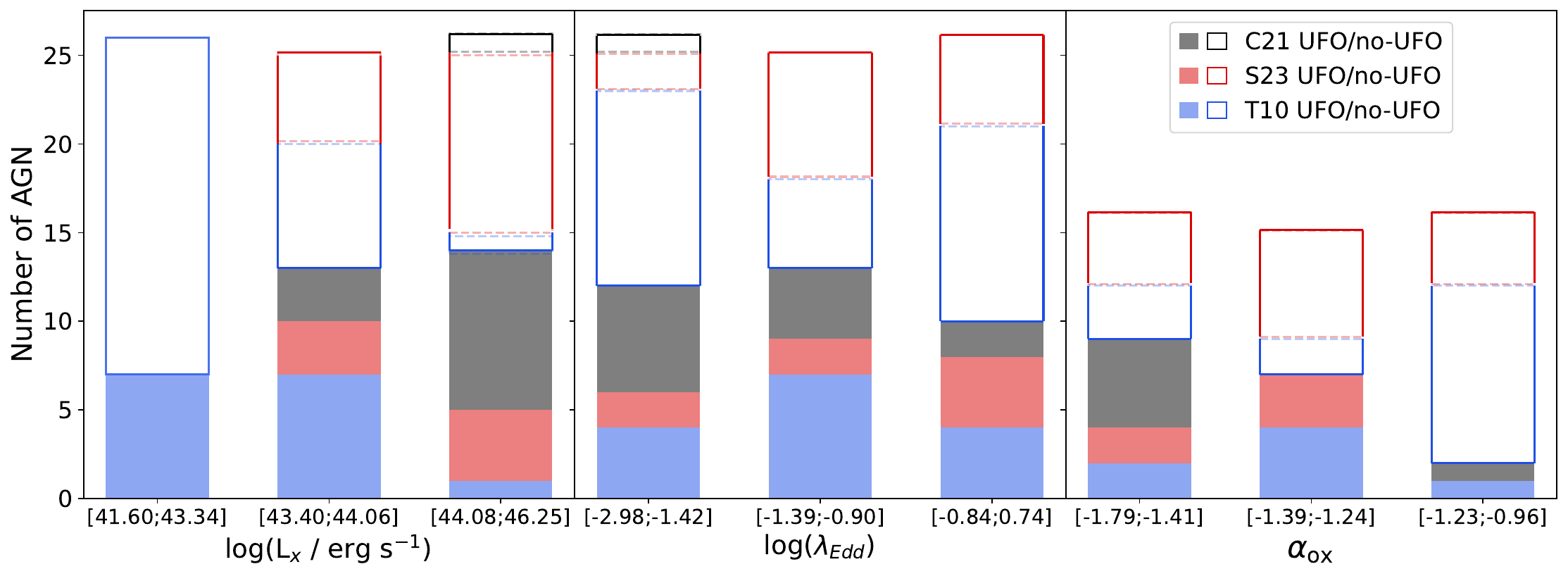}\\
    \caption[]
    {{Number of AGNs in three X-ray luminosity (left panel), Eddington ratio (central panel), and $\alpha_{\mathrm{ox}}$ (right panel) bins for the analyzed sample and sub-samples. In solid colors (black for C21, red for S23, and blue for T10) the UFO sub-samples are shown, while the colored (the same color palette is adopted) solid lines represent the no-UFO sub-samples. }}
    \label{fig:7}
\end{figure*}

\subsection{T10, S23, and C21 samples}\label{comparison_samples}

In the following, we present the distributions of the three samples, focusing on the global and derived parameters of AGNs and UFOs. The main differences are emphasized and a summary of the results is presented in Table \ref{tab:4}. 

Due to the adopted selection criteria, the three samples exhibit significant differences in redshift, X-ray, and bolometric luminosity (Fig.~\ref{fig:22}, panel a), although the disparity in luminosity between the S23 and C21 samples is not statistically significant. While the distributions of BH mass differ significantly, as expected, progressing to larger values from T10 to S23 to C21 (Fig.~\ref{fig:22}, panel b), only the C21 and S23 samples diverge in terms of the Eddington ratio.
The C21 and T10 datasets manifest some difference concerning their photon indices, steeper for the former sample (see Fig.~\ref{fig:4}, panel a). 
All samples significantly differ from each other with respect to the $\alpha_{\mathrm{ox}}$, with values progressively steepening from T10 to C21, consistent with the expected correlation with luminosity (see Fig.~\ref{fig:4}, panel b). On the other hand, the difference between the S23 and T10 samples in terms of $\Delta \alpha_{\mathrm{ox}}$ is due to the presence of a significant fraction of negative  values in the latter sample (i.e., weaker X-ray emission; see Fig.~\ref{fig:4}, panel b). If we adopt the threshold of $\Delta \alpha_{\mathrm{ox}} = -0.3$ proposed by \citet{pu20}, two sources from the T10 sample can be classified as X-ray weak sources (within errors; namely TON S180 and NGC 3783), none in S23, and only one in C21, i.e., HS 0810+2554, albeit slightly above the aforementioned threshold. 

As a further test we combined the low and intermediate redshift data sets, and this new sample (i.e., T10+S23) has been compared to the high-z data (i.e., C21). According to the KS tests (last line of Table \ref{tab:4}), the two samples differ in all parameters but the FWHM of H$\beta$ and $\Delta \alpha_{ox}$.

\subsection{UFO and no-UFO sub-samples}\label{ufo_vs_noufo}

We then conducted a similar comparison between the UFO and no-UFO sub-samples for each studied sample, including the two combined samples T10+S23 and T10+S23+C21. As shown in Table \ref{tab:7} and Figs.~\ref{fig:22}-\ref{fig:4}, no significant differences in the AGN properties are found between sources hosting UFOs and those without. In other words, based on the analyzed parameters (i.e., M$_{\textrm{BH}}$, L$_{x}$, L$_{bol}$, $\lambda_{\rm Edd}$, $\Gamma$, FWHM H$\beta$, $\alpha_{\mathrm{ox}}$, and $\Delta \alpha_{ox}$), there is currently no substantial evidence to suggest that AGNs hosting UFOs differ from those without. Since all the sources of the considered samples have been selected in order to have enough S/N to detect UFOs, the absence of any difference between the two sub-samples is unlikely to be due to detection issues. It might be instead related to a finite wind duty cycle, hinting to the possibility that all AGN, in fact, actually are capable to host UFOs during their lifetime.

The only marginal differences arise when all the samples are combined together (see last row of Table \ref{tab:7}). There is indeed an indication that UFOs are preferentially hosted in high mass and high luminosity sources. While this result is potentially biased by the C21 sample, which is almost completely constituted by sources with UFOs, this result deserves to be further investigated.
Therefore, to assess the UFO detection fraction in the combined sample, we divided the T10+S23+C21 sample into three luminosity (41.60-43.34, 43.40-44.06, and 44.08-46.25 erg s$^{-1}$) and $\lambda_{\mathrm{Edd}}$ (-2.98 to -1.42, -1.39 to -0.90, and -0.84 to -0.74) bins, so that the number of sources per bin is similar (26 AGNs in the first bin, 25 in the second, and 26 in the third; see Fig.~\ref{fig:7}, left and middle panels). In the case of L$_{\mathrm{x}}$, the UFO detection fraction in the first interval is 27\% (7/26), significantly lower than that of the second and third bins ($>98\%$ and $>99\%$ confidence level respectively, according to a binomial test), whose fractions are instead consistent with each other, 52\% (13/25) and 54\% (14/26). We must note, however, that the significantly lower UFO detection fraction observed in the 41.60-43.34 L$_{\mathrm{x}}$ bin (which contains only low-z T10 AGN) could be at least partially attributed to lower luminosity objects being missed (or excluded due to insufficient S/N) in the higher redshift samples (see Sect.~\ref{biases}).

The result on L$_{\mathrm{x}}$ seems to be independent on the Eddington ratio since the UFO detection fractions in the three $\lambda_{\mathrm{Edd}}$ bins are not significantly different according to a binomial test, that is, 46\% (12/26), 52\% (13/25), and 39\% (10/26), respectively (see middle panel of Fig.~\ref{fig:7}).
On the other hand, as shown in the right panel of Fig.~\ref{fig:7}, the UFO detection fractions in terms of $\alpha_{\mathrm{ox}}$ bins (-1.79 to -1.41, -1.39 to -1.24, and -1.23 to -0.96; the ranges have been adopted so that the number of sources per bin is similar) appear to drop at flatter $\alpha_{\mathrm{ox}}$ (i.e., for $\alpha_{\mathrm{ox}} >$ -1.24): from 56\% (9/16) and 47\% (7/15) in the first two bins, to 13\% (2/16) in the third bin (statistically different at $>99.9\%$ confidence level in both cases, according to a binomial test). This suggests that UFOs preferentially develop in X-ray under-luminous objects. We note that this is in agreement with the higher UFO detection rate in luminous sources found above since, as already mentioned, luminosity and $\alpha_{\mathrm{ox}}$ are strongly anticorrelated (see Sect.~\ref{x_uv_slope}).

\begin{table*}
\centering
\renewcommand{\arraystretch}{1.5}
\caption{Comparison between the UFO sub-samples.}
\label{tab:3}
\begin{tabular}{lllllllllllll}
\multicolumn{1}{c}{Sample} &log(N$_{\mathrm{H}}$) & v$_{out}$ & log($\xi$) & log(r$_{wind}$) & log(r$_{wind} / r_s$) & log($\dot{M}_{wind}$) \hspace{0.5cm} log($\dot{E}_{k}^{wind}$) \hspace{0.5cm} log($\dot{P}_{wind}$) \\
\hline
S23 vs T10 & -2.07 & \multicolumn{1}{c}{{\bf x}} & \hspace{0.2cm} {\bf x} & \hspace{0.2cm} {\bf x} & -2.32 & -2.52 \hspace{1.7cm} {\bf x} \hspace{1.45cm} -1.33 \\
S23 vs C21 & -4.40 & -2.01 & \hspace{0.3cm}{\bf x} & \hspace{0.2cm} {\bf x} & -1.77 & -4.04 \hspace{1.3cm} -1.62 \hspace{1.3cm} -1.62 \\
T10 vs C21 & -6.94 & -3.77 & -2.30 & -2.17 & -3.77 & -6.34 \hspace{1.3cm} -4.92 \hspace{1.3cm} -4.09 \\
T10+S23 vs C21 & -8.44 & -3.81 & -1.68 & -1.41 & -2.14 & -7.38 \hspace{1.3cm} -2.74 \hspace{1.3cm} -2.32 \\ \hline
\end{tabular}\par
\begin{FlushLeft}
\textbf{Notes.} The lower the $\log\mathrm{NHP}$ values, the more statistically different the compared samples become. Meanwhile, we mark ``\textbf{x}'' when the difference between two samples is below the adopted significance threshold ($\log\mathrm{NHP}> -1.30$, i.e., the compared samples are statistically the same).
\end{FlushLeft}
\end{table*}

\begin{figure*}
\centering    
\includegraphics[width=0.9\textwidth]{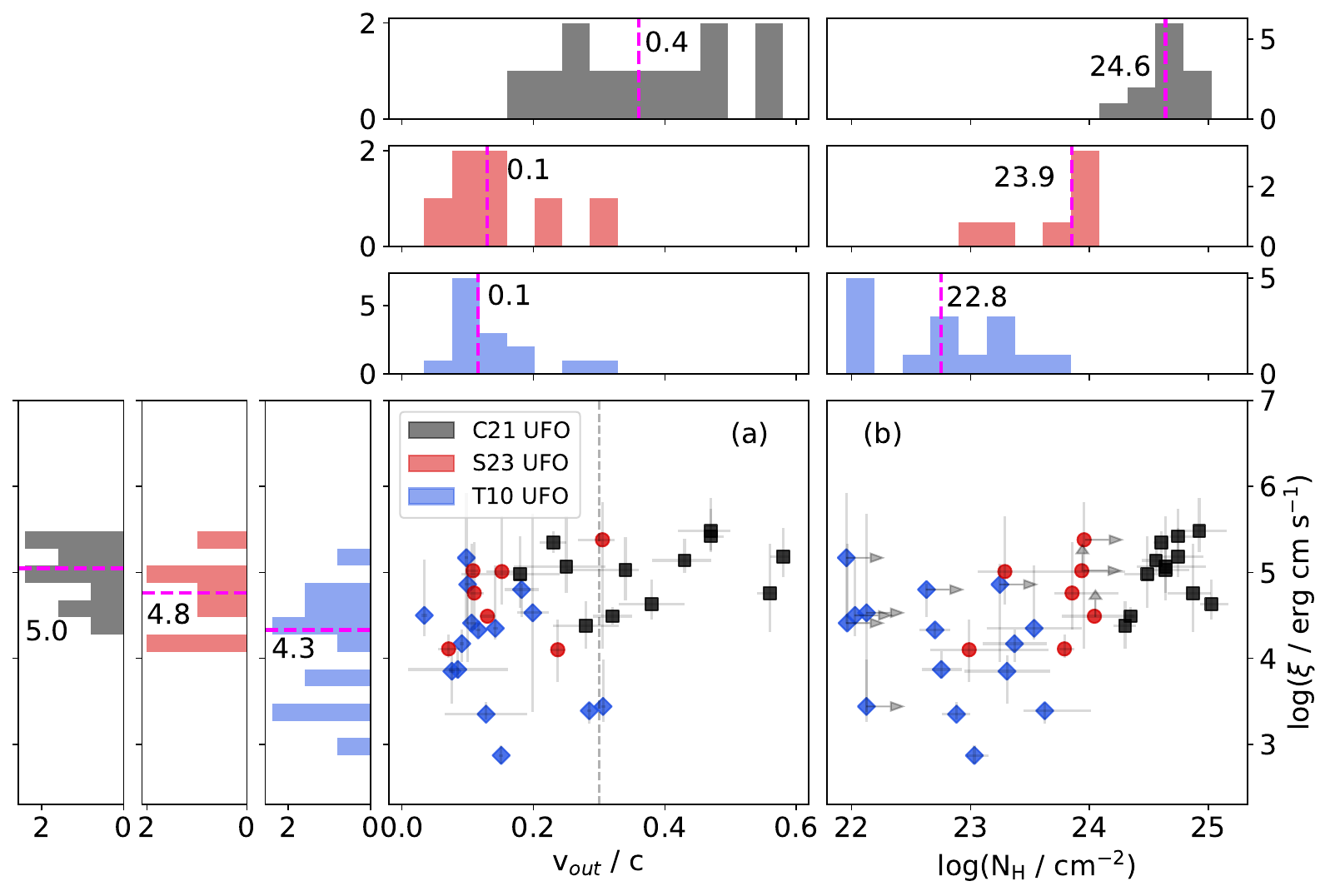}
    \caption[]
    {{Sample comparison of outflow parameters. Panel a: Outflow velocity as a function of the ionization parameter. AGNs on the right side of the dashed gray line host potentially magnetically driven winds, see text for details. Panel b: The ionization parameter versus the outflow equivalent column density. The gray arrows represent the upper/lower limits for the log($\xi$) and N$_{\rm H}$ values (see Table \ref{tab:2}).
    The S23 UFO sub-sample is shown in red circles, T10 in blue diamonds, and C21 in black squares. The dashed magenta lines show the median value of each sub-sample.}}
    \label{fig:24}
\end{figure*}

\subsection{UFO sub-samples}\label{comparison_ufos}

In this section we focus on the distributions of the outflow properties in the UFO sub-samples.  
We plot the ionization parameter versus the outflow velocity and versus the equivalent hydrogen column density of the outflow in Fig.~\ref{fig:24}, panel a and b respectively. The histograms of the observed parameters with their medians are reported in the upper and side parts of each panels, while the KS tests probability values between the samples can be found in Table \ref{tab:3}.

We note that the wind column densities are significantly different among the three samples, with T10 displaying the lowest values, and then increasing for S23 and C21. This effect may be due to an observational bias which favors the detection of low column densities (hence weaker absorption lines) in low-z and generally brighter AGNs. However, while this effect is likely significant for the high-z sources in C21, whose spectra are characterized by lower S/Ns, both T10 and S23 are selected in order to have X-ray spectra with high statistics, so their different N$_{\mathrm{H}}$ distributions should be intrinsic. The C21 sample also exhibits significantly larger outflow velocities and, although to a lesser extent, larger ionization parameters and smaller wind radii. The latter are more significantly different among the samples in terms of Schwarzschild radii, with wind radii progressively getting closer to the BH from T10 to S23 and then C21.

These results reflect in significant differences in terms of $\dot{M}_{\mathrm{wind}}$, $\dot{P}_{\mathrm{wind}}$ and $\dot{E}_k^{wind}$, typically increasing from T10 to S23 and C21 (see Fig.~\ref{fig:25} and Table \ref{tab:3}).


\section{Parameter correlations}\label{par_corr}

After the comparative analysis between the different samples shown in the previous section, we investigated for possible correlations among the AGN properties and the UFO characteristics. 
Our main diagnostic is the Spearman coefficient, whose p-value and rank respectively assess the significance and degree of monotonic relation between each parameter. In order to consider the uncertainties, we implemented the perturbation method of \citep{curran14}, available through the \texttt{python} library \textit{pymccorrelation} \citep{privon20}. We accounted for nonsymmetric uncertainties by randomly sampling among two half Gaussian distributions around the central value, and for upper (lower) limits by sampling a uniform distribution between the parameter upper (lower) limit and reasonable lower (upper) bounds\footnote{Specifically, for each given sample, we adopted as lower (upper) limit the median value of the corresponding parameter range.}. This perturbation method is also applied to compute distributions for the linear regressions of each pair of parameters, perturbed according to their uncertainties. We plotted the envelopes of the regression from the 68\% and 90\% of the line distributions, and quoted the uncertainties on the regression parameters at a 1$\sigma$ confidence level. We adopted the standard deviation to evaluate the scatter/spread of the data.

The procedure was adopted on the combined T10+S23+C21 UFO sub-samples and the results are reported in Table \ref{tab:8}. 
As expected, we find some well known relations between global AGN properties, but here we discuss only those involving at least one UFO parameter. All the investigated correlations, together with their statistical significance and corresponding plot, can be found in Appendix~\ref{allcorr}, even if not discussed here.

\subsection{UFO properties}\label{ufo_prop_corr}

\begin{figure}[h!]
\centering
    \subfloat{\includegraphics[width=.45\textwidth]{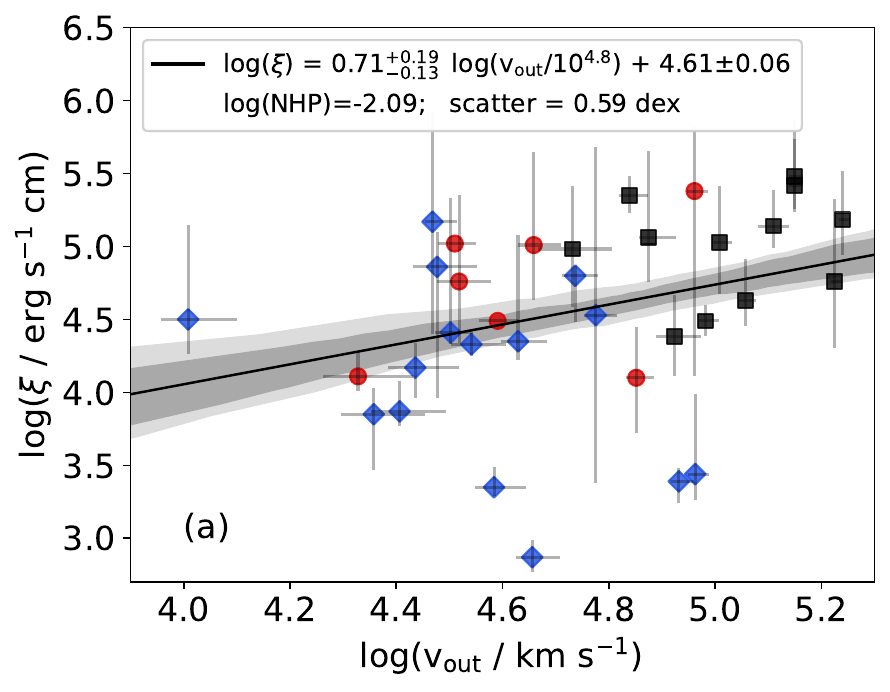}}\\
    \subfloat{\includegraphics[width=.45\textwidth]{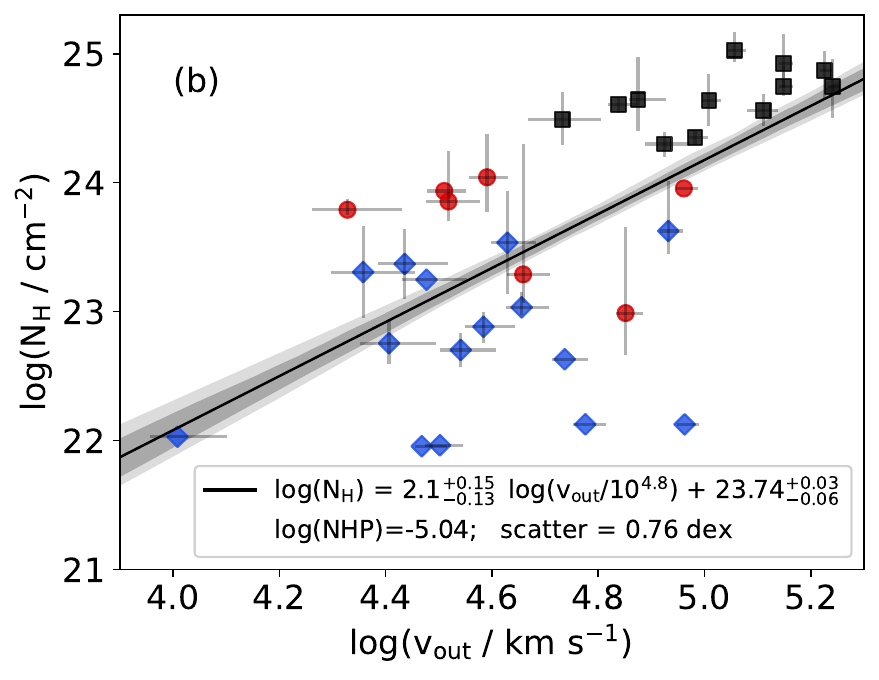}}\\
    \subfloat{\includegraphics[width=.45\textwidth]{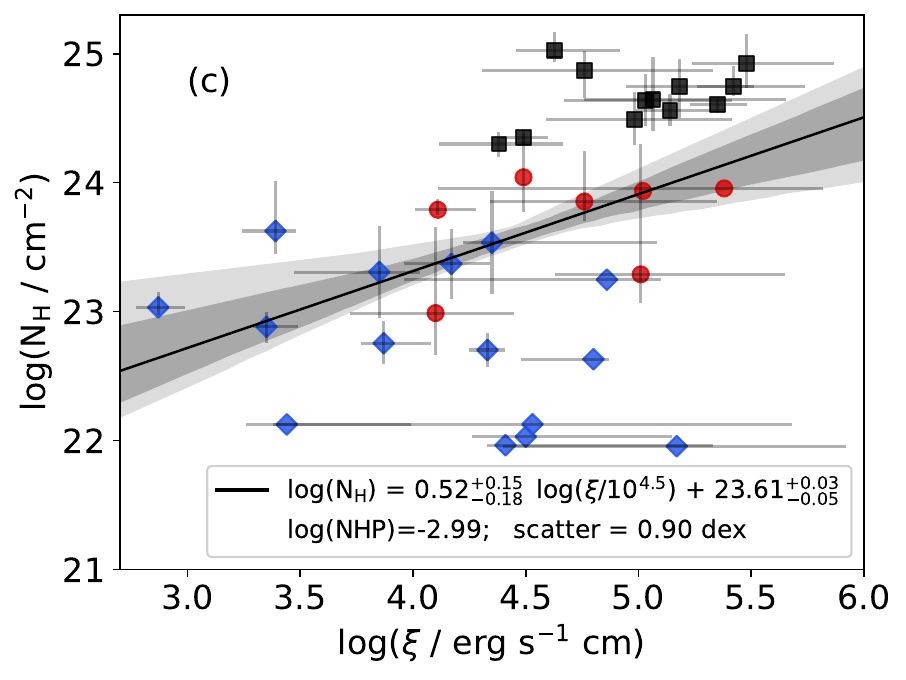}}
    \caption[]
    {{Significant correlations of observed UFO properties. Panel a: $\xi$ versus v$_{\mathrm{out}}$. Panel b: N$_{\mathrm{H}}$ versus v$_{\mathrm{out}}$. Panel c: $\xi$ versus N$_{\mathrm{H}}$. The S23 UFO sub-sample is shown in red circles, T10 in blue diamonds, and C21 in black squares. The solid lines represent the best-fitting linear correlation and the dark and light gray shadowed areas indicate the 68\% and 90\% confidence bands, respectively. In the legend, we report the best-fit coefficients, $\log\mathrm{NHP}$, and the intrinsic scatters for the correlations.}}
    \label{fig:1}
\end{figure}

The three observed outflow properties ($\xi$, N$_{\mathrm{H}}$ and v$_{\mathrm{out}}$) are significantly correlated with each other (see Table~\ref{tab:8} and Fig.~\ref{fig:1}): faster UFOs have larger ionization parameters and column densities. In particular, as already found by \citet{tombesi10}, \citet{chartas21}, and \citet{matzeu22}, $\xi$ and v$_{\mathrm{out}}$ are positively correlated (with an intrinsic scatter of 0.59 dex, see Fig.~\ref{fig:1}, panel a). We note, however, that four AGNs (PG 1211+143, NGC 4051, NGC 7582, and Ark 120) among the 34 UFOs present a lower ionization parameter with respect to the range covered by the other sources (i.e., 3.85 $\leq$ log($\xi$/erg s$^{-1}$ cm) $\leq$ 5.48), and are significantly outside the correlation. This will be further discussed in Sect. \ref{windradius}.
Interestingly, when dividing the ionization parameter by the ionizing luminosity, L$_{\rm ion}$, the positive correlation with v$_{\rm out}$ disappears. This strongly suggests that the observed $\xi$-v$_{\rm out}$ correlation is actually driven by the relation between the AGN luminosity and the outflow velocity (see discussion in Sect.~\ref{ufo_lum_corr}).

The faster outflows are also those with the largest column density since v$_{\mathrm{out}}$ is also correlated to N$_{\mathrm{H}}$ (with an intrinsic scatter of 0.76 dex; see Fig.~\ref{fig:1}, panel b). Potentially, this may result from an instrumental bias: the higher the velocity, the lower the effective area of the EPIC-pn camera where the feature can be detected, and therefore the higher the column density needed to detect an UFO. In case of dominant instrumental effects, we would expect this correlation to be more significant in the T10 sample (given that the T10 outflow velocities are smaller in comparison to those in the other samples). However, it is the C21 sample that primarily drives this correlation, and the high-z mitigates this bias since the absorption features are shifted to lower energies, where the effective area is flatter.
We also note that N$_{\mathrm{H}}$ and $\xi$ are positively correlated with each other (with an intrinsic scatter of 0.90 dex; see Fig.~\ref{fig:1}, panel c), and this is likely dominated by a well-known observed fit degeneracy between each other.

\begin{figure}[h!]
\centering
    \includegraphics[width=.45\textwidth]{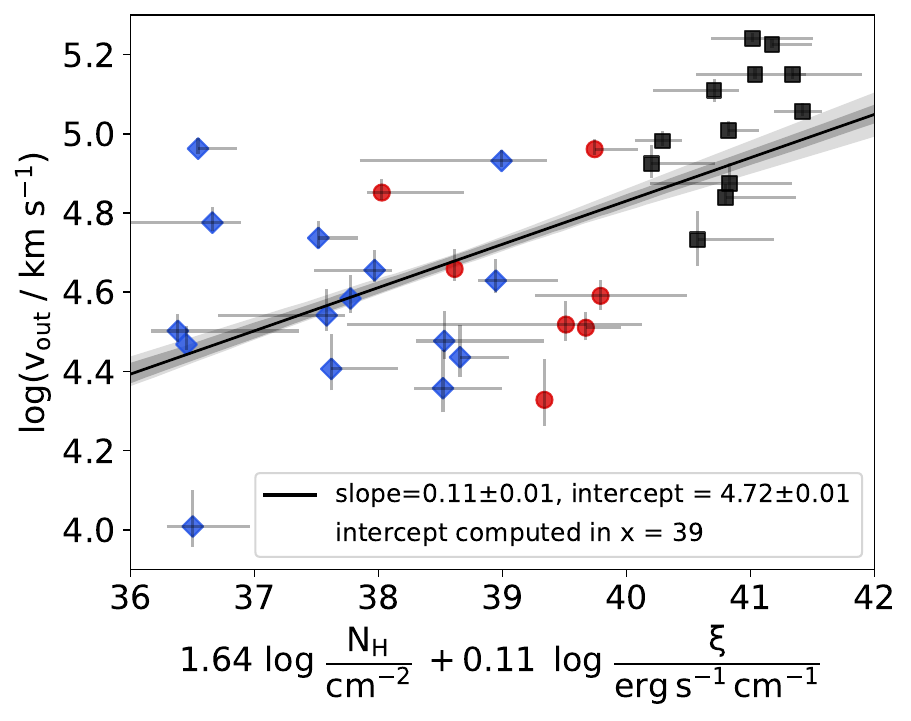}\\
    \caption[]
    {{Significant correlations: UFO universal plane. Linear combination of N$_{\mathrm{H}}$ and $\xi$ versus v$_{\rm out}$. The S23 UFO sub-sample is shown in red circles, T10 in blue diamonds, and C21 in black squares. The intrinsic scatter is 0.23 dex. The solid lines represent the best-fitting linear correlation and the dark and light gray shadowed areas indicate the 68\% and 90\% confidence bands, respectively. }}
    \label{fig:10}
\end{figure}

To further investigate the positive correlations between the three observed UFO parameters, in Fig. \ref{fig:10} we attempt to draw an “UFO universal plane”, where the outflow velocity is correlated with a linear combination (which minimize the parameters) of the ionization parameter and the column density of the wind. We find that this new relation is more significant ($\log\mathrm{NHP}$=-7.51, with an intrinsic scatter of 0.23 dex) than the individual correlations of the N$_{\mathrm{H}}$ and $\xi$ against v$_{\mathrm{out}}$ (see Table \ref{tab:8} for the respective values).

We observe that N$_{\mathrm{H}}$ and v$_{\mathrm{out}}$ correlate with $\dot{M}_{\mathrm{wind}}$, $\dot{E}_{k}^{\mathrm{wind}}$, and $\dot{P}_{\mathrm{wind}}$, as expected due to the UFO energetics derivation and the inter-correlation between the observed outflow properties. 
On the other hand, $\xi$ shows a significant positive relation only with the mass outflow rate and a marginal positive relation with $\dot{E}_{k}^{\mathrm{wind}}$, the latter is significant only if the Fe-K sub-sample is added to the UFO sources.\footnote{This sub-sample is composed of T10 AGNs with standard N$_{\mathrm{H}}$ and $\xi$ values for UFOs, albeit with a lower v$_{\mathrm{out}}$ (see Sect.~\ref{t10_sample}).}

\subsection{AGN luminosity}\label{ufo_lum_corr}

\begin{figure*}
\centering
    \subfloat[][]{\includegraphics[width=0.4635\textwidth]{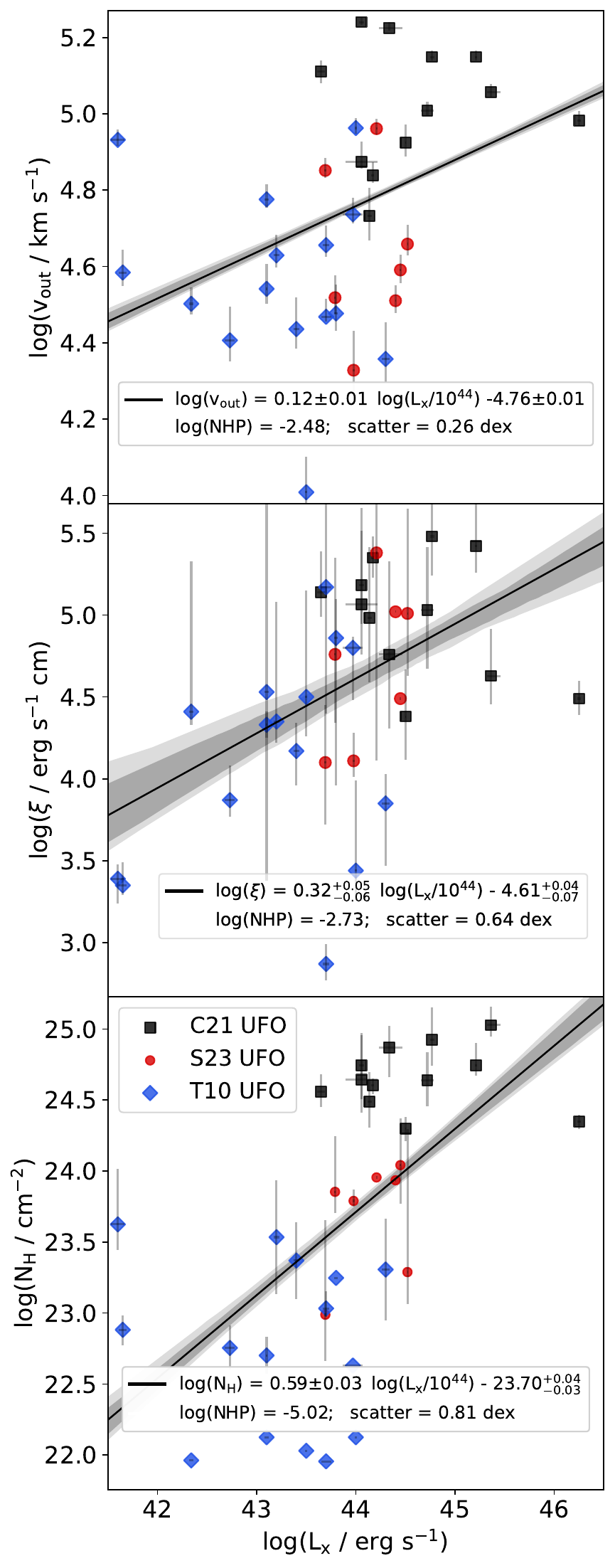}}
    \subfloat[][]{\includegraphics[width=0.45\textwidth]{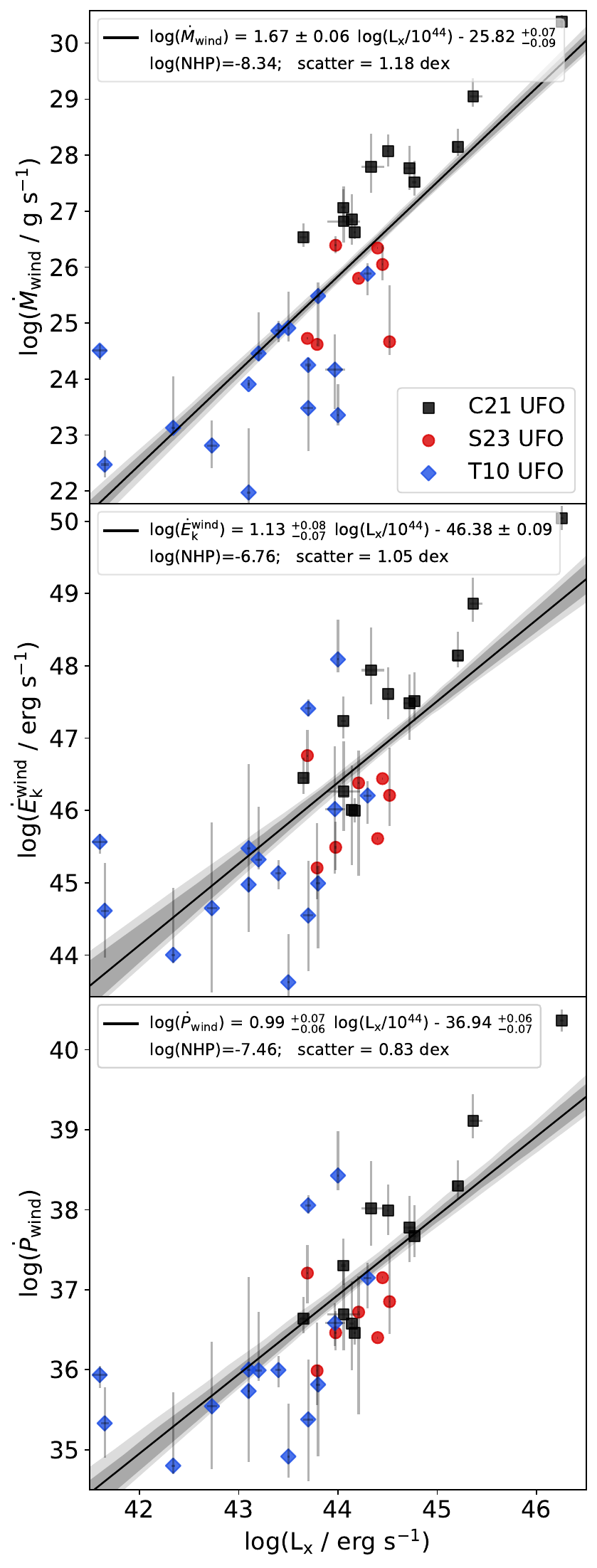}}
    \caption[]
    {{Significant correlations between luminosity and the observed UFO properties and energetics. Panel a: L$_{\mathrm{x}}$ versus v$_{\mathrm{out}}$ (upper plot), $\xi$ (middle plot), and N$_{\mathrm{H}}$ (lower plot). 
    Panel b: L$_{\mathrm{x}}$ versus $\dot{M}_{\mathrm{wind}}$ (upper plot), $\dot{E}_{\mathrm{k}}^{\mathrm{wind}}$ (middle plot), and $\dot{P}_{\mathrm{wind}}$ (lower plot).
    The S23 UFO sub-sample is shown in red circles, T10 in blue diamonds, and C21 in black squares. The solid lines represent the best-fitting linear correlation and the dark and light gray shadowed areas indicate the 68\% and 90\% confidence bands, respectively. In the legend, we report the best-fit coefficients, $\log\mathrm{NHP}$, and the intrinsic scatters for the correlations.}}
    \label{fig:2}
\end{figure*}

We observe significant correlations between the X-ray luminosity (or L$_{\mathrm{bol}}$, which is directly derived from it) and the observed outflow properties (i.e., N$_{\mathrm{H}}$, $\xi$, and v$_{\mathrm{out}}$). 
More in details, in the upper plot of Fig.~\ref{fig:2} panel a, we show a strong positive correlation between v$_{\mathrm{out}}$ and L$_{\mathrm{x}}$ (with an intrinsic scatter of 0.26 dex).
This is in agreement with the same correlation observed in different low-intermediate high-z samples and for individual AGNs (see PDS~456 in \citet{matzeu17,nardini15,reeves18}; APM~08279+5255 in \citet{chartas02,saez11}; PG~1126-041 in \citet{giustini11}; and HS 1700+6416 in \citet{lanzuisi12}). We find a slope for the correlation ($0.12\pm 0.01$ in log-log) consistent with that found by \citet{chartas21} taking into account only the T10+C21 data (0.13 $\pm$ 0.03), but flatter than the value they obtain by adding also the \citet{gofford13} sample (0.20 $\pm$ 0.03). Similarly, \citet{matzeu22} find a steeper slope (0.19 $\pm$ 0.03) considering our samples (T10, S23, C21) plus the \citet{gofford13} sample, comprehensive of the radio-loud AGNs. Our correlation coefficient (0.45, see Table \ref{tab:8}) is of the same order of that found by \citet{matzeu22} and \citet{chartas21}.
We note here that our correlation is driven by the high-z/high-velocity UFOs, and removing these specific sources (i.e., the C21 AGN) gives a nonsignificant correlation.

Significant positive correlations with the X-ray luminosity are found also for the ionization parameter (intrinsic scatter of about 0.64 dex) and the column density (intrinsic scatter of about 0.81 dex) of UFOs (middle and lower panels of Fig.~\ref{fig:2}). 
The latter relation (L$_{\mathrm{x}}$ vs N$_{\mathrm{H}}$) might be partly driven by selection effects as at higher luminosity, the gas may be more highly ionized and thus, larger columns densities are needed to detect absorption features. 
Notably, contrary to what occurs in the case of v$_{\mathrm{out}}$ versus L$_{\mathrm{x}}$, these two correlations continue to be statistically significant even after excluding the C21 sample.

It is interesting to note that the same correlations are absent or much weaker with L$_{2500\mathrm{\AA}}$.
This is not due to the lower number of sources with L$_{2500\mathrm{\AA}}$ values (the unabsorbed sample: 50/77 AGN, see Sect.~\ref{x_uv_slope}) since the correlations with L$_{\mathrm{x}}$ and L$_{\mathrm{bol}}$ are still stronger in comparison to those for L$_{2500\mathrm{\AA}}$ in this sub-sample. Instead, the weaker correlation with the UV luminosity could naturally follow from the fact that ${2500\mathrm{\AA}}$ photons hold significantly less importance in the production of highly ionized Fe than the X-ray ones.

As shown by \citet{matzeu17}, in the case of radiatively driven winds, the expected slope of the outflow velocity versus the luminosity correlation is 0.5, while the slope we find, in agreement with those by \citet{matzeu22} and \citet{chartas21}, is significantly flatter than the expected value.
Different explanations have been provided by \citet{matzeu22} to interpret the discrepancy between predicted and observed values: an increase of the slope could be reached by adding sources with outflow velocities lower than the UFO threshold \citep[as presented by][]{tombesi10}, or considering that, as the luminosity grows, the inner parts of the outflows may become over-ionized, leading to the detection of the outermost streamlines of the winds, which have lower observed velocities due to the radial dependence.
An alternative scenario is that radiation pressure alone might not supply enough kinetic power, and instead, the outflow could arise from a combination of driving mechanisms. For example, the presence of other mechanisms, such as magnetic and thermal forces, is suggested by the fact that the correlation between v$_{\textrm{out}}$ and L$_{\mathrm{x}}$ is nonsignificant for the T10+S23 sample alone. 
Hence, while radiative luminosity seems to play a key role in the formation and launch of the winds, it may not necessarily be the only driver: simply speaking, more massive SMBHs present larger luminosity, both radiative and mechanical, and thus, faster outflows.
As addressed above, the v$_{\rm out}$-L$_{\rm x}$ relation is driven by the C21 sample, which seems to steepen it, moving the correlation closer to the slope expected for radiative driving. However, as discussed in Sect.~\ref{windradius}, these sources exhibit outflow velocity within the range expected for MHD winds and only SDSS~J0921 shows an Eddington ratio consistent with radiation-driven winds. Once more, these findings suggest a combination of launching mechanisms.

When examining the correlations involving the UFO energetics (i.e., $\dot{M}_{\mathrm{wind}}$, $\dot{E}_{k}^{\mathrm{wind}}$, and $\dot{P}_{\mathrm{wind}}$) versus the X-ray (as well as the bolometric) and UV luminosities, we observe a similar behavior to that described above. This likely arises from the dependence of $\dot{M}_{\mathrm{wind}}$, $\dot{E}_{k}^{\mathrm{wind}}$, and $\dot{P}_{\mathrm{wind}}$ on the outflow velocity (see Fig.~\ref{fig:2}, panel b, for the correlation plots with L$_{\mathrm{x}}$). Similar correlations with the bolometric luminosity are obtained also by \citet{tombesi10}, \citet{gofford13}, and \citet{fiore17}, suggesting that more luminous AGNs launch more massive winds, with a substantial exchange of momentum between the radiation field and the outflow. This may be taken as an indication that UFOs may be driven by radiation pressure. However, these correlations could be driven by basic scaling relations, common to all launching mechanisms, as, for example, also in MHD models the accretion rate and mass outflow rate tend to be positively correlated \citep{fukumura18}.
Therefore, similarly to what discussed above, these results may be expected regardless of the specific launching mechanism taken into consideration.

Several theoretical models and simulations demonstrate that AGN outflows can exert a substantial influence on their surrounding environments when their mechanical power is at least $\sim$\,$10^{-3}$ of the AGN bolometric luminosity \citep[e.g.,][]{dimatteo05,king10a,ostriker10,hopkins10,gaspari19}, which consistently applies to the outflows analyzed here. Hence, these winds have the potential to contribute in removing gas from the host galaxies, as well as quenching star formation and cooling flows \citep[e.g.,][]{gaspari12,faucher12,zubovas19}.

Since the presence of more massive winds may be induced by the dependence on the BH mass, we also tried to normalize $\dot{M}_{\mathrm{wind}}$ for the BH mass (see Sect.~\ref{ufo_mass_corr}). We still obtain a significant positive correlation (coefficient 0.66 and $\log\mathrm{NHP}$ = -6.35, i.e., with significance above 4$\sigma$) with the bolometric luminosity, showing that it is not directly driven by the BH mass. Indeed, these mass-normalized mass outflow rates appear to increase from T10 to S23 to the C21 sample.

Finally, we note that similar correlations between all the UFO parameters and redshift are also present. Given that at high redshift the most luminous sources are detected, these correlations may follow from those with luminosity \citep[see][]{matzeu22}, although in some cases the correlations with z are stronger than those with the luminosity. The fact that the correlations between UFO parameters and redshift are stronger than those with luminosity potentially suggests an evolutionary effect due to the dependence of the accretion rate with redshift. In such a scenario, $\lambda_{\rm Edd}$ would exert a more pronounced influence in driving outflows than luminosity. However, the lack of significant correlations with $\lambda_{\rm Edd}$, as discussed in Sect.~\ref{sed_corr}, is against this interpretation.

\begin{figure*}
\centering
    \subfloat[][]{\includegraphics[width=0.4635\textwidth]{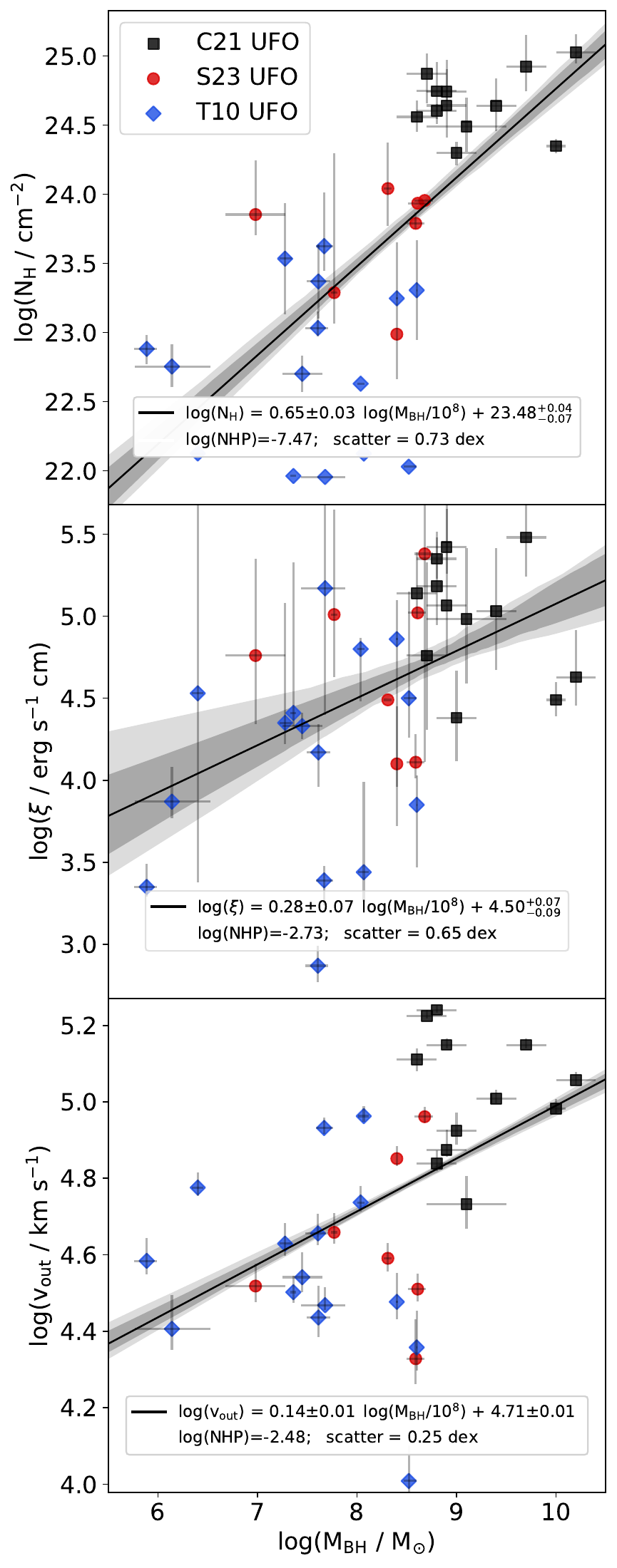}}
    \subfloat[][]{\includegraphics[width=0.45\textwidth]{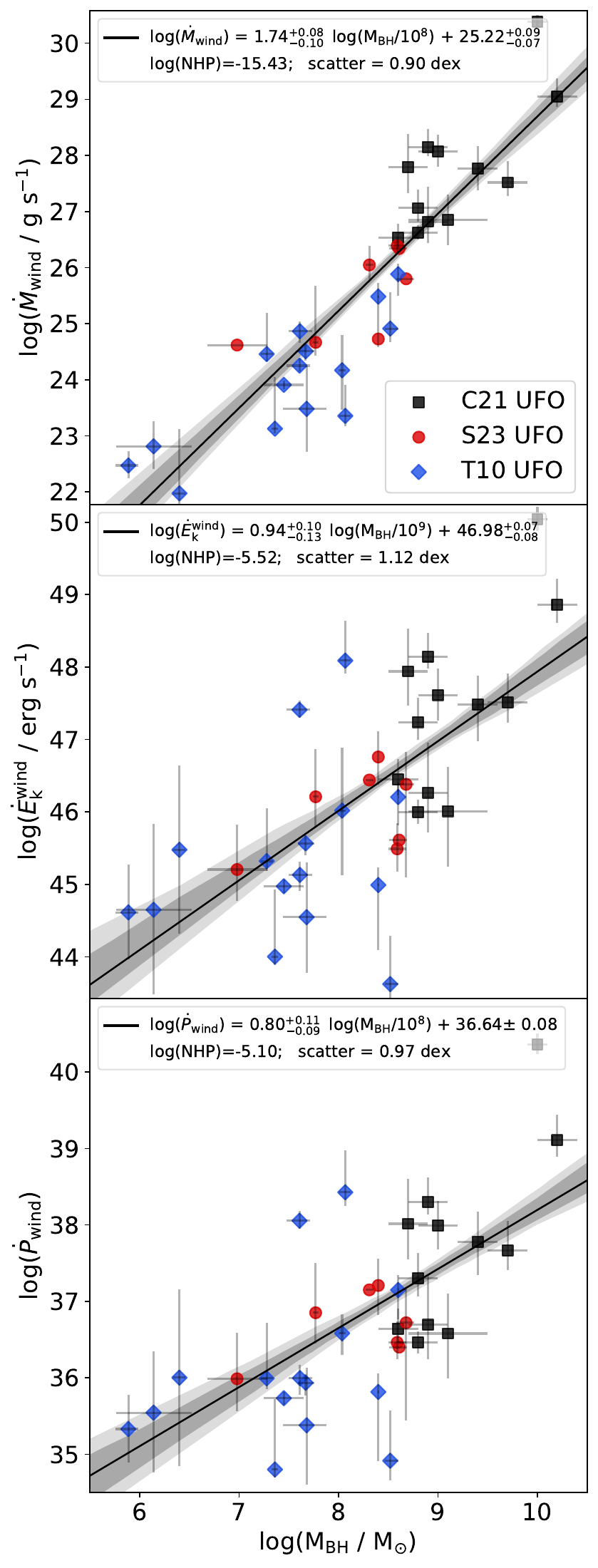}}
    \caption[]
    {{Significant correlations between the SMBH mass and the observed UFO properties and energetics. Panel a: M$_{\mathrm{BH}}$ versus N$_{\mathrm{H}}$ (upper plot), $\xi$ (middle plot), and v$_{\mathrm{out}}$ (lower plot). 
    Panel b: M$_{\mathrm{BH}}$ versus $\dot{M}_{\mathrm{wind}}$ (upper plot), $\dot{E}_{\mathrm{k}}^{\mathrm{wind}}$ (middle plot), and $\dot{P}_{\mathrm{wind}}$ (lower plot).  
    The S23 UFO sub-sample is shown in red circles, T10 in blue diamonds, and C21 in black squares. The solid lines represent the best-fitting linear correlation and the dark and light gray shadowed areas indicate the 68\% and 90\% confidence bands, respectively. In the legend, we report the best-fit coefficients, $\log\mathrm{NHP}$, and the intrinsic scatters for the correlations. }}
    \label{fig:5}
\end{figure*}

\subsection{SMBH mass}\label{ufo_mass_corr}

The same correlations discussed above for the X-ray luminosity and redshift (i.e., with UFO observed and derived properties), are found with respect to the SMBH mass (see Fig.~\ref{fig:5}). Except for the correlations involving $\dot{E}_{k}^{\mathrm{wind}}$ and $\dot{P}_{\mathrm{wind}}$, the NHPs are higher than the values obtained for L$_{\mathrm{x}}$, with the relation between the SMBH mass and $\dot{M}_{\mathrm{wind}}$ (with $\log\mathrm{NHP}$ = -15.43, i.e, significance above 8$\sigma$, and an intrinsic scatter of 0.90 dex; see upper plot panel b in Fig. \ref{fig:5}) is the strongest relation observed in our analysis and similar to what has been reported by \citet{mizumoto2019}.
In particular, we observed that the energetics of the wind all positively correlate with the SMBH mass. Notably, the correlation with the mass outflow rate is steeper ($1.73^{+0.07}_{-0.08}$) than those with $\dot{P}_{\mathrm{wind}}$ and $\dot{E}_{k}^{\mathrm{wind}}$ ($0.88^{+0.07}_{-0.08}$ and slope $1.05^{+0.09}_{-0.11}$, respectively). All of these positive correlations can be explained by the fact that SMBHs with higher masses are present in more massive and hotter halos, hence requiring stronger feedback to achieve self-regulation (e.g., \citealt{beifiori12,gaspari19,bassini19}).

\begin{figure}
\centering
    \subfloat{\includegraphics[width=.5\textwidth]{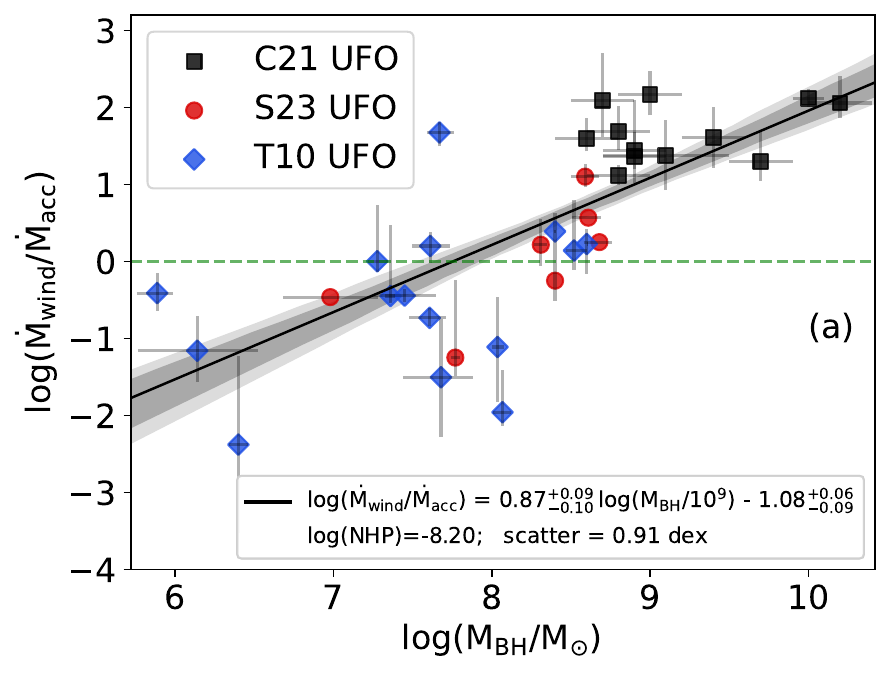}}\\
    \subfloat{\includegraphics[width=.5\textwidth]{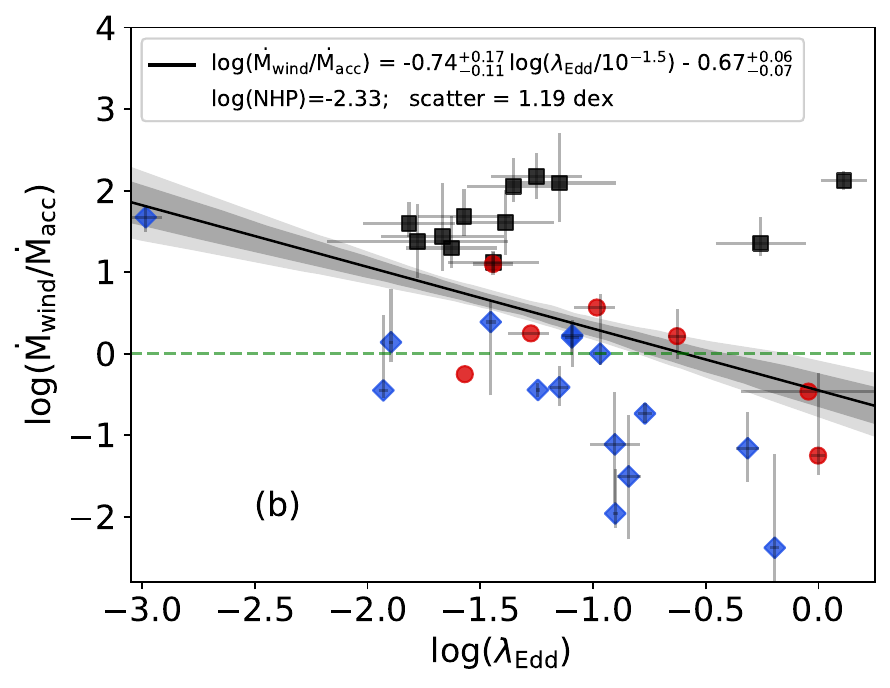}}
    \caption[]
    {{Significant correlations of observed UFO properties. Panel a: M$_{\mathrm{BH}}$ versus $\dot{M}_{\mathrm{wind}}$ normalized to $\dot{\mathrm{M}}_{\mathrm{acc}}$. Panel b: $\lambda_{\mathrm{Edd}}$ versus $\dot{M}_{\mathrm{wind}} / \dot{\mathrm{M}}_{\mathrm{acc}}$. The dashed green lines correspond to a ratio of 1. 
    The S23 UFO sub-sample is shown in red circles, T10 in blue diamonds, and C21 in black squares. The solid lines represent the best-fitting linear correlation and the dark and light gray shadowed areas indicate the 68\% and 90\% confidence bands, respectively. In the legend, we report the best-fit coefficients, $\log\mathrm{NHP}$, and the intrinsic scatters for the correlations.}}
    \label{fig:9}
\end{figure}

By normalizing the mass outflow rate to the AGN mass accretion rate, $\dot{\mathrm{M}}_{\mathrm{acc}} = \mathrm{L}_{\mathrm{bol}}/ \eta c^2$ \citep[where $\eta = 0.1 $ is the average radiative accretion efficiency assumed for the global population, e.g.,][]{peterson97,yu2002,barger05,davis11}, we observe a strong positive correlation with the SMBH mass (above 6$\sigma$, Spearman correlation coefficient of 0.77 and intrinsic scatter 0.91 dex; see Fig.~\ref{fig:9} panel a). This relation, in addition to that between M$_{\mathrm{BH}}$ and $\dot{M}_{\mathrm{wind}}$, suggests that more massive SMBHs present higher wind mass-losses, which decrease the accretion of matter onto the BH. Wind feedback is indeed thought to play an important role in the evolution of AGNs where to compensate the removal of matter through the wind, a mass accretion rate reduction can be expected \citep[e.g.,][and references therein]{crenshaw12,gaspari17,kraemer18, qiu21}.
Meanwhile, we report a weak negative relation (with Spearman correlation coefficient of -0.48 and intrinsic scatter of 1.19 dex) between $\frac{\dot{M}_{\mathrm{wind}}}{\dot{\mathrm{M}}_{\mathrm{acc}}}$ and the Eddington ratio (see Fig.~\ref{fig:9} panel b). The lower significance of this correlation in comparison with that mentioned above, suggests that the main driver is indeed the SMBH mass.
From Fig.~\ref{fig:9} (panels a and b), we also observe that the majority of AGNs hosting UFOs present $\frac{\dot{M}_{\mathrm{wind}}}{\dot{\mathrm{M}}_{\mathrm{acc}}} \geq 1$ (within errors), indicating that the outflow mass rate prevails (or it is comparable to) the mass accretion rate. As suggested in \citet{luminari20}, the outflow may have a limited duration, i.e., at a certain point the accretion disk becomes exhausted and unable to support the wind \citep[e.g.,][]{belloni97}. 

It can be then expected that, as the AGN luminosity and the BH mass increase, the wind has the power to expel a larger amount of matter from the accretion disk \citep{king03,king05,zubovas16}. For this reason, we attempted to delineate possible 3D space correlations by incorporating the BH mass to the significant correlations between the wind energetics and L$_{\mathrm{bol}}$. However, the addition does not improve the significance of any correlation. 

\subsection{Spectral energy distribution}\label{sed_corr}

From our analysis we find only marginal correlations with the parameters related to the SED of the sources, such as $\alpha_{ox}$, $\Delta \alpha_{ox}$ and $\Gamma$. In particular, $\alpha_{ox}$ anticorrelates with the column density of the ionized gas and positively correlates with $\dot{M}_{\mathrm{wind}}$, $\dot{E}_{\mathrm{k}}^{\mathrm{wind}}$, and $\dot{P}_{\mathrm{wind}}$. The relations with the energetics of the winds could be linked to their significant correlations with the X-ray (as well as bolometric) luminosity (see Sect.~\ref{ufo_lum_corr}). Moreover, all three parameters are derived using the ionizing luminosity, which is directly derived from L$_{\mathrm{bol}}$. 
As we discussed in Sect.~\ref{x_uv_slope}, the majority of AGNs in the ``unabsorbed sample'' exhibit an $\alpha_{ox}$ within the -1.8 to -1.2 range. SDSS~J0921+2854 (C21 sample) shows the highest value, that is, $\alpha_{ox} = -0.96$. We note that, if we disregard this source, a significant negative correlation (with coefficient -0.64 and $\log\mathrm{NHP}$ = -2.36) appears between $\alpha_{ox}$ and the outflow velocity of the winds. This result relates to the findings in Sect.~\ref{ufo_vs_noufo}, suggesting that X-ray weak AGNs not only have a higher probability of hosting UFOs but also exhibit faster outflows.

\begin{figure}
\centering
    \includegraphics[width=.51\textwidth]{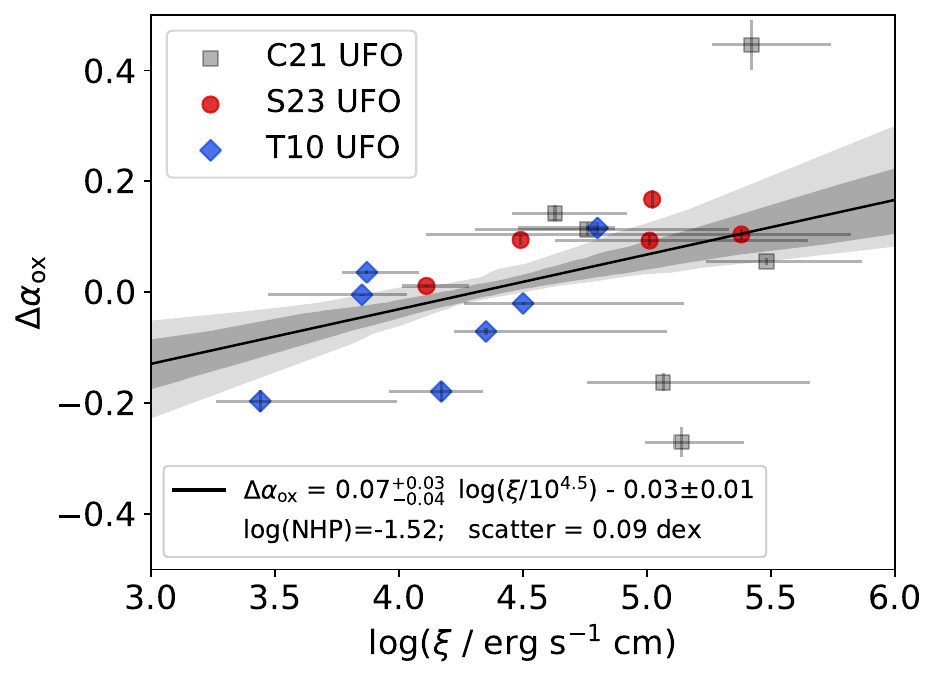}
    \caption[]
    {{ Significant correlation of T10+S23 sample. $\Delta \alpha_{\mathrm{ox}}$ versus $\xi$ for the UFOs sources. The S23 UFO sub-sample is shown in red circles and the T10 in blue diamonds. The C21 sub-sample is shown with light black squares as, if considered, the correlation is not significant. The solid line represents the best-fitting linear correlation and the dark and light gray shadowed areas indicate the 68\% and 90\% confidence bands, respectively. In the legend, we report the best-fit coefficients, $\log\mathrm{NHP}$ and the intrinsic scatter for the T10+S23 sample.}}
    \label{fig:8}
\end{figure}

While no significant correlations with the X-ray-weakness factor ($\Delta \alpha_{ox}$) emerge for the T10+S23+C21 sample\footnote{A marginal correlation can be found by adding the AGNs of the Fe-K sub-sample (see Fig.~\ref{fig:ef19})}. After the addition, the X-ray-weakness factor also weakly correlates with N$_{\mathrm{H}}$ (see Fig.~\ref{fig:ef19}), a positive correlation between $\Delta \alpha_{ox}$ and $\xi$ is seen at low/intermediate-z (i.e., for the T10+S23 sample; see Fig.~\ref{fig:8}).
It appears that AGNs with a weaker X-ray emission show a lower ionization parameter of the wind, as it can be expected since X-ray photons are indeed the main source of photoionization of the gas responsible for the UFOs.
Moreover, weak positive correlations with $\Gamma$ are found for $\dot{E}_{\mathrm{k}}^{\mathrm{wind}}$ and $\dot{P}_{\mathrm{wind}}$, suggesting that AGNs with flatter X-ray photon indices are less efficient in accelerating winds, as would be expected in the case of line-driven (but not in continuum-driven) winds since the gas would tend to be over-ionized. 
Furthermore, N$_{\mathrm{H}}$ and v$_{\mathrm{out}}$ exhibit positive and negative\footnote{To be noted that the best-fitting slope of the v$_{\mathrm{out}}$-FWHM H$\beta$ relation is consistent, within the errors, with zero (see Fig.~\ref{fig:ef8}).}, respectively, correlations with the FWHM of the H$\beta$, which is known to be related to the SED and the accretion state of AGNs \citep{marziani18}, although they may be significantly affected by turbulence, as discussed in more detail in the next section.

\begin{figure*}
\centering
    \subfloat[][]
    {\includegraphics[width=.955\textwidth]{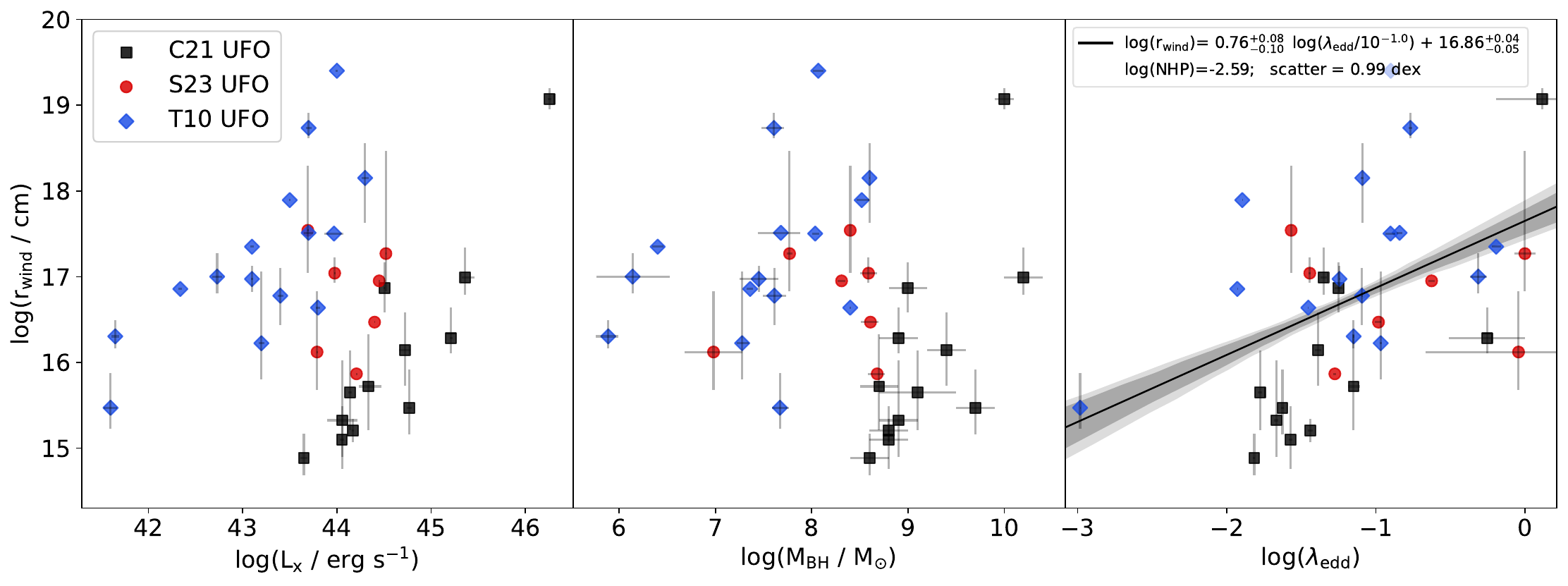}} \\
    \subfloat[][]
    {\includegraphics[width=.95\textwidth]{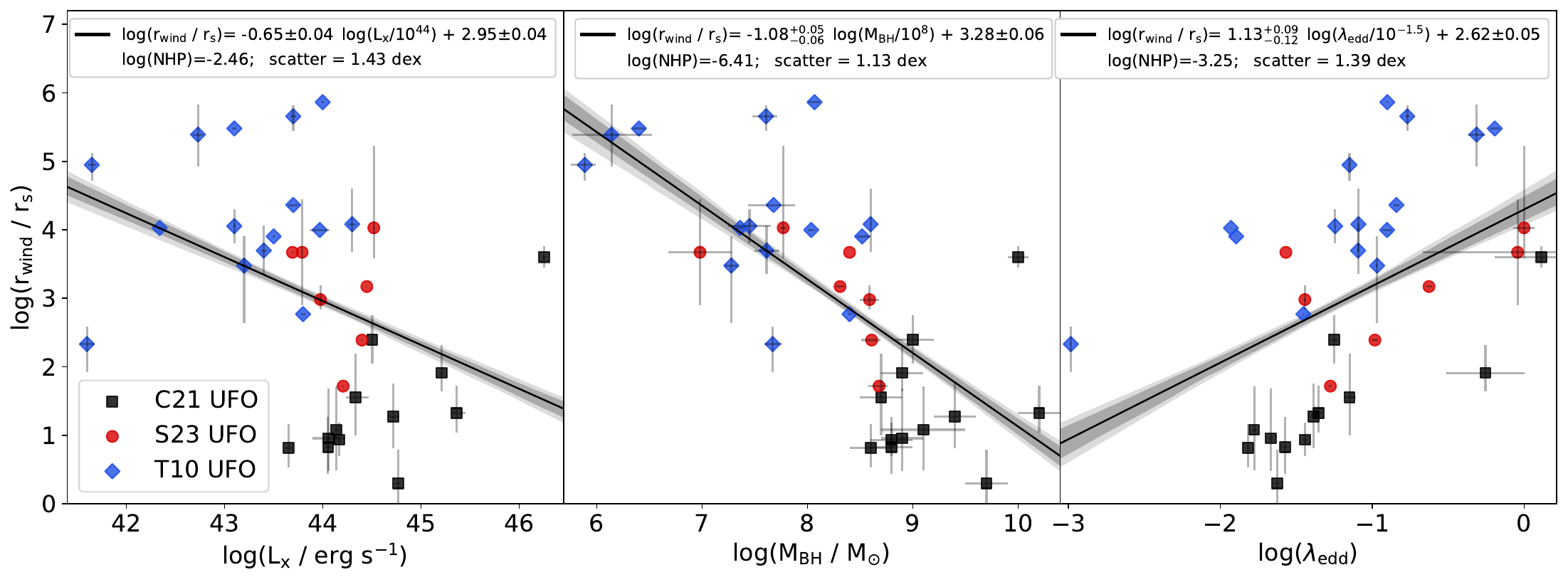}}
    \caption[]
    {{Correlations between the X-ray luminosity, SMBH mass, Eddington ratio, and the wind radii. Panel a: Correlations considering r$_{\mathrm{wind}}$. 
    Panel b: Correlations considering r$_{\mathrm{wind}}$ normalized for r$_s$.    
    The S23 UFO sub-sample is shown in red circles, T10 in blue diamonds, and C21 in black squares. The solid lines represent the best-fitting linear correlation and the dark and light gray shadowed areas indicate the 68\% and 90\% confidence bands, respectively. In the legend, we report the best-fit coefficients, $\log\mathrm{NHP}$, and the intrinsic scatters for the correlations. The correlations are not significant where the best-fit and the confidence bands are not reported.}}
    \label{fig:3}
\end{figure*}

\subsection{Wind radius and driving mechanisms}\label{windradius}

While in thermal and radiation-driven winds we would expect the launching radii to scale with the SMBH mass and the X-ray luminosity, we find that r$_{\mathrm{wind}}$ does not correlate with these parameters, as it can be seen in Fig.~\ref{fig:3} panel a (left and middle plots). Interestingly, we observe instead a significant positive correlation between the Eddington ratio and r$_{\mathrm{wind}}$ (see Fig.~\ref{fig:3} panel a right plot), suggesting that the increase of the accretion rate has some impact on the global spatial scale of the wind.
We further obtain an anticorrelation between r$_{\mathrm{wind}}$ and the UFO outflow velocity. This could be simply interpreted by the fact that a more compact wind, produced closer to the black hole, is expected to exhibit higher velocities, gradually transitioning from a fast to a slower outflow as it expands across larger scales.

We also find a positive correlation between the H$\beta$ FWHM and r$_{\mathrm{wind}}$ (and v$_{\mathrm{out}}$ as mentioned in the previous Section), which is interesting from a kinematical point of view. While line broadening in the BLR is generally attributed to virialization in the SMBH potential well, feedback and feeding processes may overcome pure gravitational effects. Hydro-dynamical simulations show that volume-filling turbulence is the irreducible by-product of the self-regulated AGN feeding/feedback cycle, with conversion efficiencies beyond 1\%, due to stretching, compressive, and baroclinic motions in a stratified medium (\citealt{wittor20,wittor23}). Even if only a small 1\% of the related feedback kinetic energy were transferred into chaotic turbulent energy ($\propto \sigma_v^2$, the gas velocity dispersion), this would overcome the virial velocity, $(GM/r)^{1/2}$ ($\sim10^3$ km\,s$^{-1}$ for $M=10^8$ M$_{\odot}$ and $r=0.1$ pc). Overall, the above-mentioned FWHM positive correlations would then be consistent with an increased turbulence driven by stronger AGN feedback (larger v$_{\rm out}$ and r$_{\rm wind}$).
On the other hand, the absence of a significant correlation between the mechanical power of the wind, $\dot{\mathrm{E}}_{\mathrm{k}}^{\mathrm{wind}}$, and r$_{\rm wind}$ could be related to the energetic properties of the outflows when it expands away.  There are indeed several hints that this mechanical power is conserved from close to the black hole to the distant molecular cloud scales, following a scale-independent energy-conserving scenario (e.g., \citealt{tombesi15,faucher12,stern2016,gaspari20}).

Further, we normalized the wind radii to the Schwarzschild radius. As expected from the lack of correlation with respect to r$_{\mathrm{wind}}$, we now obtain an anticorrelation with $M_{BH}$ and $L_x$ (see Fig.~\ref{fig:3} panel b). We observe that the anticorrelation between the X-ray luminosity and r$_{\mathrm{wind}}/r_s$ is mainly driven by the C21 sample and disappears when considering only the T10+S23 sample. On the other hand, the anticorrelation between r$_{\mathrm{wind}}/r_s$ and $M_{BH}$ remains significant at low/intermediate-z. 
While the positive correlations with the Eddington ratio and the FWHM of the H$\beta$, as well as the anticorrelation with v$_\mathrm{out}$, are still observed for r$_{\mathrm{wind}}/r_s$, the latter presents negative correlations with $\dot{\mathrm{M}}_{\mathrm{wind}}$ and $\dot{\mathrm{E}}_{\mathrm{k}}^{\mathrm{wind}}$, and a marginal negative relation with $\dot{\mathrm{P}}_{\mathrm{wind}}$ appears if the Fe-K sub-sample is included. 
Given that $\dot{\mathrm{E}}_{\mathrm{k}}^{\mathrm{wind}}$ does not correlate with r$_{\mathrm{wind}}$ (see before), the observed anticorrelation between $\dot{\mathrm{E}}_{\mathrm{k}}^{\mathrm{wind}}$ and r$_{\mathrm{wind}}/r_{\rm s}$ is certainly related to the correlation between $\dot{\mathrm{E}}_{\mathrm{k}}^{\mathrm{wind}}$ and $M_{\rm BH}$.

If we consider a thermally driven wind, at radii where the sound speed overcomes the escape velocity, the pressure gradient, which has been built up by the X-ray emission, leads to the expansion of the layer that has been heated up to the Compton temperature T$_{\mathrm{IC}}$ \citep{begelman83, done18}. Thus, the gas of the becomes unbound at the so called Compton radius, R$_{\mathrm{IC}}$\footnote{R$_{\mathrm{IC}} = 6.4 \times 10^4$ / T$_{\mathrm{IC,8}} r_g$, where T$_{\mathrm{IC,8}}$ is the Compton temperature in units of 10$^8$ K \citep[see][]{done18}.}. If we estimate a fiducial Compton radius adopting a Compton temperature of T$_{\mathrm{IC}} = 2 \times 10^7$ K  \citep[as found in][for an average quasar spectral distribution]{sazonov04}, log(R$_{\mathrm{IC}}$ / r$_{\mathrm{s}}) = 5.20$, we note that only four UFOs (all in the T10 sample: PG~1211+143, Ark~120, Mrk~766, and NGC~4507), have r$_{\mathrm{wind}}$, within errors, above the Compton radius (the black dashed line in Fig.~\ref{fig:radii}) and are then consistent with thermal launching. 
Interestingly, PG~1211+143 and Ark~120 are also outliers in the $v_{out}$ versus $\xi$ correlation reported in Sect. \ref{ufo_prop_corr}. However, both AGNs show highly variable UFOs. For instance, while T10 detects a UFO in Ark~120 during the 2003 observation, \citet{gofford13} do not find signature of absorption lines during the 2007 observation. \citet{igo20} also find no evidence of UFOs and \citet{giustini19} suggest that this face-on AGN possibly has no intersection between the wind and the line of sight.
Similarly, PG~1211+143 exhibits variability in the UFO detection across multiyear observations \citep[e.g.,][and references therein]{tombesi10, gofford13, igo20}. For example, in some cases, clear evidence of a UFO is observed, while in others, only weak or the absence of features are reported. This variability extends to NGC~4507 as well \citep[e.g.,][]{tombesi10,igo20}. The intrinsic variability of UFOs, frequently associated with changes in the observed outflow velocity, suggests that while in some observations thermal-driving may be the underlying launching mechanism (in agreement with these AGNs presenting r$_{\mathrm{wind}} > $R$_{\mathrm{IC}}$), in other a combination (or a different mechanism) could be at play.
Additionally, the outflow velocities reported in Table \ref{tab:2}, which align with the literature values, contradict with the possible thermal origin of Ark~120 and NGC~4507 outflows. Instead, they suggest the involvement of radiative or MHD winds.
Meanwhile, for all the other UFOs, in particular for the whole C21 and S23 samples, r$_{\mathrm{wind}}$ values are always inside R$_{\mathrm{IC}}$, suggesting that a simple thermal launching mechanism is unlikely and other mechanisms should be present.

\begin{figure}[h]
\centering
    \subfloat[][]{\includegraphics[width=.5\textwidth]{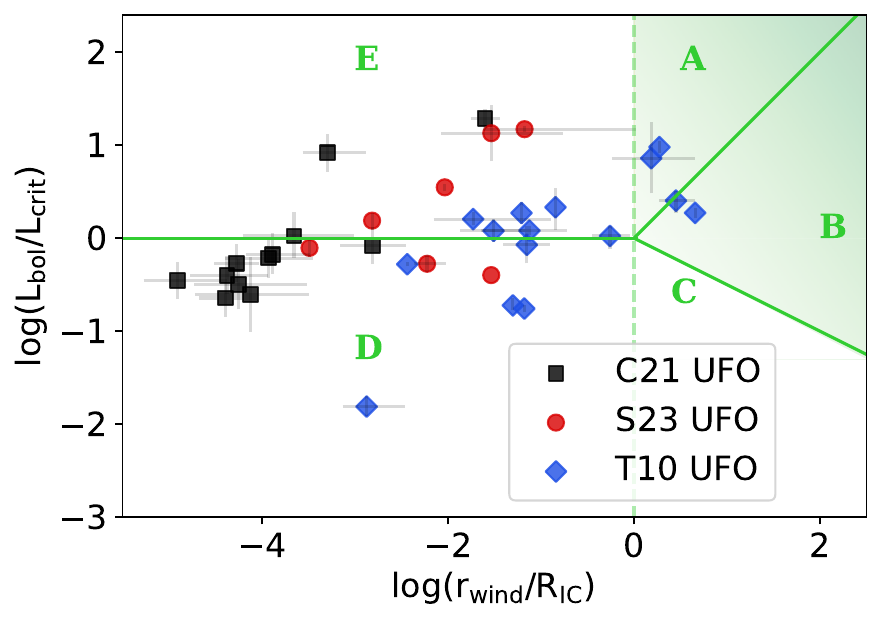}}\\
    \subfloat[][]{\includegraphics[width=.5\textwidth]{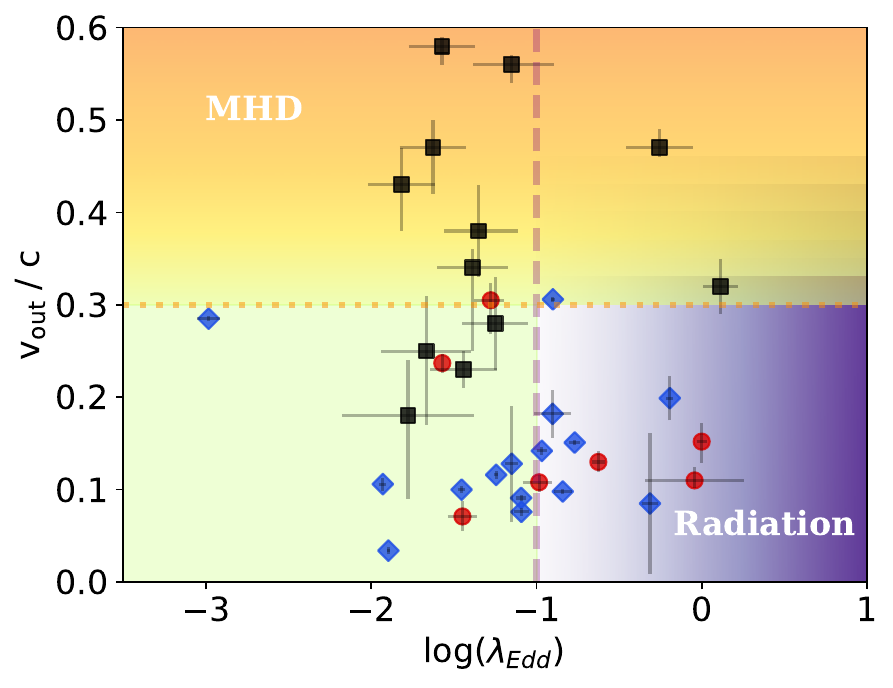}}
    \caption[]
    {{Launching mechanisms. Panel a: Parameter space for thermal winds. The green solid lines show the different wind regions boundaries described in \citet{begelman83}. In regions A and B (in shaded green), thermal winds can be launched, while this mechanism is suppressed in regions C, D, and E. Panel b: Magnetic versus radiative driving. The dashed purple line shows the possible threshold separating radiatively driven ($\lambda_{\rm Edd} \gtrsim 0.1$) winds from MHD winds, described in \citet{sadowski17}. The dotted orange line present the threshold between the two mechanisms on the basis of the outflow velocity. The shaded regions show the intersection between the two conditions. Particularly, winds in the purple shaded region are likely to be radiatively driven, while those in the orange shaded regions are possibly MHD driven.
    The S23 UFO sub-sample is shown in red circles, T10 in blue diamonds, and C21 in black squares.}}
    \label{fig:12}
\end{figure}

To further investigate the possibility of a thermal launch, we adopted the parameter space defined by \citet{begelman83}, in which five distinct regions, each showcasing various physical characteristics that can either enable or prevent the development of thermal winds, are presented (see Fig.~\ref{fig:12} panel a). On the x-axis r$_{\mathrm{wind}}$ is normalized to R$_{\mathrm{IC}}$ and on the y-axis the efficiency of the luminosity in generating a wind via Compton heating is quantified by the parameter L$_\mathrm{bol}$/L$_{\mathrm{crit}}$ \citep[where L$_{\mathrm{crit}} = 0.03 \mathrm{T}_{\mathrm{IC,8}}^{-1/2}{\rm L_{Edd}}$, see][]{begelman83}.
In region A, the thermal wind is most likely to arise as the gas is impulsively heated to the Compton temperature. As the luminosity decreases, we move into region B where the disk is steadily heated, and its material can still produce a wind as long as r$_{\mathrm{wind}}$/R$_{\mathrm{IC}}$ > L$_\mathrm{bol}$/L$_{\mathrm{crit}}$. In contrast, region C features temperatures that are too low for the material to escape. The launching of a thermal wind is even less likely in the leftward regions D and E. In region D, the entire disk is heated to a temperature below the Compton temperature, while in E, only the upper layers of the disk are heated to T$_{\mathrm{IC}}$, and the average particle velocity is lower than the escape speed. Thus, in the latter region an isothermal atmosphere with minimal wind losses can form.
In agreement with the above discussion, the majority of the T10 sample and all the S23 and C21 sources are located in the E and D region where thermal launching mechanism is not possible. Again, the same four T10 UFOs (i.e., PG~1211+143, Ark~120, Mrk~766, and NGC~4507) appear to be compatible with thermal launching, but see above discussion.
The position of the C21 and T10 AGNs at opposite ends of Fig.~\ref{fig:12}, panel a, (and Fig.~\ref{fig:radii}) is attributable to the lower column density detected in the T10 sample compared to the C21 sample. 
If we add the four AGNs of the Fe-K sub-sample to the parameter space for thermal winds, only NGC~3783 is in the B region where thermal wind can be launched, whereas the other three sources are in the E and D regions.

This evidence against thermal winds is not surprising since they are believed not to reach the mildly relativistic velocities observed in UFOs. Indeed, one can see in Fig.~\ref{fig:12}, panel b, that 12 AGNs (within errors; 1/15 in the T10, 1/7 in the S23, and 10/12 in the C21 sample) populate the upper part of the plot above 0.3c. While UV and X-ray line or radiation driving mechanisms can still accelerate the winds up to $\sim$0.3-0.4c, higher outflow velocities cannot be achieved due to the effects of radiation drag \cite[e.g.,][]{takahashi15,hagino17}. Therefore, in the case of these UFOs the involvement of a magnetic driving mechanism seems to be inescapable \citep[e.g.,][]{fukumura10}. On the other hand, radiation pressure is expected to dominate for $\lambda_{\rm Edd} \gtrsim 0.1$ (e.g., \citealt{ressler15,sadowski17}). Thus, winds in the lower right part of Fig.~\ref{fig:12}, panel b, (with v$_{\rm out} <$ 0.3c and $\lambda_{\rm Edd} \gtrsim 0.1$) could be radiatively driven.
To investigate possible differences among the sub-groups delineated by the two thresholds (i.e., MHD winds for v$_{\rm out} \gtrsim 0.3c$ and radiatively driven winds for $\lambda_{\rm Edd}\gtrsim 0.1$), we performed the two-sample KS test. As expected, we find that the two sub-groups differentiate in terms of redshift ($\log\mathrm{NHP}$ = -2.25) and BH mass ($\log\mathrm{NHP}$ = -2.25), given that the MHD sub-sample is predominantly composed of C21 AGNs (see Sect.~\ref{comparison_samples})\footnote{Additionally, we observe differences in v$_{\rm out}$ ($\log\mathrm{NHP}$ = -3.09) and Eddington ratio ($\log\mathrm{NHP}$ = -2.56), as these parameters define the two AGN sub-groups.}. Furthermore, the distinction in column density observed for the UFO sub-samples in Sect.~\ref{comparison_ufos}, persists between the MHD and radiatively driven winds ($\log\mathrm{NHP}$ = -2.26). The dependence on v$_{\rm out}$ leads to differences in two (out of three) winds energetics (i.e., $\dot{\mathrm{M}}_{\mathrm{wind}}$ with $\log\mathrm{NHP}$ = -2.10 and $\dot{\mathrm{E}}_{\mathrm{k}}^{\mathrm{wind}}$ with $\log\mathrm{NHP}$ = -1.80). A significant difference also emerges between the two driving mechanism sub-groups when considering the ratio r$_{\mathrm{wind}}/r_{\rm s}$ ($\log\mathrm{NHP}$ = -2.10), likely driven by the dependence of r$_{\rm s}$ on the BH mass.

\section{Conclusions}\label{conclusions}

We carried out an extensive statistical analysis in order to uncover connections between AGN and host galaxy properties with respect to the presence of UFOs and their characteristics. 
Our study is based on the SUBWAYS sample, which we expanded by incorporating two additional samples \citep{tombesi10,chartas21} covering different and complementary ranges of redshift and luminosity.

While our results suggest that more luminous AGNs with steeper $\alpha_{\mathrm{ox}}$ are more likely to host UFOs (i.e., with $\alpha_{\mathrm{ox}} <$ -1.24; see Sect.~\ref{ufo_vs_noufo}), we do not observe any other definitive distinctions between AGNs that exhibit UFOs and those that do not on the basis of the studied parameters (BH mass; X-ray, bolometric, and UV luminosity; Eddington ratio; X-ray photon index; FWHM of the H$\beta$; $\alpha_{\mathrm{ox}}$; and $\Delta \alpha_{ox}$; see Sect.~\ref{ufo_vs_noufo} and Table \ref{tab:7}). This is consistent with the idea that all AGNs have the potential to host UFOs with a characteristic duty cycle, which determines whether they are detectable in a specific epoch or not.

The key findings of our extensive correlation analysis (see Appendix~\ref{allcorr} for a complete view of all the correlation plots) are summarized as follows:

\begin{itemize}
    \item[\textbullet] Faster UFOs have larger ionization parameters and column densities (see Section~\ref{ufo_prop_corr}). However, the positive $\xi$-v$_{\rm out}$ correlation seems to follow from the outflow velocity and luminosity relation (see \ref{ufo_lum_corr}).
    \item[\textbullet] The correlation between outflow velocity and luminosity, however, has a significantly flatter slope (0.12; see Fig.~\ref{fig:2}) compared to the expected value for radiatively driven winds (0.5). This suggests a combination of launching mechanisms, including magnetically, radiatively, and thermally driven processes. 
    \item[\textbullet] X-ray radiation seems to play a more crucial role in driving highly ionized winds compared to UV. We find that all parameters and energetics of the observed UFOs are strongly correlated with X-ray (and bolometric) luminosity but show weaker or no correlation with the UV luminosity (see Sect.~\ref{ufo_lum_corr}).
    \item[\textbullet] More massive SMBHs suffer larger wind mass losses, thus suppressing accretion of matter onto the BH. 
    In particular, the outflow mass rate in the majority of the studied AGNs either prevails or is comparable to the mass accretion rate, suggesting a potential limitation in the duration of the outflow: When the accretion disk is depleted, it might lose the capacity to sustain the outflow (see Fig.~\ref{fig:9} panel b and Sect.~\ref{ufo_mass_corr}).
    \item[\textbullet] The UFO launching radius does not appear to correlate with either with the luminosity or with the BH mass (as instead expected in thermal and radiation driven winds) unless it is normalized by $r_{\rm s}$. However, it is always positively correlated with the Eddington ratio (Sect.~\ref{windradius}). 
\end{itemize}

In terms of the wind launching mechanism, our analysis does not unequivocally point to a single phenomenon. Instead, it suggests that multiple mechanisms may be involved in ejecting UFOs (magnetic, radiative, and thermal driving).
While thermal launching seems to be disfavored, as only a small fraction of the sources in our samples exhibit UFO properties compatible with it (Sect.~\ref{windradius}), radiation-driven winds may account for several observed correlations. However, most results display significant deviations from expectations in this context, indicating a key role played by MHD winds (Fig.~\ref{fig:12} panel b).

It is insightful to understand our observational findings in the context of the AGN feeding-feedback self-regulation.
Current theories favor a global AGN duty cycle based on the multiphase condensation of the turbulent gaseous halos around SMBHs 
(\citealt{gaspari12,McCourt:2012,Sharma:2012,McNamara:2016,Voit:2017}), which is also supported by a wide range of observational constraints 
(e.g., \citealt{Tremblay:2018,Temi:2018,Storchi-Bergmann:2019,McKinley:2021,Maccagni:2021,Olivares:2022}).
This infalling "rain" (formally known as chaotic cold accretion, or CCA; \citealt{gaspari13,gaspari20,Prasad:2017,Voit:2018}) is expected to recurrently and efficiently trigger the AGN, generating UFOs via varying launching mechanisms depending on the feeding rate.
Over the long term, the feeding rain and mechanical feedback act as a cosmic thermostat regulating galaxies and groups of galaxies.
When comparing our scaling relations with theoretical expectations (cf.~\citealt{gaspari17,gaspari19}), we find several key consistencies with a CCA-driven feedback.
Specifically, CCA theory and simulations predict 
(i) stronger outflows in more massive SMBHs due to a stronger condensation rain (Sect.~\ref{ufo_mass_corr}); 
(ii) comparable feeding and feedback mass outflow rates (Sect.~\ref{ufo_mass_corr}), as most of the raining mass is reejected back near the SMBH horizon;
(iii) outflow velocities that decrease with radius (Sect.~\ref{windradius}), as the outflow is slowed by the interaction with the multiphase atmosphere;
(iv) enhanced turbulence reflected by positive FWHM correlations (Sect.~\ref{windradius}), as faster feedback enhances the
chaotic motions seeding the CCA instabilities;
and (v) conservation of the outflow energy rate from the micro (sub-pc) scale to macro (pc-kpc) scale (Sect.~\ref{windradius}), which is vital to establishing efficient self-regulation and global quasi-thermal equilibrium.

By examining large samples of AGN, population studies play a pivotal role in advancing our understanding of UFOs. They provide valuable insights into the presence of these outflows across different AGN sub-classes and their correlations with various AGN properties. Additionally, population studies are one the most powerful tools to try to understand the role of UFOs in AGN feedback processes, their evolutionary implications for galaxy formation and evolution, and the unique characteristics of AGNs with and without UFOs. Ultimately, these comprehensive investigations contribute to a deeper understanding of the physical processes driving these outflows and their significance in shaping the cosmic landscape of AGNs and their host galaxies.
These investigations, when coupled with the goals of future X-ray missions, such as the X-Ray Imaging Spectroscopy Mission \citep[XRISM;][]{xrism20} and the Advanced Telescope for High-ENergy Astrophysics \citep[{\it Athena};][]{barret18}, are expected to provide further and unparalleled advancements in this field. 
Thanks to the unprecedented high-spectral resolution and sensitivity of these instruments, they will allow us to resolve the UFO line profiles, which hold the promise of unveiling the dominant launching mechanism; study the dynamical behavior of the inner portion of the wind; and extend the UFO search to lower luminosities and higher redshifts.

\begin{acknowledgements}
 All the Italian co-authors acknowledge support and fundings from Accordo Attuativo ASI-INAF n. 2017-14-H.0. SB, MB, VEG, MP, FU acknowledge support from PRIN MUR 2017 ``Black hole winds and the baryon life cycle of galaxies: the stone-guest at the galaxy evolution supper''. VEG, POP, MP acknowledge financial support from the High Energy National Programme (PNHE) of the  national french research agency (CNRS) and from the french spatial agency (CNES). SB and FR acknowledge funding from PRIN MUR 2022 SEAWIND 2022Y2T94C, supported by European Union - Next Generation EU. SB acknowledges support from INAF LG 2023 BLOSSOM. GM thanks the support of grant n. PID2020-115325GB-31 funded by MICIN/AEI/10.13039/501100011033. MG acknowledges partial support by NASA HST GO-15890.020/023-A and the \textit{BlackHoleWeather} program. AL acknowledge support from the HORIZON-2020 grant “Integrated Activities for the High Energy Astrophysics Domain" (AHEAD-2020), G.A. 871158. EB is supported by a Center of Excellence of the Israel Science Foundation (grant 1937/19). BDM acknowledges support via a Ramón y Cajal Fellowship (RYC2018-025950-I), the Spanish MINECO grant PID2020-117252GB-I00, and the AGAUR/Generalitat de Catalunya grant SGR-386/2021. MB acknowledges support from INAF under project 1.05.12.04.01 - MINI-GRANTS di RSN1 "Mini-feedback". GP acknowledges funding from the European Research Council (ERC) under the European Union’s Horizon 2020 research and innovation programme (grant agreement No 865637), support from Bando per il Finanziamento della Ricerca Fondamentale 2022 dell’Istituto Nazionale di Astrofisica (INAF): GO Large program and from the Framework per l'Attrazione e il Rafforzamento delle Eccellenze (FARE) per la ricerca in Italia (R20L5S39T9). YK acknowledges support from PAPIIT DGAPA gran IN 102023. RS acknowledges support from  the agreement ASI-INAF eXTP Fase B - 2020-3-HH.1-2021 and the INAF-PRIN grant “A Systematic Study of the largest reservoir of baryons and metals in the Universe: the circumgalactic medium of galaxies” (No. 1.05.01.85.10). LZ acknowledges financial support from the Bando Ricerca Fondamentale INAF 2022 Large Grant “Toward an holistic view of the Titans: multi-band observations of z $>$ 6 QSOs powered by greedy supermassive black-holes. FP acknowledges financial support from the Bando Ricerca Fondamentale INAF 2023 "Exploring the origin of radio emission in Radio Quiet AGN". MP acknowledges grant PID2021-127718NB-I00 funded by the Spanish Ministry of Science and Innovation/State Agency of Research (MICIN/AEI/ 10.13039/501100011033). The research leading to these results has received funding from the European Union’s Horizon 2020 Programme under the AHEAD2020 project (grant agreement n. 871158).
\end{acknowledgements}


\begin{landscape}
\renewcommand{\arraystretch}{1.8}
\begin{table}
\caption{Correlation analysis results: T10+S23+C21 UFOs sub-sample.}
\label{tab:8}
\begin{tabular}{llllllllllllllllllrrrrr}
\multicolumn{1}{c}{} & M$_{BH}$ & z & L$_x$ & L$_{bol}$ & L$_{2500 \AA}$ & H$\beta$ & $\Gamma$ & $\lambda _{\rm Edd}$ & $\alpha _{ox}$ & $\Delta \alpha _{ox}$ & \cellcolor[HTML]{DAF7A6}$\xi$ & \cellcolor[HTML]{DAF7A6}N$_{\mathrm{H}}$ & \cellcolor[HTML]{DAF7A6}v$_{out}$ & \cellcolor[HTML]{DAF7A6}r$_{wind}$ & \cellcolor[HTML]{DAF7A6}r$_{wind} / r_s$ & \cellcolor[HTML]{DAF7A6}$\dot{M} _{wind}$ & \cellcolor[HTML]{DAF7A6}$\dot{E} _{k}^{wind}$ & \cellcolor[HTML]{DAF7A6}$\dot{P}_{wind}$ \\ \hline 
\multicolumn{1}{l|}{M$_{BH}$} & - & \cellcolor[HTML]{fcaa8d}\textcolor{white}{\makecell{ 0.90 \\ -13.48 }} & \cellcolor[HTML]{fdcab5}\makecell{ 0.79 \\ -8.97 } & \cellcolor[HTML]{fdcab5}\makecell{ 0.79 \\ -8.97 } & \cellcolor[HTML]{ffede5}\makecell{ 0.40 \\ -2.25 } & x & x & \cellcolor[HTML]{d8e7f5}\makecell{ -0.42 \\ -1.90 } & \cellcolor[HTML]{e2edf8}\makecell{ -0.34 \\ -1.32 } & x & \cellcolor[HTML]{fee9df}\makecell{ 0.52 \\ -3.15 } & \cellcolor[HTML]{fdcebb}\makecell{ 0.79 \\ -8.31 } & \cellcolor[HTML]{fee7dc}\makecell{ 0.57 \\ -3.78 } & x & \cellcolor[HTML]{5ba3d0}\makecell{ -0.76 \\ -6.62 } & \cellcolor[HTML]{fc9c7d}\textcolor{white}{\makecell{ 0.93 \\ -15.43 }} & \cellcolor[HTML]{fee1d3}\makecell{ 0.68 \\ -5.52 } & \cellcolor[HTML]{fee2d5}\makecell{ 0.65 \\ -5.10 } \\ 
\arrayrulecolor{black}\cdashline{2-21}[0.5pt/5pt]
\multicolumn{1}{l|}{z} & \cellcolor[HTML]{fcaa8d}\textcolor{white}{\makecell{ 0.90 \\ -13.48 }} & - & \cellcolor[HTML]{fdcab5}\makecell{ 0.79 \\ -9.04 } & \cellcolor[HTML]{fdcab5}\makecell{ 0.79 \\ -9.03 } &  \cellcolor[HTML]{ffede5}\makecell{ 0.39 \\ -2.16 }  & x & x & x & \cellcolor[HTML]{deebf7}\makecell{  -0.37 \\ -1.53 } & \cellcolor[HTML]{ffefe8}\makecell{ 0.21 \\ -1.65 } & \cellcolor[HTML]{fee8dd}\makecell{ 0.54 \\ -3.64 } & \cellcolor[HTML]{fdc9b3}\makecell{ 0.82 \\ -9.15 } & \cellcolor[HTML]{fee7dc}\makecell{ 0.49 \\  -3.70 } & \cellcolor[HTML]{e2edf8}\makecell{  -0.34 \\ -1.32 } & \cellcolor[HTML]{65aad4}\makecell{ -0.74 \\  -6.25 } & \cellcolor[HTML]{fcb095}\textcolor{white}{\makecell{ 0.89 \\ -12.63 }} & \cellcolor[HTML]{fedecf}\makecell{ 0.70 \\ -6.09 } & \cellcolor[HTML]{fedfd0}\makecell{ 0.70 \\ -5.86 } \\ 
\arrayrulecolor{black}\cdashline{2-21}[0.5pt/5pt]
\multicolumn{1}{l|}{L$_x$}  & \cellcolor[HTML]{fdcab5}\makecell{ 0.79 \\ -8.97 } & \cellcolor[HTML]{fdcab5}\makecell{ 0.79 \\ -9.04 } & - & \cellcolor[HTML]{67000d}\textcolor{white}{\makecell{ 0.98 \\ -44.64 }} & \cellcolor[HTML]{fee8dd}\makecell{ 0.42 \\ -3.63 } & x & x & x & \cellcolor[HTML]{dfebf7}\makecell{ -0.40 \\ -1.50 } & \cellcolor[HTML]{fee9df}\makecell{ 0.45 \\ -3.19 } & \cellcolor[HTML]{ffebe2}\makecell{ 0.47 \\ -2.73 } & \cellcolor[HTML]{fedfd0}\makecell{ 0.69 \\ -5.78 } & \cellcolor[HTML]{ffece3}\makecell{ 0.45 \\ -2.48 } & x & \cellcolor[HTML]{cee0f2}\makecell{ -0.49 \\ -2.53} & \cellcolor[HTML]{fdcebb}\makecell{ 0.80 \\ -8.34 } & \cellcolor[HTML]{fed9c9}\makecell{ 0.74 \\ -6.76 } & \cellcolor[HTML]{fdd4c2}\makecell{ 0.76 \\ -7.46 }  \\ 
\arrayrulecolor{black}\cdashline{2-21}[0.5pt/5pt]
\multicolumn{1}{l|}{L$_{bol}$}  & \cellcolor[HTML]{fdcab5}\makecell{ 0.79 \\ -8.97 } & \cellcolor[HTML]{fdcab5}\makecell{ 0.79 \\ -9.03 } & \cellcolor[HTML]{67000d}\textcolor{white}{\makecell{ 0.98 \\ -44.64 }} & - & \cellcolor[HTML]{fee8dd}\makecell{ 0.42 \\ -3.63 } & x & x & x & \cellcolor[HTML]{dfebf7}\makecell{ -0.40 \\ -1.50 }  & \cellcolor[HTML]{fee9df}\makecell{ 0.45 \\ -3.19 } & \cellcolor[HTML]{ffebe2}\makecell{ 0.47 \\ -2.73 } & \cellcolor[HTML]{fedfd0}\makecell{ 0.69 \\ -5.79 } & \cellcolor[HTML]{ffece3}\makecell{ 0.45 \\ -2.48 } & x & \cellcolor[HTML]{cee0f2}\makecell{ -0.49 \\  -2.53 } & \cellcolor[HTML]{fdcebb}\makecell{ 0.80 \\ -8.34 } & \cellcolor[HTML]{fed9c9}\makecell{ 0.74 \\ -6.76 } & \cellcolor[HTML]{fdd4c2}\makecell{ 0.76 \\ -7.46 }  \\ 
\arrayrulecolor{black}\cdashline{2-21}[0.5pt/5pt]
\multicolumn{1}{l|}{L$_{2500 \AA}$}  & \cellcolor[HTML]{ffede5}\makecell{ 0.40 \\ -2.25 } & \cellcolor[HTML]{ffede5}\makecell{ 0.39 \\ -2.16 } & \cellcolor[HTML]{fee8dd}\makecell{ 0.42 \\ -3.63 } & \cellcolor[HTML]{fee8dd}\makecell{ 0.42 \\ -3.63 } & - & x & \cellcolor[HTML]{fff0e8}\makecell{ 0.30 \\ -1.54 } & x & \cellcolor[HTML]{084082}\textcolor{white}{\makecell{ -0.88 \\ -11.36}} & \cellcolor[HTML]{ffefe8}\makecell{ 0.36 \\ -1.62 } & x & \cellcolor[HTML]{ffefe8}\makecell{ 0.36 \\ -1.61 } & x & x & x & \cellcolor[HTML]{ffeee7}\makecell{ 0.39 \\ -1.85 } & \cellcolor[HTML]{ffeee7}\makecell{ 0.39 \\ -1.87 } & \cellcolor[HTML]{ffeee6}\makecell{ 0.40 \\ -2.03 }   \\ 
\arrayrulecolor{black}\cdashline{2-21}[0.5pt/5pt]
\multicolumn{1}{l|}{H$\beta$}  & x & x & x & x & x & - & x & x & x & x & x & \cellcolor[HTML]{ffeee6}\makecell{ 0.43 \\ -1.98 } & \cellcolor[HTML]{ffede5}\makecell{ -0.45 \\ -2.10 } & \cellcolor[HTML]{ffece3}\makecell{ 0.46 \\ -2.49 } & \cellcolor[HTML]{fff0e8}\makecell{ 0.35 \\ -1.55 }  & x & x & x  \\ 
\arrayrulecolor{black}\cdashline{2-21}[0.5pt/5pt]
\multicolumn{1}{l|}{$\Gamma$}  & x & x & x & x & \cellcolor[HTML]{fff0e8}\makecell{ 0.30 \\ -1.54 } & x & - & x & \cellcolor[HTML]{dae8f6}\makecell{ -0.40 \\ -1.75 } & x & x & x & x & x & x & x & \cellcolor[HTML]{fff0e8}\makecell{ 0.36 \\ -1.56 } & \cellcolor[HTML]{fff0e8}\makecell{ 0.32 \\ -1.46 } \\ 
\arrayrulecolor{black}\cdashline{2-21}[0.5pt/5pt]
\multicolumn{1}{l|}{$\lambda _{\rm Edd}$}  & \cellcolor[HTML]{d8e7f5}\makecell{ -0.42 \\ -1.90 } & x & x & x  & x & x & x & - & x & \cellcolor[HTML]{ddeaf7}\makecell{ -0.01 \\ -1.58 } & x & x& x & \cellcolor[HTML]{ffece3}\makecell{ 0.48 \\ -2.59 } & \cellcolor[HTML]{fee9df}\makecell{ 0.52 \\ -3.22 } & x & x & x \\ 
\arrayrulecolor{black}\cdashline{2-21}[0.5pt/5pt]
\multicolumn{1}{l|}{$\alpha _{ox}$}  & \cellcolor[HTML]{e2edf8}\makecell{ -0.34 \\ -1.32 } & \cellcolor[HTML]{deebf7}\makecell{  -0.37 \\ -1.53 } & \cellcolor[HTML]{dfebf7}\makecell{ -0.40 \\ -1.50 } & \cellcolor[HTML]{dfebf7}\makecell{ -0.40 \\ -1.50 }& \cellcolor[HTML]{084082}\textcolor{white}{\makecell{ -0.88 \\ -11.36 }} & x & \cellcolor[HTML]{dae8f6}\makecell{  -0.40 \\ -1.75 } & x & - & x & x & \cellcolor[HTML]{e1edf8}\makecell{  -0.35 \\ -1.34 } & x & x & x & \cellcolor[HTML]{e1edf8}\makecell{  -0.35 \\ -1.37 } & \cellcolor[HTML]{dfebf7}\makecell{  -0.37 \\ -1.50 } & \cellcolor[HTML]{dfebf7}\makecell{ -0.35 \\ -1.52 } \\
\arrayrulecolor{black}\cdashline{2-21}[0.5pt/5pt]
\multicolumn{1}{l|}{$\Delta \alpha _{ox}$}  & x & \cellcolor[HTML]{ffefe8}\makecell{ 0.21 \\ -1.65 } & \cellcolor[HTML]{fee9df}\makecell{ 0.45 \\ -3.19 } & \cellcolor[HTML]{fee9df}\makecell{ 0.45 \\ -3.19 } &  \cellcolor[HTML]{ffefe8}\makecell{ 0.36 \\ -1.62 }  & x & x & \cellcolor[HTML]{ddeaf7}\makecell{ -0.01 \\ -1.58 } & x & - & x & x & x & x & x & x & x & x \\ 
\arrayrulecolor{black}\cdashline{2-21}[0.5pt/5pt]
\multicolumn{1}{l|}{\cellcolor[HTML]{DAF7A6}$\xi$}  & \cellcolor[HTML]{fee9df}\makecell{ 0.52 \\ -3.15 } & \cellcolor[HTML]{fee8dd}\makecell{ 0.54 \\ -3.64 } & \cellcolor[HTML]{ffebe2}\makecell{ 0.47 \\ -2.73 } & \cellcolor[HTML]{ffebe2}\makecell{ 0.47 \\ -2.73 }  & x & x & x & x & x & x & - & \cellcolor[HTML]{fee8dd}\makecell{ 0.54 \\ -3.61 } & \cellcolor[HTML]{ffeee6}\makecell{ 0.29 \\ -2.09 } & \cellcolor[HTML]{c1d9ed}\makecell{ -0.56 \\  -3.22 } & \cellcolor[HTML]{91c3de}\makecell{ -0.67 \\  -4.92 } & \cellcolor[HTML]{ffebe2}\makecell{ 0.49 \\ -2.73 } & x & x  \\ 
\arrayrulecolor{black}\cdashline{2-21}[0.5pt/5pt]
\multicolumn{1}{l|}{\cellcolor[HTML]{DAF7A6}N$_{\mathrm{H}}$}  & \cellcolor[HTML]{fdcebb}\makecell{ 0.79 \\ -8.31 } & \cellcolor[HTML]{fdc9b3}\makecell{ 0.82 \\ -9.15 } & \cellcolor[HTML]{fedfd0}\makecell{ 0.69 \\ -5.78 } & \cellcolor[HTML]{fedfd0}\makecell{ 0.69 \\ -5.79 } & \cellcolor[HTML]{ffefe8}\makecell{ 0.36 \\ -1.61 } & \cellcolor[HTML]{ffeee6}\makecell{ 0.43 \\ -1.98 } & x & x & \cellcolor[HTML]{e1edf8}\makecell{ -0.35 \\ -1.34 } & x & \cellcolor[HTML]{fee8dd}\makecell{ 0.54 \\ -3.61 } & - & \cellcolor[HTML]{fee1d3}\makecell{ 0.67 \\ -5.52} & \cellcolor[HTML]{94c4df}\makecell{-0.67 \\ -4.85} & \cellcolor[HTML]{08306b}\textcolor{white}{\makecell{ -0.89 \\ -12.10 }} & \cellcolor[HTML]{fcb398}\textcolor{white}{\makecell{ 0.88 \\ -12.28 }} & \cellcolor[HTML]{fee2d5}\makecell{ 0.66 \\ -5.16 } & \cellcolor[HTML]{fee6da}\makecell{ 0.59 \\ -4.04 }  \\ 
\arrayrulecolor{black}\cdashline{2-21}[0.5pt/5pt]
\multicolumn{1}{l|}{\cellcolor[HTML]{DAF7A6}v$_{out}$}  & \cellcolor[HTML]{fee7dc}\makecell{ 0.57 \\ -3.78 } & \cellcolor[HTML]{fee7dc}\makecell{ 0.49 \\ -3.70 } & \cellcolor[HTML]{ffece3}\makecell{ 0.45 \\ -2.48 } & \cellcolor[HTML]{ffece3}\makecell{ 0.45 \\ -2.48 } & x & \cellcolor[HTML]{ffede5}\makecell{ -0.45 \\ -2.10 } & x & x & x & x & \cellcolor[HTML]{ffeee6}\makecell{ 0.29 \\  -2.09 } & \cellcolor[HTML]{fee1d3}\makecell{ 0.67 \\ -5.52 } & - & \cellcolor[HTML]{d6e5f4}\makecell{ -0.44 \\  -2.07 } & \cellcolor[HTML]{b4d3e9}\makecell{ -0.60 \\ -3.73 } & \cellcolor[HTML]{feeae0}\makecell{ 0.54 \\ -3.10 } & \cellcolor[HTML]{fdcdb9}\makecell{ 0.80 \\ -8.42 } & \cellcolor[HTML]{fee0d2}\makecell{ 0.69 \\ -5.71 } \\
\arrayrulecolor{black}\cdashline{2-21}[0.5pt/5pt] 
\multicolumn{1}{l|}{\cellcolor[HTML]{DAF7A6}r$_{wind}$}  & x & \cellcolor[HTML]{e2edf8}\makecell{ -0.34 \\ -1.32 }  & x & x & x & \cellcolor[HTML]{ffece3}\makecell{ 0.46 \\ -2.49 } & x & \cellcolor[HTML]{ffece3}\makecell{  0.48 \\  -2.59 } & x & x  & \cellcolor[HTML]{c1d9ed}\makecell{ -0.56 \\  -3.22 } & \cellcolor[HTML]{94c4df}\makecell{ -0.67 \\ -4.85 } & \cellcolor[HTML]{d6e5f4}\makecell{ -0.44 \\  -2.07} & - & \cellcolor[HTML]{fdd1be}\makecell{0.78 \\ -7.92 } & \cellcolor[HTML]{dce9f6}\makecell{ -0.40 \\ -1.69} & x & x \\ 
\arrayrulecolor{black}\cdashline{2-21}[0.5pt/5pt]  
\multicolumn{1}{l|}{\cellcolor[HTML]{DAF7A6}r$_{wind} / r_s$} & \cellcolor[HTML]{5ba3d0}\makecell{-0.76 \\ -6.62} & \cellcolor[HTML]{65aad4}\makecell{ -0.74 \\  -6.25 } & \cellcolor[HTML]{cee0f2}\makecell{ -0.49 \\  -2.53} &\cellcolor[HTML]{cee0f2}\makecell{ -0.49 \\  -2.53} & x & \cellcolor[HTML]{fff0e8}\makecell{ 0.35 \\  -1.55}  & x & \cellcolor[HTML]{fee9df}\makecell{ 0.52 \\ -3.22 } & x & x & \cellcolor[HTML]{91c3de}\makecell{ -0.67 \\  -4.92 } & \cellcolor[HTML]{08306b}\textcolor{white}{\makecell{ -0.89 \\ -12.10 }} & \cellcolor[HTML]{b4d3e9}\makecell{  -0.60 \\  -3.73 } & \cellcolor[HTML]{fdd1be}\makecell{ 0.78 \\  -7.92 } & - & \cellcolor[HTML]{3b8bc2}\makecell{ -0.80 \\ -7.89 } & \cellcolor[HTML]{d7e6f5}\makecell{ -0.43 \\ -1.98} & x \\ 
\arrayrulecolor{black}\cdashline{2-21}[0.5pt/5pt]
\multicolumn{1}{l|}{\cellcolor[HTML]{DAF7A6}$\dot{M} _{wind}$} & \cellcolor[HTML]{fc9c7d}\textcolor{white}{\makecell{ 0.93 \\ -15.43 }} & \cellcolor[HTML]{fcb095}\textcolor{white}{\makecell{ 0.89 \\ -12.63 }} & \cellcolor[HTML]{fdcebb}\makecell{ 0.80 \\ -8.34 } & \cellcolor[HTML]{fdcebb}\makecell{ 0.80 \\ -8.34  } & \cellcolor[HTML]{ffeee7}\makecell{ 0.39 \\ -1.85 } & x & x & x & \cellcolor[HTML]{e1edf8}\makecell{  -0.35 \\ -1.37 }  & x & \cellcolor[HTML]{ffebe2}\makecell{ 0.49 \\ -2.73 } & \cellcolor[HTML]{fcb398}\textcolor{white}{\makecell{ 0.88 \\ -12.28 }} & \cellcolor[HTML]{feeae0}\makecell{ 0.54 \\ -3.10 } & \cellcolor[HTML]{dce9f6}\makecell{ -0.40 \\ -1.69} & \cellcolor[HTML]{3b8bc2}\makecell{ -0.80 \\ -7.89 } & - & \cellcolor[HTML]{fee1d4}\makecell{ 0.66 \\ -5.34 } & \cellcolor[HTML]{fee3d6}\makecell{ 0.63 \\ -5.01 } \\ 
\arrayrulecolor{black}\cdashline{2-21}[0.5pt/5pt]
\multicolumn{1}{l|}{\cellcolor[HTML]{DAF7A6}$\dot{E} _{k}^{wind}$}  & \cellcolor[HTML]{fee1d3}\makecell{ 0.68 \\ -5.52 } & \cellcolor[HTML]{fedecf}\makecell{ 0.70 \\ -6.09} & \cellcolor[HTML]{fed9c9}\makecell{ 0.74 \\ -6.76 } & \cellcolor[HTML]{fed9c9}\makecell{ 0.74 \\ -6.76 } & \cellcolor[HTML]{ffeee7}\makecell{ 0.39 \\ -1.87 } & x & \cellcolor[HTML]{fff0e8}\makecell{ 0.36 \\ -1.56 } & x & \cellcolor[HTML]{dfebf7}\makecell{  -0.37 \\ -1.50 } & x & x & \cellcolor[HTML]{fee2d5}\makecell{ 0.66 \\ -5.16 } & \cellcolor[HTML]{fdcdb9}\makecell{ 0.80 \\ -8.42 } & x & \cellcolor[HTML]{d7e6f5}\makecell{ -0.43 \\ -1.98 } & \cellcolor[HTML]{fee1d4}\makecell{ 0.66 \\ -5.34 } & - & \cellcolor[HTML]{f75c41}\textcolor{white}{\makecell{ 0.97 \\ -23.99 }} \\ 
\arrayrulecolor{black}\cdashline{2-21}[0.5pt/5pt]
\multicolumn{1}{l|}{\cellcolor[HTML]{DAF7A6}$\dot{P}_{wind}$}  & \cellcolor[HTML]{fee2d5}\makecell{ 0.65 \\ -5.10 } & \cellcolor[HTML]{fedfd0}\makecell{ 0.70 \\ -5.86 } & \cellcolor[HTML]{fdd4c2}\makecell{ 0.76 \\ -7.46  } & \cellcolor[HTML]{fdd4c2}\makecell{ 0.76 \\ -7.46 } & \cellcolor[HTML]{ffeee6}\makecell{ 0.40 \\ -2.03 } & x & \cellcolor[HTML]{fff0e8}\makecell{ 0.32 \\ -1.46 } & x & \cellcolor[HTML]{deebf7}\makecell{ -0.35 \\ -1.52 } & x & x & \cellcolor[HTML]{fee6da}\makecell{ 0.59 \\ -4.04 } & \cellcolor[HTML]{fee0d2}\makecell{ 0.69 \\ -5.71 } & x & x & \cellcolor[HTML]{fee3d6}\makecell{ 0.63 \\ -5.01 } & \cellcolor[HTML]{f75c41}\textcolor{white}{\makecell{ 0.97 \\ -23.99 }} & - \\
\arrayrulecolor{black}\cdashline{2-21}[0.5pt/5pt]
\hline
\hline
\end{tabular}\par
\smallskip
\begin{FlushLeft}
\textbf{Notes.}
For each pair, the Spearman correlation coefficient is reported (top), along with the logarithm of the \textit{NHP} (bottom). If the latter is larger than –1.30, we consider the correlation with low significance and an “X” is shown. Cells displaying positive correlations are shaded in red, with darker shades indicating more significant correlations. Conversely, cells representing negative correlations are shaded in blue, following the same shading scheme. Additionally, cells highlighted in green within the first row and column indicate correlations involving at least one UFO parameter.
\end{FlushLeft}
\end{table}
\end{landscape}

%
   \bibliographystyle{aa} 
   \bibliography{biblio} 

\vspace{1cm}

\noindent \textbf{Affiliations:} \\
\noindent
\textit{
     $^{1}$Université Grenoble Alpes, CNRS, IPAG, 38000 Grenoble, France \\
     $^{2}$Dipartimento di Matematica e Fisica, Università degli Studi Roma Tre, Via della Vasca Navale 84, 00146 Roma, Italy \\
     $^{3}$Department of Physics and Astronomy (DIFA), University of Bologna, Via Gobetti, 93/2, I-40129 Bologna, Italy\\
     $^{4}$INAF-Osservatorio di Astrofisica e Scienza dello Spazio di Bologna, Via Gobetti, 93/3, I-40129 Bologna, Italy\\
     $^{5}$European Space Agency (ESA), European Space Astronomy Centre (ESAC), E-28691 Villanueva de la Cañada, Madrid, Spain\\
     $^{6}$Department of Physics and Astronomy, College of Charleston, Charleston, SC, 29424, USA\\
     $^{7}$Physics Department, The Technion, 32000 Haifa, Israel \\
     $^{8}$Dipartimento di Fisica, Universitá di Trieste, Sezione di Astronomia, Via G.B. Tiepolo 11, I-34131 Trieste, Italy \\
     $^{9}$SRON Netherlands Institute for Space Research, Niels Bohrweg 4, 2333 CA Leiden, The Netherlands  \\
     $^{10}$INAF – Osservatorio Astrofisico di Arcetri, Largo Enrico Fermi 5, 50125 Firenze, Italy  \\
     $^{11}$Departament de Física, EEBE, Universitat Politècnica de Catalunya, Av. Eduard Maristany 16, 08019 Barcelona, Spain  \\
     $^{12}$INAF – Osservatorio Astronomico di Trieste, Via G. B. Tiepolo 11, 34143 Trieste, Italy  \\
     $^{13}$Department of Astrophysical Sciences, Princeton University, 4 Ivy Lane, Princeton, NJ 08544-1001, USA  \\
     $^{14}$Centro de Astrobiologia (CAB), CSIC-INTA, Camino Bajo del Castillo s/n, Campus ESAC, 28692, Villanueva de la Ca\~nada, Madrid, Spain \\
     $^{15}$ESA – European Space Research and Technology Centre (ESTEC), Keplerlaan 1, 2201 AZ Noordwijk, The Netherlands \\
     $^{16}$Department of Physics \& Astronomy, University of Leicester, Leicester LE1 7RH, UK \\
     $^{17}$Astronomical Institute Anton Pannekoek, University of Amsterdam , Science Park 904, NL-1098 XH The Netherlands \\
     $^{18}$Department of Physics, Institute for Astrophysics and Computational Sciences, The Catholic University of America, Washington, DC 20064, USA \\
     $^{19}$Space Telescope Science Institute, 3700 San Martin Drive, Baltimore, MD 21218, USA  \\
     $^{20}$Instituto de Astronomía, Universidad Nacional Autónoma de México, Circuito Exterior, Ciudad Universitaria, Ciudad de México 04510, México\\
     $^{21}$INAF – Istituto di Astrofisica e Planetologia Spaziali, Via Fosso del Cavaliere, 00133 Roma, Italy  \\
     $^{22}$INAF – Osservatorio Astronomico di Roma, Via Frascati 33, 00078 Monte Porzio Catone (Roma), Italy \\
     $^{23}$Dipartimento di Fisica e Astronomia, Università di Firenze, via G. Sansone 1, 50019 Sesto Fiorentino, Firenze, Italy \\
     $^{24}$Department of Astronomy, The Ohio State University, 140 West 18th Avenue, Columbus, OH 43210, USA  \\
     $^{25}$Center for Cosmology and Astroparticle Physics, 191 West Woodruff Avenue, Columbus, OH 43210, USA  \\
     $^{26}$Space Telescope Science Institute, 3700 San Martin Drive, Baltimore, MD 21218, USA \\
     $^{27}$Space Science Data Center – ASI, Via del Politecnico s.n.c., 00133 Roma, Italy  \\
     $^{28}$INAF – Osservatorio Astronomico di Brera, Via Bianchi 46, 23807 Merate (LC), Italy  \\
     $^{29}$Department of Physics, University of Rome ‘Tor Vergata’, Via della Ricerca Scientifica 1, 00133 Rome, Italy  \\
     $^{30}$INFN - Rome Tor Vergata, Via della Ricerca Scientifica 1, 00133 Rome, Italy \\
     $^{31}$Department of Astronomy, University of Maryland, College Park, MD 20742, USA \\
     $^{32}$NASA/Goddard Space Flight Center, Code 662, Greenbelt, MD 20771, USA \\
     $^{33}$Max-Planck-Institut für extraterrestrische Physik, Giessenbachstra{\ss}e 1, 85748 Garching bei München, Germany \\
     $^{34}$Cavendish Laboratory, University of Cambridge, 19 J.J. Thomson Avenue, Cambridge CB3 0HE, UK \\
     $^{35}$Kavli Institute for Cosmology, University of Cambridge, Madingley Road, Cambridge CB3 0HA, UK }

\begin{appendix} 
\onecolumn
\section{AGN and UFO parameters}\nopagebreak
\begin{landscape}
\footnotesize\setlength{\tabcolsep}{2.6pt}
\renewcommand{\arraystretch}{1.5}
\begin{longtable}{llllllllllll}
\caption{AGN global parameters.}\\
\label{tab:1}
& Source & z & log(M$_{\textrm{BH}}$) & log(L$_X$) & FWHM H$\beta$ & $\Gamma$ & N$_{\mathrm{H}}^{\mathrm{neutral}}$ & $\alpha_{\mathrm{ox}}$ & $\Delta \alpha_{ox}$ & log(L$_{bol}$) & $\lambda_{\rm Edd}$ \\ 
\hline
\endfirsthead
\caption{continued.}\\
\hline\hline
& Source & z & log(M$_{\textrm{BH}}$) & log(L$_X$) & FWHM H$\beta$ & $\Gamma$ & N$_{\mathrm{H}}^{\mathrm{neutral}}$ & $\alpha_{\mathrm{ox}}$ & $\Delta \alpha_{ox}$ & log(L$_{bol}$) & $\lambda_{\rm Edd}$ \\ 
\hline
\endhead
\endfoot
\endlastfoot
\hline
S23$^{1}$:
&&&&&& UFO sub-sample && \\
&LBQS 1338-0038 & 0.23745 & 7.77 $\pm$ 0.03 & 44.52 $\pm$ 0.01 & 2369 $\pm$ 5 $^{a}$ & 1.71 $\pm$ 0.03 && -1.33 & 0.10 & 45.87 $\pm$ 0.01 & 0.99 \\ 
&PG 0804+761 * & 0.100 & 8.31 $\pm$ 0.04 & 44.45 $\pm$ 0.01 & 3053 $\pm$ 38 $^{e}$ & 1.97 $^{+ 0.05}_{-0.06}$ && -1.75 & -0.22 & 45.78 $\pm$ 0.03 & 0.24 \\ 
&PG 0947+396 * & 0.20553 & 8.68 $^{+0.08}_{- 0.10}$ & 44.21 $\pm$ 0.01 & 4340 $\pm$ 651 $^{c}$ & 1.72 $^{+0.05}_{- 0.06}$ && -1.37 & 0.11 & 45.50 $\pm$ 0.01 & 0.05 \\ 
&PG 1114+445 * & 0.144 & 8.59 $\pm$ 0.09 & 43.98 $\pm$ 0.01 & 4825 $\pm$ 723 $^{c}$ & 1.81 $^{+ 0.22}_{- 0.10}$ && -1.43 & 0.02 & 45.24 $\pm$ 0.01 & 0.04 \\ 
&PG 1202+281 & 0.16501 & 8.61 $^{+0.08}_{- 0.10}$ & 44.40 $\pm$ 0.01 & 4950 $\pm$ 742 $^{c}$ & 1.64 $\pm$ 0.01 & & -1.22 & 0.19 & 45.73 $\pm$ 0.02 & 0.10 \\ 
&2MASX J105144+3539 & 0.159 & 8.40 $\pm$ 0.30 $\dagger$ & 43.69 $\pm$ 0.01 & & 1.56 $\pm$ 0.02 &  67.60 $\pm$ 10.00 $^{(a)}$ & x & x & 44.93 $\pm$ 0.02 & 0.03 \\ 
&2MASX J165315+2349 & 0.103 & 6.98 $\pm$ 0.30 $\dagger$ & 43.79 $\pm$ 0.01 &  & 1.60$^{+0.09}_{-0.08}$ & 1778.20 $\pm$ 20.00 $^{(a)}$ & x & x & 45.04 $\pm$ 0.02 & 0.90 \\ 
&&&&&& no-UFO sub-sample && \\
&HB 89-1257+286 * & 0.091 & 7.46 $\pm$ 0.30 $\dagger$ & 43.55 $\pm$ 0.01 & 2676 $\pm$ 10 $^{a}$ & 1.81$^{+ 0.07}_{- 0.08}$ && -1.30 & -0.02 & 44.78 $\pm$ 0.02 & 0.18  \\ 
&HB 89-1529+050 & 0.21817 & 8.75 $\pm$ 0.30 $\dagger$ & 44.22 $\pm$ 0.01 & 4261 $\pm$ 15 $^{a}$ & 1.75 $\pm$ 0.05 && -1.20 & 0.18 & 45.52 $\pm$ 0.01 & 0.05 \\
&PG 0052+251 & 0.15445 & 8.57 $\pm$ 0.09 $^{g}$ & 44.61 $\pm$ 0.01 & 4165 $\pm$ 381 $^{g}$ & 1.75 $\pm$ 0.02 & & -1.22 & 0.24 & 45.97 $\pm$ 0.03 & 0.20 \\ 
&PG 0953+414 & 0.23410 & 8.27 $^{+0.06}_{-0.09}$ & 44.60 $\pm$ 0.01 & 3071 $\pm$ 27 $^{b}$ & 1.90 $\pm$ 0.01 && -1.49 & 0.14 & 45.96 $\pm$ 0.02 & 0.39 \\ 
&PG 1216+069 & 0.33130 & 9.20 $^{+0.09}_{- 0.11}$ & 44.77 $\pm$ 0.01 & 5190 $\pm$ 1020 $^{f}$ & 1.74 $\pm$ 0.01 & & -1.46 & 0.15 & 46.17 $\pm$ 0.01 & 0.08 \\  
&PG 1307+085 & 0.15384 & 8.64 $\pm$ 0.12 $^{g}$ & 44.31 $\pm$ 0.01 & 5059 $\pm$ 133 $^{e}$  & 1.82 $\pm$ 0.02 && -1.62 & -0.09  & 45.62 $\pm$ 0.01 & 0.08 \\ 
&PG 1352+183 & 0.15147 & 8.42 $^{+0.08}_{- 0.10}$ & 43.89 $\pm$ 0.01 & 4210 $\pm$ 631 $^{c}$ & 1.92 $\pm$ 0.03 && -1.34 & 0.04 & 45.14 $\pm$ 0.03 & 0.04 \\ 
&PG 1402+261 * & 0.164 & 7.94 $^{+0.08}_{- 0.10}$ & 44.03 $\pm$ 0.01 & 2100 $\pm$ 315 $^{c}$ & 2.08$^{+0.03}_{-0.34}$ && -1.37 & 0.14 & 45.30 $\pm$ 0.01 & 0.18 \\ 
&PG 1416-129 & 0.129 & 9.05 $^{+0.08}_{- 0.10}$ & 44.17 $\pm$ 0.01 & 3766 $\pm$ 377 $^{f}$ & 1.60 $\pm$ 0.02 && -1.02 & 0.24 & 45.46 $\pm$ 0.02 & 0.02 \\ 
&PG 1425+267 & 0.36361 & 9.22 $\pm$ 0.30 $\dagger$ & 44.82 $\pm$ 0.01 & 9875 $\pm$ 1481 $^{c}$ & 1.71 $\pm$ 0.11 & & -1.50 & 0.12 & 46.24 $\pm$ 0.01 & 0.08 \\ 
&PG 1435-067 & 0.12900 & 8.37 $^{+0.08}_{- 0.10}$ & 43.68 $\pm$ 0.01 & 3180 $^{d}$ & 1.68 $^{+0.06}_{-0.07}$ && -1.39 & 0.06 & 44.92 $\pm$ 0.02 & 0.03 \\ 
&PG 1626+554 & 0.13170 & 8.50 $^{+0.08}_{- 0.10}$ & 44.08 $\pm$ 0.01 & 4390 $\pm$ 658 $^{c}$ & 1.89 $\pm$ 0.03 && -1.43 & 0.07 & 45.35 $\pm$ 0.01 & 0.06 \\ 
&SDSS J144414+0633 & 0.20768 & 8.10 $\pm$ 0.30 $\dagger$ & 44.47 $\pm$ 0.01 & 4083 $\pm$ 18 $^{a}$ & 1.72 $\pm$ 0.03 && -1.30 & 0.14 & 45.81 $\pm$ 0.03 & 0.40 \\ 
&WISE J053756.30-024513.1 + & 0.11 & 7.73 $\pm$ 0.30 $\dagger$ & 43.69 $\pm$ 0.01 & 2938 $\pm$ 10 $^{a}$ & 1.65 $\pm$ 0.06 && x & x & 44.93 $\pm$ 0.02 & 0.12 \\ 
&2MASX J0220-0728 & 0.21343 & 8.42 $\pm$ 0.30 $\dagger$ & 44.21 $\pm$ 0.01 & 6891 $\pm$ 20 $^{a}$ & 1.67$^{+0.05}_{-0.06}$ && -1.02 & 0.17 & 45.51 $\pm$ 0.03 & 0.10 \\ 
\hdashline
T10$^{2}$:
&&&&&& UFO sub-sample && \\
&Ark 120 + & 0.0327 & 8.07 $^{+0.05}_{-0.06}$ $^{(p)}$ & 43.96 $\pm$ 0.01 & 5850 $\pm$ 480 $^{(p)}$ & 1.86 $\pm$ 0.01 & 0.1 $^{(d)}$ & x & x & 45.27 $\pm$ 0.01 & 0.13 \\
&IC 4329A & 0.0161 & 8.11 $^{+0.33}_{-0.10}$ $^{(q)}$ & 43.70 $\pm$ 0.05 & 6431 $\pm$ 532 $^{(q)}$ & 1.65 $\pm$ 0.01 & & -1.57 & -0.20 & 44.93 $\pm$ 0.05 & 0.14 \\
&MCG-5-23-16* & 0.00849 & 7.45 $\pm$ 0.20 $^{(h)}$ & 43.10 $\pm$ 0.01 & & 1.53 $\pm$ 0.01 & 185 $\pm$ 30 $^{(b)}$ & x & x & 44.31 $\pm$ 0.01 & 0.06 \\
&Mrk 79* & 0.0222 & 7.61 $^{+0.11}_{-0.14}$ $^{(p)}$ & 43.40 $\pm$ 0.03 & 4219 $\pm$ 262 $^{(p)}$ & 1.55 $\pm$ 0.02 & 0.6 $^{(d)}$ & -1.38 & -0.20 & 44.62 $\pm$ 0.03 & 0.08 \\
&Mrk 205* & 0.07085 & 8.40 $\pm$ 0.01 $^{(i)}$ & 43.80 $\pm$ 0.02 & & 1.71 $\pm$ 0.04 & 8$^{+2}_{-1}$ $^{(c)}$ & x & x & 45.05 $\pm$ 0.02 & 0.04 \\ 
&Mrk 290* & 0.03022 & 7.28 $\pm$ 0.06 $^{(p)}$ & 43.20 $\pm$ 0.03 & 4270 $\pm$ 157 $^{(p)}$ & 1.61 $\pm$ 0.05 & 1.8 $^{(a)}$ & -1.31 & -0.09  & 44.41 $\pm$ 0.03 & 0.11 \\
&Mrk 509* & 0.0344 & 8.04 $\pm$ 0.04 $^{(p)}$ & 43.97 $\pm$ 0.10 & 2715 $\pm$ 101 $^{(p)}$ & 1.67 $\pm$ 0.60 & 0.5  $^{(d)}$ & -1.05 & 0.13 & 45.23 $\pm$ 0.11  & 0.13 \\
&Mrk 766* & 0.0129 & 6.82 $^{+0.05}_{-0.06}$ $^{(p)}$ & 42.73 $\pm$ 0.05  & 2423 $\pm$ 59 $^{(p)}$ & 1.94 $\pm$ 0.02 & 0.4 $^{+0.2}_{-0.4}$ $^{(d)}$ & -1.03 & 0.04  & 43.93 $\pm$ 0.05  & 0.49 \\ 
&Mrk 841* & 0.0364 & 8.52 $^{+0.08}_{-0.05}$ $^{(t)}$ & 43.50 $\pm$ 0.01 & 4958 $\pm$ 87 $^{(t)}$ & 1.61 $\pm$ 0.04 & 1.3 $^{(d)}$ & -1.26 & -0.02 & 44.72 $\pm$ 0.01 & 0.01 \\ 
&NGC 4051* & 0.00234 & 5.89 $^{+0.08}_{-0.15}$ $^{(p)}$ & 41.65 $\pm$ 0.04 & 1499 $\pm$ 35 $^{(p)}$ & 1.96 $\pm$ 0.02 & 410 $\pm$ 60 $^{(b)}$ & x & x & 42.84 $\pm$ 0.04  & 0.07 \\ 
&NGC 4151* & 0.0332 & 7.36 $\pm$ 0.03 $^{(p)}$ & 42.34 $\pm$ 0.02 & 4393 $\pm$ 110 $^{(p)}$ & 1.58 $\pm$ 0.03 & 1100 $\pm$ 40 $^{(b)}$ & x & x & 43.53 $\pm$ 0.02  & 0.01 \\ 
&NGC 4507 & 0.0118 & 6.40 $\pm$ 0.06 $^{(u)}$ & 43.10 $\pm$ 0.02 & & 1.50 $\pm$ 0.06 & 3800 $\pm$ 500 $^{(b)}$ & x & x & 44.31 $\pm$ 0.02 & 0.64 \\ 
&NGC 7582* & 0.00525 & 7.67 $^{+0.09}_{-0.08}$ $^{(u)}$ & 41.60 $\pm$ 0.07 & & 0.80 $\pm$ 0.06 & 1900 $\pm$ 240  $^{(b)}$ & x & x & 42.79 $\pm$ 0.07 & 0.001 \\ 
&PG 1211+143* & 0.0809 & 7.61 $_{-0.13}^{+0.10}$ $^{(r)}$ & 43.70 $\pm$ 0.03 & 1832 $\pm$ 81 $^{(s)}$ & 2.8 $\pm$ 0.12 & 1170 $\pm$ 150 $^{(b)}$ & x & x & 44.94 $\pm$ 0.03 & 0.17 \\ 
&1H 0419-577* & 0.104 & 8.60 $\pm$ 0.01 $^{(u)}$ & 44.30 $\pm$ 0.04 & 4700 $\pm$ 400 $^{(o)}$ & 1.21 $\pm$ 0.05  & 0.3 $^{+0.1}_{-0.2}$  $^{(d)}$ & -1.46  & -0.02  & 45.61 $\pm$ 0.05 & 0.08 \\ 
&&&&&&  Fe-K sub-sample  \\
&ESO 323-G077 & 0.015 & 7.40 $\pm$ 0.01 $^{(u)}$ & 43.00 $\pm$ 0.01 & & 2.36 $\pm$ 0.09 & 1200 $\pm$ 100 $^{(b)}$ & x & x & 44.20 $\pm$ 0.01 & 0.05 \\ 
&Mrk 279 & 0.03  & 7.45 $^{+0.10}_{-0.13}$ $^{(p)}$ & 42.78 $\pm$ 0.01 & 3385 $\pm$ 349 $^{(p)}$ & 1.71 $\pm$ 0.02  & 2.2 $^{+6.9}_{-6.0}$ $^{(d)}$ & -1.25 & 0.05 & 43.98 $\pm$ 0.01 & 0.03 \\ 
&NGC 3516* & 0.009 & 7.40 $^{+0.04}_{-0.06}$ $^{(p)}$ & 43.10 $\pm$ 0.02 & 5295 $\pm$ 100 $^{(p)}$ & 1.99 $\pm$ 0.03 & 478 $\pm$ 40  $^{(b)}$ & x & x & 44.27 $\pm$ 0.02 & 0.06 \\
&NGC 3783* & 0.01 & 7.34 $^{+0.05}_{-0.06}$ $^{(p)}$ & 43.10 $\pm$ 0.02 & 4728 $\pm$ 676 $^{(p)}$ & 1.77 $\pm$ 0.03 & 0.1 $^{(d)}$ & -1.45 & -0.27 & 44.35 $\pm$ 0.02 & 0.08 \\ 
&&&&&&  no-UFO sub-sample &  & & \\
&Ark 564 & 0.025 & 7.99 $^{+0.09}_{-0.08}$ $^{(l)}$ & 43.50 $\pm$ 0.04 & & 2.01 $\pm$ 0.07 & & -1.05 & 0.15 & 44.67 $\pm$ 0.04 & 0.04 \\ 
&ESO 198-G024 & 0.046 & 8.09 $^{+0.12}_{-0.34}$ $^{(o)}$ & 43.70 $\pm$ 0.01 & & 1.62 $\pm$ 0.02 & 0.4 $^{(d)}$ & -1.16 & 0.12 & 44.94 $\pm$ 0.01 & 0.06 \\ 
&ESO 511-G030 & 0.022 & 7.23 $\pm$ 0.05 $^{(m)}$ & 43.50 $\pm$ 0.01 & & 1.71 $\pm$ 0.01 & 0.1 $^{(d)}$ & -1.19 & 0.07 & 44.75 $\pm$ 0.01 & 0.26 \\ 
&Fairall 9 & 0.047 & 8.32$^{+0.08}_{-0.12}$ $^{(p)}$ & 44.60 $\pm$ 0.01 & 6901 $\pm$ 707 $^{(p)}$ & 1.64 $\pm$ 0.03 & & -1.22 & 0.09 & 45.99 $\pm$ 0.01 & 0.37 \\ 
&H 557-385 & 0.034 & 7.14 $^{+0.52}_{-0.16}$ $^{(o)}$ & 44.04 $\pm$ 0.08 & & 1.17 $\pm$ 0.07 & 2100 $\pm$ 200 $^{(b)}$ & x & x & 45.31 $\pm$ 0.09 & 1.19 \\ 
&IRAS 5078+1626 + & 0.018 & 7.65 $\pm$ 0.17 $^{(o)}$ & 41.60 $\pm$ 0.02 & 4549 $\pm$ 608 $^{(o)}$ & 1.55 $\pm$ 0.02 & & x & x & 42.82 $\pm$ 0.02 & 0.001 \\
&MCG-6-30-15* & 0.008 & 6.29 $^{+0.16}_{-0.24}$ $^{(p)}$ & 42.80 $\pm$ 0.05 & 1422 $\pm$ 416 $^{(p)}$ & 2.21 $\pm$ 0.03 & 370 $\pm$ 30 $^{(b)}$ & x & x & 43.97 $\pm$ 0.05 & 0.37  \\
&MCG+8-11-11 & 0.02 & 7.45 $\pm$ 0.05 $^{(p)}$ & 43.80 $\pm$ 0.01 & 4475 $\pm$ 274 $^{(p)}$ & 1.57 $\pm$ 0.01 & & -1.07 & 0.19 & 45.03 $\pm$ 0.01 & 0.30 \\ 
&Mrk 110 & 0.035 & 7.29 $\pm$ 0.10 $^{(p)}$ & 43.77 $\pm$ 0.06 & 3333 $\pm$ 21 $^{(p)}$ & 1.68 $\pm$ 0.02 & 0.2 $^{(d)}$ & -1.20 & 0.12 & 45.02 $\pm$ 0.07 & 0.42 \\ 
&Mrk 335 & 0.026 & 7.23$^{+0.04}_{-0.04}$ $^{(p)}$ & 43.30 $\pm$ 0.01 & 1418 $\pm$ 118 $^{(p)}$ & 1.99 $\pm$ 0.02 & & -1.44 & -0.11 & 44.56 $\pm$ 0.01 & 0.17  \\ 
&Mrk 590 & 0.026 & 7.56 $^{+0.06}_{-0.07}$ $^{(p)}$ & 42.90 $\pm$ 0.05 & 5403 $\pm$ 130 $^{(p)}$ & 1.52 $\pm$ 0.03 & 0.2 $^{(d)}$ & -1.13 & 0.02 & 44.20 $\pm$ 0.05 & 0.04 \\
&Mrk 704 & 0.029 & 7.57 $^{+0.07}_{-0.10}$ $^{(p)}$ & 43.30 $\pm$ 0.02 & 3406 $\pm$ 275 $^{(p)}$ & 1.70 $\pm$ 0.10 & 500 $\pm$ 120 $^{(b)}$ & x & x & 44.48 $\pm$ 0.02 & 0.07 \\
&NGC 526A & 0.019 & 8.11 $\pm$ 0.65 $^{(n)}$ & 43.30 $\pm$ 0.02 & & 1.43 $\pm$ 0.02 & 65.00 $^{+4.7}_{-3.2}$ $^{(a)}$ & x & x & 44.52 $\pm$ 0.02 & 0.03 \\ 
&NGC 2110 & 0.008 & 8.43 $^{+0.56}_{-0.34}$ $^{(t)}$  & 42.60 $\pm$ 0.04 & & 1.59 $\pm$  0.04 & 221 $\pm$ 11 $^{(d)}$ & x & x & 43.79 $\pm$ 0.04 & 0.002 \\ 
&NGC 3227* & 0.004 & 6.69 $^{+0.08}_{-0.10}$ $^{(p)}$ & 42.10 $\pm$ 0.08 & 3837 $\pm$ 94 $^{(p)}$ & 1.53 $\pm$ 0.03 & 7.2 $\pm$ 1.7 $^{(h)}$ & x & x & 43.32 $\pm$ 0.08 & 0.03 \\ 
&NGC 4593 & 0.009 & 6.91 $^{+0.07}_{-0.07}$ $^{(p)}$ & 42.20 $\pm$ 0.10 & 3597 $\pm$ 72 $^{(p)}$ & 1.68 $\pm$ 0.01 & & -1.15 & -0.06 & 43.41 $\pm$ 0.10 & 0.03 \\ 
&NGC 5506* & 0.006 & 6.71 $^{+0.19}_{-0.10}$ $^{(u)}$ & 43.19 $\pm$ 0.01 & & 1.84 $\pm$ 0.04 & 182 $\pm$ 40 $^{(b)}$ & x & x & 44.40 $\pm$ 0.01 & 0.39 \\
&NGC 5548* & 0.017 & 7.68 $\pm$ 0.02 $^{(p)}$ & 42.90 $\pm$ 0.01 & 7736 $\pm$ 76 $^{(p)}$ & 1.63 $\pm$ 0.01 & 0.7 $^{(d)}$ & -1.22 & -0.01 & 44.20 $\pm$ 0.01 & 0.03 \\ 
&NGC 7172* & 0.009 & 8.65 $\pm$ 0.10 $^{(u)}$ & 42.90 $\pm$ 0.02 & & 1.71 $\pm$ 0.05 & 760 $\pm$ 20 $^{(c)}$ & x & x & 44.11 $\pm$ 0.02 & 0.002 \\ 
&NGC 7213 & 0.006 & 6.83 $^{+0.84}_{-0.33}$ $^{(t)}$ & 43.10 $\pm$ 0.05 & & 1.71 $\pm$ 0.02 & 0.8$^{+0.4}_{-0.3}$ $^{(d)}$ & -1.07 & -0.10 & 44.34 $\pm$ 0.05 & 0.26  \\ 
&NGC 7314 & 0.005 & 5.94 $\pm$ 0.23 $^{(u)}$ & 42.40 $\pm$ 0.07 & & 1.97 $\pm$ 0.04 & 73 $\pm$ 2 $^{(c)}$ & x & x & 43.62 $\pm$ 0.07 & 0.38 \\ 
&NGC 7469* & 0.016 & 6.99 $\pm$ 0.05 $^{(p)}$ & 42.30 $\pm$ 0.01 & 3148 $\pm$ 346 $^{(p)}$ & 1.74 $\pm$ 0.02 & 0.9 $^{(d)}$ & -1.26 & -0.05 & 43.45 $\pm$ 0.01 & 0.02 \\
&TON S180* & 0.06 & 6.08 $^{+0.11}_{-0.15}$ $^{(u)}$ & 43.70 $\pm$ 0.04 & & 2.14 $\pm$ 0.05 & & -1.72 & -0.28 & 44.93 $\pm$ 0.04 & 5.52  \\
\hdashline
C21$^{3}$: \\
&&&&&&  UFO sub-sample && \\
&APM 08279+5255* & 3.910 & 10.00 $\pm$ 0.10 & 46.25 $\pm$ 0.02 & 6990 $\pm$ 460 $^{(v)}$ & 1.75 $\pm$ 0.03 & 626 $^{+18}_{-22}$$^{(f)}$ & x & x & 48.21 $\pm$ 0.03 & 1.30 \\ 
&HS 0810+2554* & 1.510 & 8.60 $\pm$ 0.20 & 45.66 $\pm$ 0.04 & 4400 $\pm$ 60 $^{(v)}$ & 2.28 $\pm$ 0.04 & & -1.68$^{(3)}$ & -0.27 & 44.88 $\pm$ 0.04 & 0.02 \\
&HS 1700+6416 & 2.735 & 10.20 $\pm$ 0.20 & 45.36 $^{+0.09}_{-0.04}$ & & 1.83 $\pm$ 0.16 & & -1.79$^{(3)}$ & 0.14 & 46.95 $^{+0.13}_{-0.05}$ & 0.04  \\
&MG J0414+0534 + & 2.640 & 9.00 $\pm$ 0.20 & 44.50 $\pm$ 0.01 & & 1.77 $\pm$ 0.07 & & x & x & 45.85 $\pm$ 0.01 & 0.06 \\
&PG 1115+080* & 1.720 & 8.80 $\pm$ 0.20 & 44.17 $\pm$ 0.01 & & 1.83 $\pm$ 0.04 & 27 $\pm$ 5$^{(g)}$ & x & x & 45.46 $\pm$ 0.02 & 0.04\\
&Q 2237+0305 & 1.695 & 9.10 $\pm$ 0.40 & 44.14 $\pm$ 0.02 & 3800 $\pm$ 1400 $^{(v)}$ & 1.62 $\pm$ 0.24 & 142 $^{+155}_{-134}$ $^{(e)}$ & x & x & 45.42 $\pm$ 0.02 & 0.02\\
&SDSS J0921+2854* & 1.410 & 8.90 $\pm$ 0.20 & 45.81 $\pm$ 0.01 && 1.69 $\pm$ 0.03 & & -0.96$^{(3)}$ & 0.45 & 46.75 $\pm$ 0.02 & 0.56 \\
&SDSS J1029+2623 & 2.197 & 8.80 $\pm$ 0.20 & 44.05 $\pm$ 0.03 & & 1.50 $\pm$ 0.04 & < 40 $^{(e)}$ & x & x & 45.33 $\pm$ 0.03 & 0.03 \\
&SDSS J1128+2402 & 1.608 & 8.70 $\pm$ 0.20 & 44.34 $^{+0.13}_{-0.10}$ & & 1.78 $\pm$ 0.10 & & -1.59$^{(3)}$ & 0.11 & 45.65 $^{+0-16}_{-0.12}$ & 0.07\\
&SDSS J1353+1138 & 1.627 & 9.40 $\pm$ 0.20 & 44.72 $\pm$ 0.06 & & 2.10 $\pm$ 0.08 & 61 $^{+30}_{26}$ $^{(e)}$ & x & x & 46.11 $\pm$ 0.08 & 0.04\\
&SDSS J1442+4055 & 2.593 & 9.70 $\pm$ 0.20 & 44.77 $\pm$ 0.02 & & 1.81 $\pm$ 0.09 & & -1.48$^{(3)}$ & 0.06 & 46.17 $\pm$ 0.03 & 0.02 \\
&SDSS J1529+1038 & 1.984 & 8.90 $\pm$ 0.20 & 44.06 $\pm$ 0.16 & & 1.93 $\pm$ 0.05 & & -1.67$^{(3)}$ & -0.16 & 45.33 $\pm$ 0.18 & 0.02 \\
&&&&&&  no-UFO sub-sample && \\
&SDSS J0904+1512 & 1.826 & 9.30 $\pm$ 0.20 & 44.25 $\pm$ 0.08 & & 1.97 $^{+0.14}_{-0.13}$ & 25 $^{+36}_{-25}$ $^{(e)}$ & x & x & 45.55 $\pm$ 0.09 & 0.01 \\
\end{longtable}
\par
\smallskip
\scriptsize{
\textbf{Notes.} \\
    \textit{Columns}: (1) Source name; (2) redshift; (3) black hole mass in unit of M$_{\odot}$; (4) intrinsic luminosity in the 2-10 keV energy band, in units of erg s$^{-1}$; (5) FWHM of the H$\beta$ broad emission line, in units of km s$^{-1}$; (6) X-ray photon indices: (7) neutral absorber column density in units of 10$^{20}$ cm$^{-2}$; (8) observed $\alpha_{\mathrm{ox}}$; (9) difference between the observed $\alpha_{\mathrm{ox}}$ and the value expected based on the UV luminosity of the AGN \citep[e.g.,][]{lusso10}{}; (10) bolometric luminosity calculated from the 2-10 keV luminosities using X-ray bolometric correction factor (see Sect.~\ref{bol_lum}) in units of erg s$^{-1}$; (11) Eddington ratio. \\
    * multi-epochs sources. We report the mean values of each parameter. In case of a multi-epochs AGNs hosting UFO, the means refer only to  observations with detected ultra-fast outflows. \\
    + AGN with intrinsic N$_{\mathrm{H}}$ below the adopted threshold (i.e., $5 \times 10^{20}$ cm$^{-2}$; see Sect.~~\ref{x_uv_slope}), but with N$_{\mathrm{H}}^{\rm Gal}$ above it. \\
    $^{1}$ redshift, black hole masses, 2-10 keV luminosities, X-ray photon indices and neutral column densities of the S23 sample have been taken from \citet{matzeu22} and references therein, unless stated otherwise. $\dagger$ BH masses were taken from \citet{perna20} and \citet{matzeu22}. However, as the uncertainties are not present in the papers, we opted to apply an uncertainty of 0.3 dex. This accounts for the dependence of the BH mass on the square of FWHM, alongside the uncertainties in luminosity (whether line or continuum) and zero point. \\
    $^{2}$ redshift, 2-10 keV luminosities and X-ray photon indices of the T10 sample have been taken from \citet{tombesi10} and references therein. As the X-ray luminosity uncertainties were not reported in the paper, we adopted those documented in the XMM archive for the 2-10 keV flux, and we applied them as a percentage to the luminosity. \\
    $^{3}$ redshift, black hole masses, 2-10 keV luminosities, X-ray photon indices and $\alpha_{\mathrm{ox}}$ of the C21 sample have been taken from \citet{chartas21} and references therein. All the reported luminosities are corrected for lensing magnification. \\
    $^{(a)}$ values from the automatic fit in the SDSS; $^{(b)}$ \citet{ricci17}; $^{(c)}$ \citet{runnoe13}; $^{(d)}$ \citet{sani10} ; $^{(e)}$ \citet{ho14}; $^{(f)}$ \citet{vestegaard02}; $^{(g)}$ \citet{peterson04}; $^{(h)}$ \citet{ponti12}; $^{(i)}$ \citet{kelly07}; $^{(l)}$ \citet{botte04}; $^{(m)}$ \citet{marin16}; $^{(n)}$ \citet{middleton08}; $^{(o)}$ \citet{wang07}; $^{(p)}$ \citet{bentz15}; $^{(q)}$ \citet{markowitz09}; $^{(r)}$ \citet{peterson04}; $^{(s)}$ \citet{Danehkar2018}; $^{(t)}$ \citet{woo02}; $^{(u)}$ \citet{mckernan10}; $^{(v)}$ \citet{assef11}.\\
    \textit{N$_{\mathrm{H}}$ references:} $^{(a)}$ \citet{matzeu22}, $^{(b)}$ \citet{tombesi10}, $^{(c)}$ \citet{laha19}, $^{(d)}$ \citet{winter12}, $^{(e)}$ \citet{chartas21}, $^{(f)}$ \citet{chartas09}, $^{(g)}$ \citet{chartas03}, $^{(h)}$ \citet{gondoin03}.\\
    The reported $\alpha_{\rm ox}$ and $\Delta \alpha_{\rm ox}$ values refer only to the unabsorbed sample defined in Sect.~\ref{x_uv_slope}.\\
    We note that all the reported uncertainties are taken from the literature and no systematic errors are taken into account.}
\end{landscape}

\begin{landscape}
\begin{ThreePartTable}
\footnotesize\setlength{\tabcolsep}{2.6pt}
\begin{TableNotes}
\item[\textbf{Notes.}]
    \item[\textit{Columns}: (1) Source name; (2) ionized absorber column density in units of cm$^{-2}$; (3) ionization parameter in unit of erg cm s$^{-1}$; (4) outflow velocity in units of c; (5) wind location in units of cm;]
    \item[(6) outflow mass rate in units of g s$^{-1}$: (7) mechanical power in units of erg s$^{-1}$; (8) outflow momentum rate; (9) momentum flux of the radiation field; (10) correction factor for relativistic effects,]
    \item[to be applied on the N$_{\mathrm{H}}$ and r$_\textrm{wind}$ values reported in the table.] 
    \item[$\dagger$ detected UFO in Obs ID 0841481001.]
    \item[$^+$ detected UFO in Obs ID 0102040401.]
    \item[* multi-epoch source with detected UFO in Obs ID: 0651330101 and 0651330301; the reported values are means.]
    \item[N$_{\mathrm{H}}$, $\xi$ and v$_{out}$ values of each sample are taken from the respective papers.]
    \item[We note that all the reported uncertainties are taken from the literature and no systematic errors are taken into account.]
\end{TableNotes}
\renewcommand{\arraystretch}{1.5}
\begin{longtable}{lllllllllllrrr}
\caption{UFO global parameters. }\\
\label{tab:2}
& Source & log(N$_{\mathrm{H}}$) & log($\xi$) & v$_{out}$ & log($\textrm{r}_\textrm{wind}$) & log($\dot{\textrm{M}}_{\textrm{wind}}$) & log($\dot{\textrm{E}}_k^{\textrm{wind}}$) & log($\dot{\textrm{P}}_{\textrm{wind}}$) & $\Psi$ \\ 
\hline
\endfirsthead
\caption{continued.}\\
\hline\hline
& Source & log(N$_{\mathrm{H}}$) & log($\xi$) & v$_{out}$ & log($\textrm{r}_\textrm{wind}$) & log($\dot{\textrm{M}}_{\textrm{wind}}$) & log($\dot{\textrm{E}}_k^{\textrm{wind}}$) & log($\dot{\textrm{P}}_{\textrm{wind}}$) & $\Psi$ \\ 
\hline
\endhead
\endfoot
\insertTableNotes
\endlastfoot
\hline
S23: \\
& LBQS 1338-0038 & 23.16 $^{+1.01}_{-0.22}$ & 5.01$^{+ 0.64}_{- 0.38}$ & 0.15 $\pm$ 0.02 & 17.40$_{-0.44}^{+1.19}$ & 26.97$_{-0.39}^{+0.64}$ & 45.98$_{-0.43}^{+0.66}$ & 36.63$_{-0.40}^{+0.65}$ & 1.36\\ 
& PG 0804+761 $^+$ & 23.93 $^{+0.33}_{-0.27}$ & > 4.49 & 0.13 $\pm$ 0.01 & > 17.07  & > 27.33 & > 46.21 & > 36.93 & 1.30\\
& PG 0947+396 $\dagger$ & > 23.68 & 5.38$^{+0.44}_{-1.27}$ & 0.31$^{+0.02}_{-0.04}$ & > 16.14 & 26.53$_{-1.27}^{+0.44}$ & 46.16$_{-1.28}^{+0.45}$ & 36.50$_{-1.27}^{+0.44}$ & 1.88\\
& PG 1114+445 * & 23.73 $^{+0.08}_{-0.04}$ & 4.65$^{+0.77}_{-0.44}$ & 0.07 $\pm$ 0.02 & 17.10$_{-0.11}^{+0.19}$ & 26.91$_{-0.14}^{+0.20}$ & 45.26$_{-0.31}^{+0.36}$  & 36.24$_{-0.22}^{+0.27}$ & 1.15 \\
&PG 1202+281 & > 23.84 & > 5.02 & 0.11 $\pm$ 0.01 & > 16.56  & > 26.67   & > 45.39   & > 36.18 & 1.24 \\ 
&2MASX J105144+3539 & 22.78$^{+0.67}_{-0.33}$ & 4.10$^{+0.35}_{-0.38}$ & 0.24 $\pm$ 0.01 & 17.75$_{-0.50}^{+0.75}$ & 27.13$_{-0.38}^{+0.35}$ & 46.53$_{-0.39}^{+0.35}$ & 36.98$_{-0.38}^{+0.35}$ & 1.62\\
&2MASX J165315+2349 & 23.76$^{+0.39}_{-0.15}$ & 4.76$^{+0.59}_{-0.42}$ &  0.11$^{+0.02}_{-0.01}$ & 16.22$_{-0.45}^{+0.71}$ & 26.24$_{-0.42}^{+0.59}$ & 44.98$_{-0.43}^{+0.62}$ & 35.76$_{-0.43}^{+0.60}$ & 1.25 \\
\hdashline
T10: \\
&Ark 120  & > 21.85 &	3.44 $^{+0.55}_{-0.18}$ & 0.31 $\pm$ 0.01 & > 19.68 & 28.24 $^{+0.55}_{-0.18}$ & 47.86 $^{+0.55}_{-0.18}$ & 38.20 $^{+0.55}_{-0.18}$ & 1.88 \\
&IC 4329A & > 21.87 &5.17 $^{+0.75}_{-0.77}$ & 0.10 $\pm$ 0.01 & > 17.60  & 25.69 $^{+0.75}_{-0.77}$ & 44.32 $^{+0.75}_{-0.77}$ & 35.15 $^{+0.75}_{-0.77}$ & 1.22 \\
&MCG-5-23-16 & 22.60 $\pm$ 0.13 & 4.33 $\pm$ 0.08 & 0.12 $\pm$ 0.01 & 17.08 $\pm$ 0.15 & 25.97 $\pm$ 0.08  & 44.75 $\pm$ 0.09 & 35.51 $\pm$ 0.09 & 1.26 \\
& Mrk 79  &	23.29 $\pm$ 0.27 & 4.17$^{+0.17}_{-0.21}$ & 0.09 $\pm$ 0.01 & 16.86 $^{+0.34}_{-0.32}$ & 26.33 $^{+0.17}_{-0.21}$ & 44.91 $^{+0.18}_{-0.22}$ & 35.77 $^{+0.18}_{-0.22}$ & 1.20\\
&Mrk 205  &	> 23.16 &	4.86$^{+0.14}_{-0.90}$ & 0.10 $\pm$ 0.01 & > 16.73 & 26.11 $^{+0.24}_{-0.90}$ & 44.77 $^{+0.25}_{-0.90}$ & 35.59 $^{+0.24}_{-0.90}$ & 1.22 \\
&Mrk 290  &	23.41 $\pm$ 0.40 &	4.35$^{+0.73}_{-0.13}$ & 0.14 $\pm$ 0.01 & 16.35 $^{+0.83}_{-0.42}$ & 26.14 $^{+0.73}_{-0.13}$ & 45.10 $^{+0.73}_{-0.14}$ & 35.77 $^{+0.73}_{-0.14}$ & 1.33 \\
& Mrk 509  & > 22.47 & 4.80$^{+0.07}_{-0.32}$ & 0.18 $\pm$ 0.03 & > 17.66 & 26.62$_{- 0.71}^{+ 0.63}$ & 45.79$_{-0.89}^{+ 0.87}$ & 36.36$_{- 0.27}^{+0.25}$ & 1.44 \\
& Mrk 766 &	22.68 $^{+0.16}_{0.15}$ & 3.87$^{+0.21}_{-0.10}$ & 0.09 $\pm$ 0.07 & 17.08$_{- 0.19}^{+ 0.27}$  & 25.91$_{- 0.40}^{+ 0.44}$  & 44.42$_{- 1.17}^{+ 1.19}$ & 35.32$^{+0.79}_{-0.81}$ & 1.19 \\
&Mrk 841 & > 22.00 &	4.50 $^{+0.65}_{-0.24}$ & 0.03 $\pm$ 0.01 & > 17.92  & 25.68 $^{+0.65}_{-0.25}$ & 43.40 $^{+0.67}_{-0.28}$ & 34.69 $^{+0.66}_{-0.26}$ & 1.07 \\
& NGC 4051 & 22.77 $^{+0.10}_{-0.11}$ & 3.35$^{+0.14}_{-0.06}$ &	0.13 $\pm$ 0.06 & 16.42$_{-0.13}^{+0.18}$  & 25.52$_{-0.23}^{+0.26}$  & 44.39$_{-0.65}^{+ 0.66}$  & 35.11$_{-0.43}^{+ 0.45}$ & 1.29 \\ 
& NGC 4151 & > 21.87 & 4.41 $^{+0.90}_{-0.08}$ & 0.11 $\pm$ 0.01 & > 16.95 & 25.07$_{-0.09}^{+0.92}$ & 43.78 $_{-0.12}^{+ 0.92}$ & 34.58$_{-0.10}^{+0.92}$ & 1.24 \\ 
& NGC 4507 & > 21.95 & 4.53 $\pm$ 1.15 & 0.20 $\pm$ 0.02 & > 17.53 & 26.00 $\pm$ 1.15 & 45.25 $\pm$ 1.16 & 35.78 $\pm$ 1.16 & 1.50\\
&NGC 7582  & 23.37$^{+0.39}_{-0.18}$ & 3.39 $^{+0.09}_{-0.15}$ & 0.29 $\pm$ 0.01 & 15.73 $^{+0.24}_{-0.41}$ & 25.78 $^{+0.11}_{-0.17}$ & 45.34 $^{+0.11}_{-0.17}$ & 35.71 $^{+0.11}_{-0.17}$ & 1.80 \\
& PG 1211+143 &	22.90 $^{+0.12}_{-0.06}$ &	2.87$^{+0.12}_{-0.10}$ & 0.15 $\pm$ 0.01 & 18.87 $^{+0.12}_{-0.17}$ & 28.17 $^{+0.12}_{-0.11}$ & 47.18 $^{+0.13}_{-0.11}$ & 37.83 $^{+0.13}_{-0.11}$ & 1.36 \\
& 1H 0419577 & 23.24 $\pm$ 0.36 & 3.85 $^{+0.18}_{-0.38}$ & 0.08 $\pm$ 0.01 & 18.22 $^{+0.53}_{-0.41}$ & 27.57 $^{+0.19}_{-0.38}$ & 45.98 $^{+0.20}_{-0.38}$ & 36.92 $^{+0.19}_{-0.39}$ & 1.17 \\
\hdashline
C21: \\
&APM 082795255 & 24.06 $\pm$ 0.05 & 4.49 $_{-0.10}^{+0.11}$ & 0.32$\pm$ 0.03 & 19.36 $\pm$ 0.12 & 30.38 $_{-0.11}^{+0.12}$ & 50.04 $_{-0.16}^{+0.17}$ & 40.36 $_{-0.13}^{+0.14}$ & 1.94 \\
& HS 08102554 & 24.16 $_{-0.11}^{+0.12}$ & 5.14 $_{-0.15}^{+0.25}$ & 0.43 $_{-0.05}^{+0.04}$ & 15.28 $_{-0.19}^{+0.28}$ & 26.53$_{-0.16}^{+0.26}$ & 46.45 $_{-0.22}^{-0.28}$ & 36.64 $_{-0.19}^{+0.27}$ & 2.51\\ 
& HS 17006416 & 24.68 $_{-0.08}^{+0.13}$ & 4.63 $_{-0.17}^{+0.29}$ & 0.38 $\pm$ 0.05 & 17.34 $_{-0.20}^{+0.34}$ & 29.05 $_{-0.19}^{+0.32}$ & 48.86 $_{-0.25}^{+0.36}$ & 39.11 $_{-0.21}^{+0.34}$ & 2.23 \\
& MG J0414053 & 24.05 $_{-0.09}^{+0.08}$ & 4.38 $_{-0.27}^{+0.29}$ & 0.28 $\pm$ 0.05 & 17.12 $_{-0.28}^{+0.30}$ & 28.07 $_{-0.28}^{+0.30}$ & 47.61 $_{-0.35}^{0.37}$ & 37.99 $_{-0.31}^{+0.33}$ & 1.78 \\
& PG 1115080 & 24.40 $_{-0.06}^{+0.05}$ & 5.35 $_{-0.12}^{+0.13}$ & 0.23 $\pm$ 0.02 & 15.41 $\pm$ 0.14 & 26.62 $_{-0.13}^{+0.14}$ & 46.00 $\pm$ 0.17  & 36.46 $_{-0.14}^{+0.15}$ & 1.60\\
& Q 22370305 & 24.33 $_{-0.18}^{+0.21}$ & 4.98 $_{-0.40}^{+0.43}$ & 0.18 $_{-0.09}^{+0.06}$ & 15.81 $_{-0.44}^{+0.48}$ & 26.85 $_{-0.45}^{+0.46}$ & 46.01 $_{-0.76}^{+0.62}$ & 36.58 $_{-0.59}^{+0.52}$ & 1.44 \\
& SDSS J09212854 & 24.30 $_{-0.07}^{+0.16}$ & 5.42 $_{-0.16}^{+0.32}$ & 0.47 $_{-0.01}^{+0.02}$ & 16.72$_{-0.18}^{+0.36}$ & 28.15 $_{-0.16}^{+0.32}$ & 48.15 $_{-0.16}^{+0.33}$ & 38.30 $_{-0.16}^{+0.32}$ & 2.77\\
& SDSS J10292623 & 24.17 $_{-0.24}^{+0.21}$ & 5.18 $_{-0.24}^{+0.33}$ & 0.58 $_{-0.02}^{+0.01}$ & 15.68 $_{-0.34}^{+0.39}$ & 27.06$_{-0.24}^{+0.33}$ & 47.24$_{-0.24}^{+0.33}$ & 37.30 $_{-0.24}^{+0.33}$ & 3.76 \\
& SDSS J11282402 & 24.32 $_{-0.21}^{+0.15}$ & 4.76$_{-0.45}^{+0.57}$ & 0.56 $_{-0.02}^{+0.01}$ & 16.27 $_{-0.51}^{+0.61}$ & 27.79$_{-0.47}^{+0.59}$ & 47.94$_{-0.47}^{+0.59}$ & 38.02 $_{-0.47}^{+0.59}$ & 2.03\\
& SDSS J13531138 & 24.33 $_{-0.18}^{+0.20}$ & 5.03 $_{-0.36}^{+0.38}$ & 0.34 $_{-0.09}^{+0.02}$ & 16.45$_{-0.41}^{+0.44}$ & 27.77$_{-0.38}^{+0.39}$ & 47.48 $_{-0.50}^{+0.40}$ & 37.78 $_{-0.43}^{+0.40}$ & 1.44 \\
& SDSS J14424055 & 24.48 $_{-0.18}^{+0.23}$ & 5.48 $_{-0.24}^{+0.39}$ & 0.47 $_{-0.05}^{+0.03}$ & 15.91$_{-0.30}^{+0.45}$ & 27.52$_{-0.25}^{+0.39}$ & 47.51 $_{-0.28}^{+0.40}$ & 37.67 $_{-0.26}^{+0.39}$ & 2.77 \\
& SDSS J15291038 & 24.42 $_{-0.23}^{+0.33}$ & 5.06 $_{-0.31}^{+0.59}$ & 0.25 $_{-0.08}^{+0.06}$ & 15.55 $_{-0.42}^{+0.70}$ & 26.82$_{-0.38}^{+0.63}$ & 46.27 $_{-0.55}^{+0.69}$ & 36.69 $_{-0.45}^{+0.65}$ & 1.67 \\
\end{longtable}
\end{ThreePartTable}
\end{landscape}

\section{$\alpha_{\mathrm{ox}}$ methodological figures}\nopagebreak

\begin{figure}[h!]
\centering
    \subfloat[][]
    {\includegraphics[width=.5\textwidth]{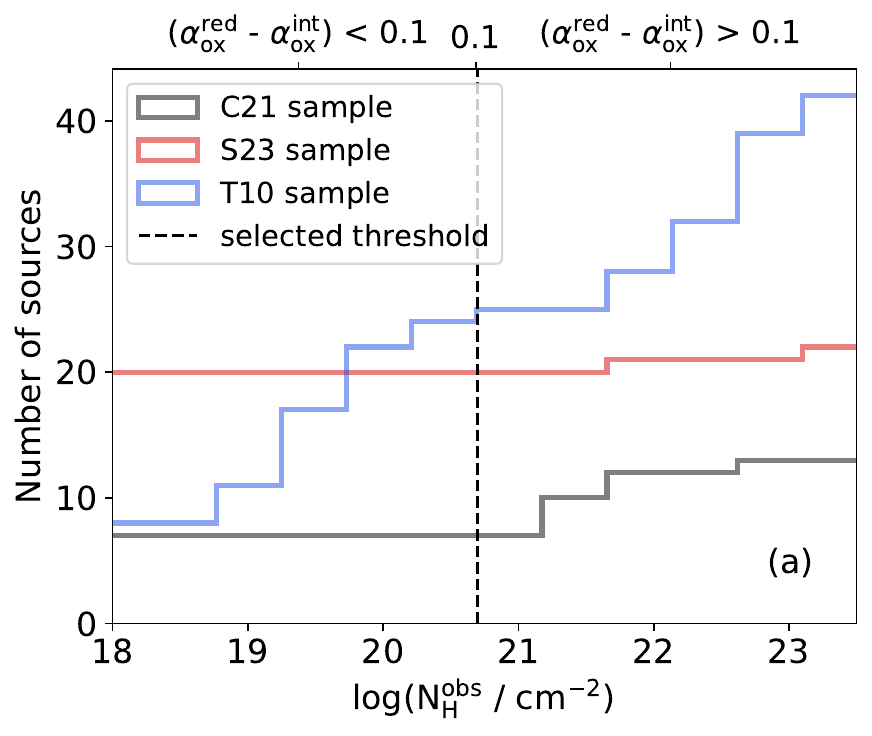}}
    \subfloat[][]
    {\includegraphics[width=.51\textwidth]{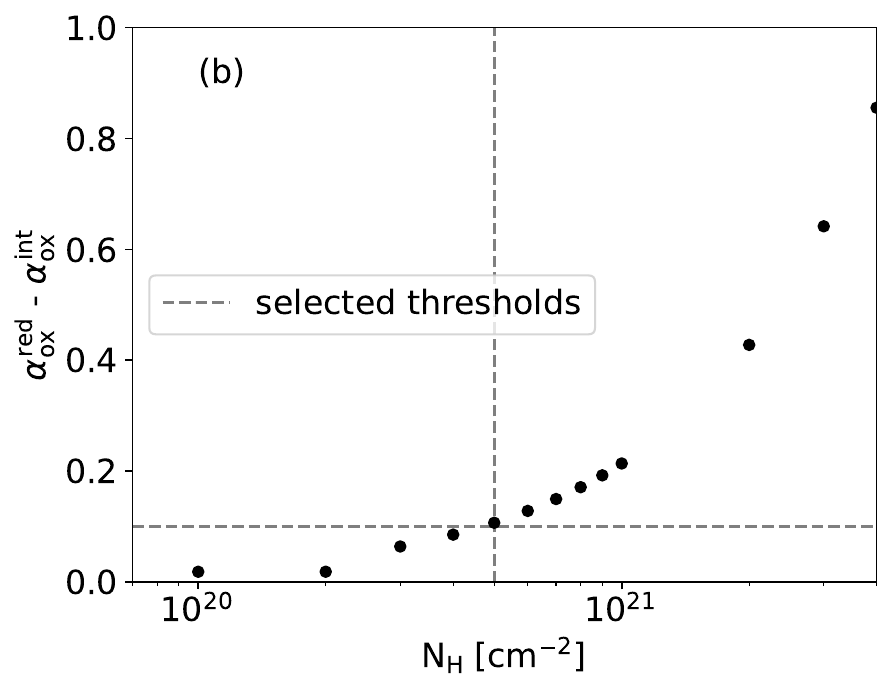}}
    \caption[]{{Unabsorbed sample.
    As reported in Sect.~\ref{x_uv_slope}, the UV and the X-ray intrinsic luminosities can be affected by the presence of gas and dust along the line of sight. To improve the reliability of our analysis, we thus identify a threshold in the neutral absorber column density (i.e., N$_{\mathrm{H}} < 5 \times 10^{20}$ cm$^{-2}$), which ensures that the AGN observed $\alpha_{\mathrm{ox}}$ (i.e., potentially affected by absorption and reddening) are closer to their intrinsic values.
    Panel a: cumulative curves of the neutral column density observed for the SUBWAYS (in red), T10 (in blue) and C21 (in black) samples. The dashed line indicates the adopted threshold above which the $\alpha_{\mathrm{ox}}$ is considered significantly affected by absorption and reddening. Panel b: expected deviation from $\alpha_{\mathrm{ox}}^{\mathrm{int}}$ = -1.5 as a function of the neutral absorber equivalent hydrogen column density, N$_{\mathrm{H}}$. 
    This plot highlights how absorption and reddening affect the observed $\alpha_{\mathrm{ox}}$. The dashed lines highlight the adopted thresholds, which helps define a maximum neutral N$_{\mathrm{H}}$ and the corresponding expected deviation from the intrinsic $\alpha_{\mathrm{ox}}$, above which the observed $\alpha_{\mathrm{ox}}$ cannot be considered reliable.
    }}
    \label{fig:14}
\end{figure} 

\section{$\alpha_{\mathrm{ox}}$ methodological figures}\nopagebreak

\begin{figure}[h!]
\centering
    \subfloat
    {\includegraphics[width=.5\textwidth]{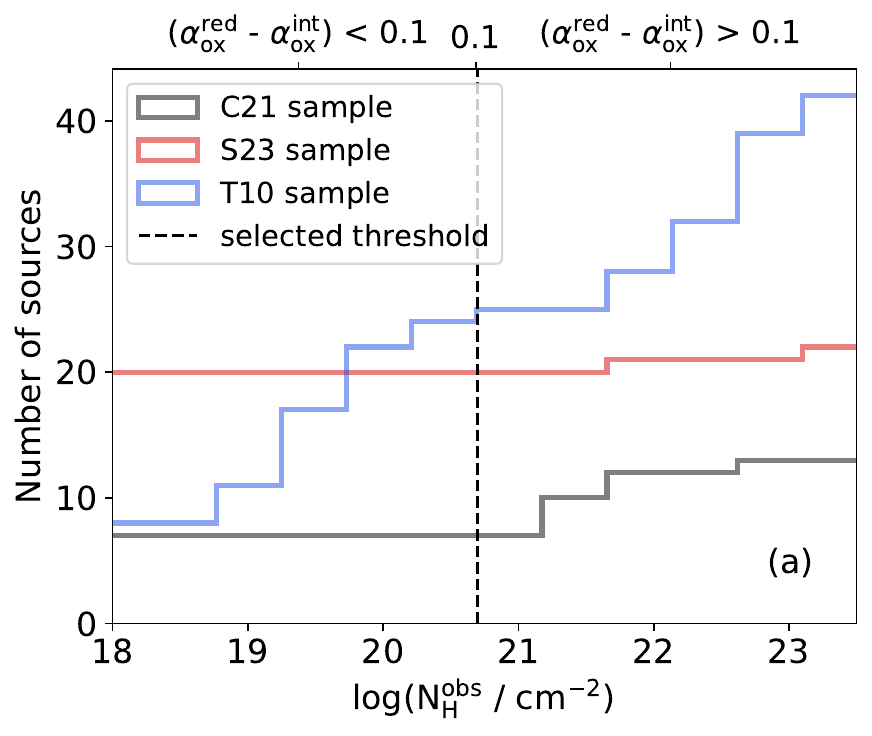}}
    \subfloat
    {\includegraphics[width=.51\textwidth]{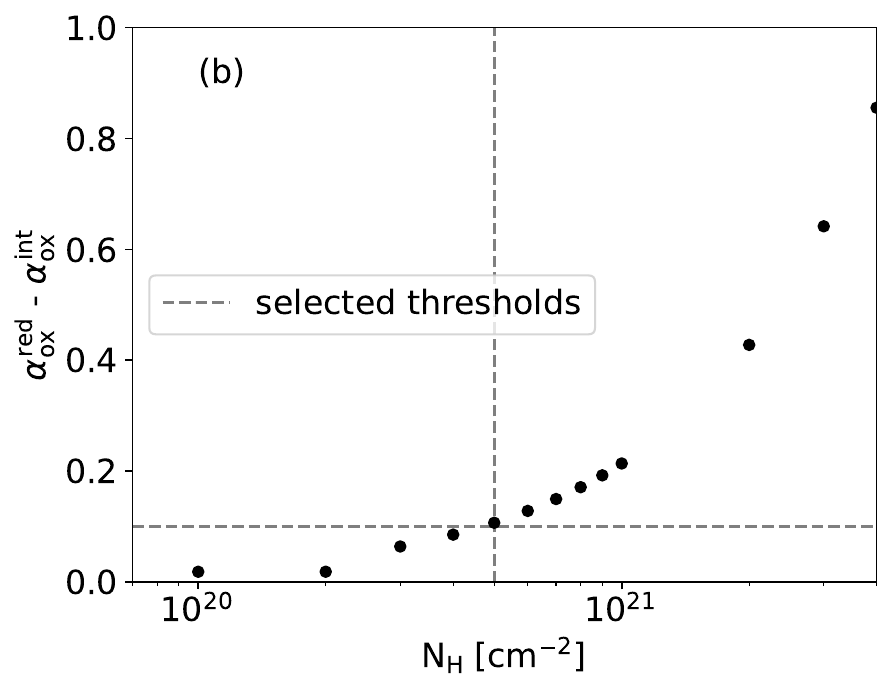}}
    \caption[]{{Unabsorbed sample.
    As reported in Sect.~\ref{x_uv_slope}, the UV and the X-ray intrinsic luminosities can be affected by the presence of gas and dust along the line of sight. To improve the reliability of our analysis, we thus identify a threshold in the neutral absorber column density (i.e., N$_{\mathrm{H}} < 5 \times 10^{20}$ cm$^{-2}$), which ensures that the AGN observed $\alpha_{\mathrm{ox}}$ (i.e., potentially affected by absorption and reddening) are closer to their intrinsic values.
    Panel a: Cumulative curves of the neutral column density observed for the SUBWAYS (in red), T10 (in blue) and C21 (in black) samples. The dashed line indicates the adopted threshold above which the $\alpha_{\mathrm{ox}}$ is considered significantly affected by absorption and reddening. Panel b: Expected deviation from $\alpha_{\mathrm{ox}}^{\mathrm{int}}$ = -1.5 as a function of the neutral absorber equivalent hydrogen column density, N$_{\mathrm{H}}$. 
    This plot highlights how absorption and reddening affect the observed $\alpha_{\mathrm{ox}}$. The dashed lines highlight the adopted thresholds, which help define a maximum neutral N$_{\mathrm{H}}$ and the corresponding expected deviation from the intrinsic $\alpha_{\mathrm{ox}}$, above which the observed $\alpha_{\mathrm{ox}}$ cannot be considered reliable.
    }}
    \label{fig:14}
\end{figure} 

\section{Parameter distributions}\nopagebreak

\begin{table}[h!]
\centering
\renewcommand{\arraystretch}{1.5}
\caption{Comparison between UFO and no-UFO sub-samples. }
\label{tab:7}
\begin{tabular}{lllllllllll}
\multicolumn{1}{c}{\makecell{UFO vs no-UFO \\ sub-samples}} & log(M$_{\textrm{BH}}$) & log(L$_{x}$) & log(L$_{bol}$) & $\lambda _{\rm Edd}$ &$\Gamma$ & FWHM H$\beta$ & $\alpha_{\mathrm{ox}}$ & $\Delta \alpha_{ox}$\\
\hline
\multicolumn{1}{c}{S23} & \multicolumn{1}{c}{{\bf x}} & \multicolumn{1}{c}{{\bf x}} & \multicolumn{1}{c}{{\bf x}} & \multicolumn{1}{c}{{\bf x}} & \multicolumn{1}{c}{{\bf x}} & \multicolumn{1}{c}{{\bf x}} & \multicolumn{1}{c}{{\bf x}} & \multicolumn{1}{c}{{\bf x}}  \\
\multicolumn{1}{c}{T10} & \multicolumn{1}{c}{{\bf x}} & \multicolumn{1}{c}{{\bf x}} & \multicolumn{1}{c}{{\bf x}} & \multicolumn{1}{c}{{\bf x}} & \multicolumn{1}{c}{{\bf x}} & \multicolumn{1}{c}{{\bf x}} & \multicolumn{1}{c}{{\bf x}} & \multicolumn{1}{c}{{\bf x}} \\
\multicolumn{1}{c}{C21} & \multicolumn{1}{c}{{\bf x}} & \multicolumn{1}{c}{{\bf x}} & \multicolumn{1}{c}{{\bf x}} & \multicolumn{1}{c}{{\bf x}} & \multicolumn{1}{c}{{\bf x}} & \multicolumn{1}{c}{/} & \multicolumn{1}{c}{/} & \multicolumn{1}{c}{/} \\
\multicolumn{1}{c}{T10+S23} & \multicolumn{1}{c}{{\bf x}} & \multicolumn{1}{c}{{\bf x}} & \multicolumn{1}{c}{{\bf x}} & \multicolumn{1}{c}{{\bf x}} & \multicolumn{1}{c}{{\bf x}} & \multicolumn{1}{c}{{\bf x}} & \multicolumn{1}{c}{{\bf x}} & \multicolumn{1}{c}{{\bf x}} \\
T10+S23+C21 & -2.00 & -1.70 & \multicolumn{1}{c}{{\bf x}} & \multicolumn{1}{c}{{\bf x}} & \multicolumn{1}{c}{{\bf x}} & \multicolumn{1}{c}{/} & \multicolumn{1}{c}{/} & \multicolumn{1}{c}{/} \\ \hline
\end{tabular}\par
\smallskip
\begin{FlushLeft}
\textbf{Notes.} Alongside Fig.~\ref{fig:22} and \ref{fig:4}, we obtained no substantial evidence indicating differences between AGNs hosting UFOs and those without. This suggests that all AGNs might be capable of hosting these outflows during their lifetime and their observability is linked to the wind duty cycle (see Sect.~\ref{ufo_vs_noufo}).
The lower the $\log\mathrm{NHP}$ values, the more statistically different the compared samples become. Meanwhile, we mark ``\textbf{x}'' when the difference between two samples is below the adopted significance threshold ($\log\mathrm{NHP}> -1.30$, i.e., the compared samples are statistically the same).
The C21 sample has limited data and a comparison between sub-samples could not be performed, we thus adopt ``\textbf{/}'' for the respective comparisons.
\end{FlushLeft}
\end{table}

\begin{figure}[h!]
\centering
    \includegraphics[width=.5\textwidth]{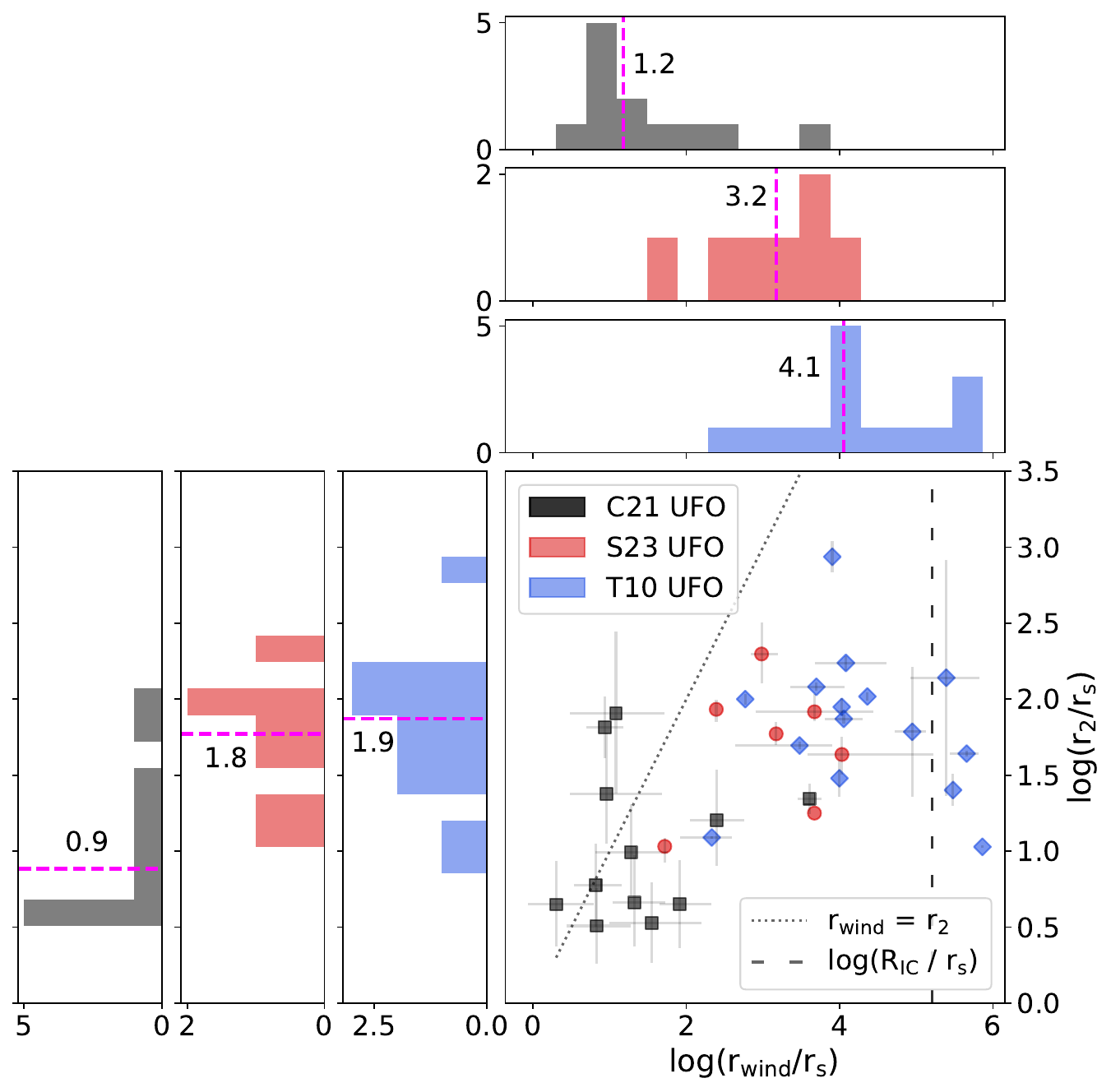}
    \caption[]
    {Estimates of the distance between the wind and the SMBH in terms of the Schwarzschild radius, r$_\mathrm{s}$. The S23 UFO sub-sample is shown in red circles, T10 in blue diamonds, and C21 in black squares. The black dotted line shows r$_\mathrm{wind}$ = r$_\mathrm{2}$. The black dashed line represents the ratio between R$_\mathrm{IC}$ and r$_\mathrm{s}$, see text for more details. The dashed magenta lines show the median value of each sub-sample. Our analysis considers only r${_\mathrm{wind}}$ as, within errors, it is always bigger than (or consistent with) r$_{\mathrm{2}}$ (see Sect.~\ref{sec:out_global}). }
    \label{fig:radii}
\end{figure}

\begin{figure}[h!]
\centering
    \subfloat{\includegraphics[width=.5\textwidth]{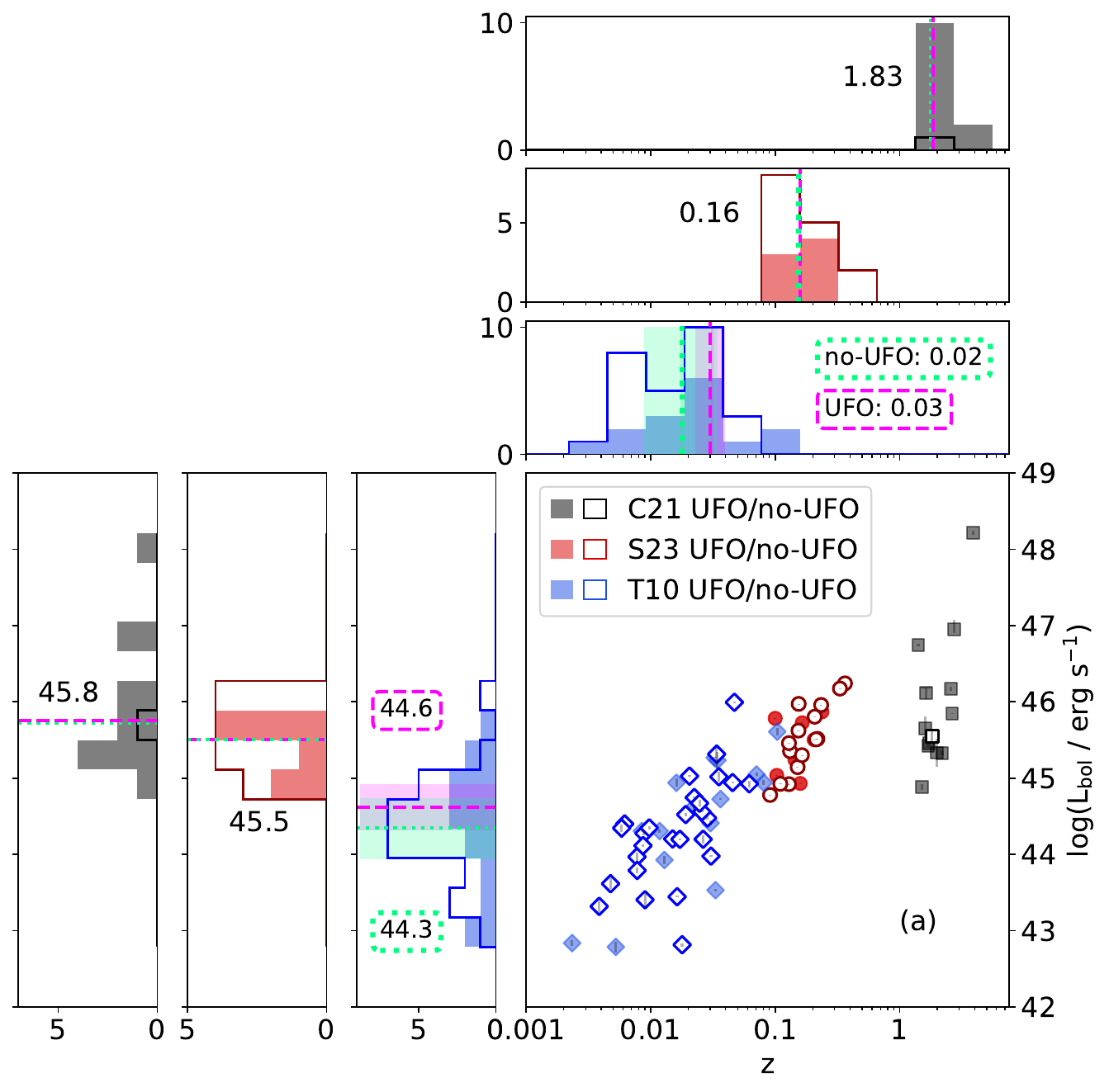}}
    \subfloat{\includegraphics[width=.5\textwidth]{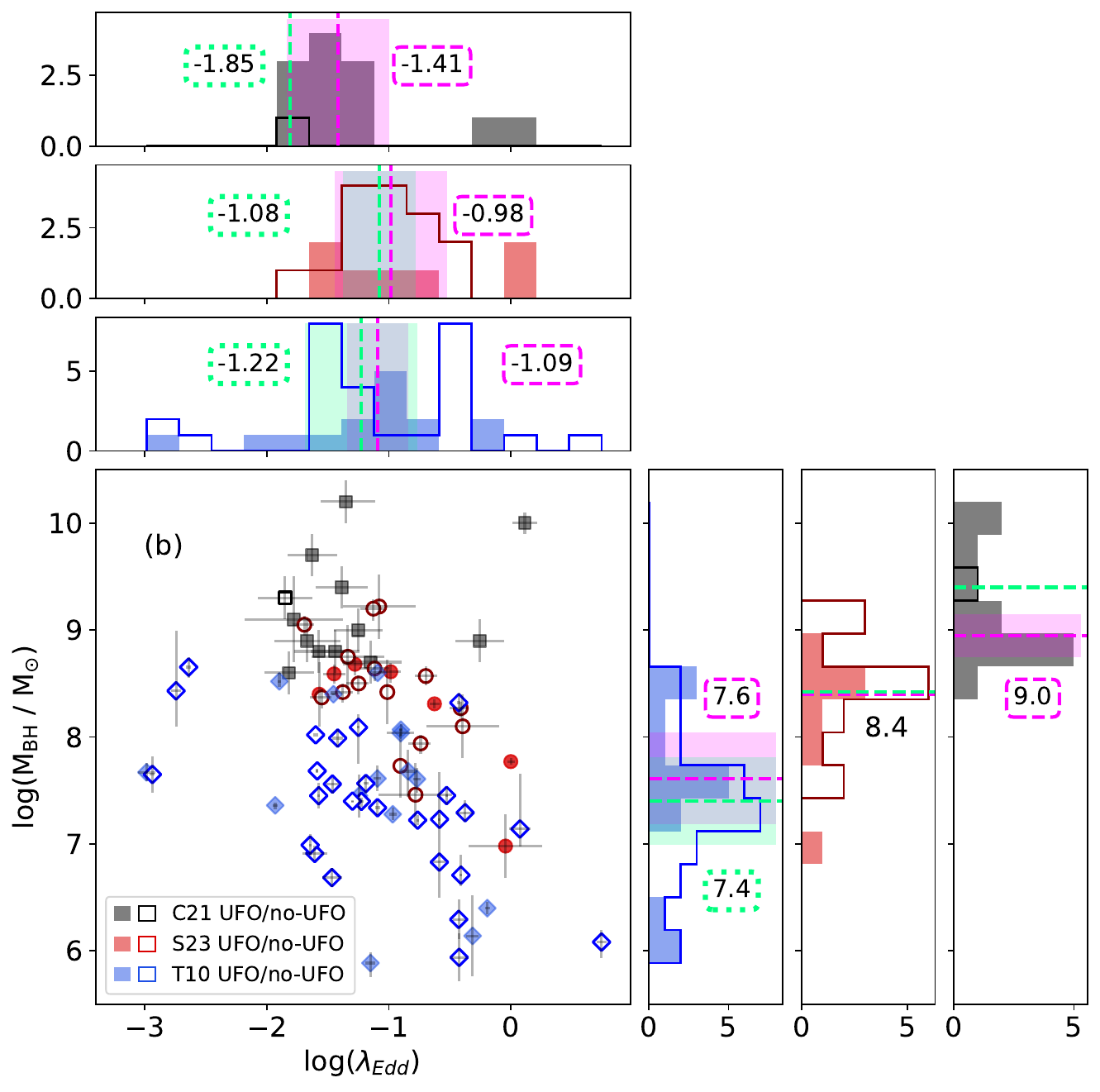}}
    \caption[]
    {{Samples comparison of AGN properties. Alongside Table \ref{tab:7} and Fig.~\ref{fig:4}, we obtained no substantial evidence indicating differences between AGNs hosting UFOs and those without. This suggests that all AGNs might be capable of hosting these outflows during their lifetime and their observability is linked to the wind duty cycle (see Sect.~\ref{ufo_vs_noufo}). Panel a: Bolometric luminosity versus redshift. Panel b: SMBH mass versus Eddington ratio. The S23 sample is shown in red circles, T10 sample in blue diamonds and C21 sample in black squares. For each distribution and scatter plot, the UFO sub-samples are represented as color filled histograms and dots, respectively. The dashed magenta and dotted green lines show the median value of each UFO and no-UFO sub-sample, respectively. We report both median values only when these are different and we include the corresponding 1$\sigma$ error-bars on the median.
    }}
    \label{fig:22}
\end{figure}

\begin{figure}[h!]
\centering
    \subfloat{\includegraphics[width=.5\textwidth]{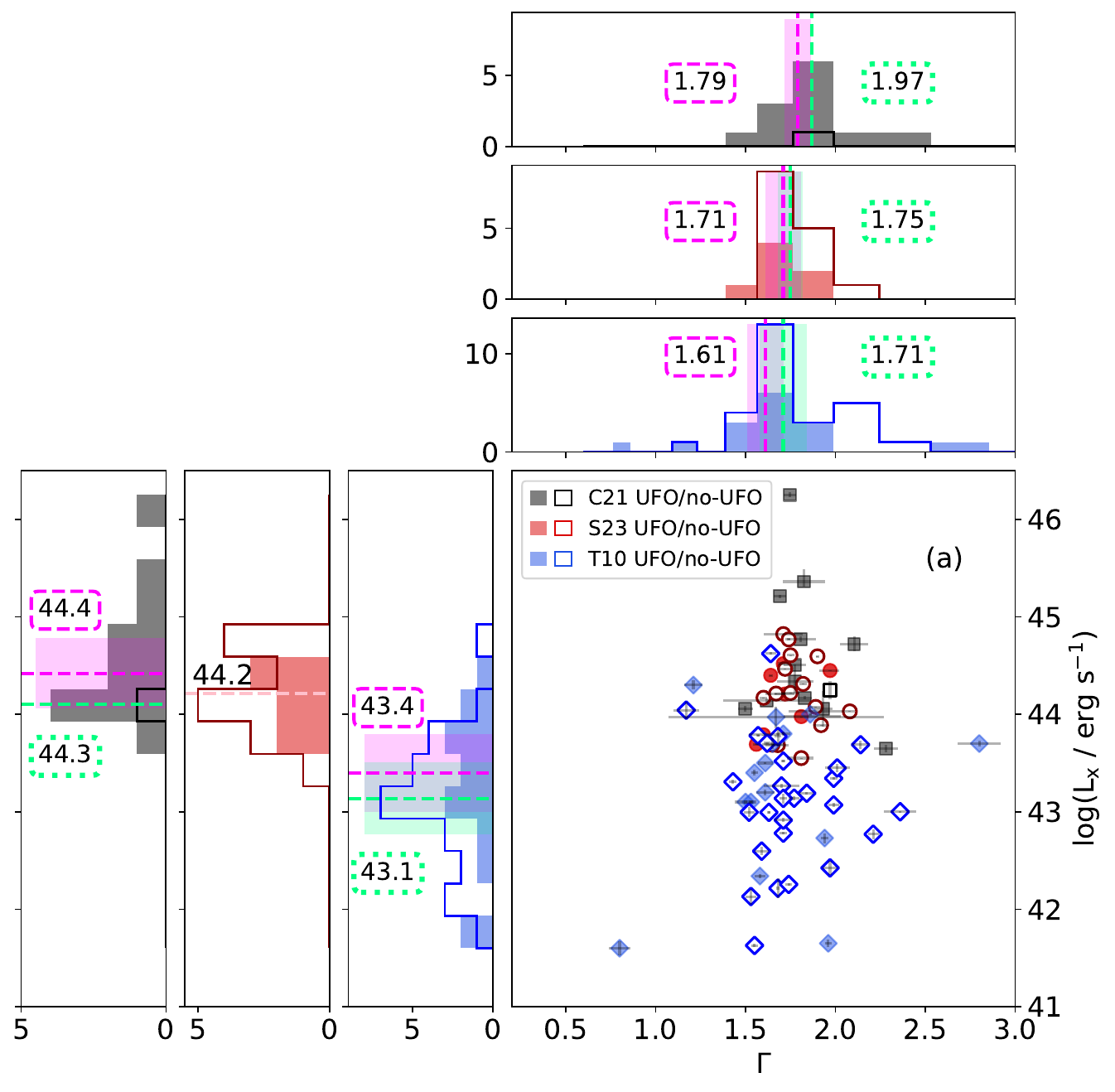}}
    \subfloat{\includegraphics[width=.5\textwidth]{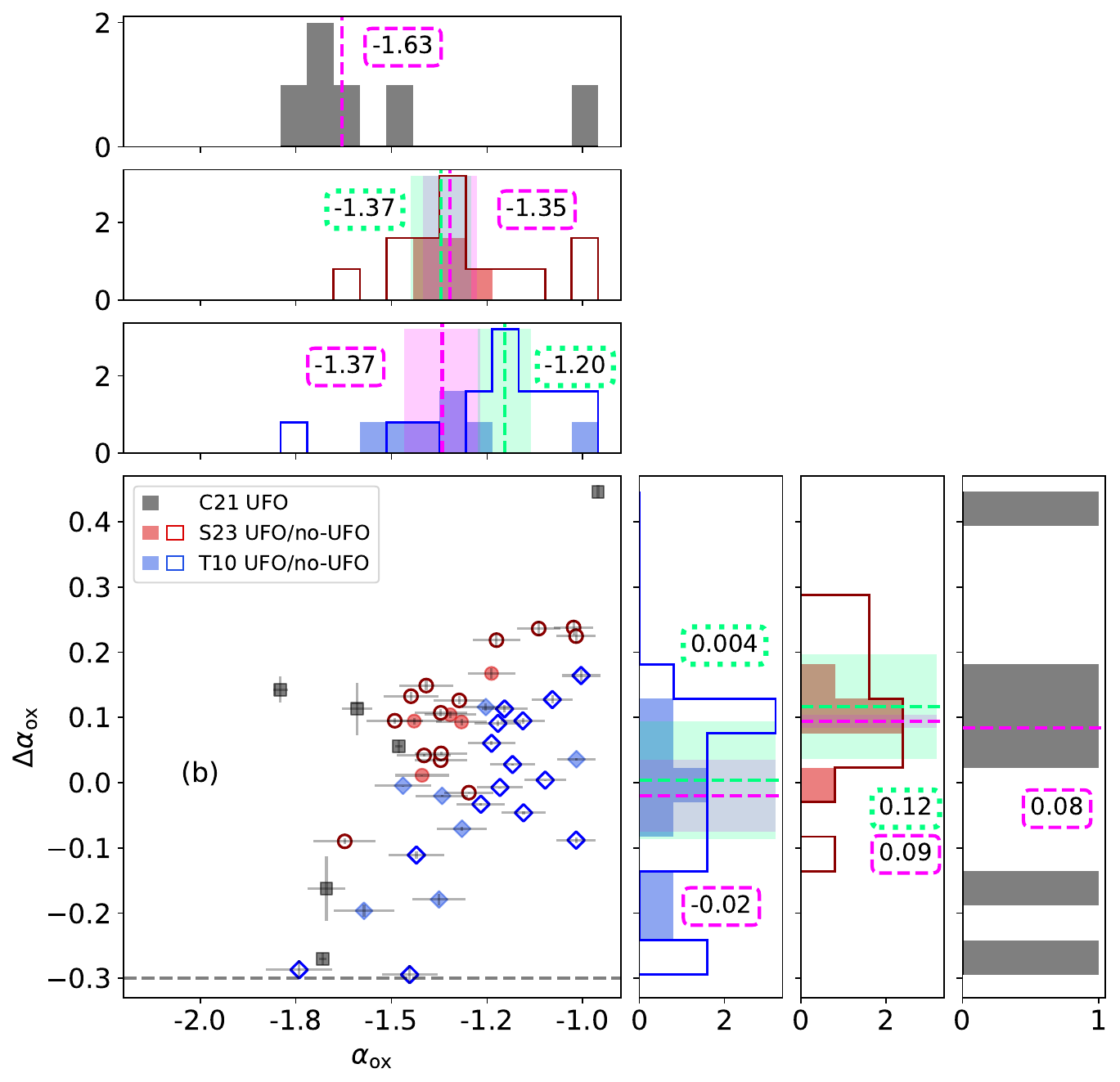}}
    \caption[]
    {{Samples comparison of AGN properties. Alongside Table \ref{tab:7} and Fig.~\ref{fig:22}, we obtained no substantial evidence indicating differences between AGNs hosting UFOs and those without. This suggests that all AGNs might be capable of hosting these outflows during their lifetime and their observability is linked to the wind duty cycle (see Sect.~\ref{ufo_vs_noufo}). Panel a: Observed X-ray spectral index versus the 2-10 keV luminosity. Panel b: Difference between the observed $\alpha_{\mathrm{ox}}$ and the expected value based on the UV luminosity of each AGN versus $\alpha_{\mathrm{ox}}$. The S23 sample is shown in red circles, T10 sample in blue diamonds and C21 sample in black squares (in the C21 no-UFO sub-sample, no sources are present as they are not part of the unabsorbed sample, see Sect.~\ref{x_uv_slope}).
    For each distribution and scatter plot, the UFO sub-samples are represented as color filled histograms and dots, respectively. The gray dashed line shows the threshold between X-ray normal and weak AGNs defined in \citet{pu20}. The dashed magenta and dotted green lines show the median value of each UFO and no-UFO sub-sample, respectively. We report both median values only when these are different and we include the corresponding 1$\sigma$ error-bars on the median.}}
    \label{fig:4}
\end{figure}

\begin{figure}[h!]
\centering
    \subfloat{\includegraphics[width=.5\textwidth]{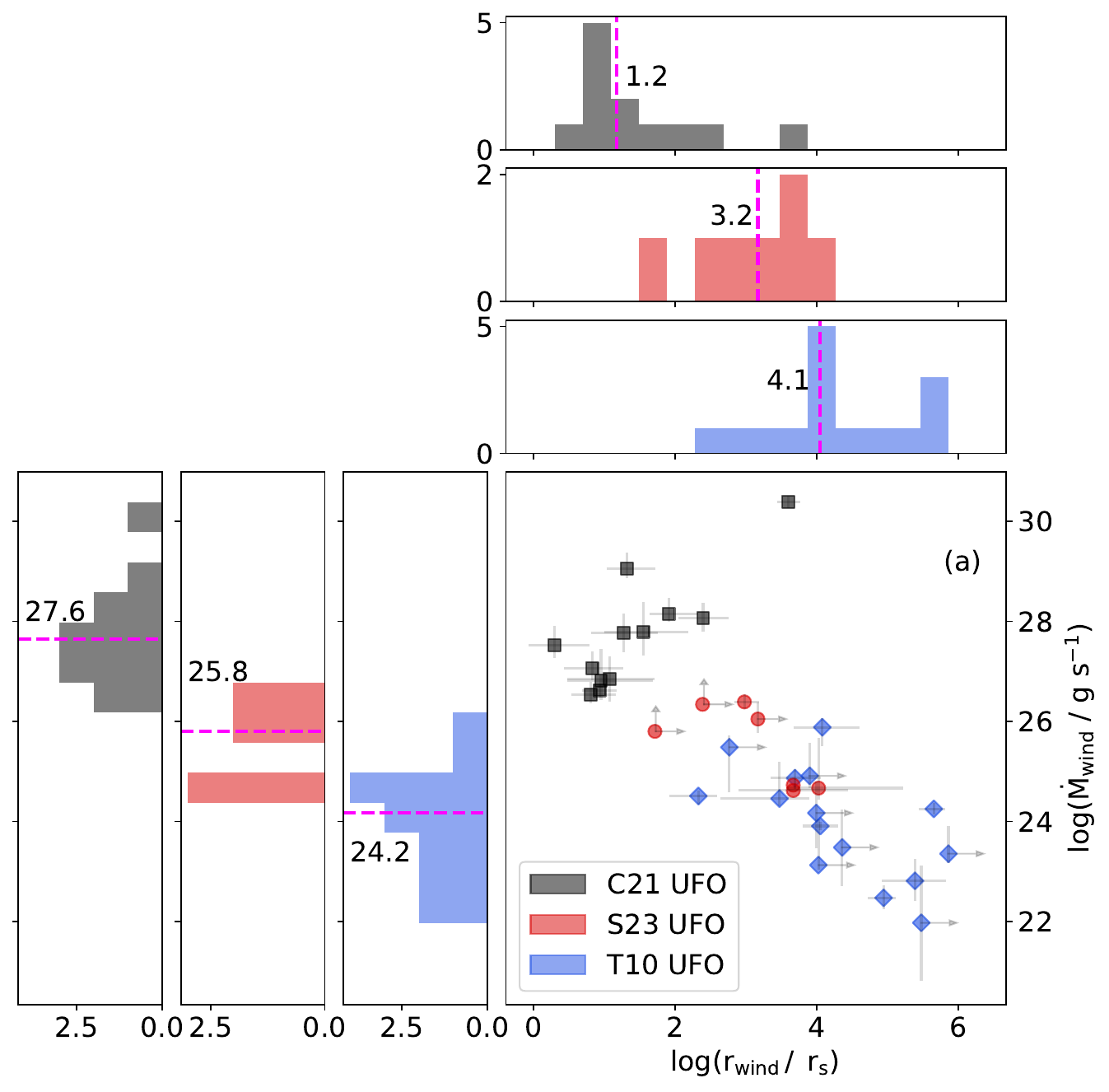}}
    \subfloat{\includegraphics[width=.5\textwidth]{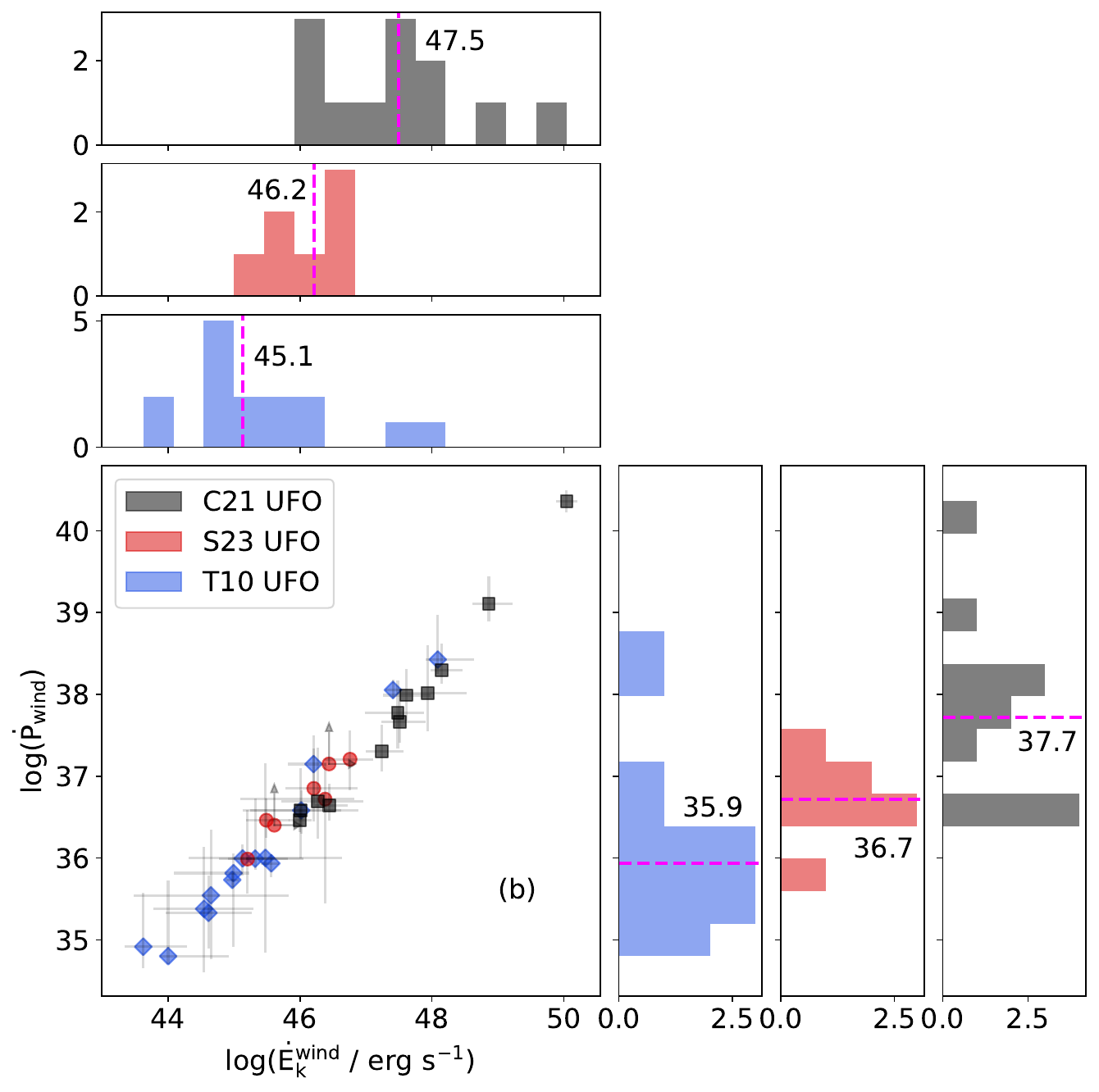}}
    \caption[]
    {{Samples comparison of outflow derived properties. Panel a: Distance between the wind and the SMBH versus mass outflow rate. Panel b: Kinetic power versus outflow momentum rate. 
    Each UFO sub-sample follows the same color/marker code as Fig.~\ref{fig:radii}. 
    The dashed magenta lines show the median values. The gray arrows represent the upper/lower limits for the log($\xi$) and N$_{\rm H}$ values (see Table \ref{tab:2}). We observe statistically significant differences for the launching radius and energetics of the wind (see Sect.~\ref{comparison_ufos}). }}
    \label{fig:25}
\end{figure} 

\twocolumn
\section{Tested correlations}\label{allcorr}

\begin{figure*}
    \centering
    \begin{subfigure}[b]{0.405\textwidth}
        \includegraphics[width=\linewidth]{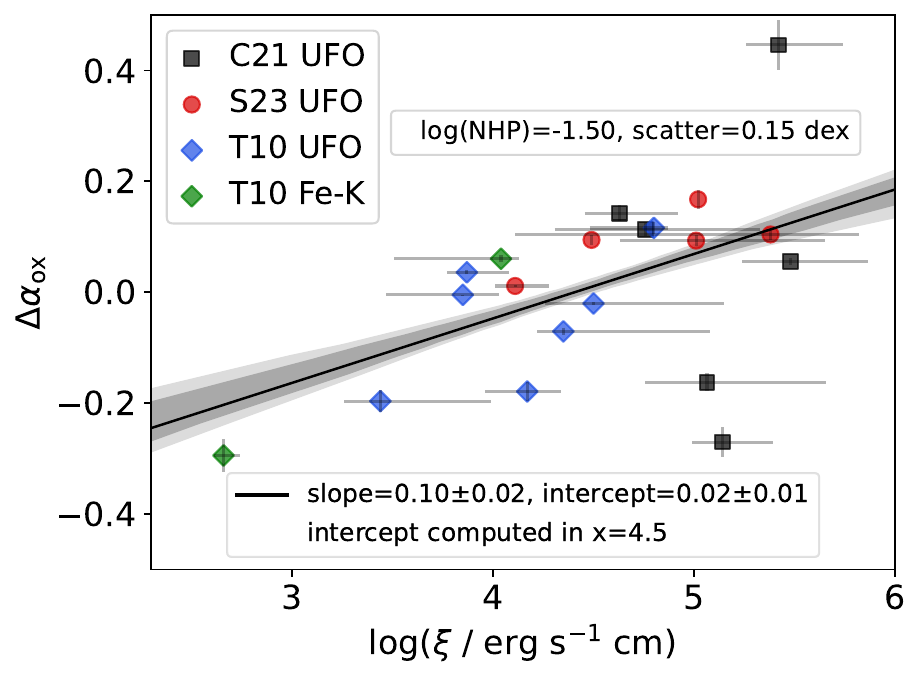}
    \end{subfigure}
    \begin{subfigure}[b]{0.4\textwidth}
        \includegraphics[width=\linewidth]{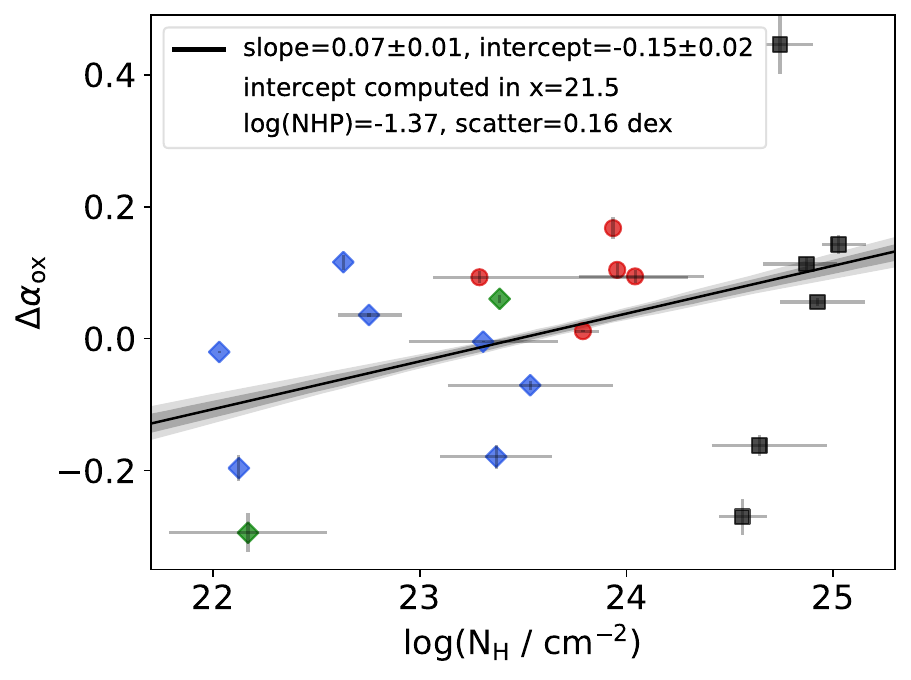}
    \end{subfigure}\\
    \begin{subfigure}[b]{0.405\textwidth}
        \includegraphics[width=\linewidth]{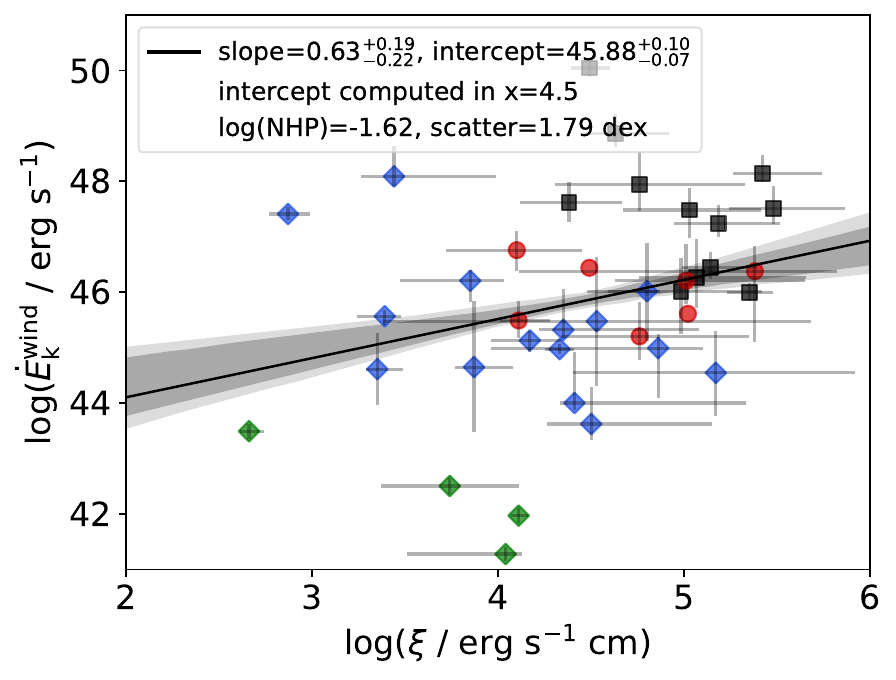}
    \end{subfigure}
    \begin{subfigure}[b]{0.4\textwidth}
        \includegraphics[width=\linewidth]{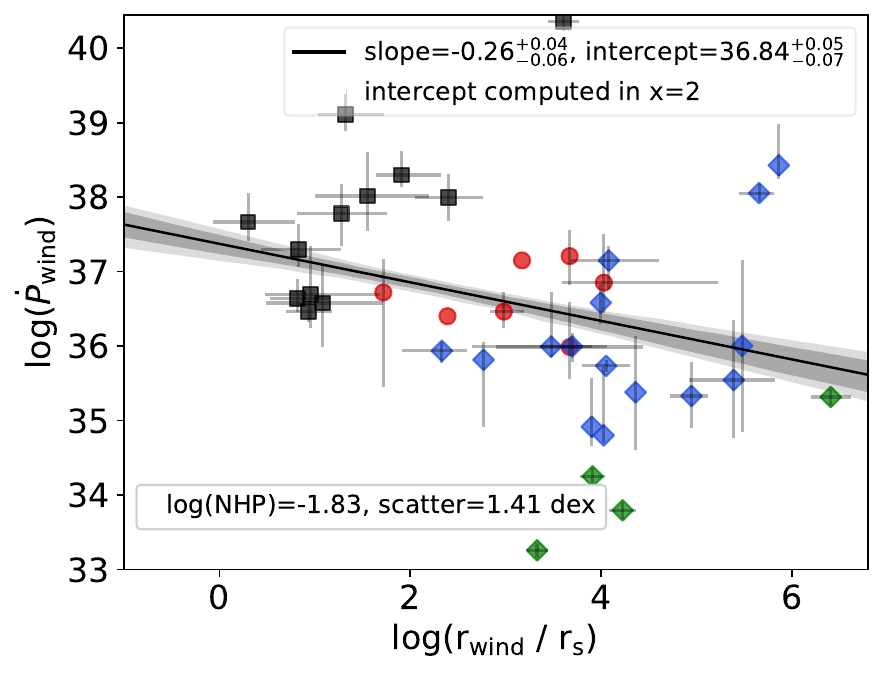}
    \end{subfigure}\\
    \caption[]{{Significant correlations after the addition of the Fe-K sub-sample. The S23 UFO sub-sample is shown in red dots, the T10 in blue diamonds, the C21 in black squares, and the Fe-K sub-sample in green diamonds. The best-fitting linear correlations, applied exclusively to statistically significant correlations, are presented by the solid black lines and the  dark and light gray shadowed areas indicate the 68\% and 90\% confidence bands, respectively. In the legend, we report the best-fit coefficients, $\log\mathrm{NHP}$, and the intrinsic scatters for the correlations.}}
    \label{fig:ef19}
\end{figure*}

In this work, we investigate potential correlations between the AGN properties and the UFO characteristics and, in the following figures, we display all pairs of tested parameters. For each significant (positive/negative) correlation, we report the best-fitting linear regressions, associated $\log\mathrm{NHP}$ values and the intrinsic scatters of the data. Each UFO sub-sample is color-coded as in the main text, that is, S23 in red circles, T10 in blue diamonds and C21 in black squares. 
Four of the reported correlations (log($\xi$)-$\Delta\alpha_{\rm{ox}}$, N$_{\rm{H}}$-$\Delta\alpha_{\rm{ox}}$, log($\xi$)-$\dot{\rm{E}}_{\rm{k}}^{\rm{wind}}$, and r$_{\rm{wind}}/\rm{r}_{\rm{s}}$-$\dot{\rm{P}}_{\rm{wind}}$ in Fig.~\ref{fig:ef19}) are only significant when the Fe-K sub-sample is added to the T10 UFO sub-sample. In these cases, AGNs from the Fe-K sub-sample are presented in green diamonds.

\begin{figure*}
\centering
    \begin{subfigure}[b]{0.4\textwidth}
        \includegraphics[width=\linewidth]{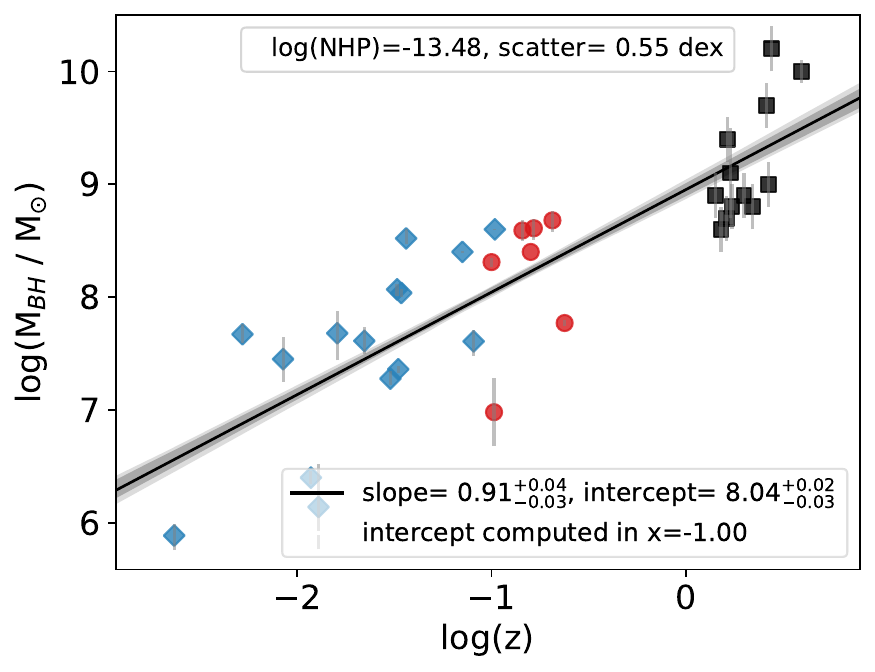}
    \end{subfigure}
    \begin{subfigure}[b]{0.4\textwidth}
        \includegraphics[width=\linewidth]{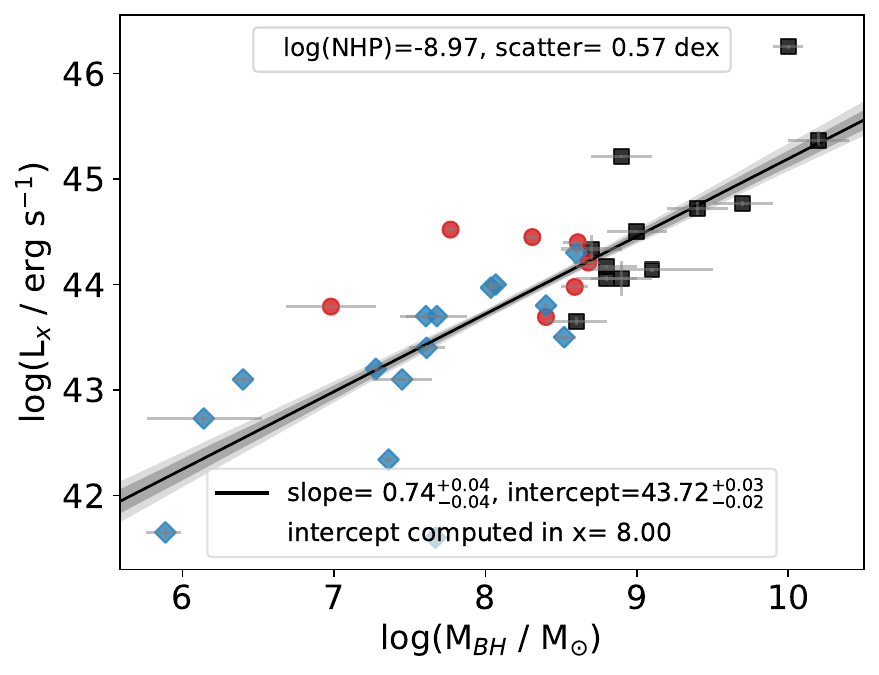}
    \end{subfigure}
    \begin{subfigure}[b]{0.4\textwidth}
        \includegraphics[width=\linewidth]{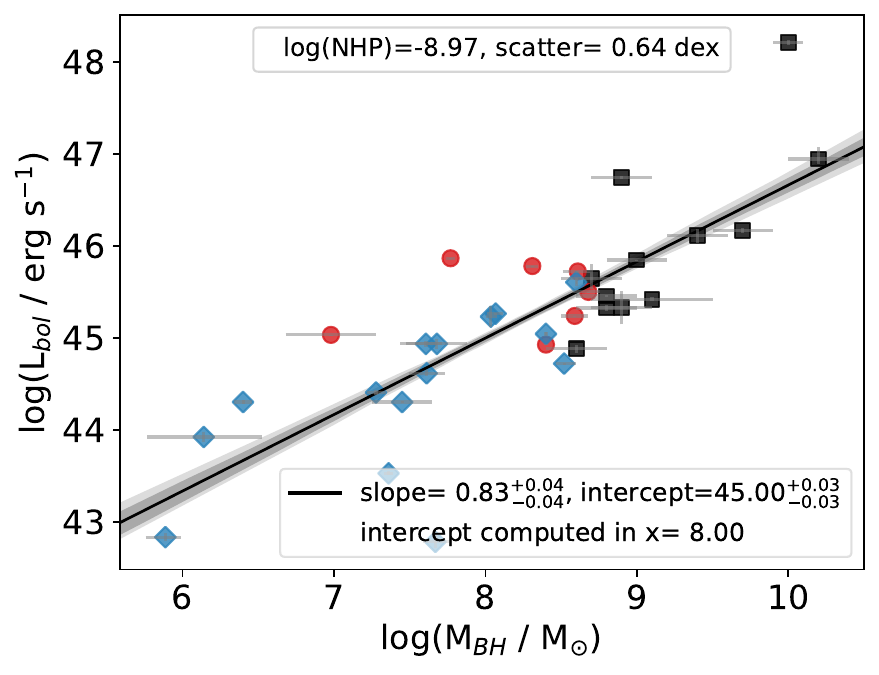}
    \end{subfigure}
    \begin{subfigure}[b]{0.4\textwidth}
        \includegraphics[width=\linewidth]{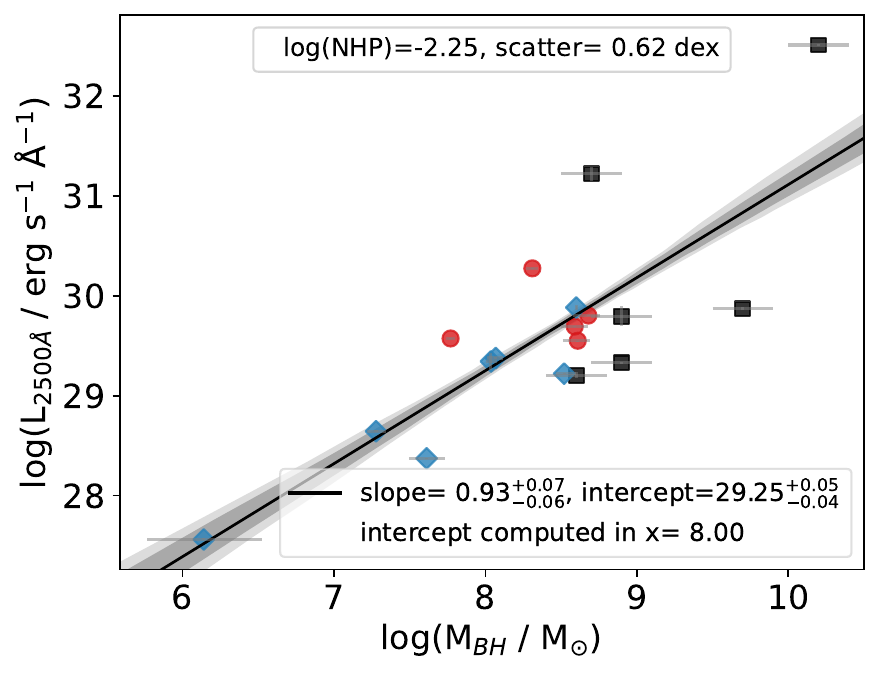}
    \end{subfigure}
    \begin{subfigure}[b]{0.4\textwidth}
        \includegraphics[width=\linewidth]{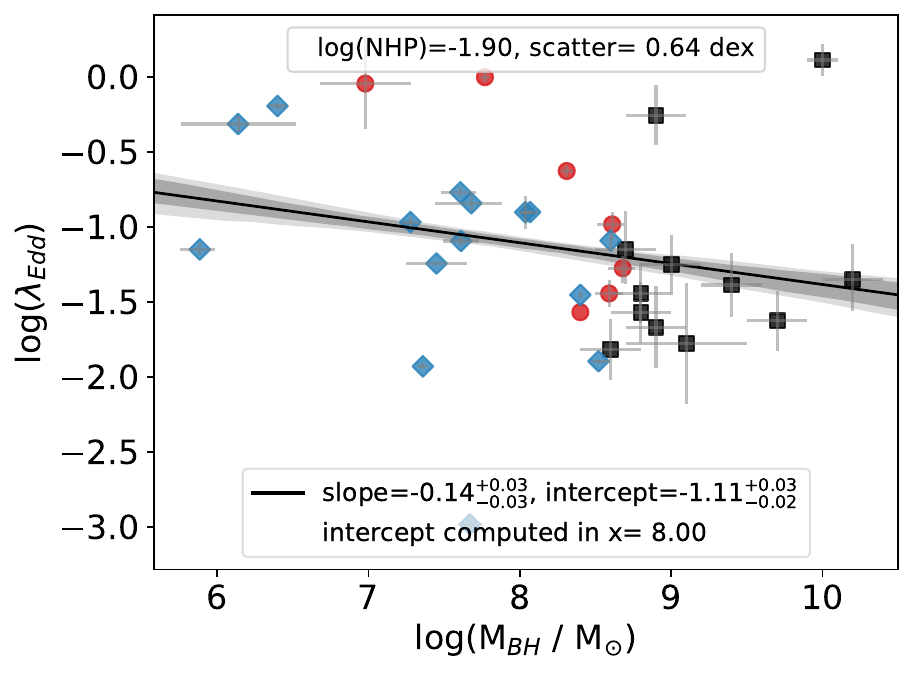}
    \end{subfigure}\\
    \begin{subfigure}[b]{0.225\textwidth}
        \includegraphics[width=\linewidth]{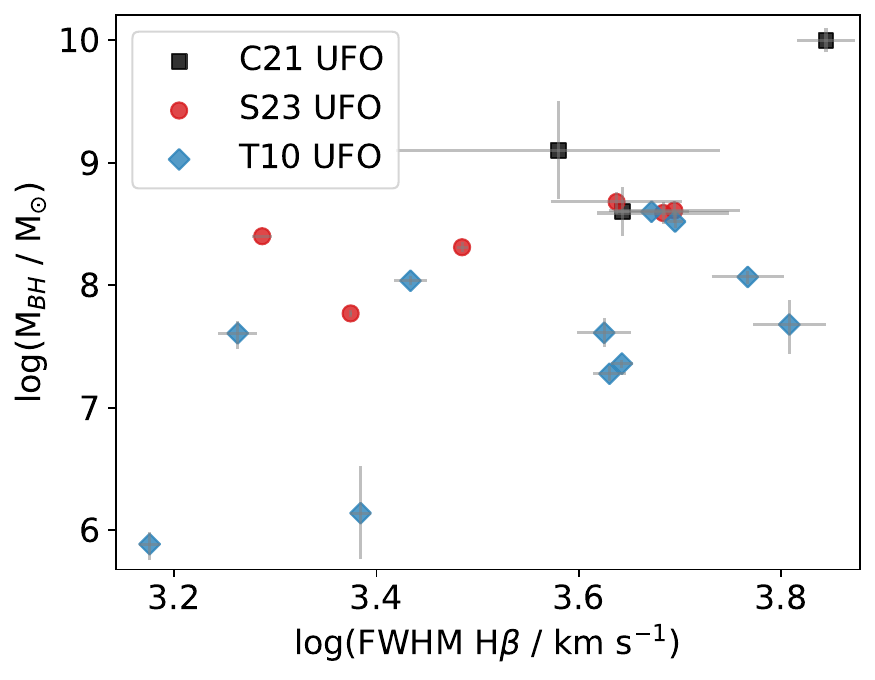}
    \end{subfigure}
    \begin{subfigure}[b]{0.235\textwidth}
        \includegraphics[width=\linewidth]{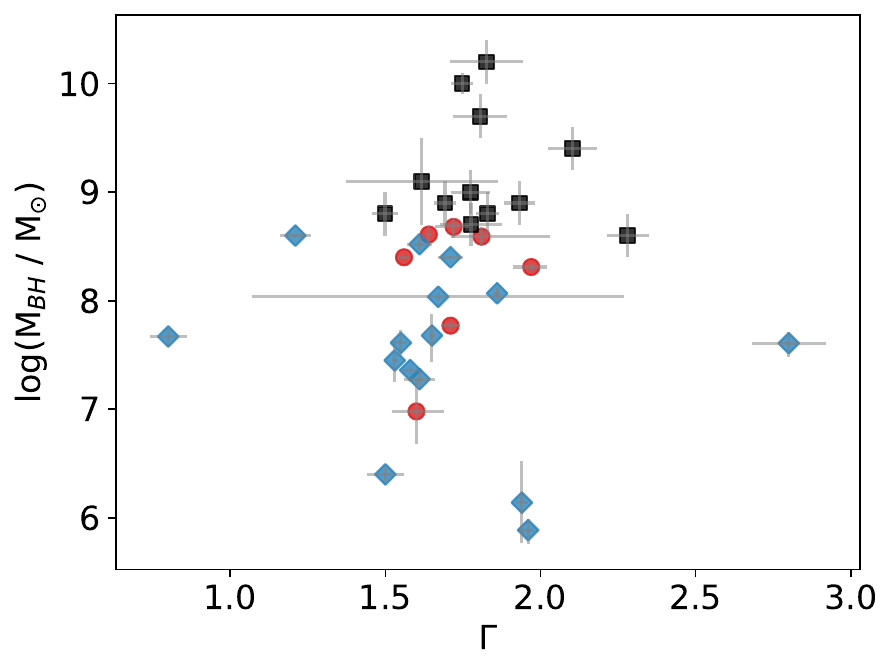}
    \end{subfigure}
    \begin{subfigure}[b]{0.23\textwidth}
        \includegraphics[width=\linewidth]{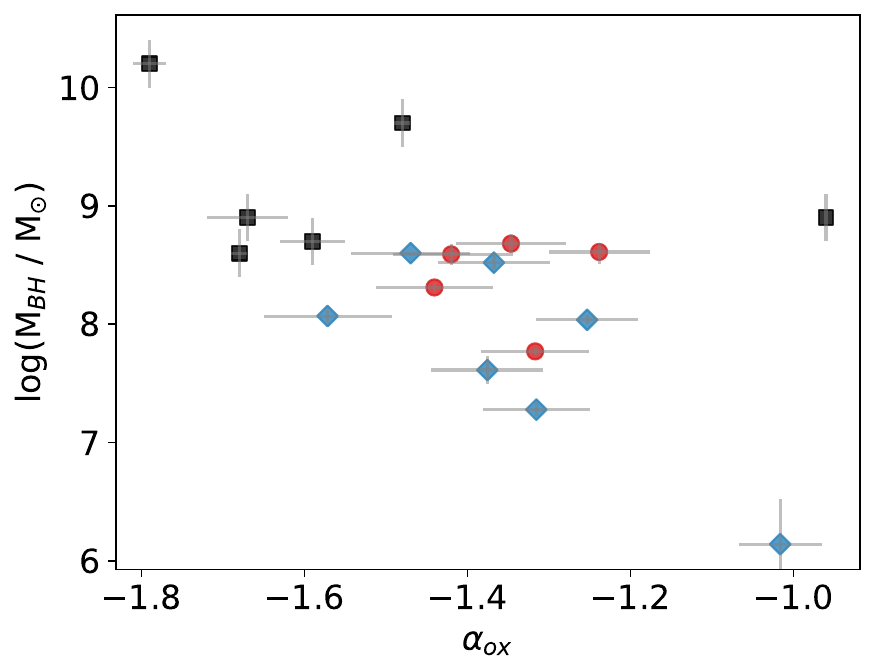}
    \end{subfigure}
    \begin{subfigure}[b]{0.23\textwidth}
        \includegraphics[width=\linewidth]{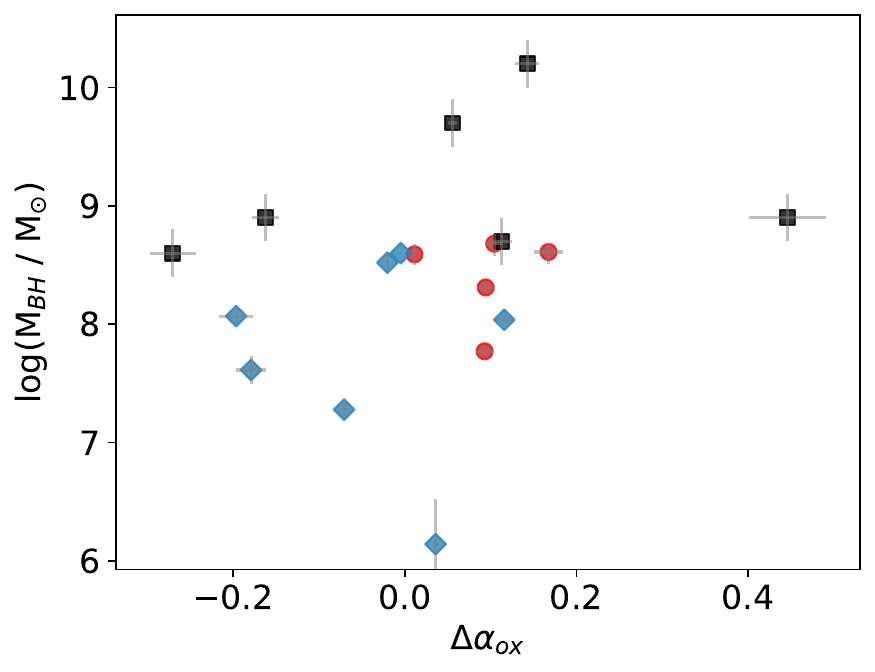}
    \end{subfigure}
    \caption{SMBH mass versus AGN parameters. Significant and nonsignificant correlations for the S23, T10, and C21 samples. The best-fitting linear correlations, applied exclusively to statistically significant correlations, are presented by the solid black lines and the dark and light gray shadowed areas indicate the 68\% and 90\% confidence bands, respectively. In the legend, we report the best-fit coefficients, $\log\mathrm{NHP}$, and the intrinsic scatters for the correlations.}
    \label{fig:ef1}
\end{figure*}

\begin{figure*}
\centering
    \begin{subfigure}[b]{0.4\textwidth}
        \includegraphics[width=\linewidth]{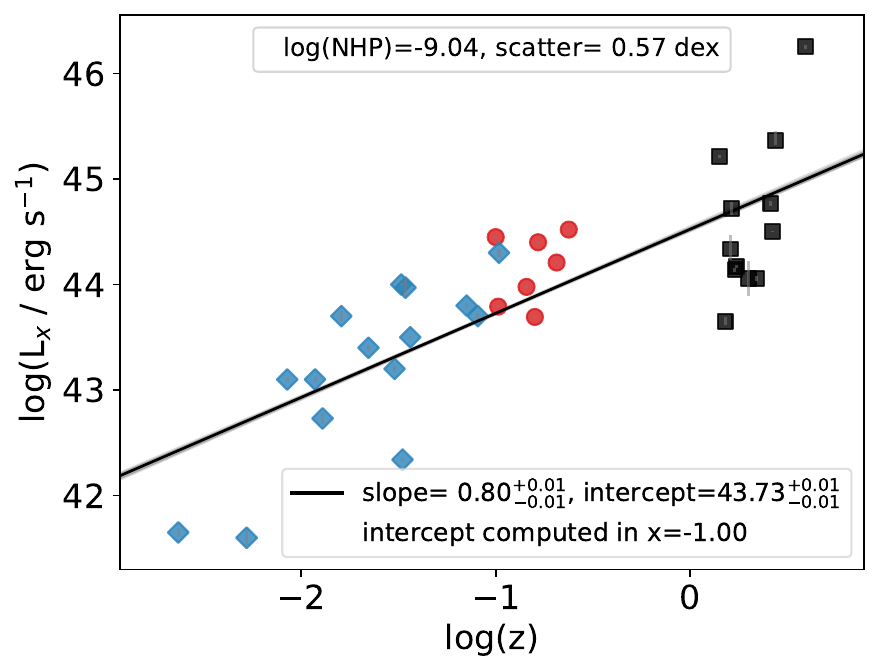}
    \end{subfigure}
    \begin{subfigure}[b]{0.4\textwidth}
        \includegraphics[width=\linewidth]{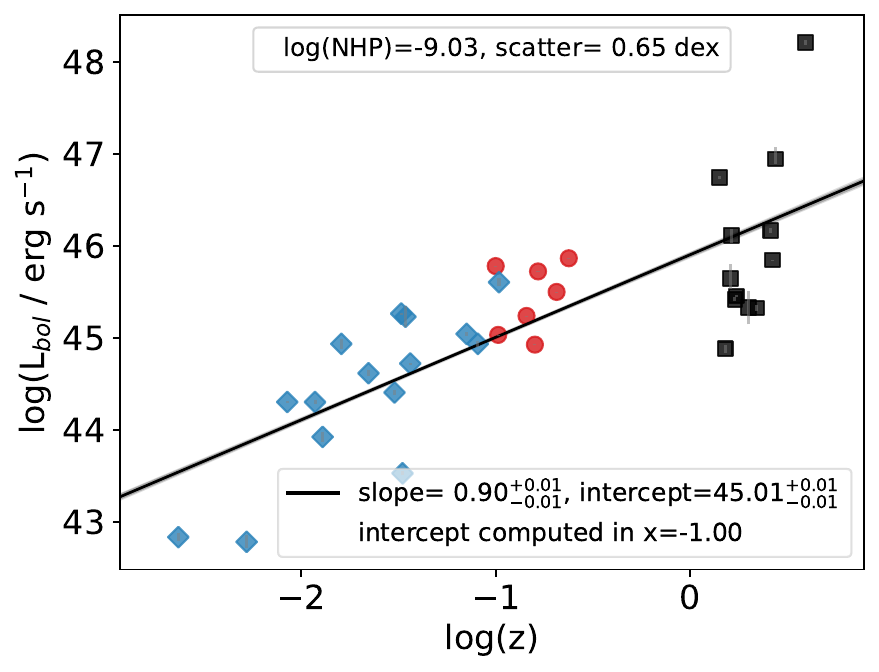}
    \end{subfigure}
    \begin{subfigure}[b]{0.4\textwidth}
        \includegraphics[width=\linewidth]{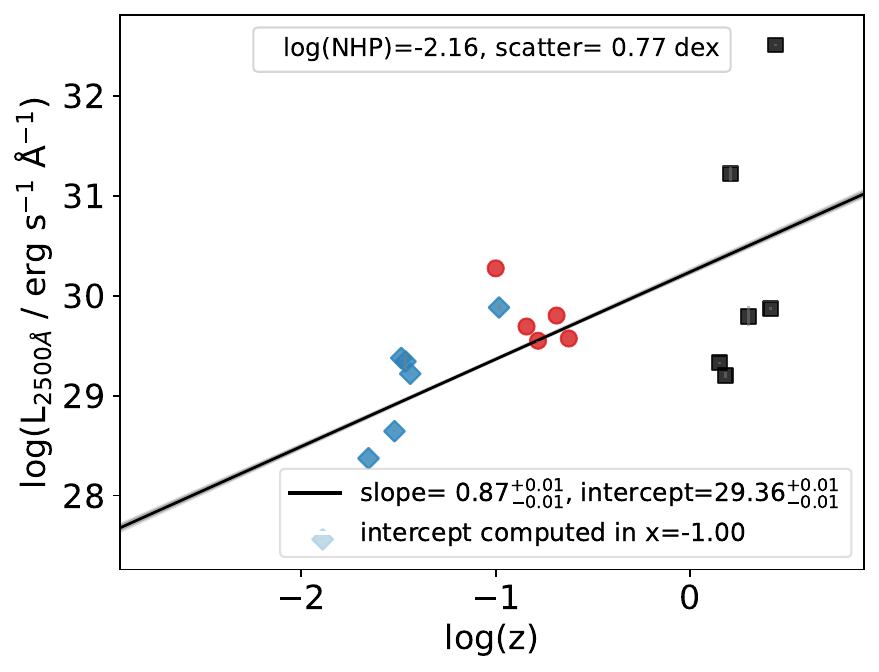}
    \end{subfigure}
    \begin{subfigure}[b]{0.4\textwidth}
        \includegraphics[width=\linewidth]{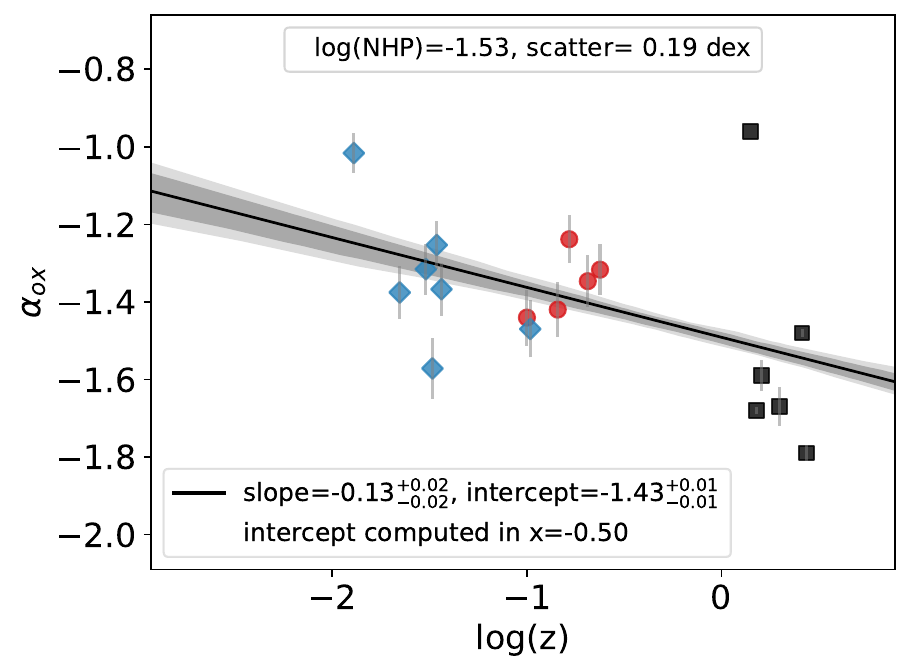}
    \end{subfigure}
    \begin{subfigure}[b]{0.4\textwidth}
        \includegraphics[width=\linewidth]{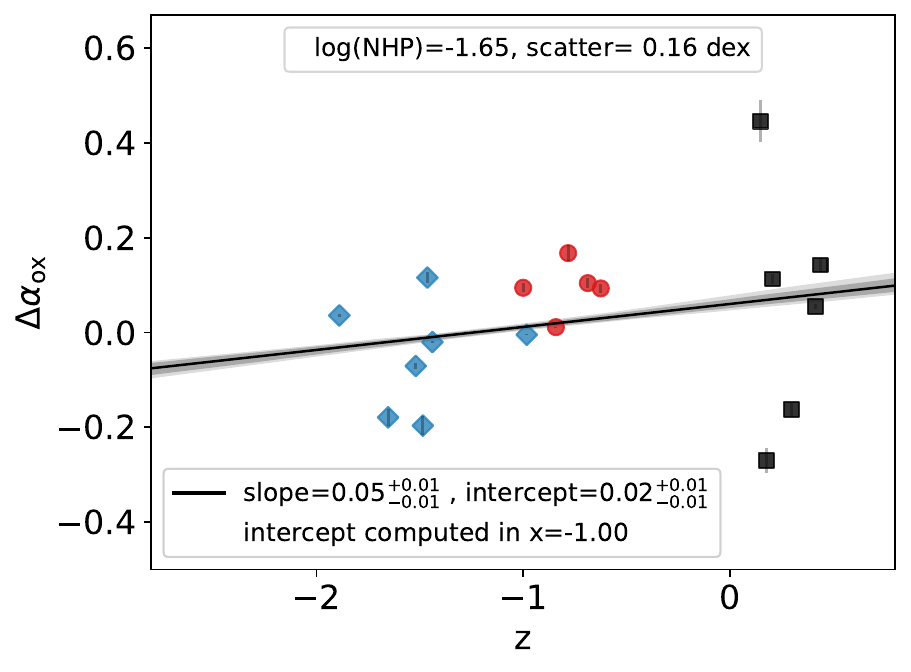}
    \end{subfigure}\\
    \begin{subfigure}[b]{0.25\textwidth}
        \includegraphics[width=\linewidth]{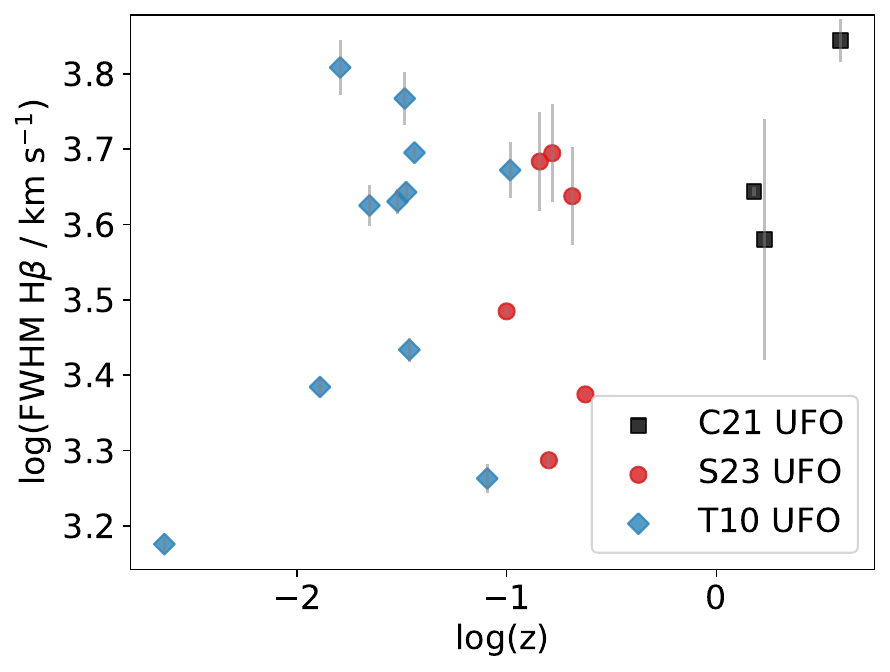}
    \end{subfigure}
    \begin{subfigure}[b]{0.25\textwidth}
        \includegraphics[width=\linewidth]{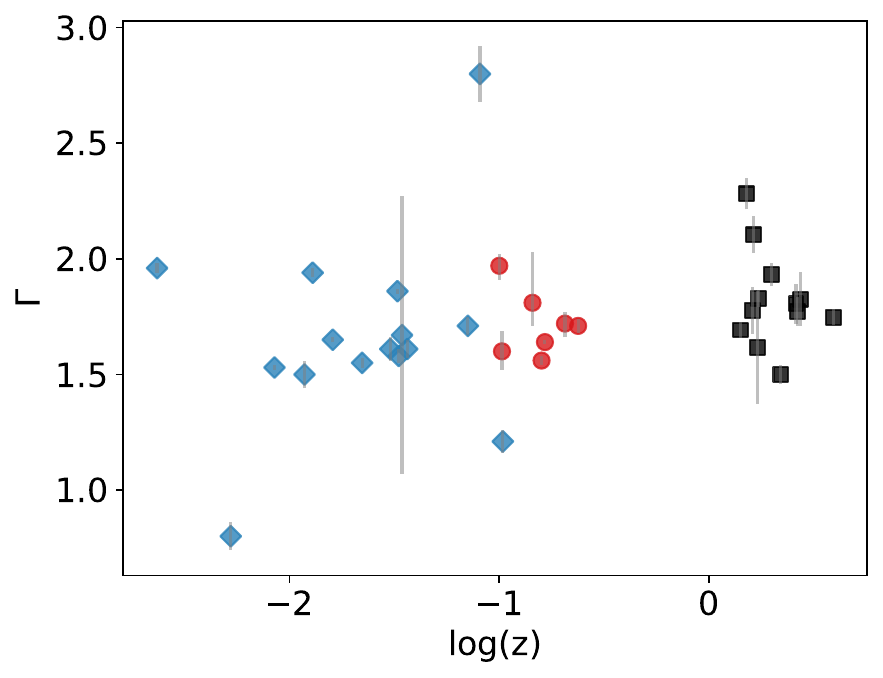}
    \end{subfigure}
    \begin{subfigure}[b]{0.25\textwidth}
        \includegraphics[width=\linewidth]{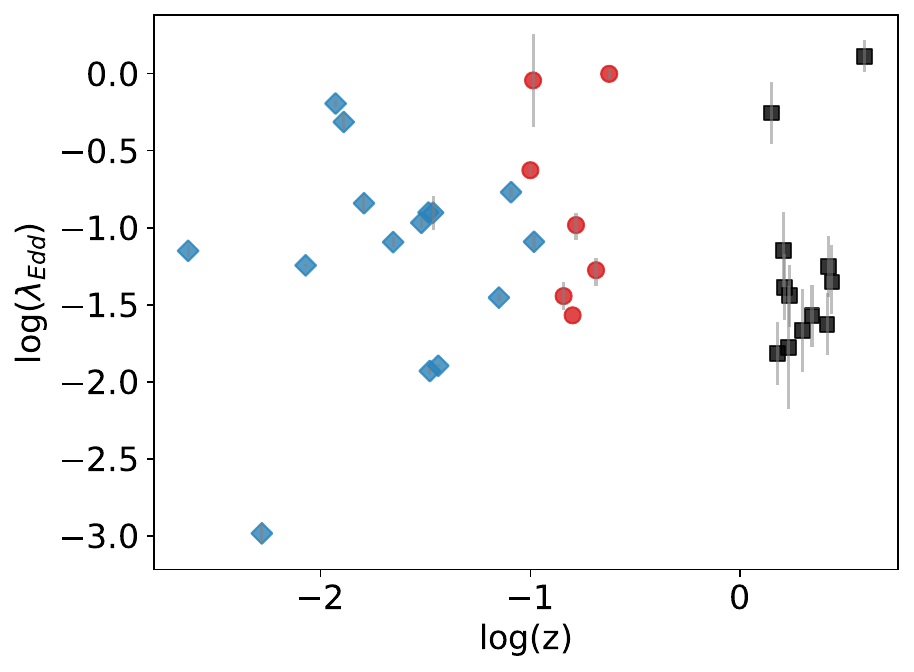}
    \end{subfigure}\\
    \caption{Redshift versus AGN parameters. Significant and nonsignificant correlations for the S23, T10, and C21 samples. The best-fitting linear correlations, applied exclusively to statistically significant correlations, are presented by the solid black lines and the dark and light gray shadowed areas indicate the 68\% and 90\% confidence bands, respectively. In the legend, we report the best-fit coefficients, $\log\mathrm{NHP}$, and the intrinsic scatters for the correlations.}
    \label{fig:ef2}
\end{figure*}

\begin{figure*}
\centering
    \begin{subfigure}[b]{0.4\textwidth}
        \includegraphics[width=\linewidth]{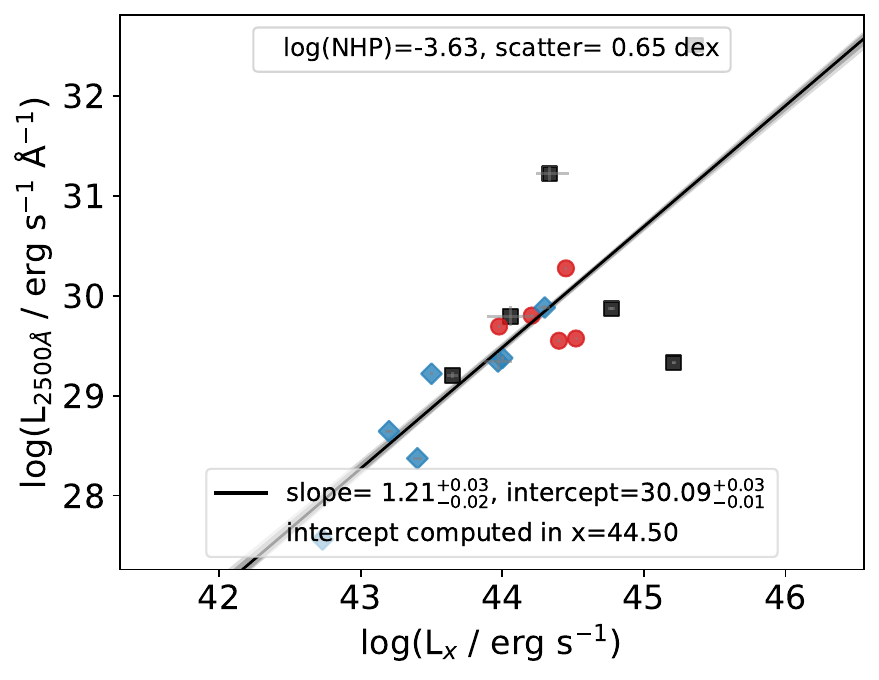}
    \end{subfigure}
    \begin{subfigure}[b]{0.4\textwidth}
        \includegraphics[width=\linewidth]{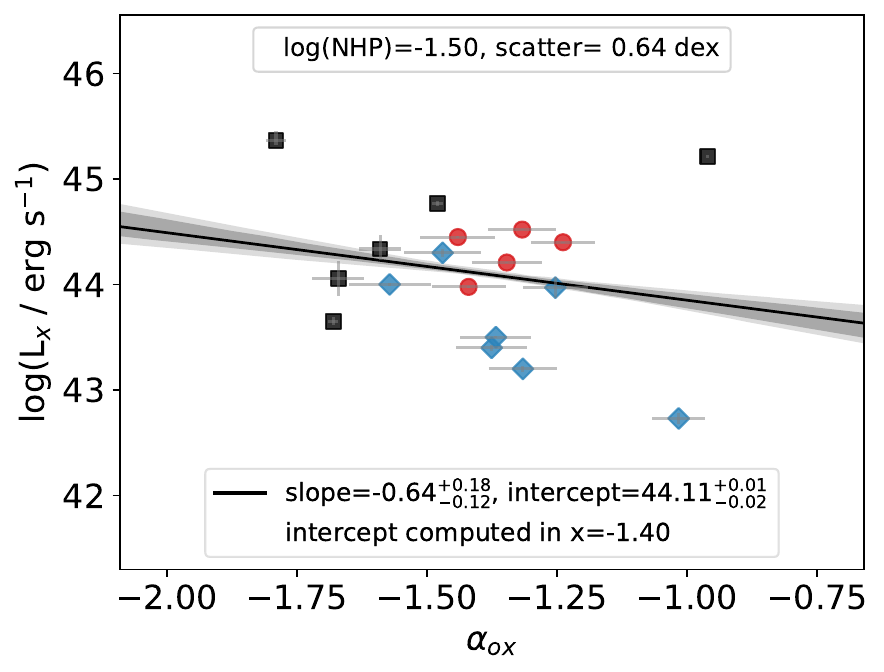}
    \end{subfigure}
    \begin{subfigure}[b]{0.4\textwidth}
        \includegraphics[width=\linewidth]{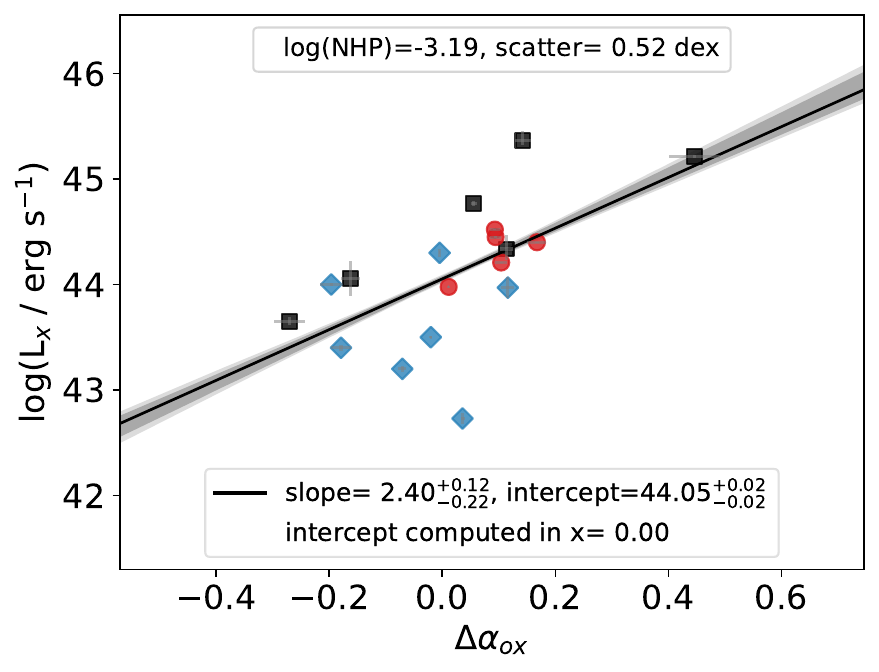}
    \end{subfigure}
    \begin{subfigure}[b]{0.4\textwidth}
        \includegraphics[width=\linewidth]{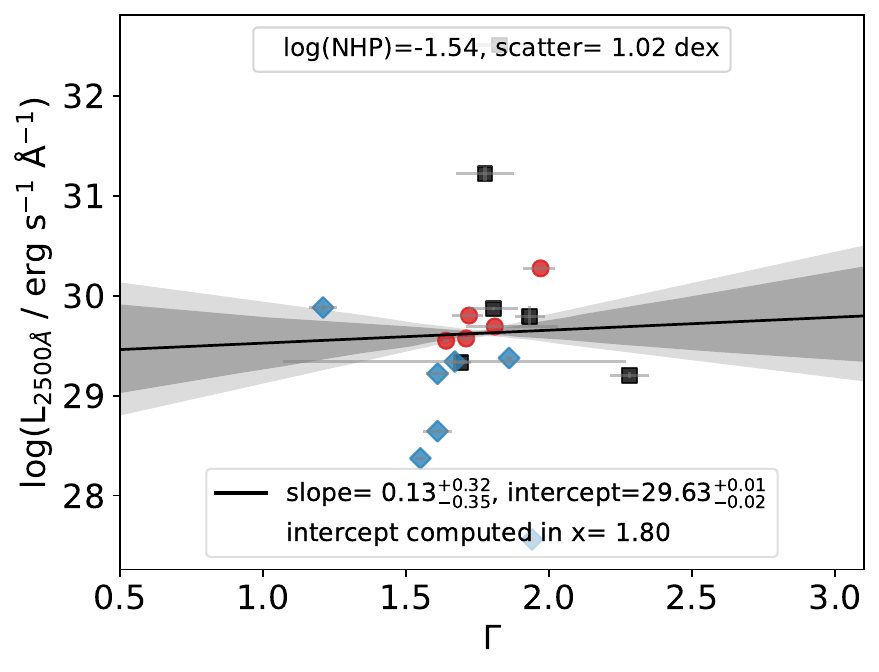}
    \end{subfigure}
    \begin{subfigure}[b]{0.4\textwidth}
        \includegraphics[width=\linewidth]{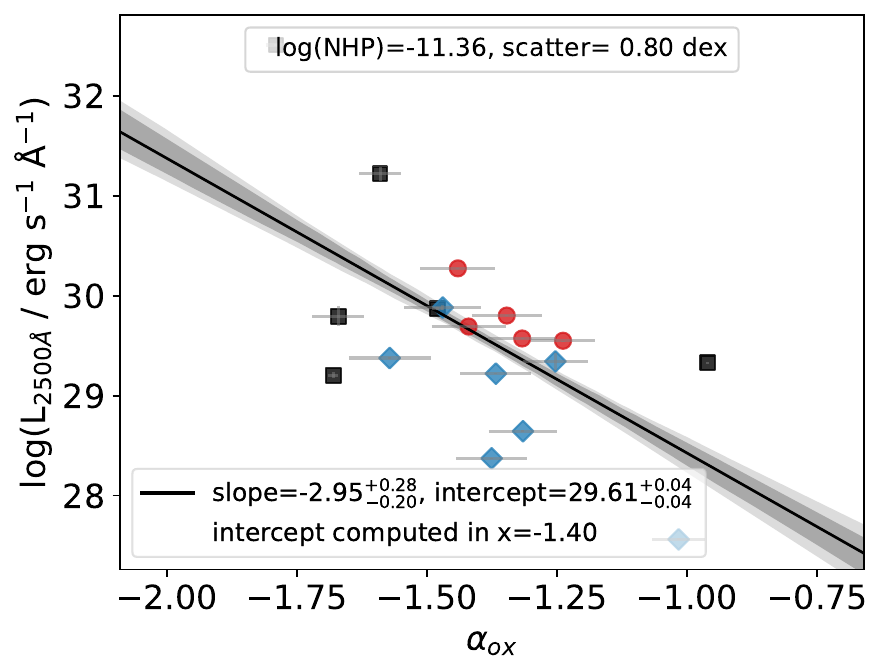}
    \end{subfigure}
    \begin{subfigure}[b]{0.4\textwidth}
        \includegraphics[width=\linewidth]{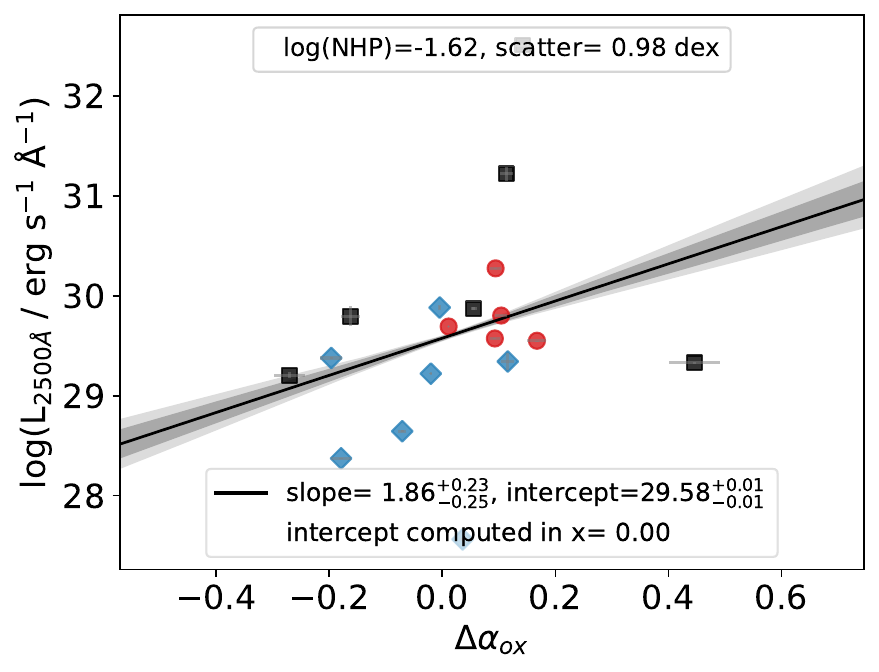}
    \end{subfigure}
    \begin{subfigure}[b]{0.225\textwidth}
        \includegraphics[width=\linewidth]{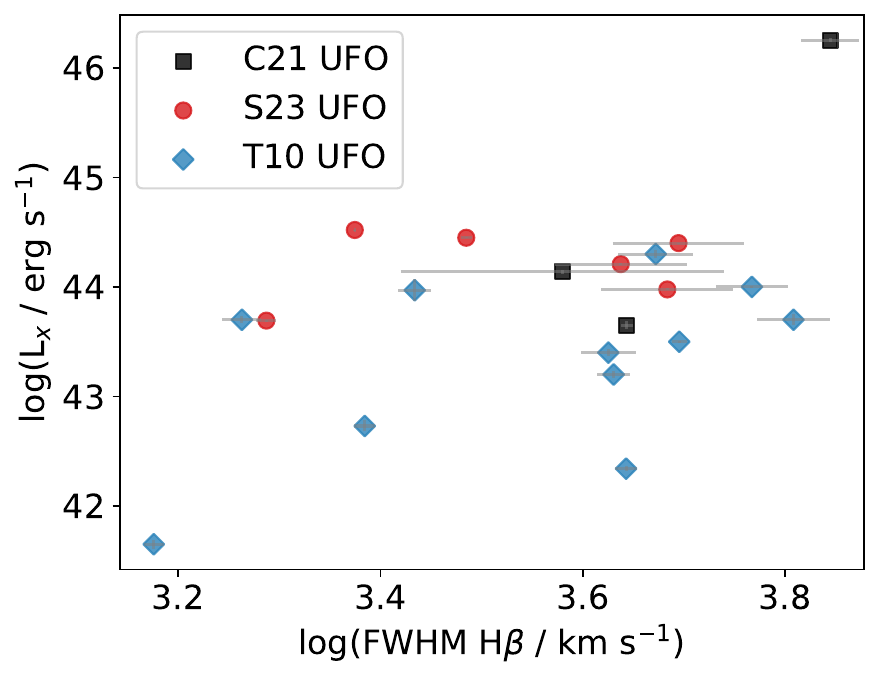}
    \end{subfigure}
    \begin{subfigure}[b]{0.23\textwidth}
        \includegraphics[width=\linewidth]{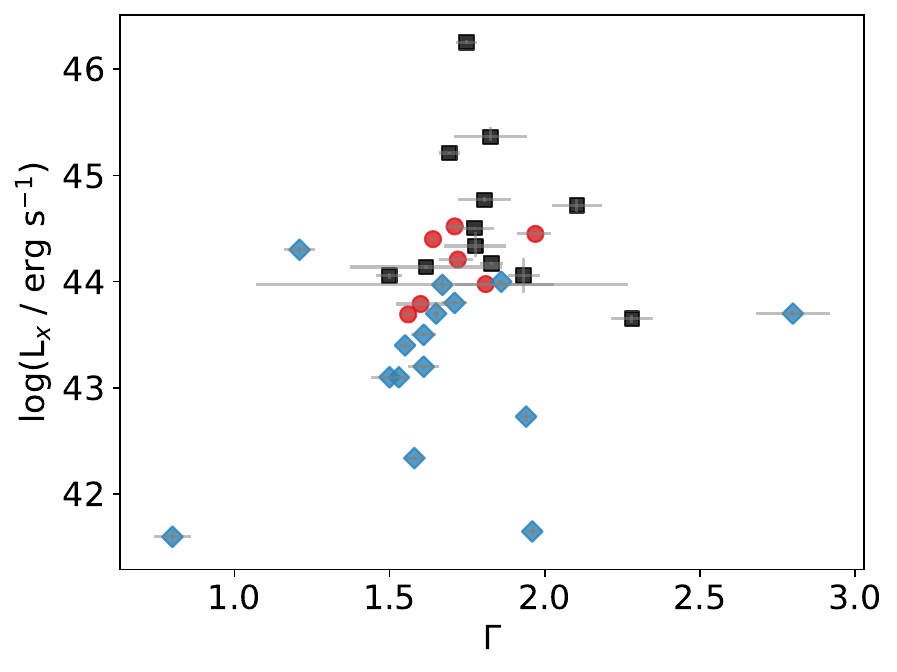}
    \end{subfigure}
    \begin{subfigure}[b]{0.23\textwidth}
        \includegraphics[width=\linewidth]{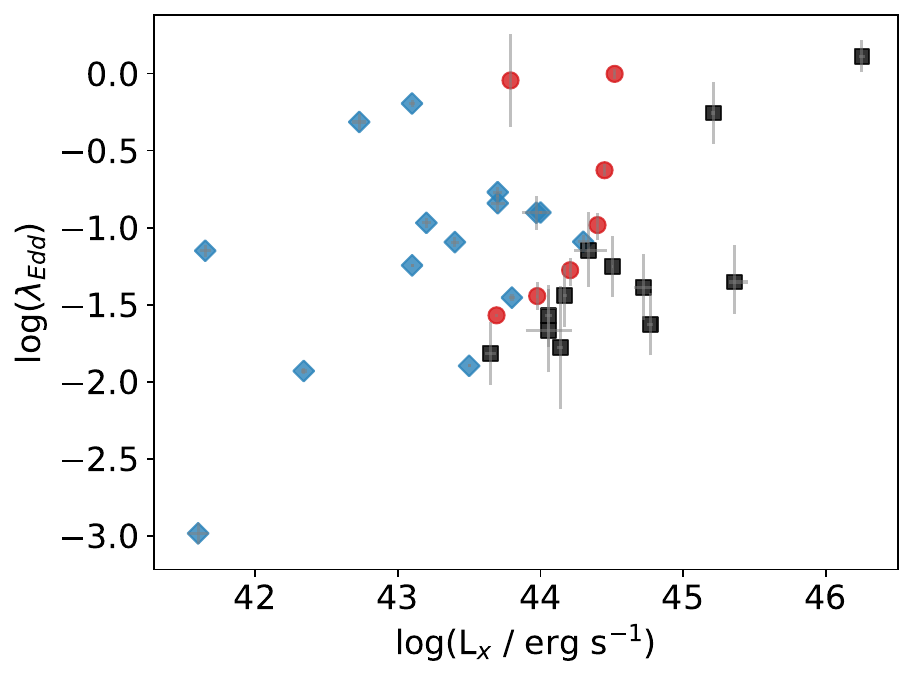}
    \end{subfigure} 
    \begin{subfigure}[b]{0.23\textwidth}
        \includegraphics[width=\linewidth]{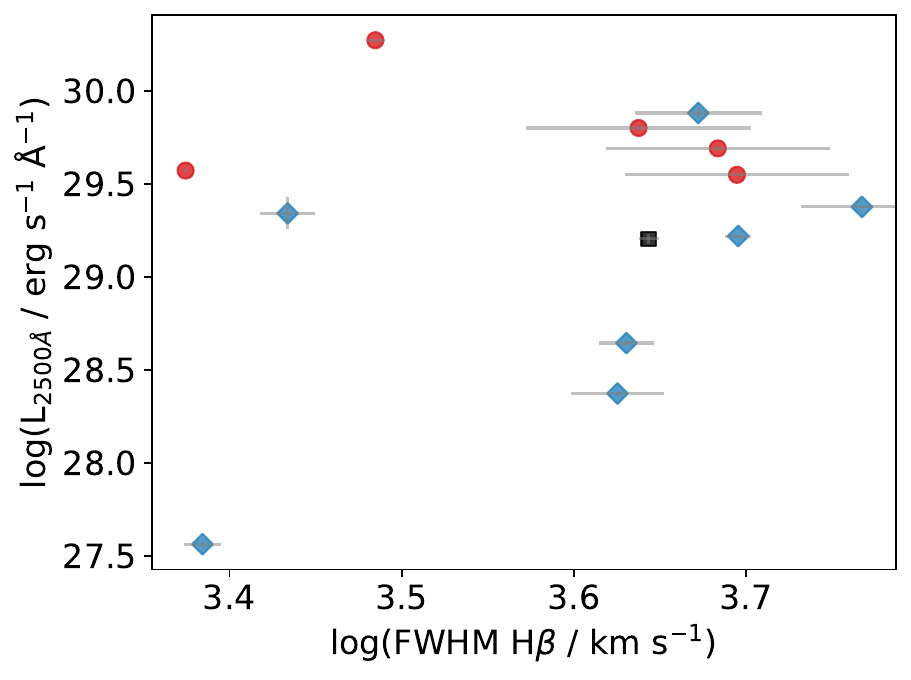}
    \end{subfigure}
        \begin{subfigure}[b]{0.23\textwidth}
        \includegraphics[width=\linewidth]{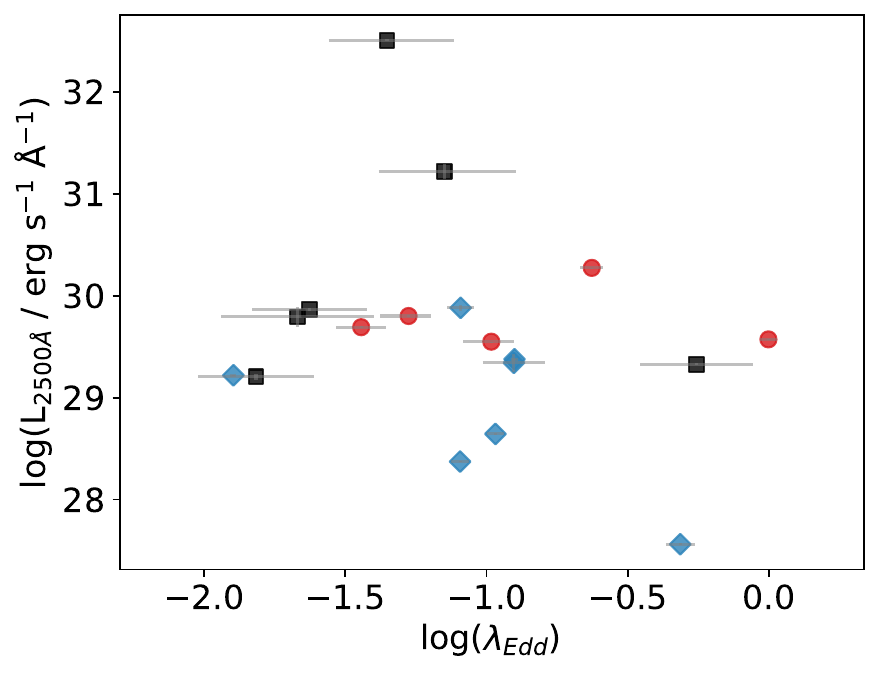}
    \end{subfigure}
    \caption{X-ray and UV luminosity versus AGN parameters. Significant and nonsignificant correlations for the S23, T10, and C21 samples. The best-fitting linear correlations, applied exclusively to statistically significant correlations, are presented by the solid black lines and the dark and light gray shadowed areas indicate the 68\% and 90\% confidence bands, respectively.}
    \label{fig:ef3}
\end{figure*}

\begin{figure*}
\centering
    \begin{subfigure}[b]{0.4\textwidth}
        \includegraphics[width=\linewidth]{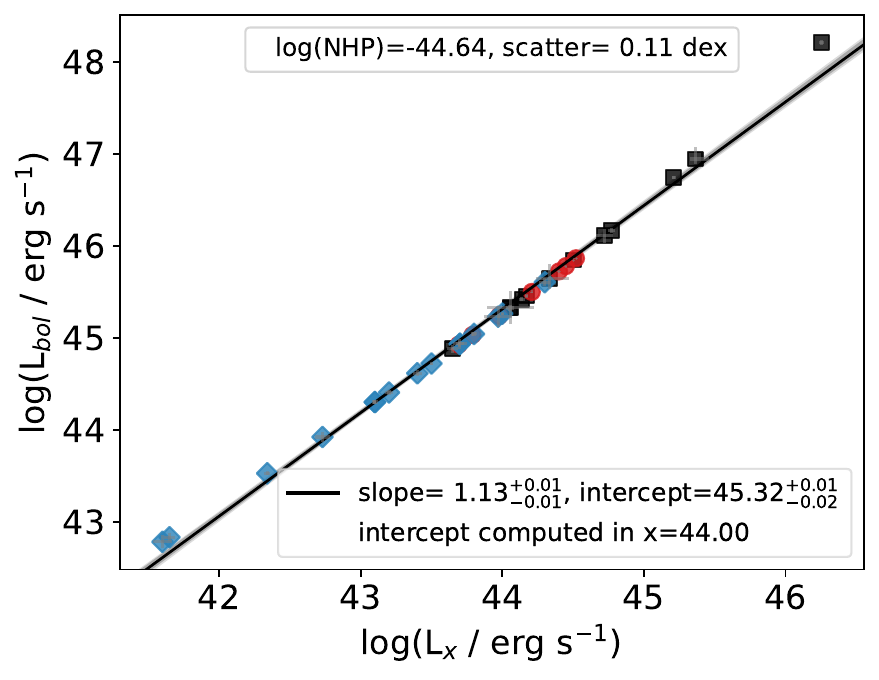}
    \end{subfigure}
    \begin{subfigure}[b]{0.4\textwidth}
        \includegraphics[width=\linewidth]{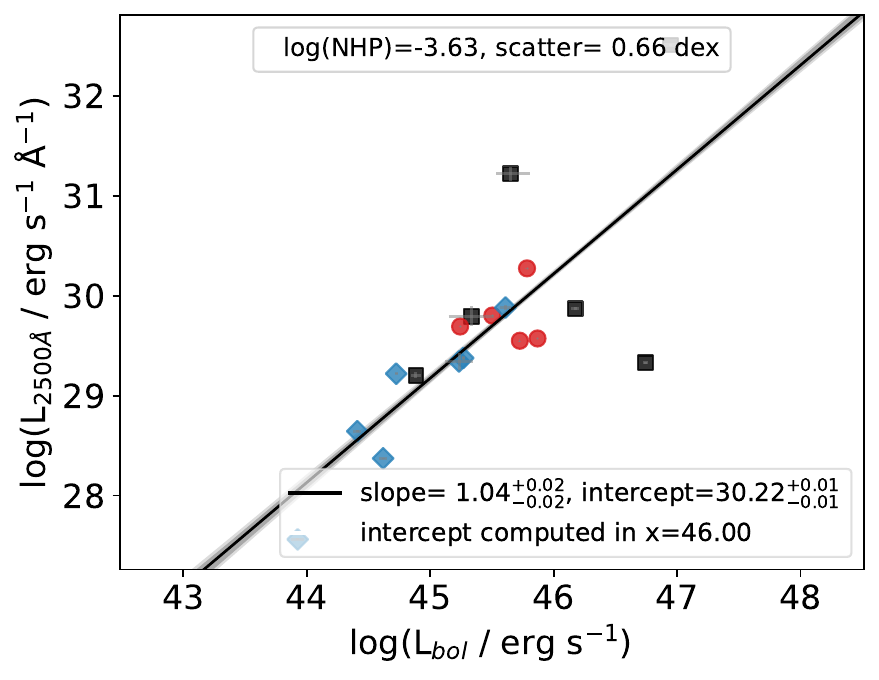}
    \end{subfigure}
    \begin{subfigure}[b]{0.4\textwidth}
        \includegraphics[width=\linewidth]{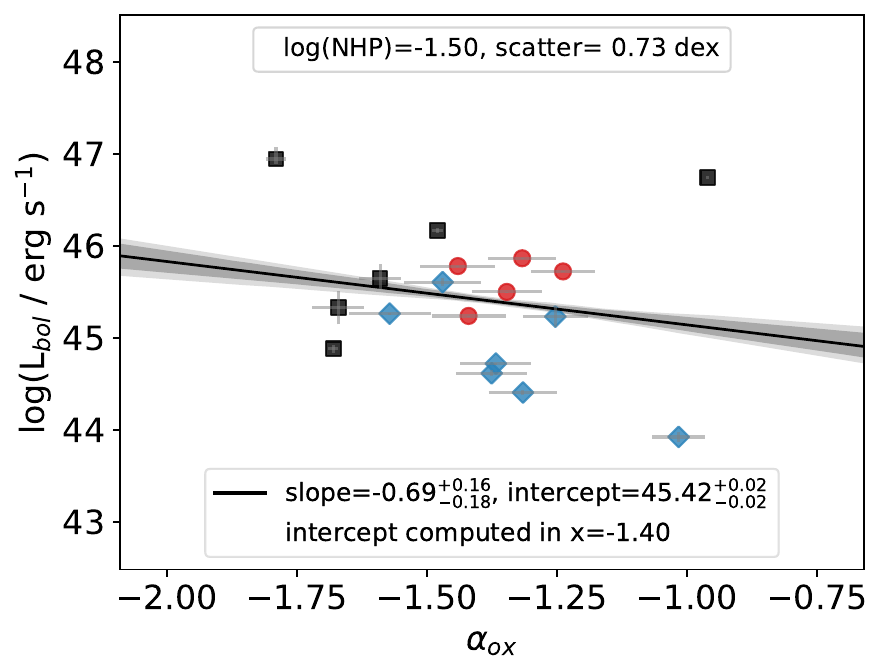}
    \end{subfigure}
    \begin{subfigure}[b]{0.4\textwidth}
        \includegraphics[width=\linewidth]{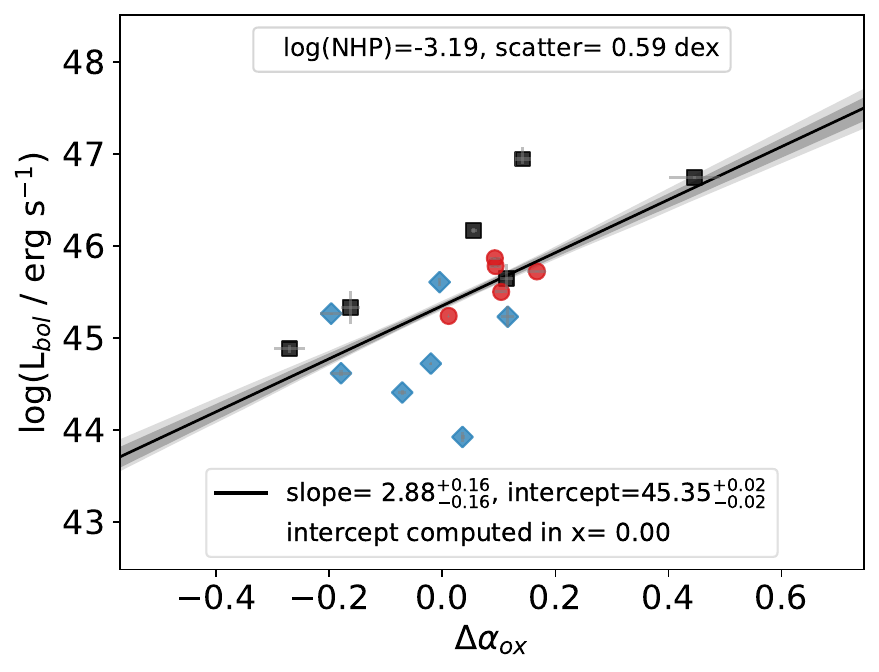}
    \end{subfigure}
    \begin{subfigure}[b]{0.25\textwidth}
        \includegraphics[width=\linewidth]{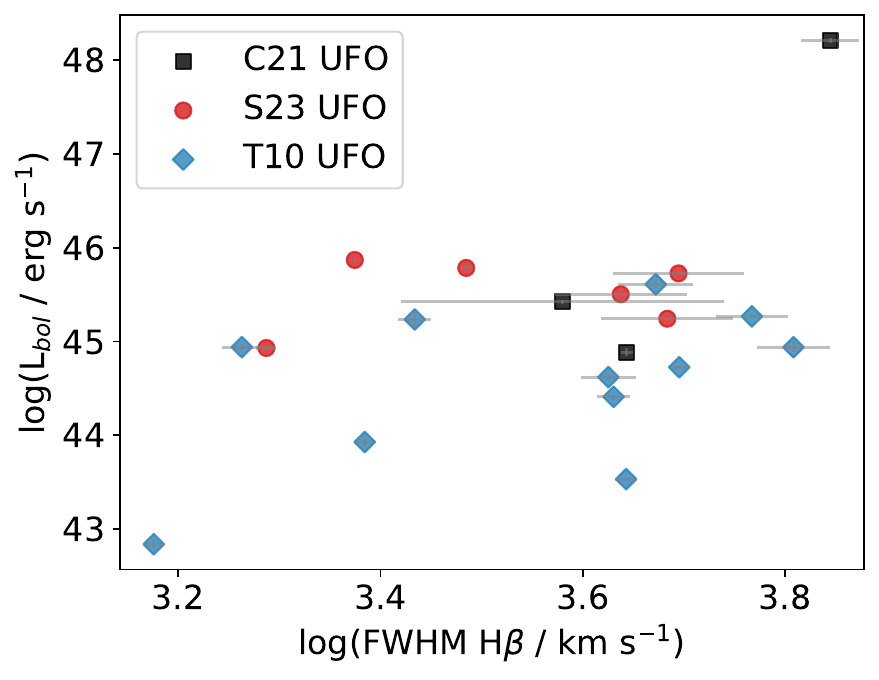}
    \end{subfigure}
    \begin{subfigure}[b]{0.255\textwidth}
        \includegraphics[width=\linewidth]{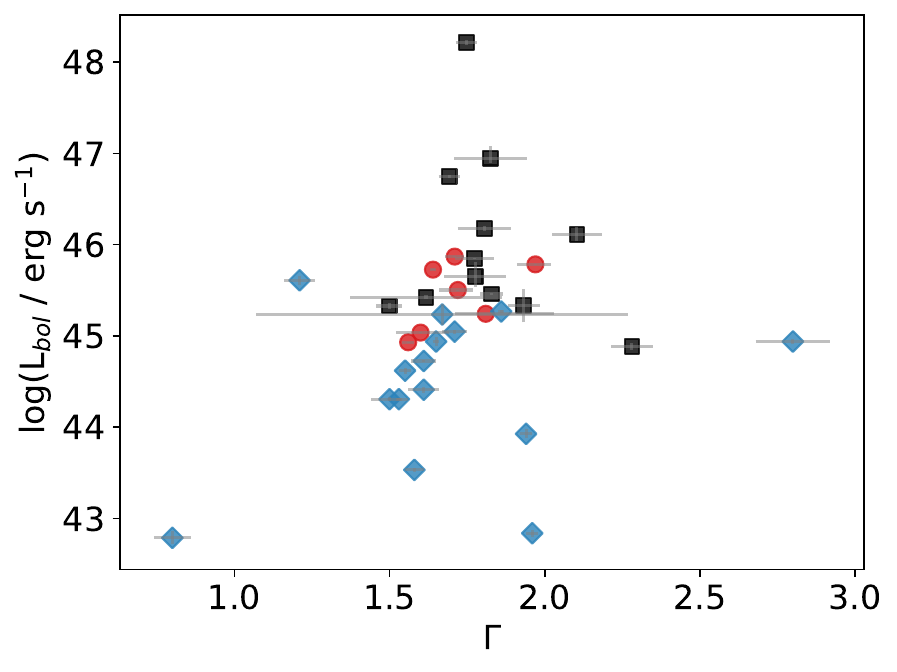}
    \end{subfigure}
    \begin{subfigure}[b]{0.255\textwidth}
        \includegraphics[width=\linewidth]{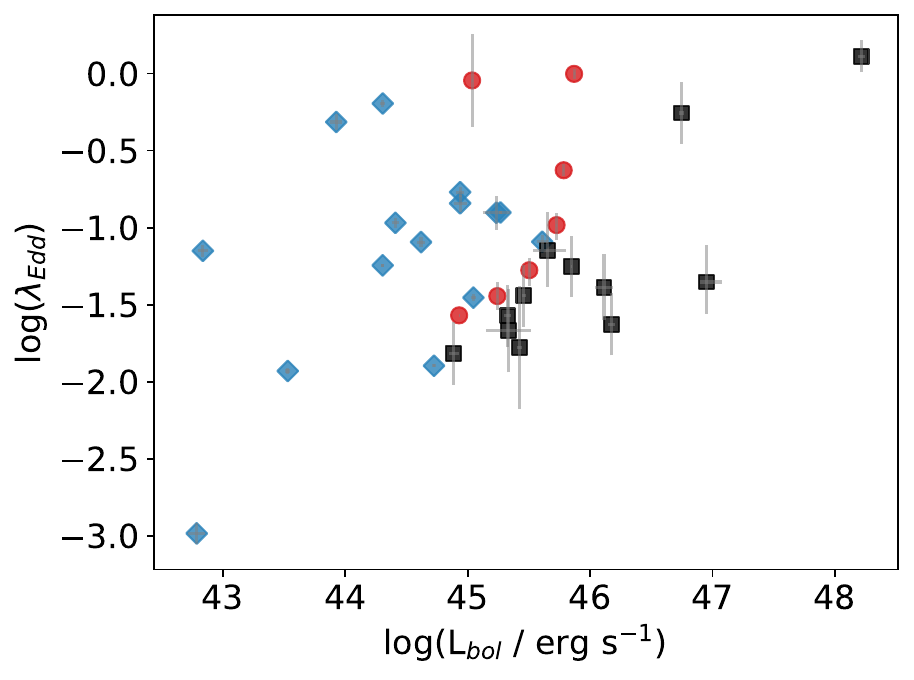}
    \end{subfigure}\\
    \caption{Bolometric luminosity versus AGN parameters. Significant and nonsignificant correlations for the S23, T10, and C21 samples. The best-fitting linear correlations, applied exclusively to statistically significant correlations, are presented by the solid black lines and the dark and light gray shadowed areas indicate the 68\% and 90\% confidence bands, respectively. In the legend, we report the best-fit coefficients, $\log\mathrm{NHP}$, and the intrinsic scatters for the correlations.}
    \label{fig:ef4}
\end{figure*}

\begin{figure*}
\centering
    \begin{subfigure}[b]{0.4\textwidth}
        \includegraphics[width=\linewidth]{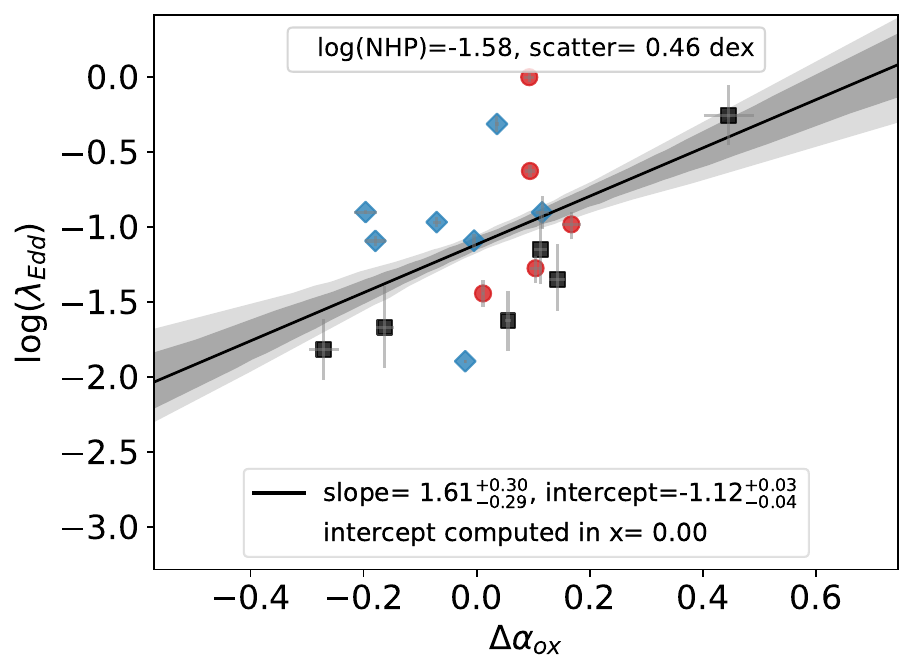}
    \end{subfigure}
    \begin{subfigure}[b]{0.4\textwidth}
        \includegraphics[width=\linewidth]{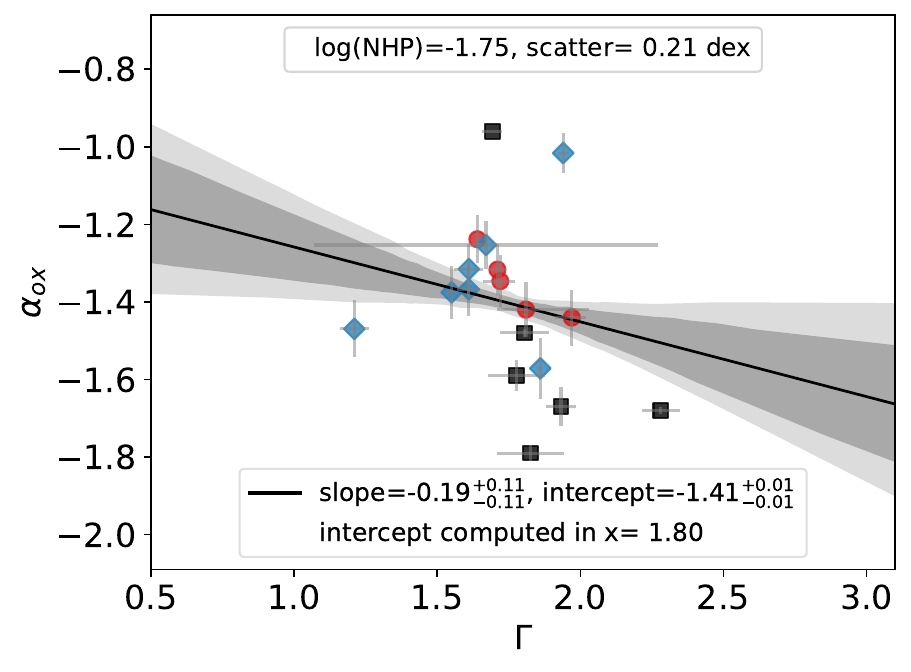}
    \end{subfigure}
    \begin{subfigure}[b]{0.25\textwidth}
        \includegraphics[width=\linewidth]{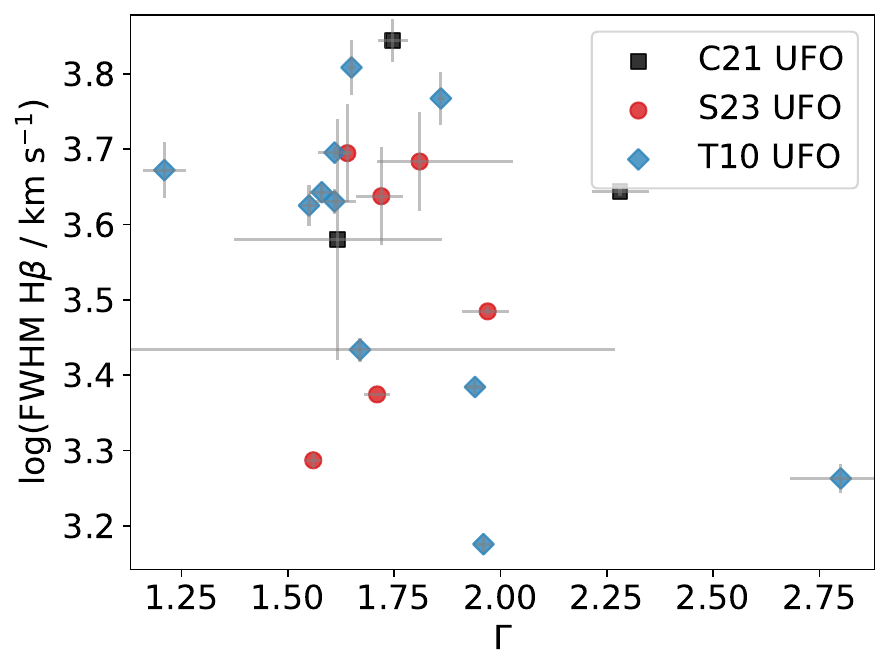}
    \end{subfigure}
    \begin{subfigure}[b]{0.25\textwidth}
        \includegraphics[width=\linewidth]{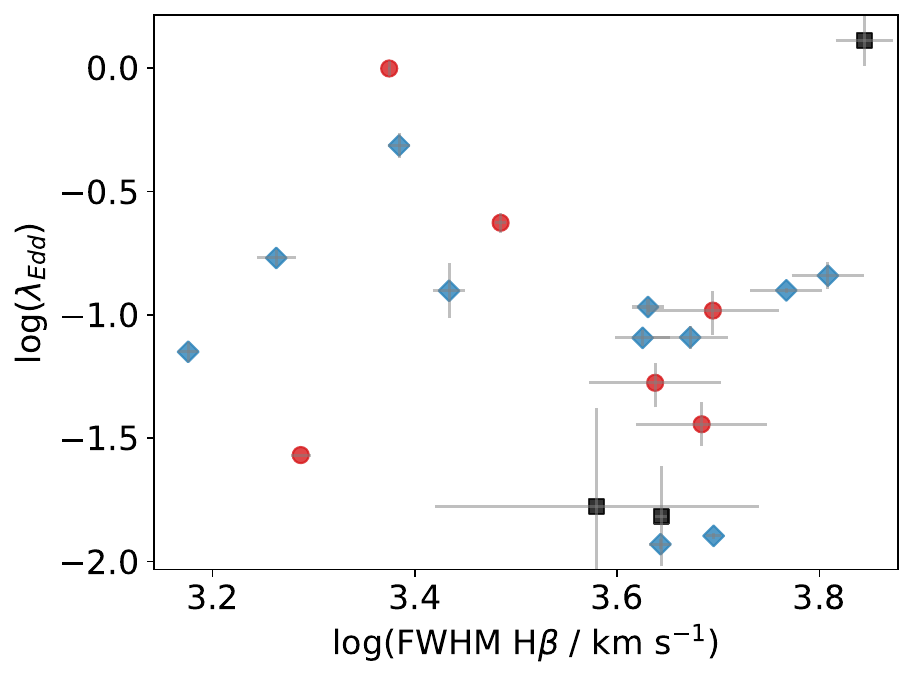}
    \end{subfigure}
    \begin{subfigure}[b]{0.25\textwidth}
        \includegraphics[width=\linewidth]{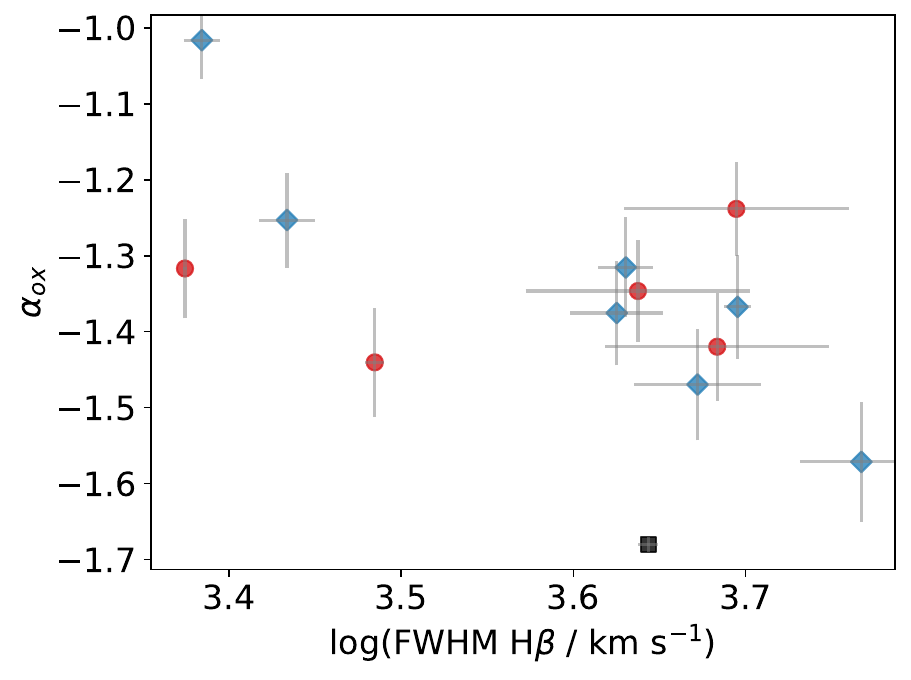}
    \end{subfigure}
    \begin{subfigure}[b]{0.25\textwidth}
        \includegraphics[width=\linewidth]{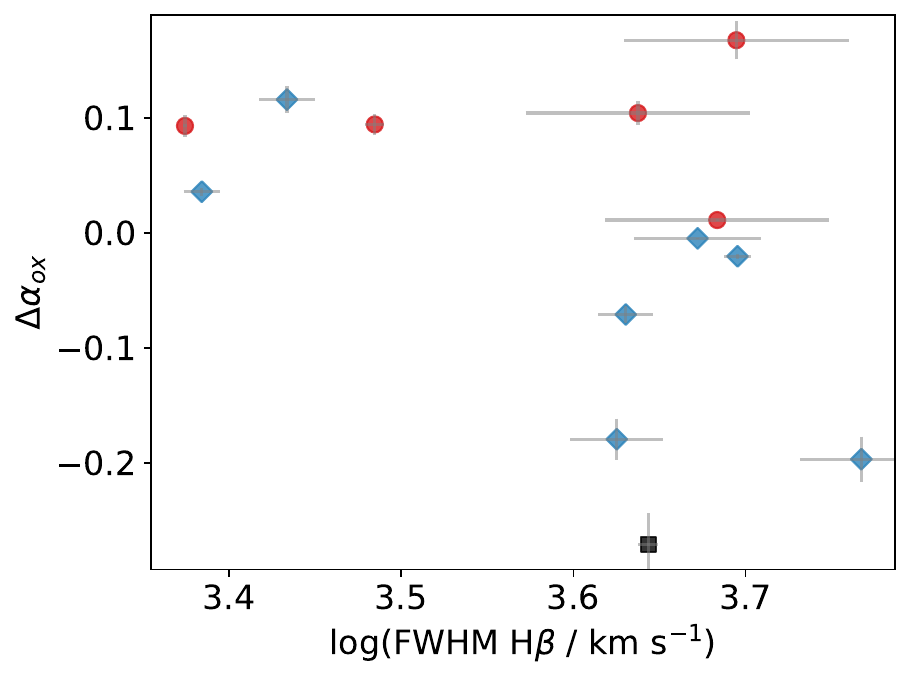}
    \end{subfigure}
    \begin{subfigure}[b]{0.25\textwidth}
        \includegraphics[width=\linewidth]{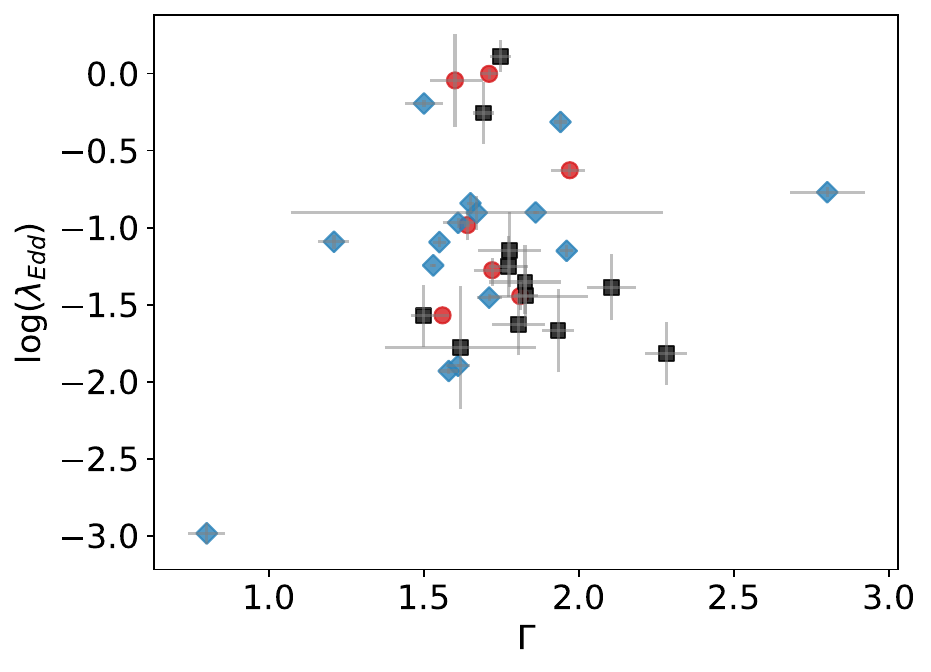}
    \end{subfigure}
    
    \begin{subfigure}[b]{0.25\textwidth}
        \includegraphics[width=\linewidth]{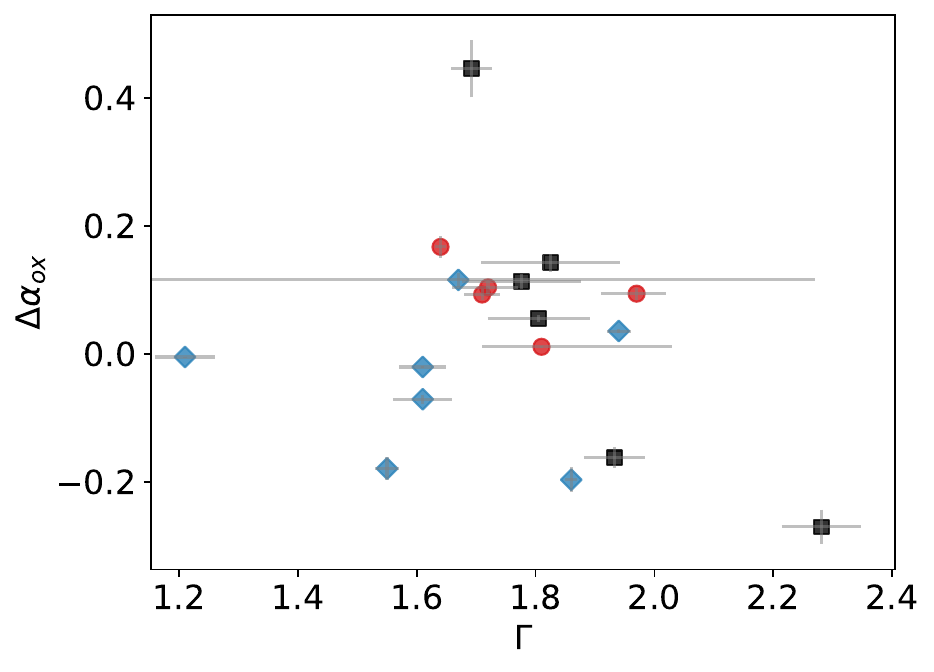}
    \end{subfigure}
    \begin{subfigure}[b]{0.25\textwidth}
        \includegraphics[width=\linewidth]{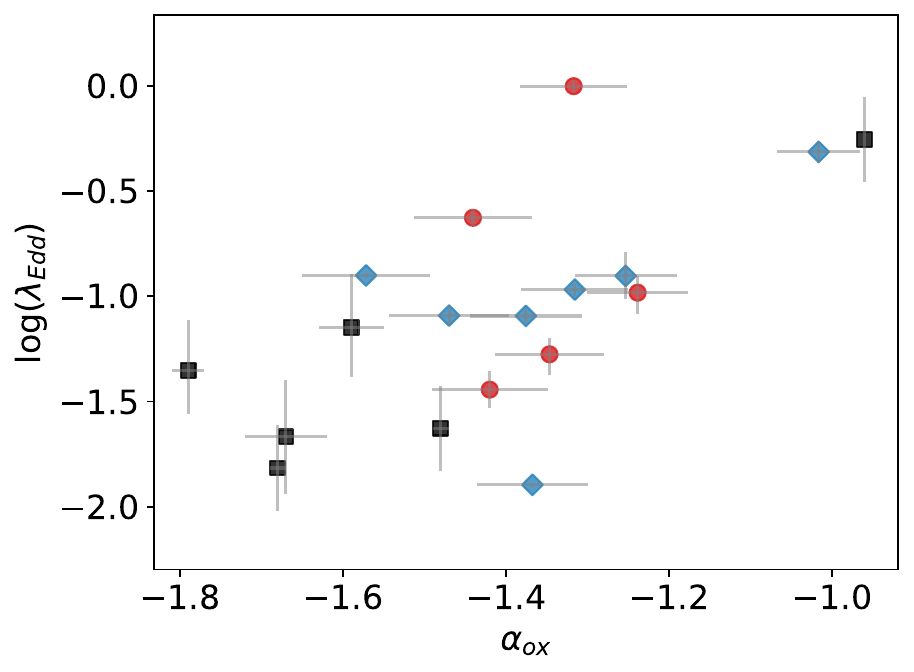}
    \end{subfigure}
    \begin{subfigure}[b]{0.25\textwidth}
        \includegraphics[width=\linewidth]{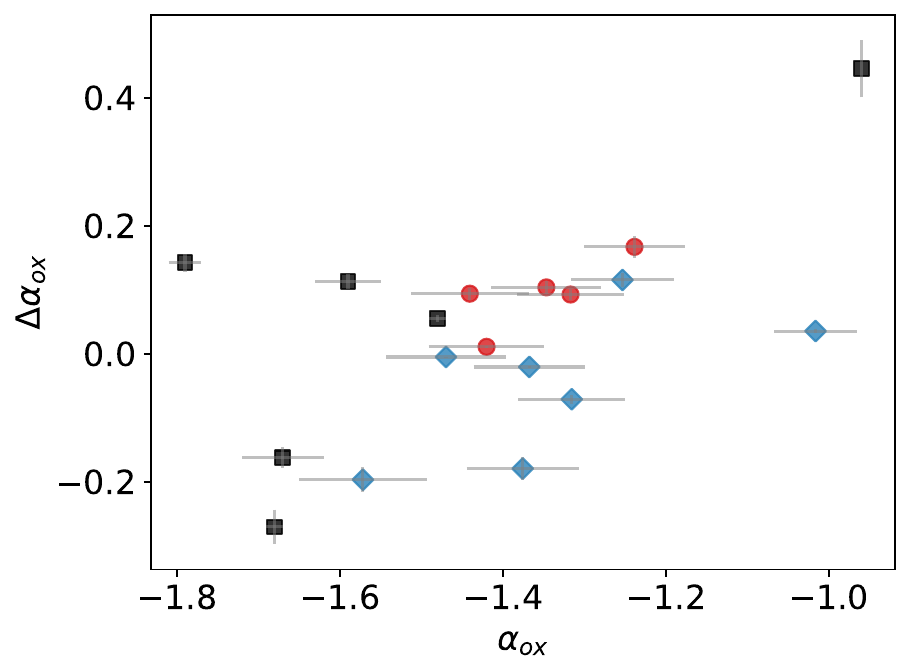}
    \end{subfigure}\\
    \caption{SED parameters versus AGN parameters. Significant and nonsignificant correlations for the S23, T10, and C21 samples. The best-fitting linear correlations, applied exclusively to statistically significant correlations, are presented by the solid black lines and the dark and light gray shadowed areas indicate the 68\% and 90\% confidence bands, respectively. In the legend, we report the best-fit coefficients, $\log\mathrm{NHP}$, and the intrinsic scatters for the correlations.}
    \label{fig:ef5}
\end{figure*}

\begin{figure*}
\centering
    \begin{subfigure}[b]{0.41\textwidth}
        \includegraphics[width=\linewidth]{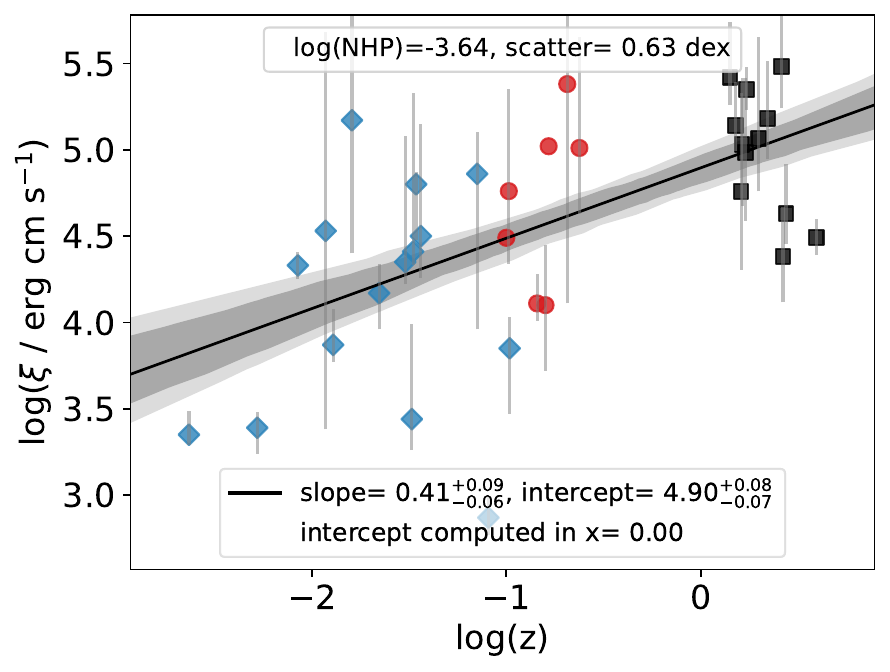}
    \end{subfigure}
    \begin{subfigure}[b]{0.4\textwidth}
        \includegraphics[width=\linewidth]{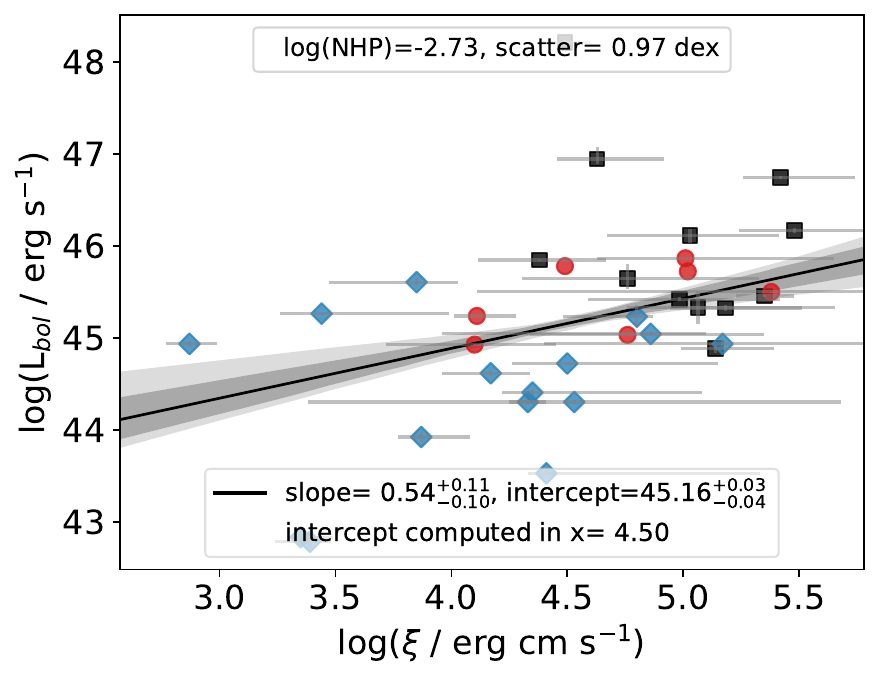}
    \end{subfigure}
    \begin{subfigure}[b]{0.25\textwidth}
        \includegraphics[width=\linewidth]{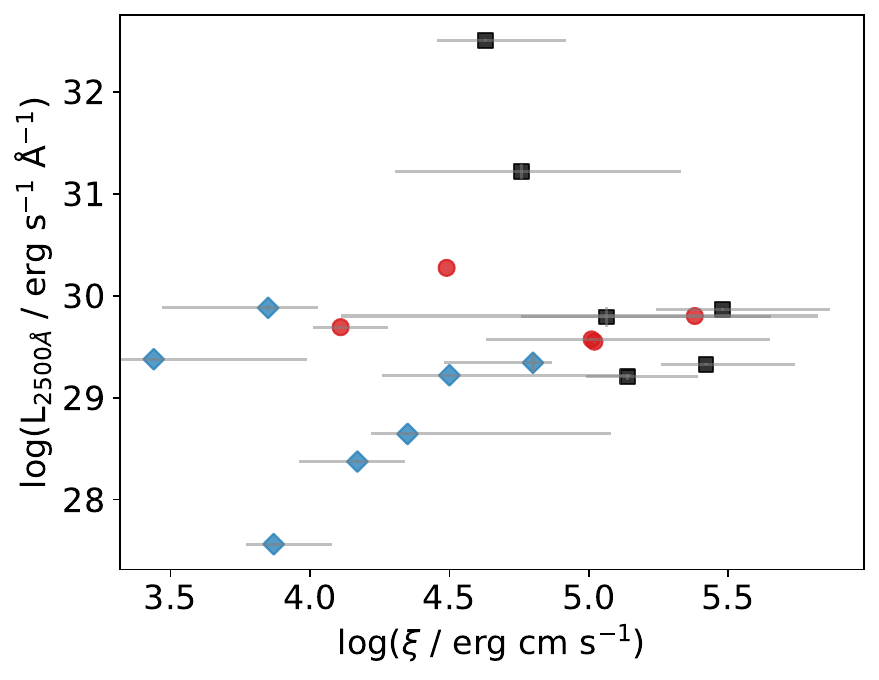}
    \end{subfigure}
    \begin{subfigure}[b]{0.25\textwidth}
        \includegraphics[width=\linewidth]{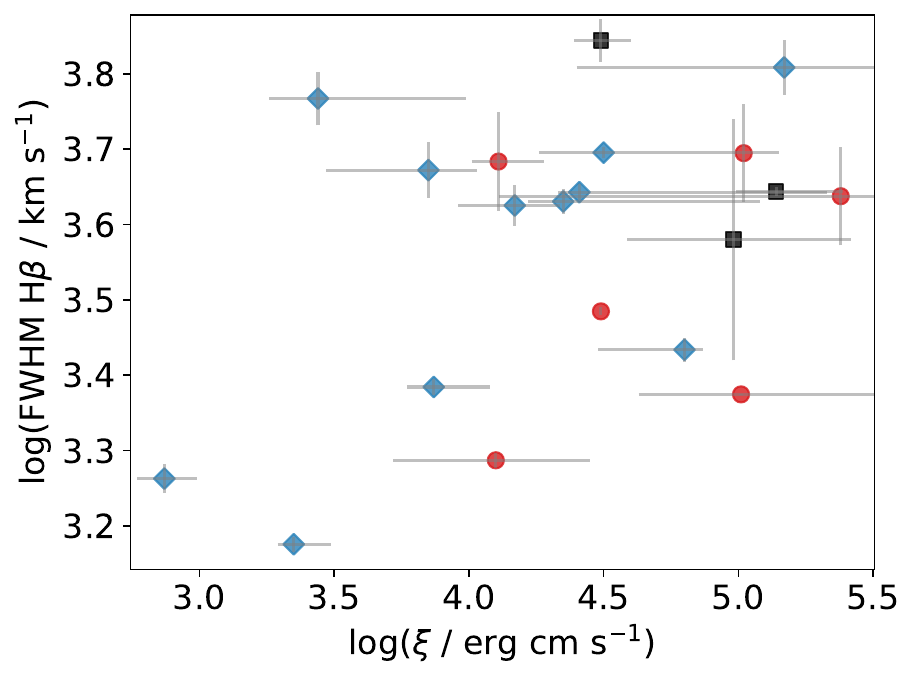}
    \end{subfigure}
    \begin{subfigure}[b]{0.25\textwidth}
        \includegraphics[width=\linewidth]{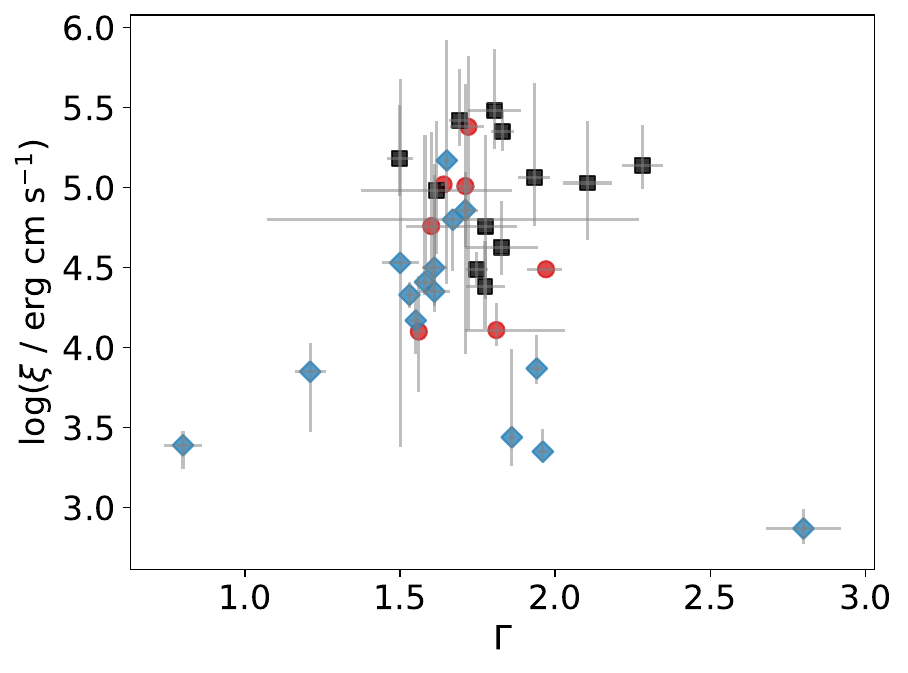}
    \end{subfigure}
    \begin{subfigure}[b]{0.25\textwidth}
        \includegraphics[width=\linewidth]{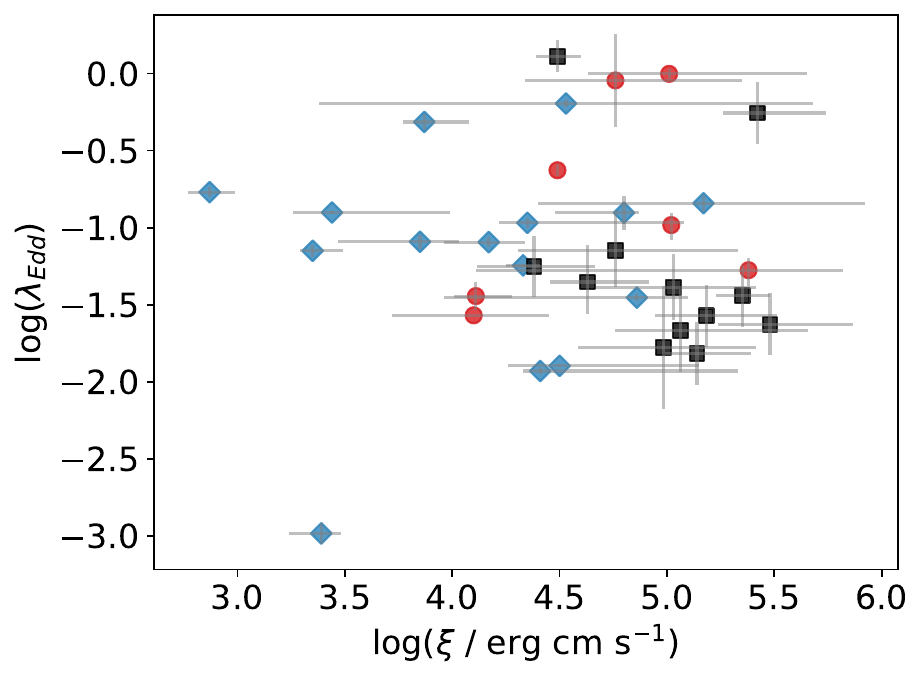}
    \end{subfigure}
    \begin{subfigure}[b]{0.25\textwidth}
        \includegraphics[width=\linewidth]{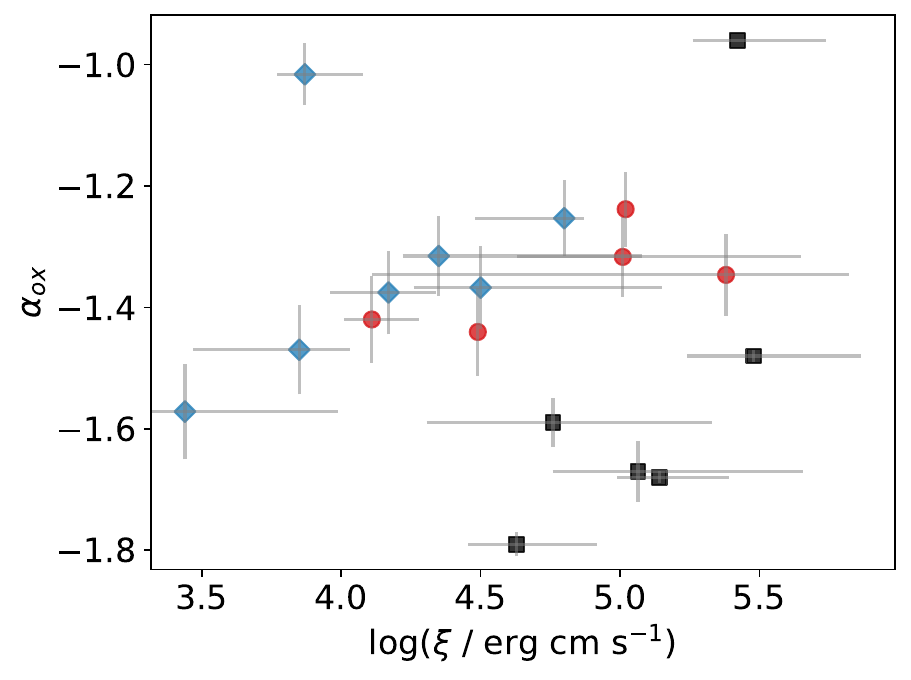}
    \end{subfigure}
    \begin{subfigure}[b]{0.25\textwidth}
        \includegraphics[width=\linewidth]{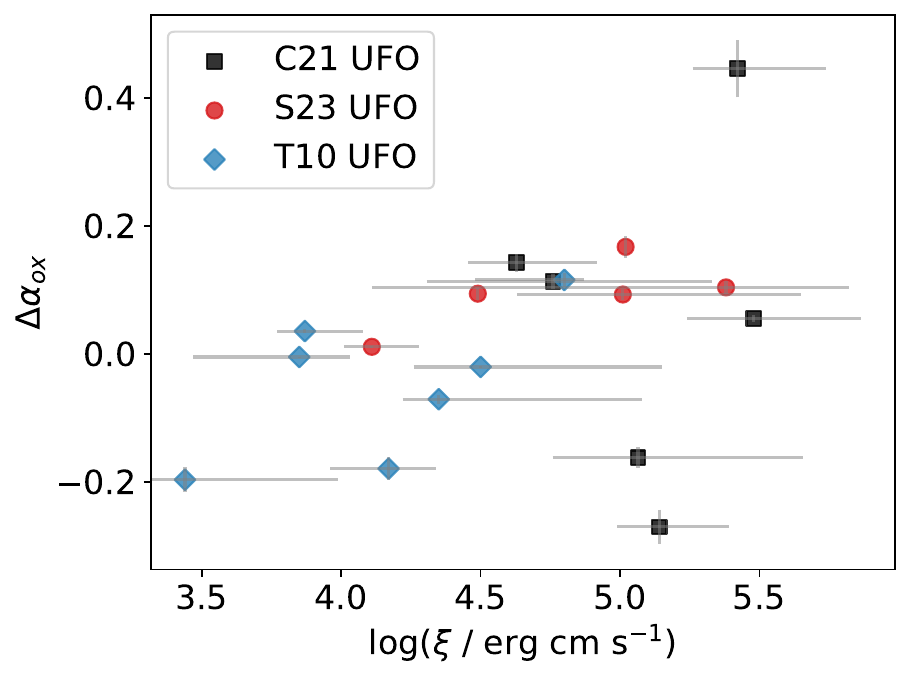}
    \end{subfigure}\\

\caption[]{{UFO ionization parameter versus AGN parameters. Significant and nonsignificant correlations for the S23, T10, and C21 samples. The best-fitting linear correlations, applied exclusively to statistically significant correlations, are presented by the solid black lines and the dark and light gray shadowed areas indicate the 68\% and 90\% confidence bands, respectively. In the legend, we report the best-fit coefficients, $\log\mathrm{NHP}$, and the intrinsic scatters for the correlations.}}
\label{fig:ef6}
\end{figure*}

\begin{figure*}
\centering
    \begin{subfigure}[b]{0.35\textwidth}
        \includegraphics[width=\linewidth]{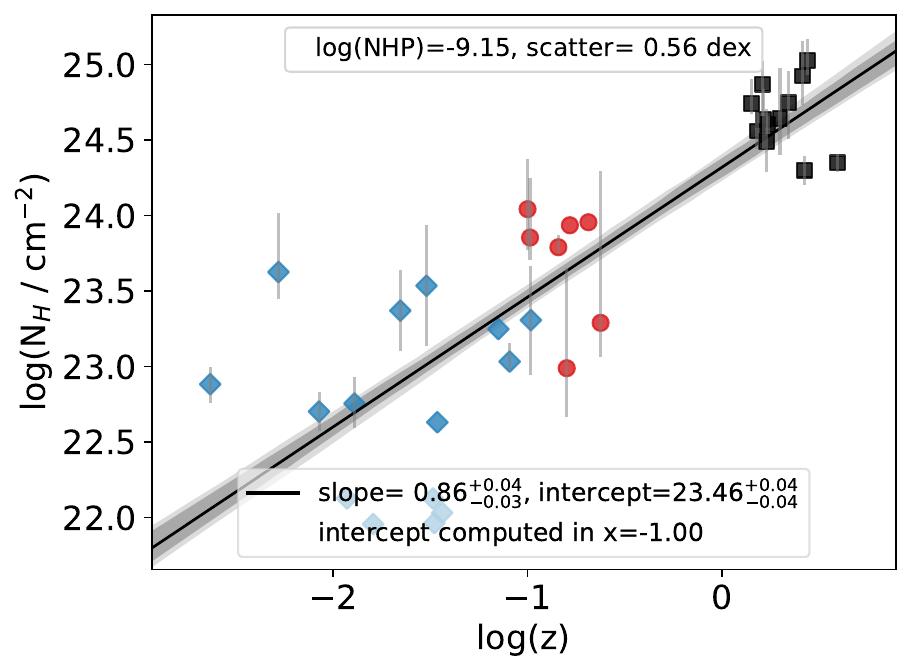}
    \end{subfigure}
    \begin{subfigure}[b]{0.35\textwidth}
        \includegraphics[width=\linewidth]{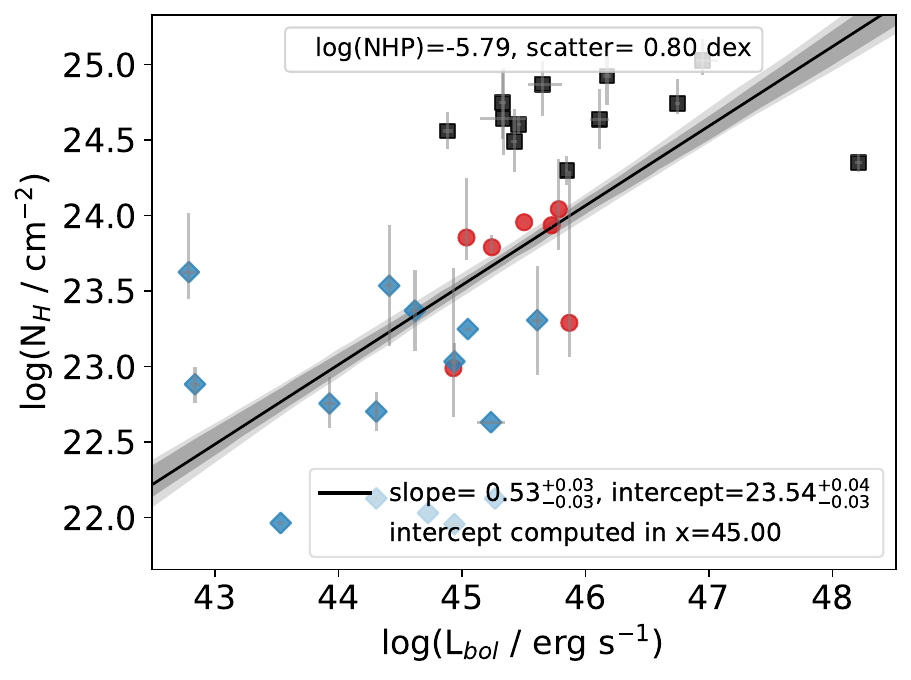}
    \end{subfigure}
    \begin{subfigure}[b]{0.35\textwidth}
        \includegraphics[width=\linewidth]{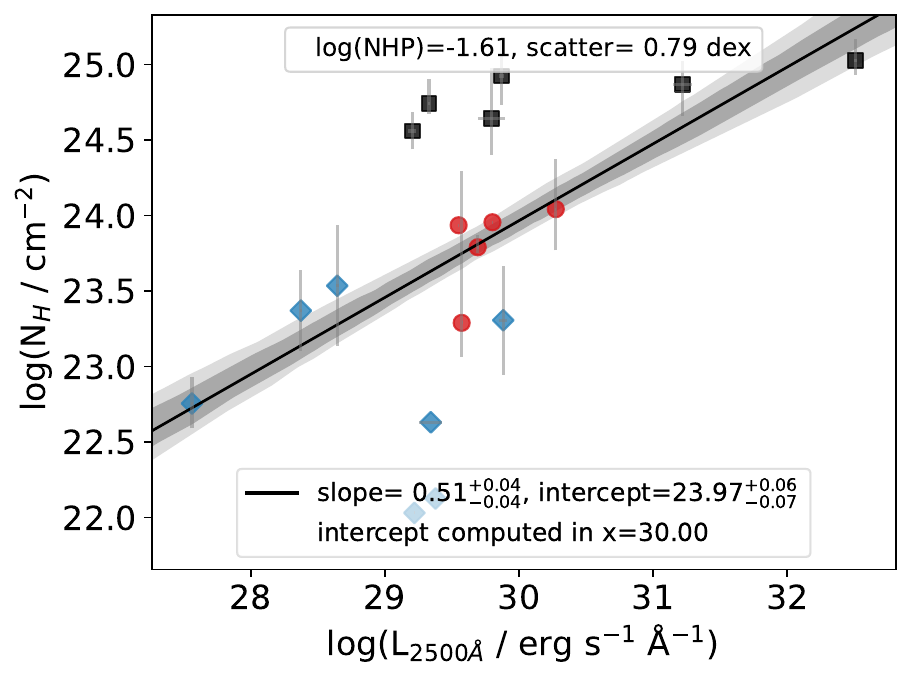}
    \end{subfigure}
    \begin{subfigure}[b]{0.35\textwidth}
        \includegraphics[width=\linewidth]{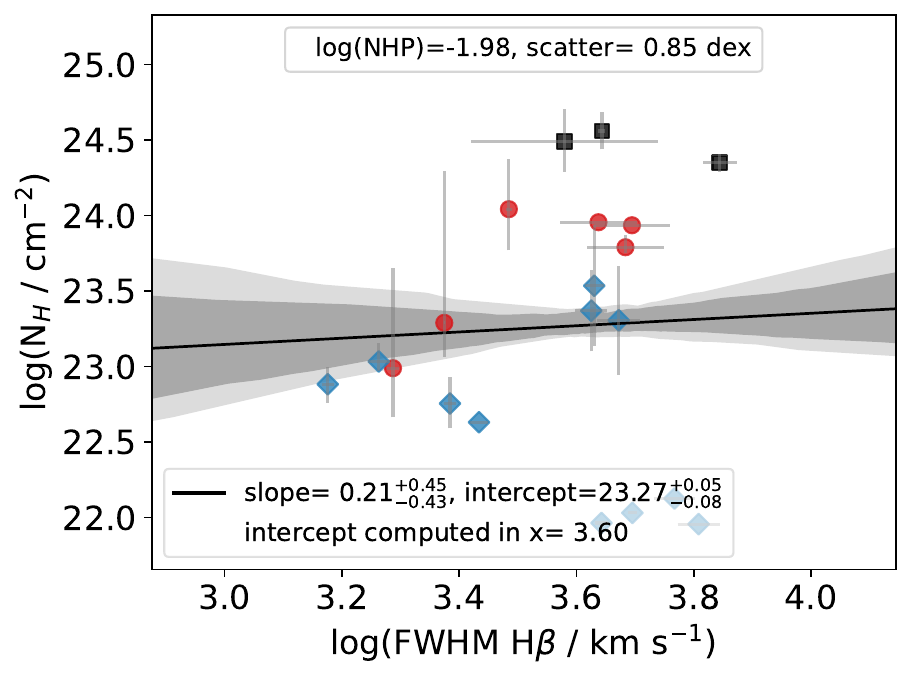}
    \end{subfigure}
    \begin{subfigure}[b]{0.35\textwidth}
        \includegraphics[width=\linewidth]{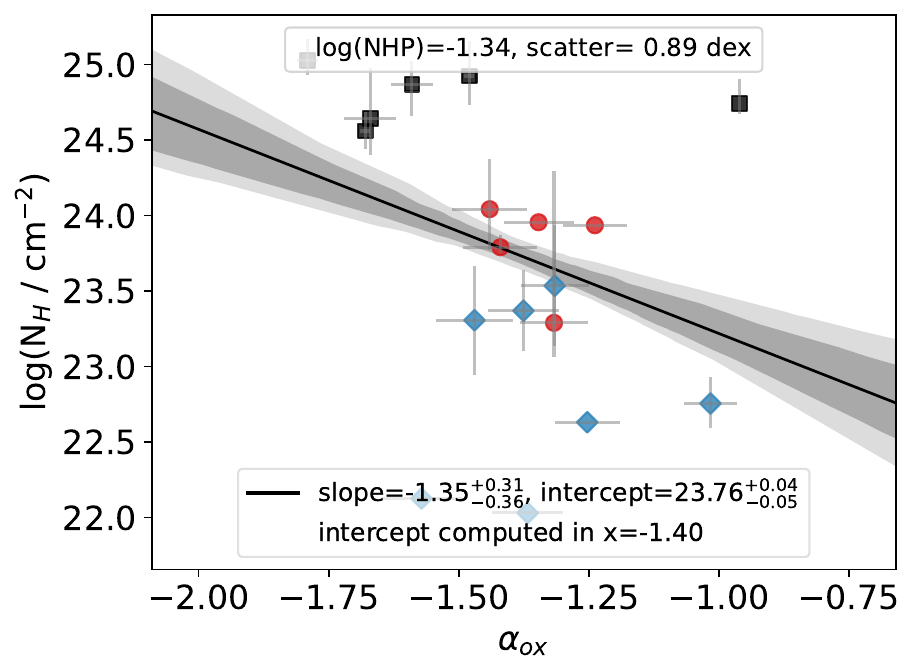}
    \end{subfigure}\\
    \begin{subfigure}[b]{0.25\textwidth}
        \includegraphics[width=\linewidth]{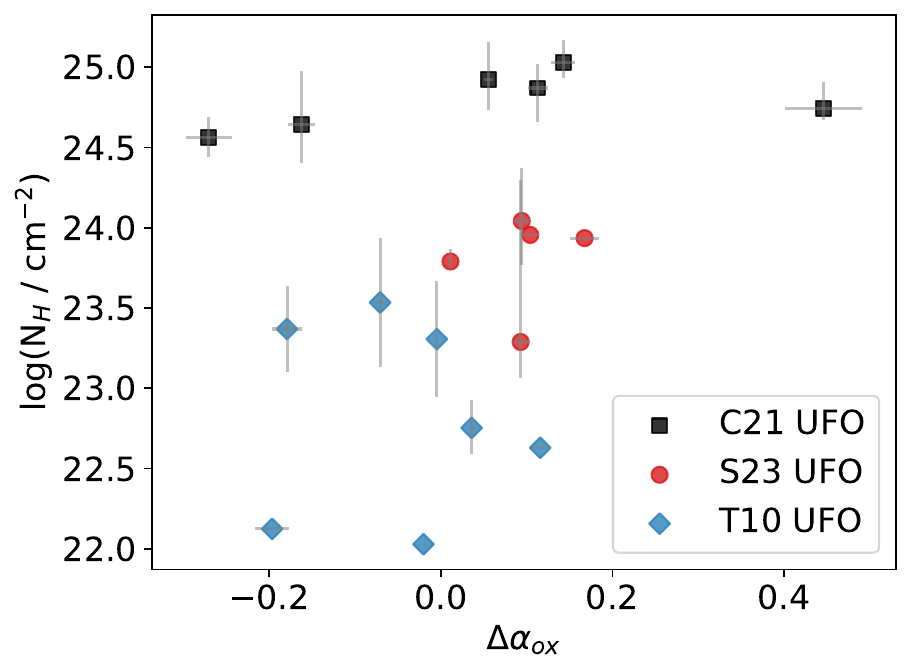}
    \end{subfigure}    
    \begin{subfigure}[b]{0.25\textwidth}
        \includegraphics[width=\linewidth]{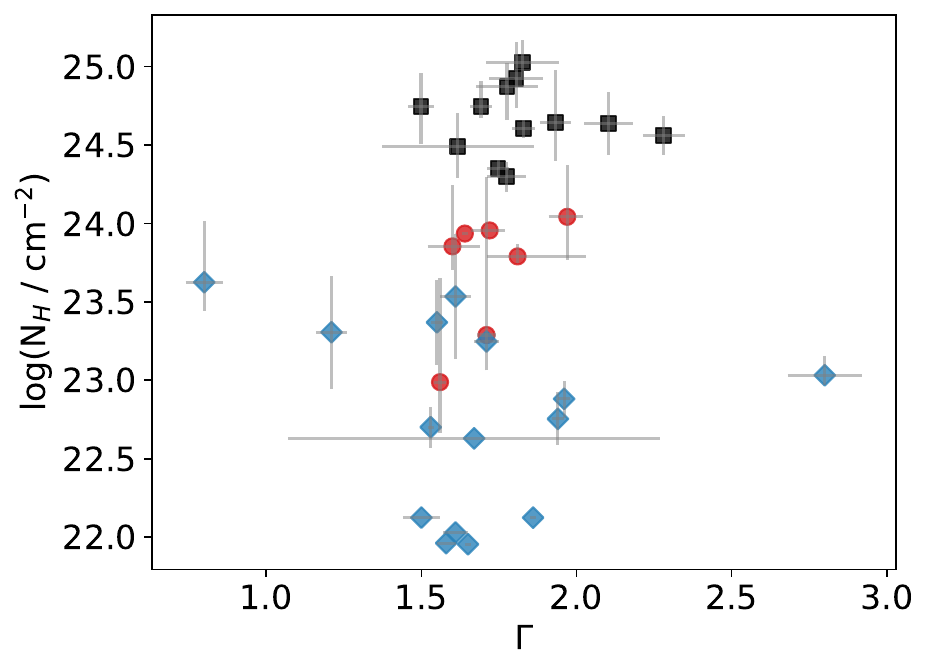}
    \end{subfigure}
    \begin{subfigure}[b]{0.25\textwidth}
        \includegraphics[width=\linewidth]{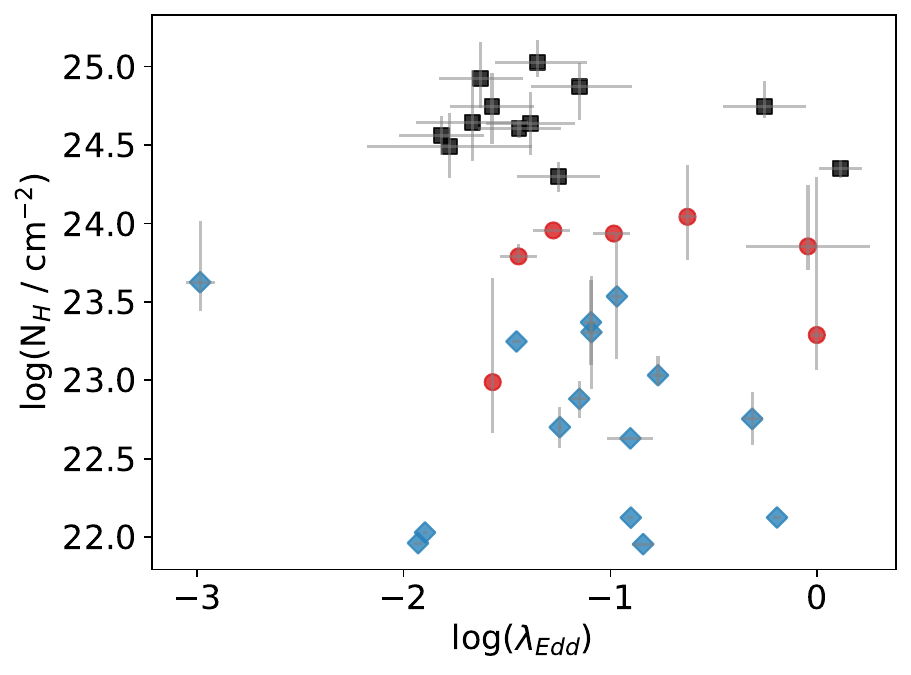}
    \end{subfigure}\\

\caption[]{{UFO column density versus AGN parameters. Significant and nonsignificant correlations for the S23, T10, and C21 samples. The best-fitting linear correlations, applied exclusively to statistically significant correlations, are presented by the solid black lines and the dark and light gray shadowed areas indicate the 68\% and 90\% confidence bands, respectively. In the legend, we report the best-fit coefficients, $\log\mathrm{NHP}$, and the intrinsic scatters for the correlations.}}
\label{fig:ef7}
\end{figure*}

\begin{figure*}
\centering
    \begin{subfigure}[b]{0.4\textwidth}
        \includegraphics[width=\linewidth]{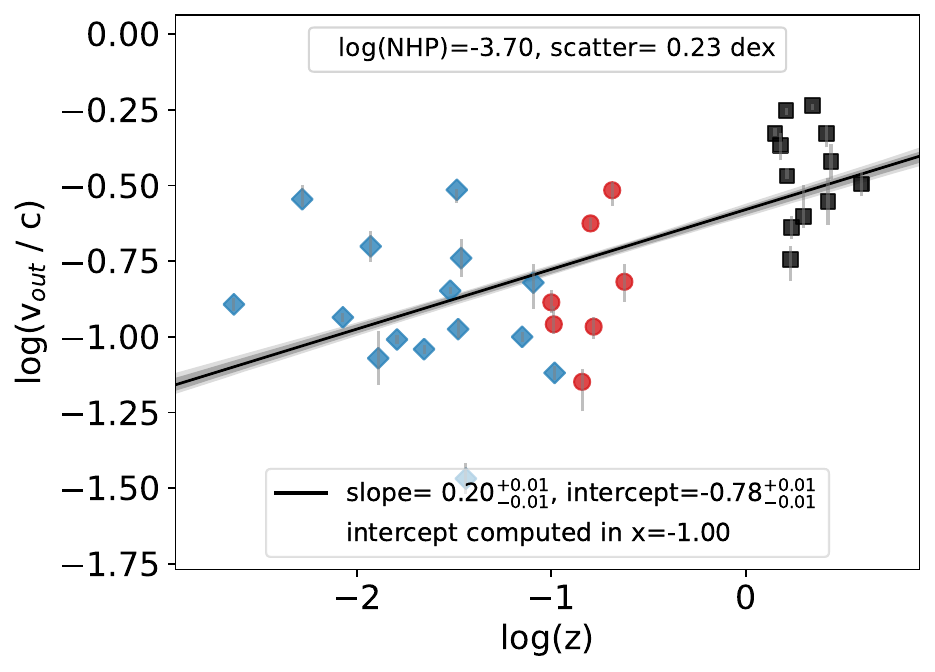}
    \end{subfigure}
    \begin{subfigure}[b]{0.375\textwidth}
        \includegraphics[width=\linewidth]{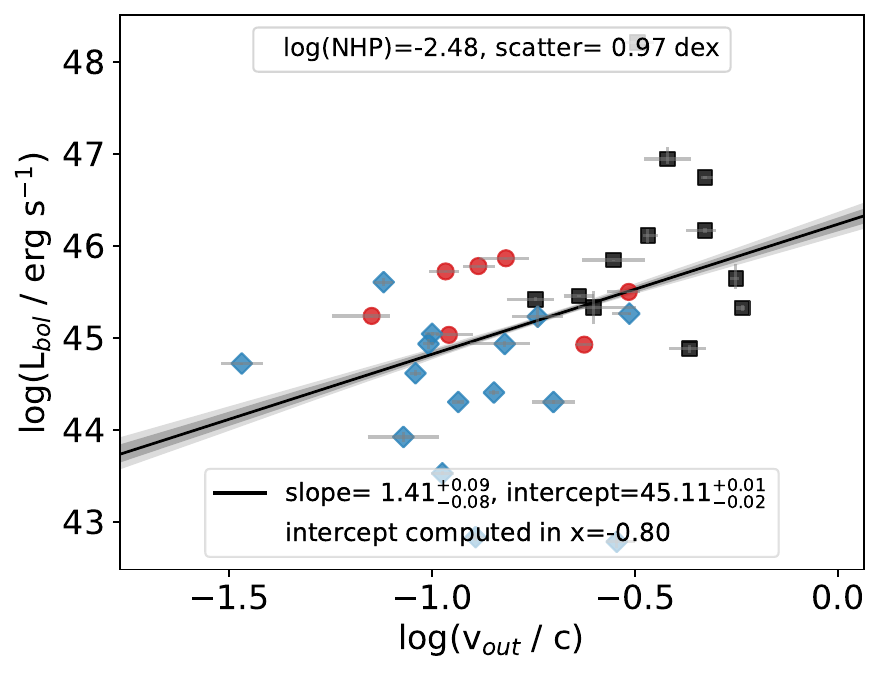}
    \end{subfigure}
    \begin{subfigure}[b]{0.4\textwidth}
        \includegraphics[width=\linewidth]{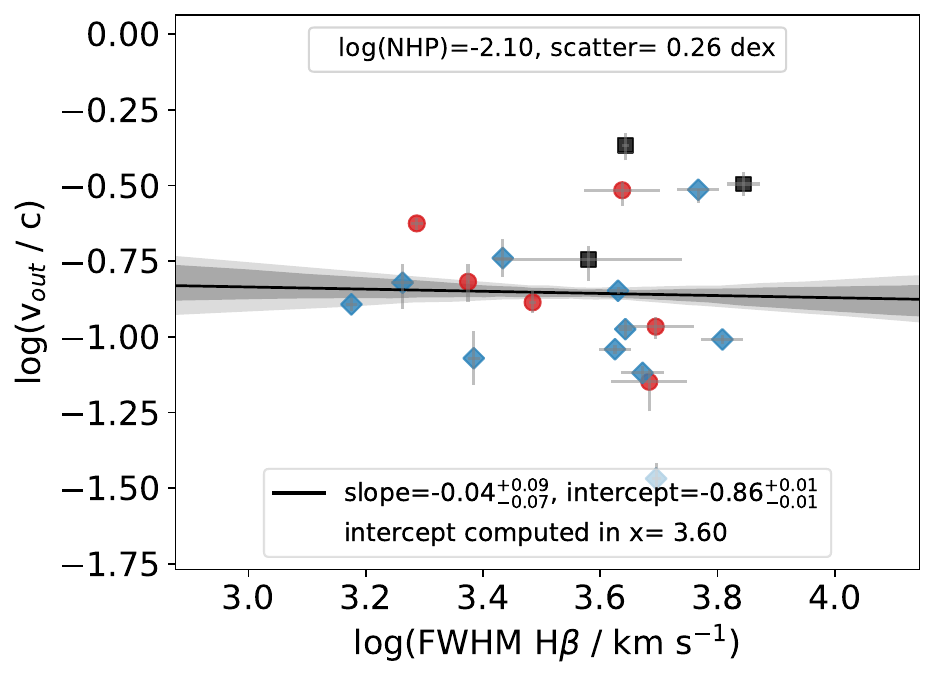}
    \end{subfigure}
    \begin{subfigure}[b]{0.25\textwidth}
        \includegraphics[width=\linewidth]{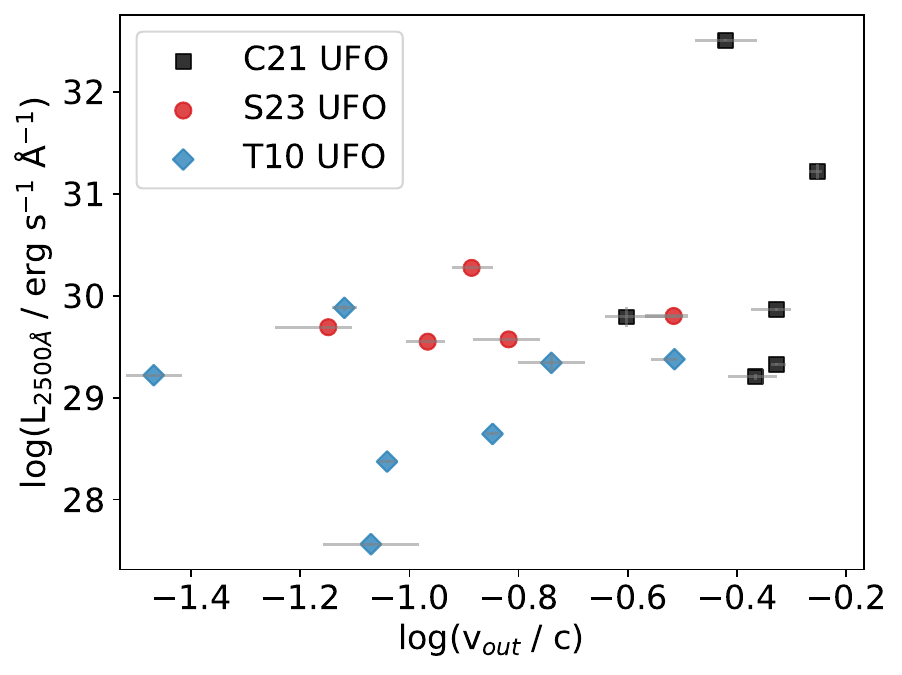}
    \end{subfigure}
    \begin{subfigure}[b]{0.25\textwidth}
        \includegraphics[width=\linewidth]{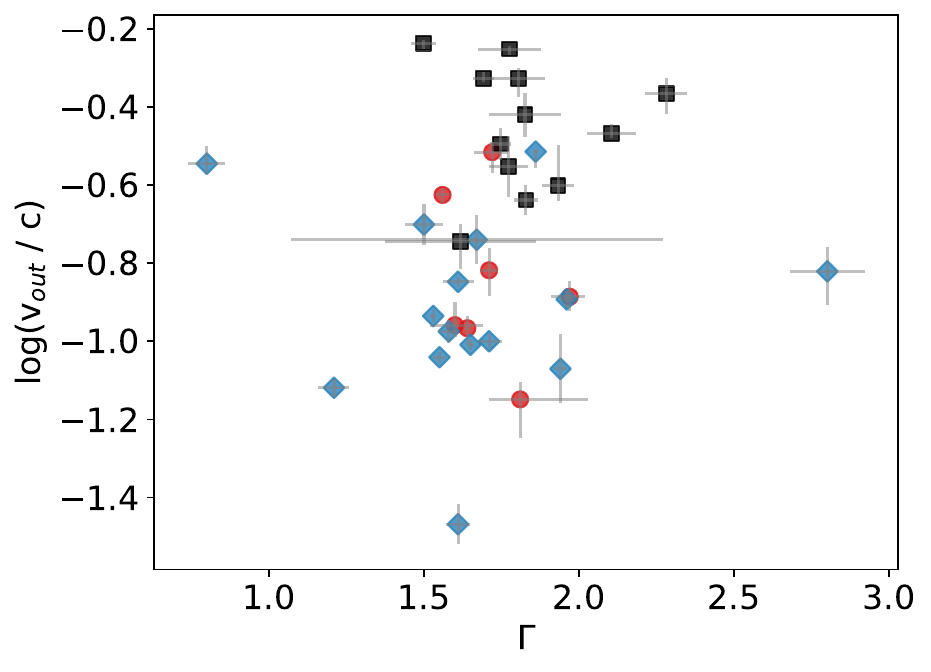}
    \end{subfigure}
    \begin{subfigure}[b]{0.25\textwidth}
        \includegraphics[width=\linewidth]{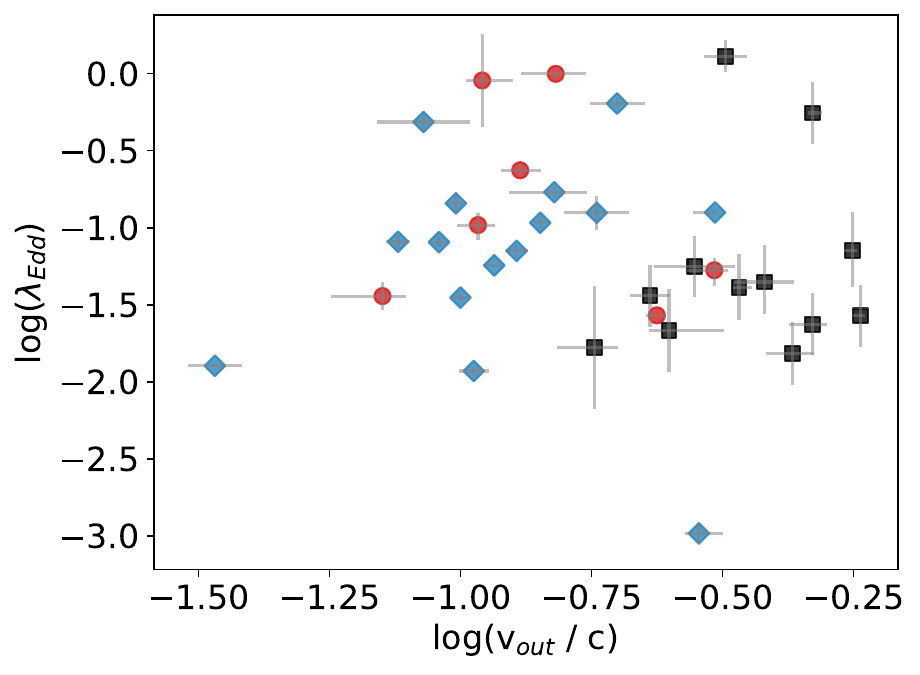}
    \end{subfigure}
    \begin{subfigure}[b]{0.25\textwidth}
        \includegraphics[width=\linewidth]{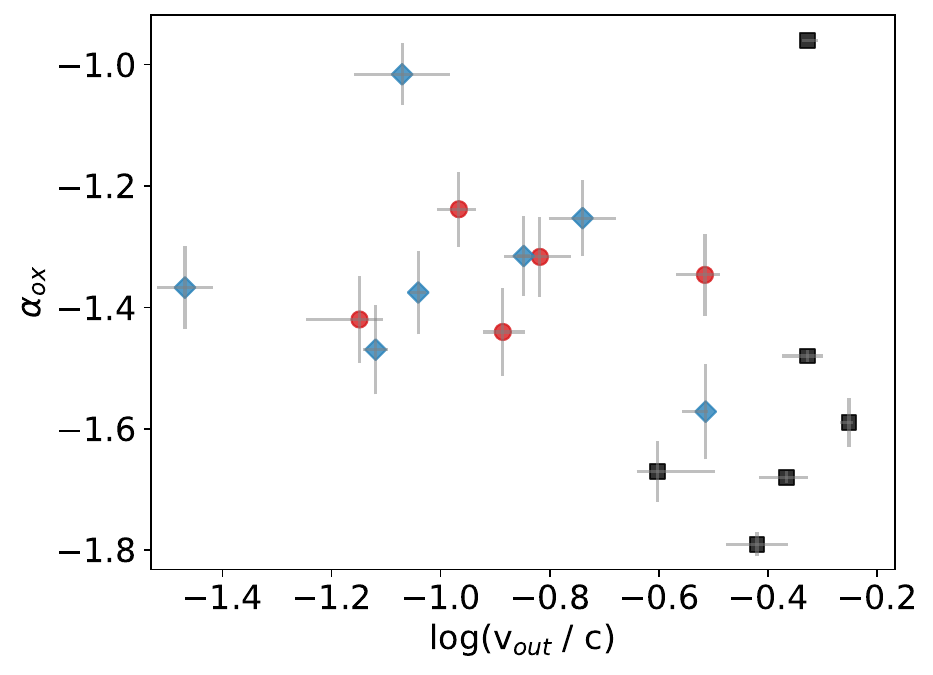}
    \end{subfigure}
    \begin{subfigure}[b]{0.25\textwidth}
        \includegraphics[width=\linewidth]{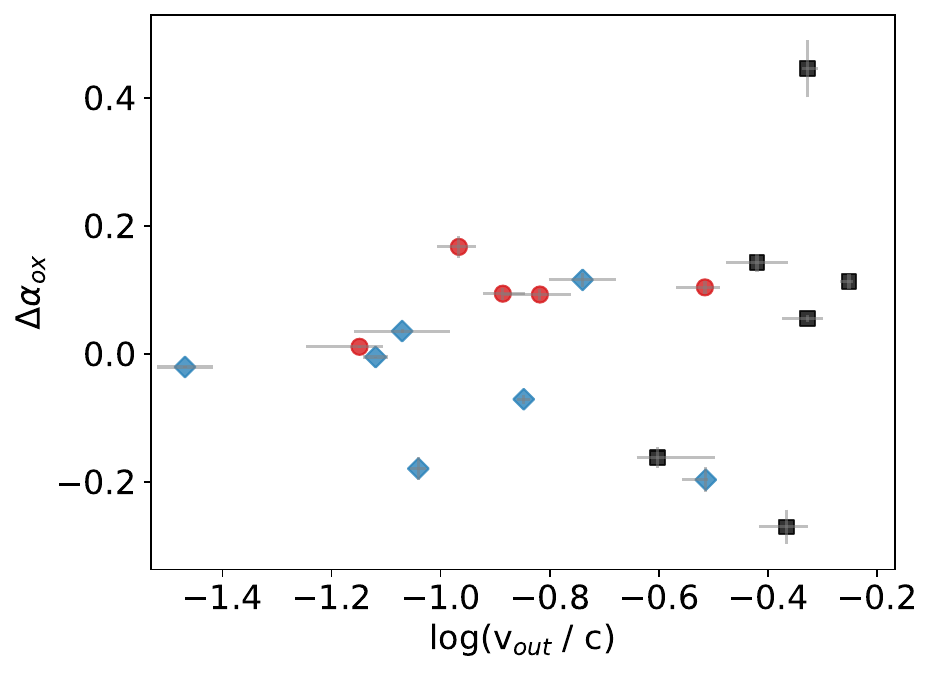}
    \end{subfigure}\\

\caption[]{{UFO outflow velocity versus AGN parameters. Significant and nonsignificant correlations for the S23, T10, and C21 samples. The best-fitting linear correlations, applied exclusively to statistically significant correlations, are presented by the solid black lines and the dark and light gray shadowed areas indicate the 68\% and 90\% confidence bands, respectively. In the legend, we report the best-fit coefficients, $\log\mathrm{NHP}$, and the intrinsic scatters for the correlations.}}
\label{fig:ef8}
\end{figure*}

\begin{figure*}
\centering
    \begin{subfigure}[b]{0.4\textwidth}
        \includegraphics[width=\linewidth]{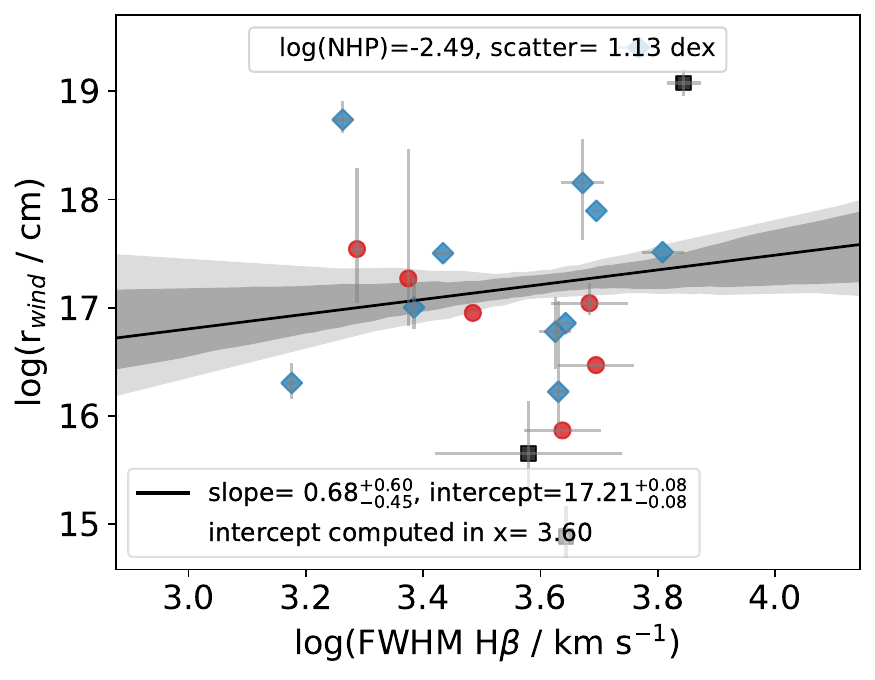}
    \end{subfigure}
    \begin{subfigure}[b]{0.4\textwidth}
        \includegraphics[width=\linewidth]{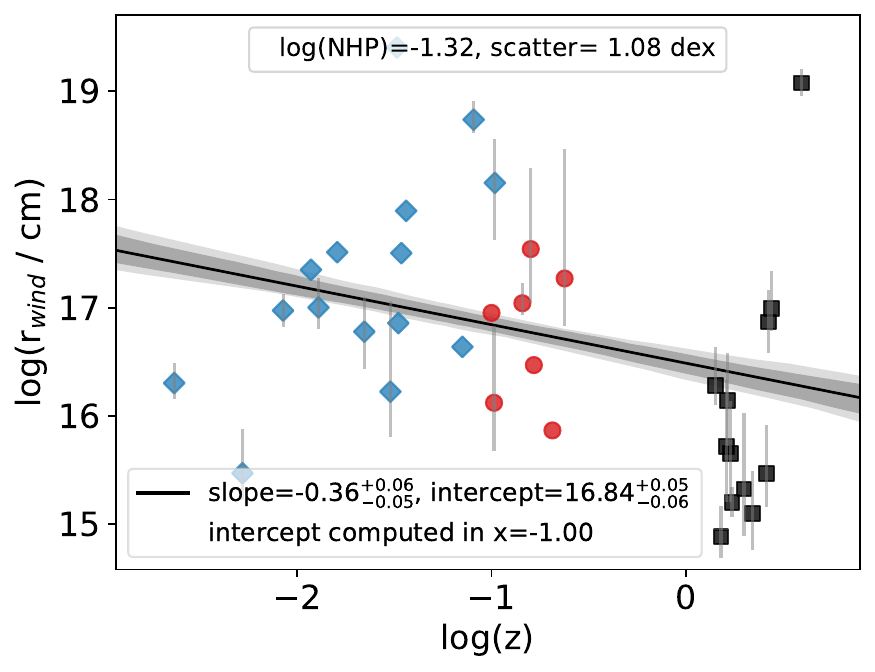}
    \end{subfigure}
    \begin{subfigure}[b]{0.25\textwidth}
        \includegraphics[width=\linewidth]{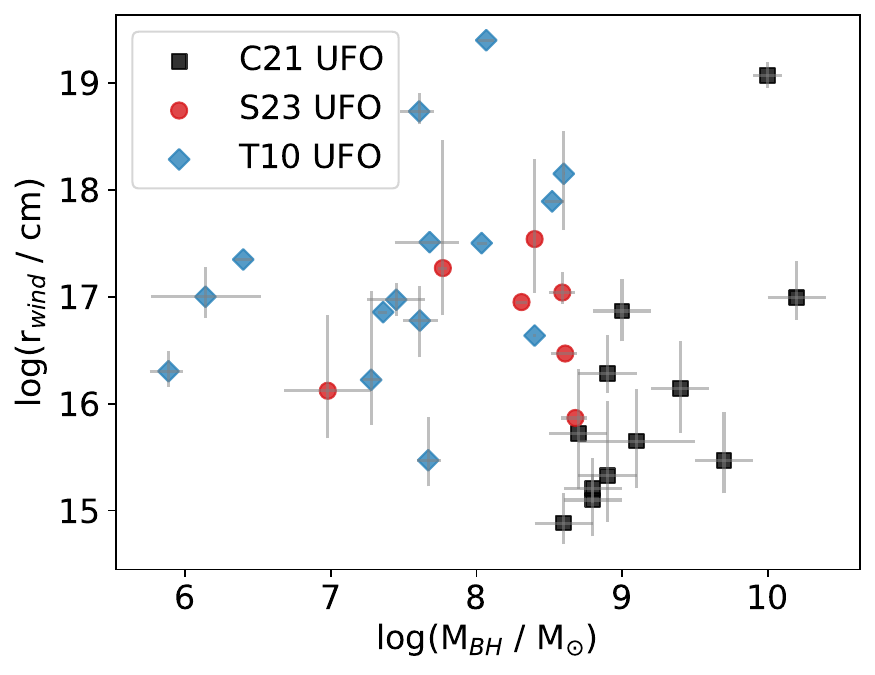}
    \end{subfigure}
    \begin{subfigure}[b]{0.25\textwidth}
        \includegraphics[width=\linewidth]{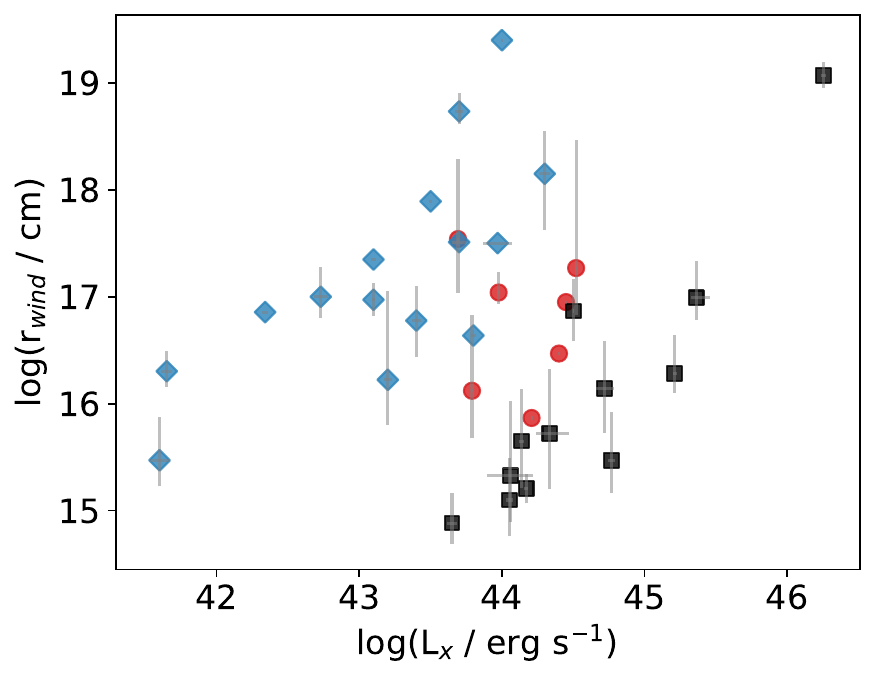}
    \end{subfigure}
    \begin{subfigure}[b]{0.25\textwidth}
        \includegraphics[width=\linewidth]{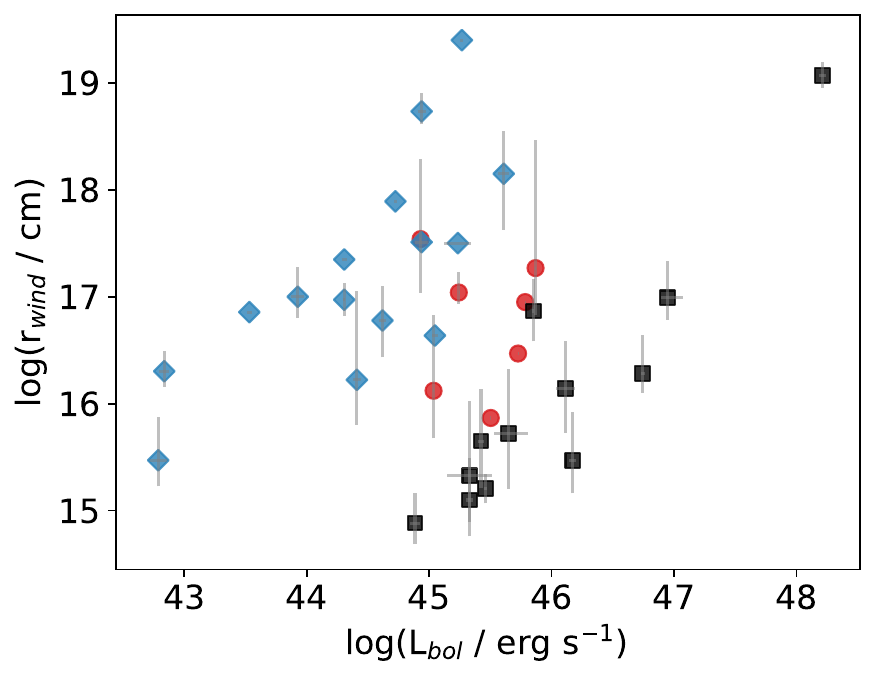}
    \end{subfigure}
    \begin{subfigure}[b]{0.25\textwidth}
        \includegraphics[width=\linewidth]{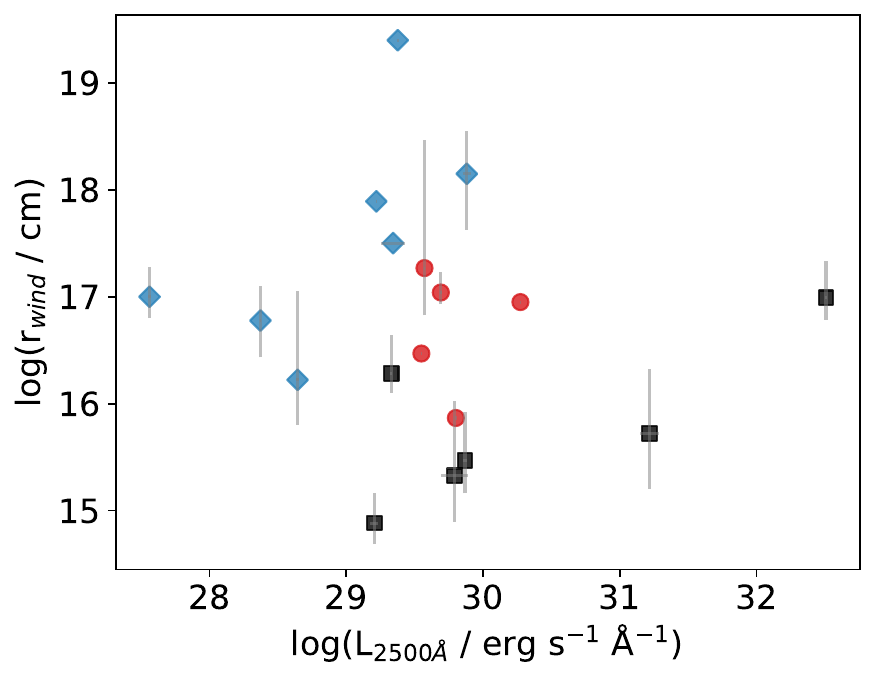}
    \end{subfigure}
    \begin{subfigure}[b]{0.25\textwidth}
        \includegraphics[width=\linewidth]{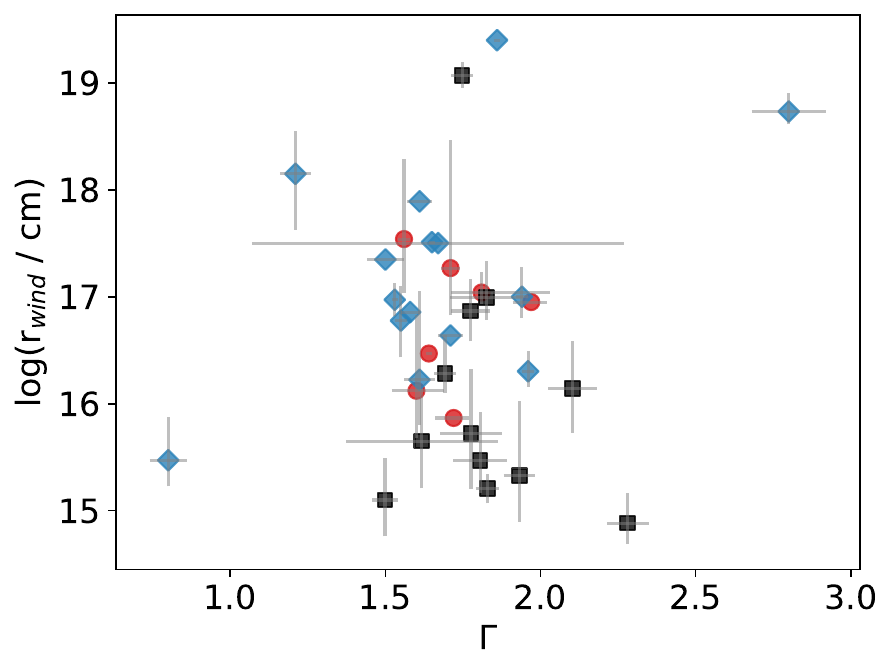}
    \end{subfigure}
    \begin{subfigure}[b]{0.25\textwidth}
        \includegraphics[width=\linewidth]{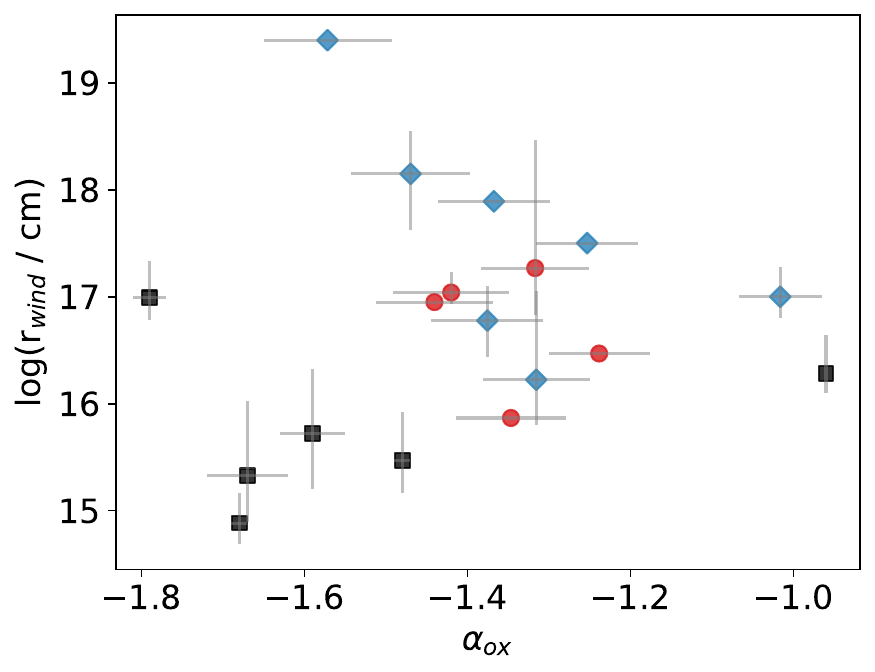}
    \end{subfigure}
    \begin{subfigure}[b]{0.25\textwidth}
        \includegraphics[width=\linewidth]{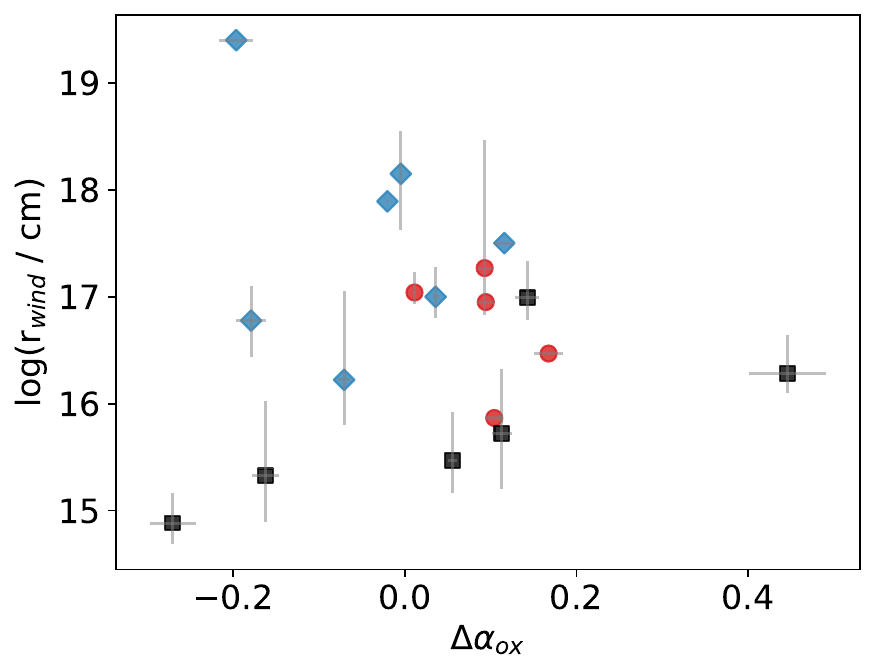}
    \end{subfigure}\\
\caption[]{{UFO launching radius versus AGN parameters. Significant and nonsignificant correlations for the S23, T10, and C21 samples. The best-fitting linear correlations, applied exclusively to statistically significant correlations, are presented by the solid black lines and the dark and light gray shadowed areas indicate the 68\% and 90\% confidence bands, respectively. In the legend, we report the best-fit coefficients, $\log\mathrm{NHP}$, and the intrinsic scatters for the correlations.}}
\label{fig:ef9}
\end{figure*}

\begin{figure*}
\centering
    \begin{subfigure}[b]{0.41\textwidth}
        \includegraphics[width=\linewidth]{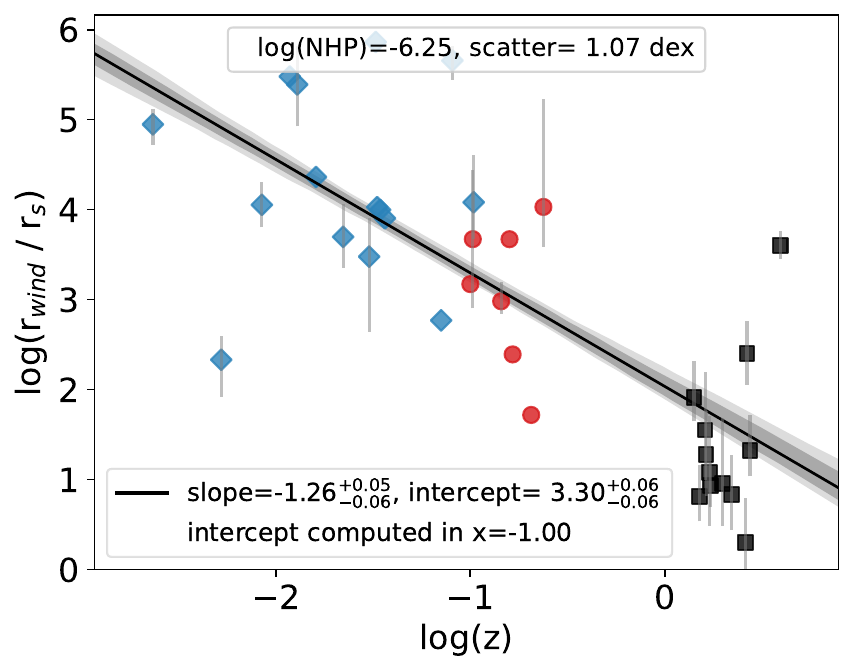}
    \end{subfigure}
    \begin{subfigure}[b]{0.4\textwidth}
        \includegraphics[width=\linewidth]{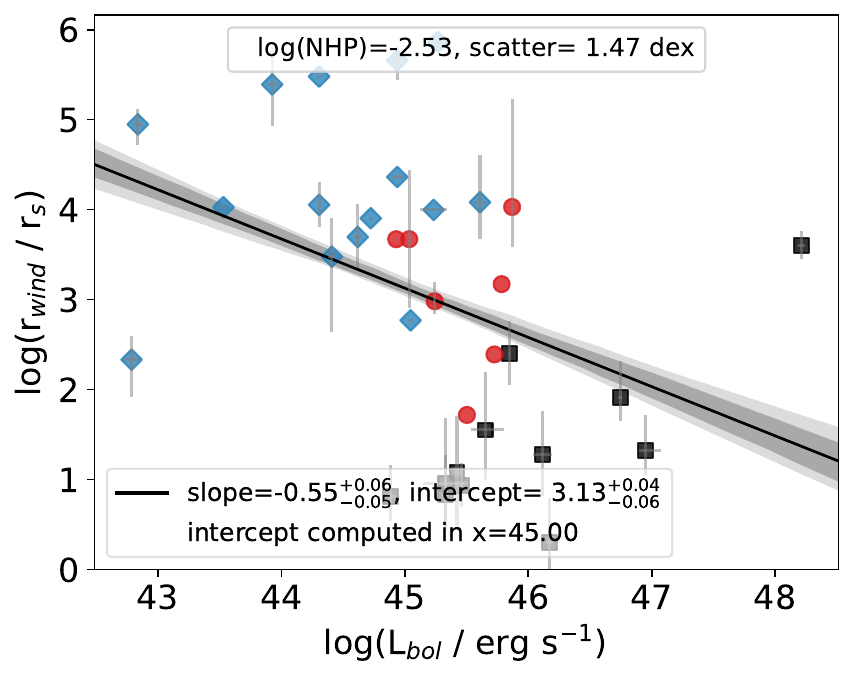}
    \end{subfigure}
    \begin{subfigure}[b]{0.4\textwidth}
        \includegraphics[width=\linewidth]{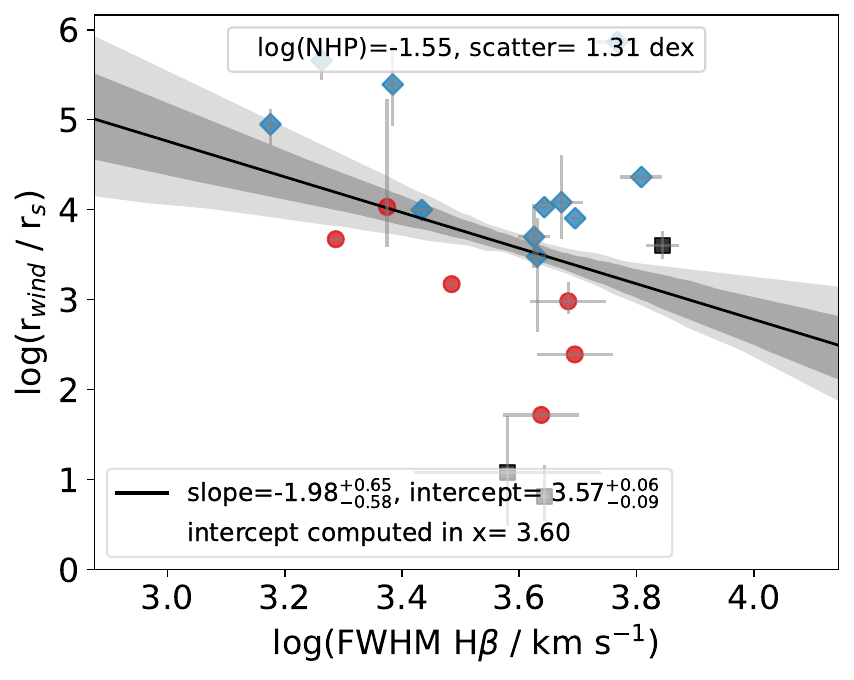}
    \end{subfigure}
    \begin{subfigure}[b]{0.25\textwidth}
        \includegraphics[width=\linewidth]{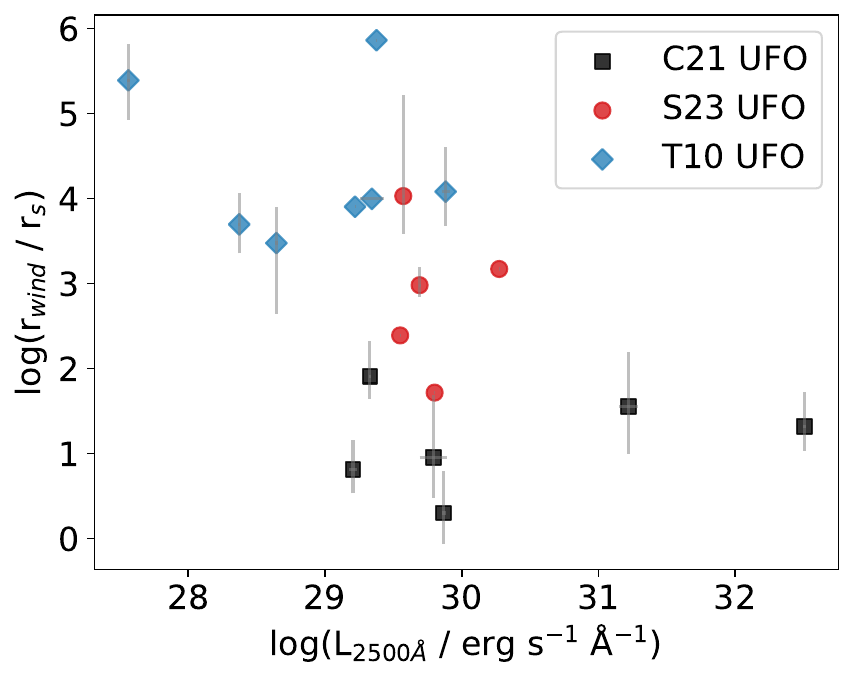}
    \end{subfigure}
    \begin{subfigure}[b]{0.25\textwidth}
        \includegraphics[width=\linewidth]{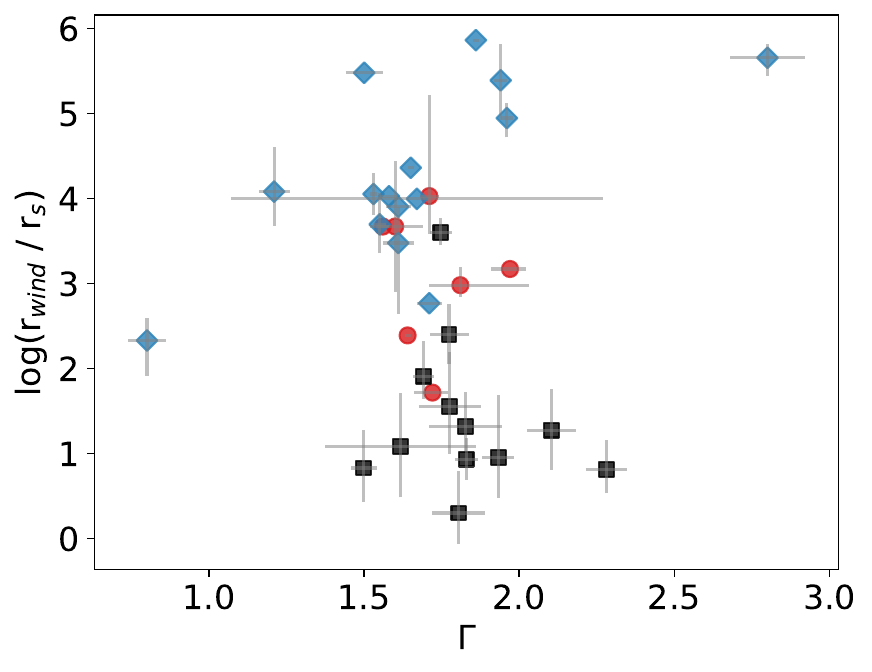}
    \end{subfigure}
    \begin{subfigure}[b]{0.25\textwidth}
        \includegraphics[width=\linewidth]{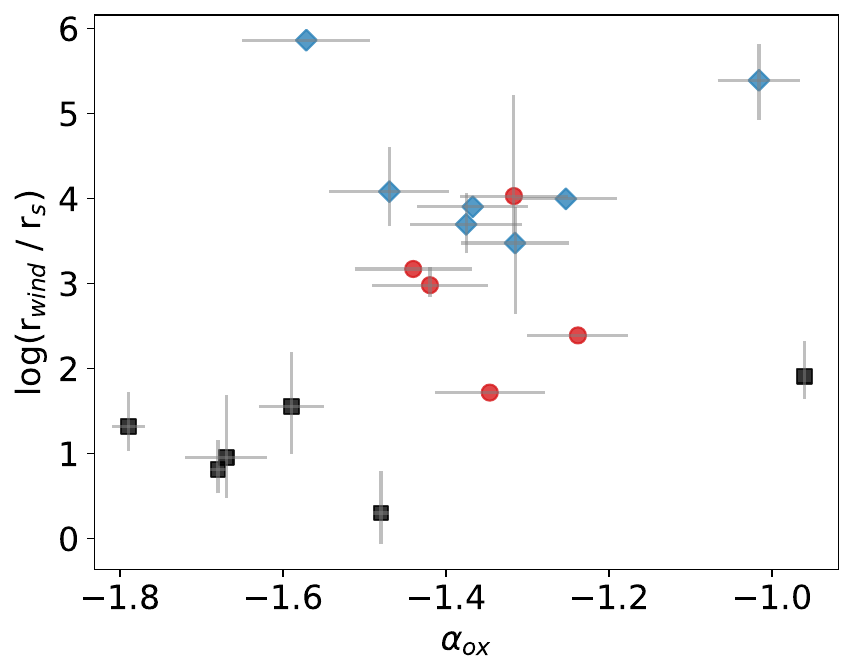}
    \end{subfigure}
    \begin{subfigure}[b]{0.25\textwidth}
        \includegraphics[width=\linewidth]{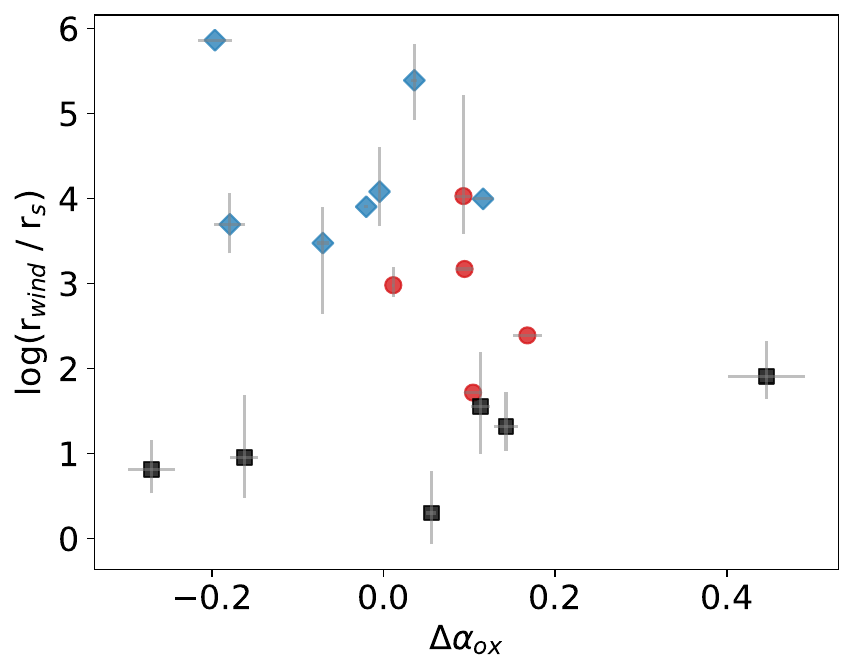}
    \end{subfigure}\\

\caption[]{{UFO launching normalized radius versus AGN parameters. Significant and nonsignificant correlations for the S23, T10, and C21 samples. The best-fitting linear correlations, applied exclusively to statistically significant correlations, are presented by the solid black lines and the dark and light gray shadowed areas indicate the 68\% and 90\% confidence bands, respectively. In the legend, we report the best-fit coefficients, $\log\mathrm{NHP}$, and the intrinsic scatters for the correlations.}}
\label{fig:ef10}
\end{figure*}

\begin{figure*}
\centering
    \begin{subfigure}[b]{0.4\textwidth}
        \includegraphics[width=\linewidth]{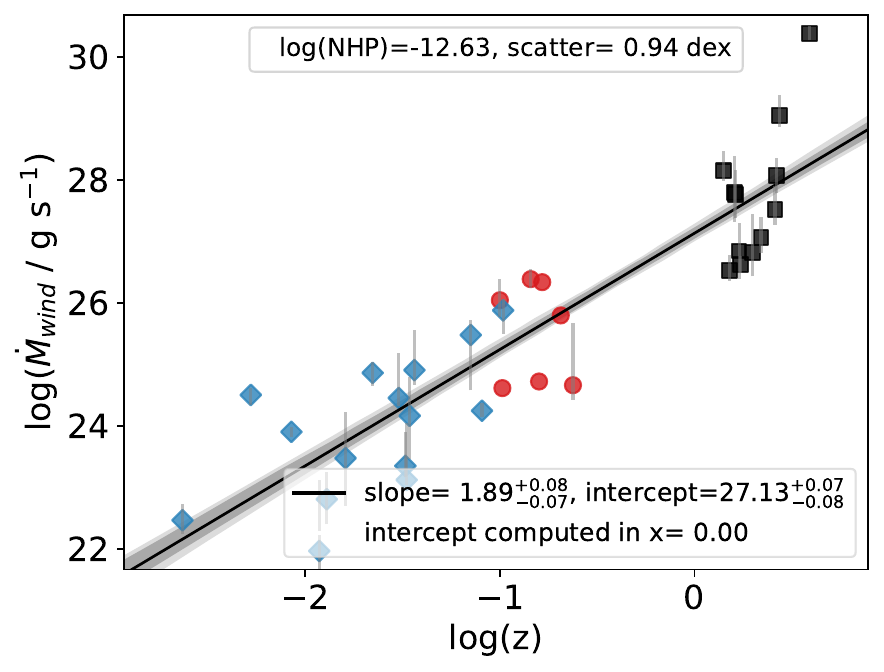}
    \end{subfigure}
    \begin{subfigure}[b]{0.4\textwidth}
        \includegraphics[width=\linewidth]{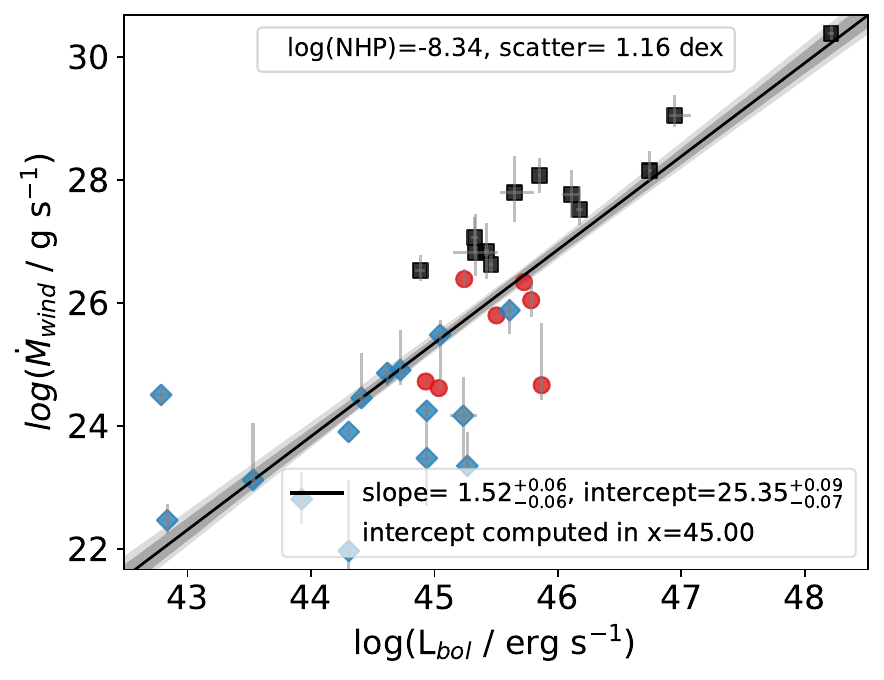}
    \end{subfigure}
    \begin{subfigure}[b]{0.4\textwidth}
        \includegraphics[width=\linewidth]{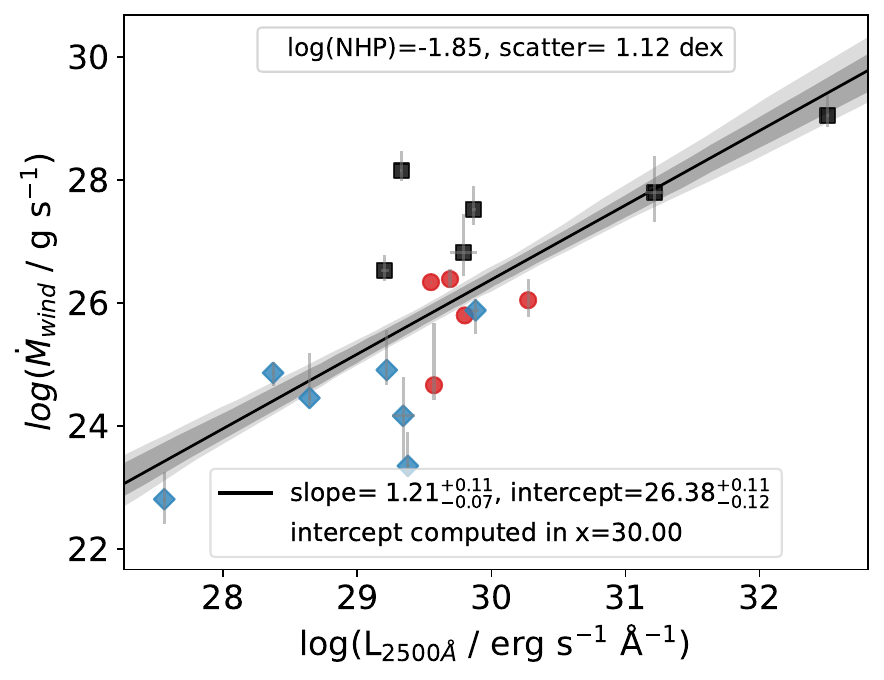}
    \end{subfigure}
    \begin{subfigure}[b]{0.4\textwidth}
        \includegraphics[width=\linewidth]{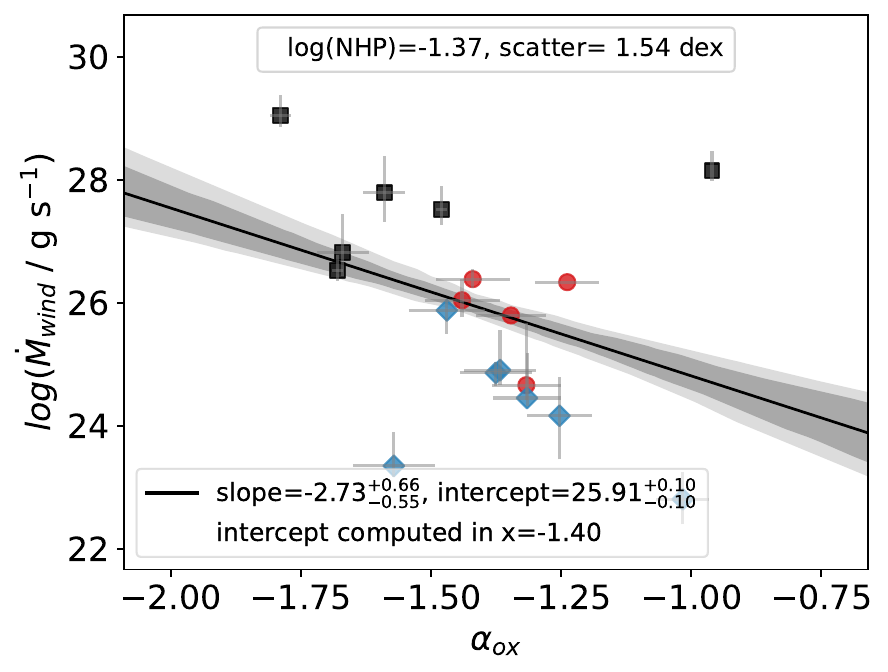}
    \end{subfigure}
    \begin{subfigure}[b]{0.25\textwidth}
        \includegraphics[width=\linewidth]{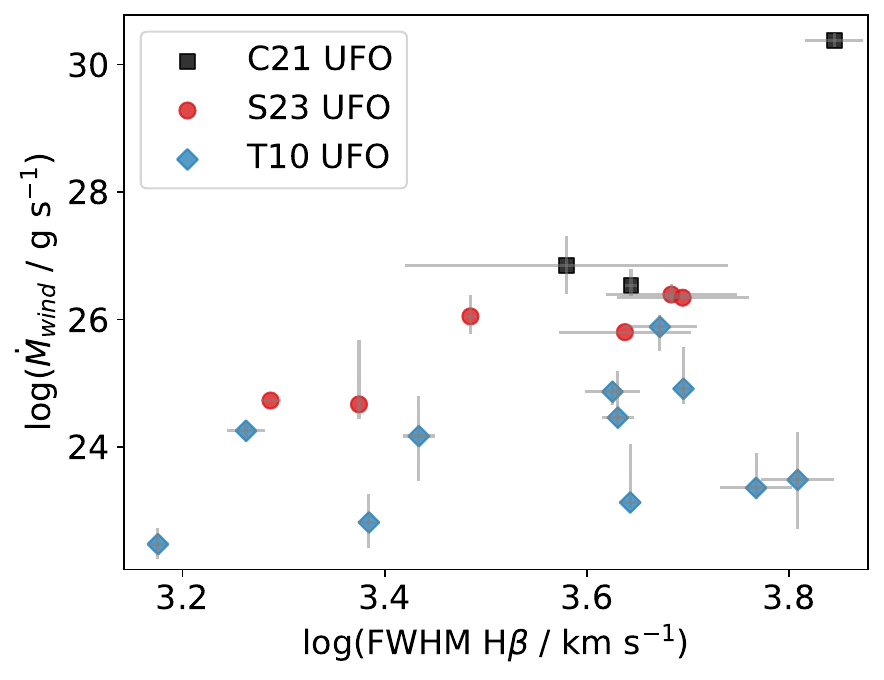}
    \end{subfigure}
    \begin{subfigure}[b]{0.25\textwidth}
        \includegraphics[width=\linewidth]{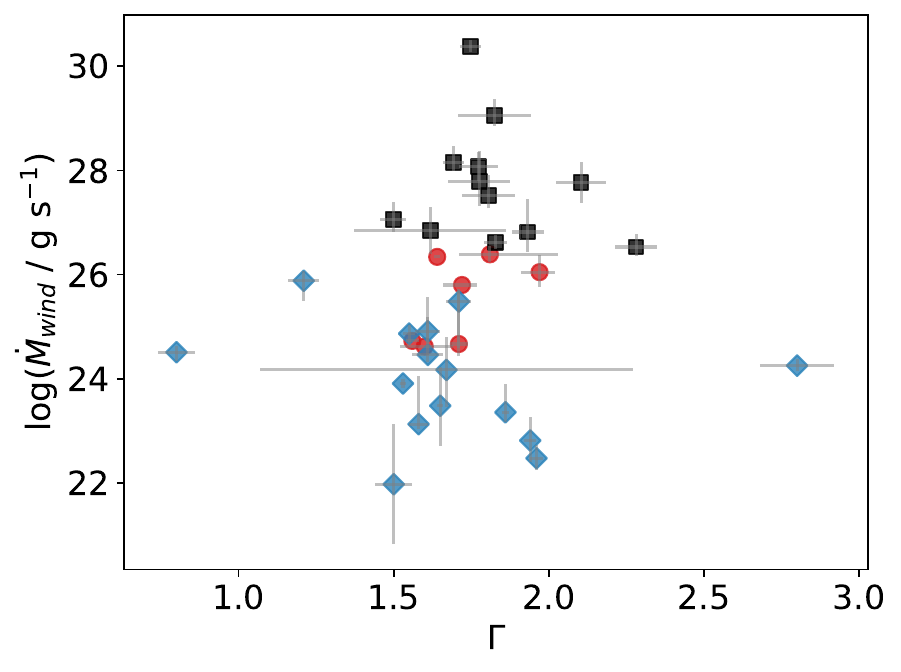}
    \end{subfigure}
    \begin{subfigure}[b]{0.25\textwidth}
        \includegraphics[width=\linewidth]{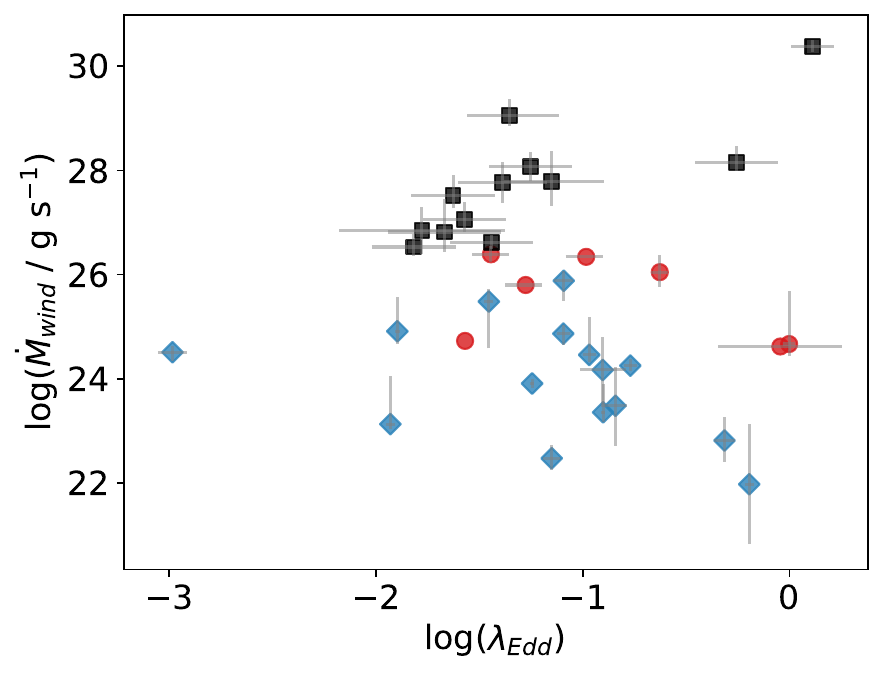}
    \end{subfigure}
    \begin{subfigure}[b]{0.25\textwidth}
        \includegraphics[width=\linewidth]{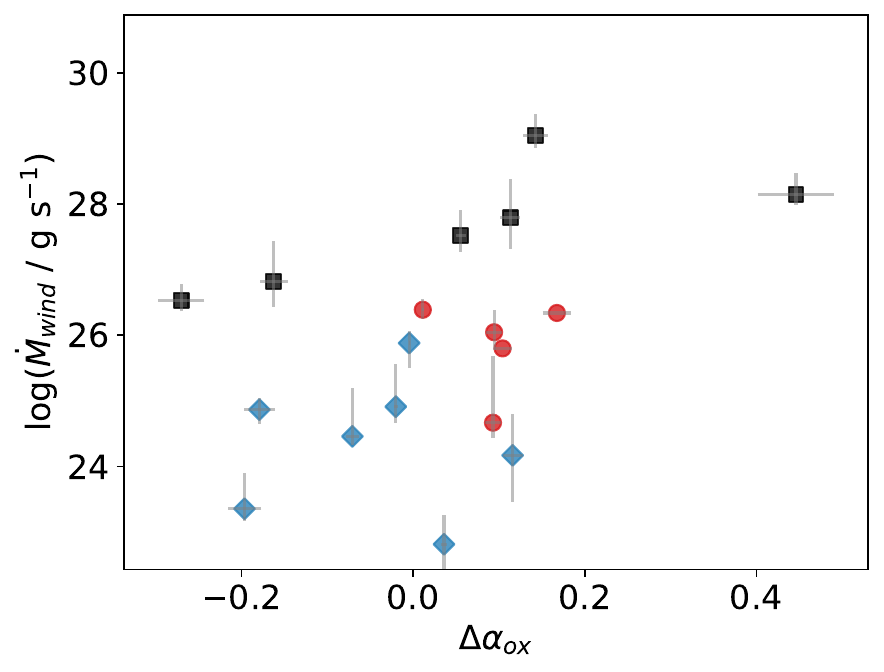}
    \end{subfigure}\\

\caption[]{{UFO mass outflow rate versus AGN parameters. Significant and nonsignificant correlations for the S23, T10, and C21 samples. The best-fitting linear correlations, applied exclusively to statistically significant correlations, are presented by the solid black lines and the dark and light gray shadowed areas indicate the 68\% and 90\% confidence bands, respectively. In the legend, we report the best-fit coefficients, $\log\mathrm{NHP}$, and the intrinsic scatters for the correlations.}}
\label{fig:ef11}
\end{figure*}

\begin{figure*}
\centering
    \begin{subfigure}[b]{0.35\textwidth}
        \includegraphics[width=\linewidth]{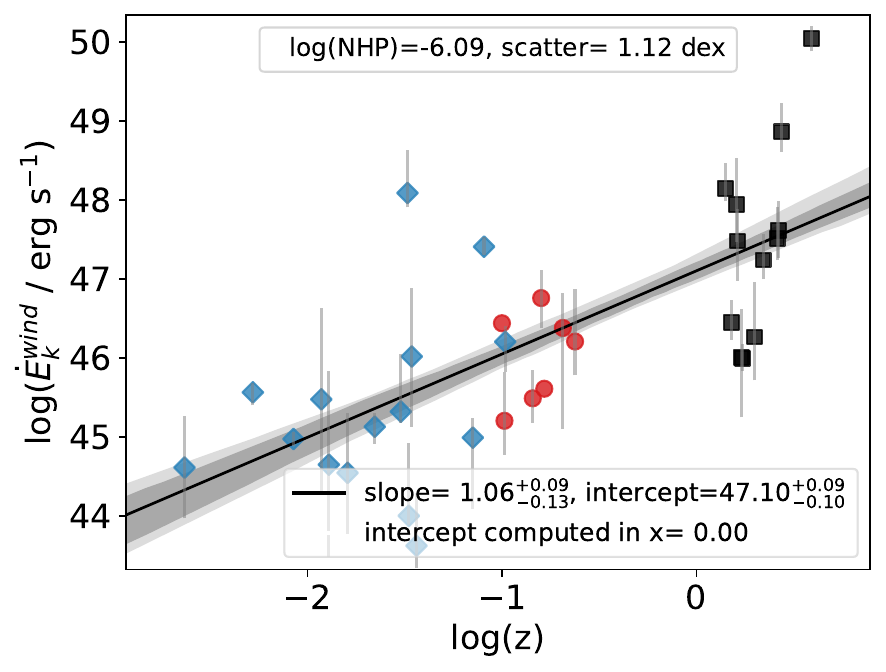}
    \end{subfigure}
    \begin{subfigure}[b]{0.35\textwidth}
        \includegraphics[width=\linewidth]{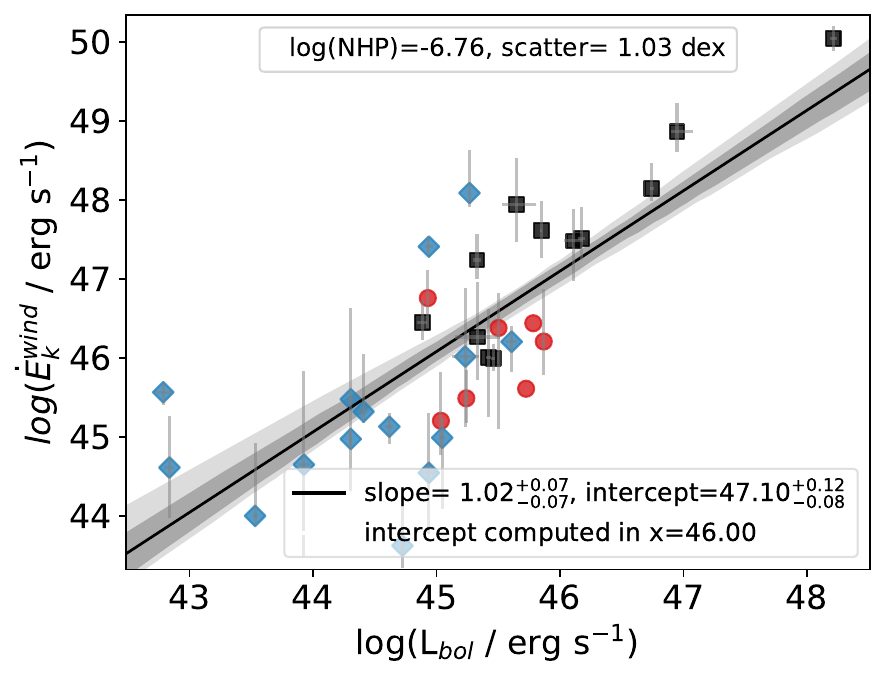}
    \end{subfigure}
    \begin{subfigure}[b]{0.35\textwidth}
        \includegraphics[width=\linewidth]{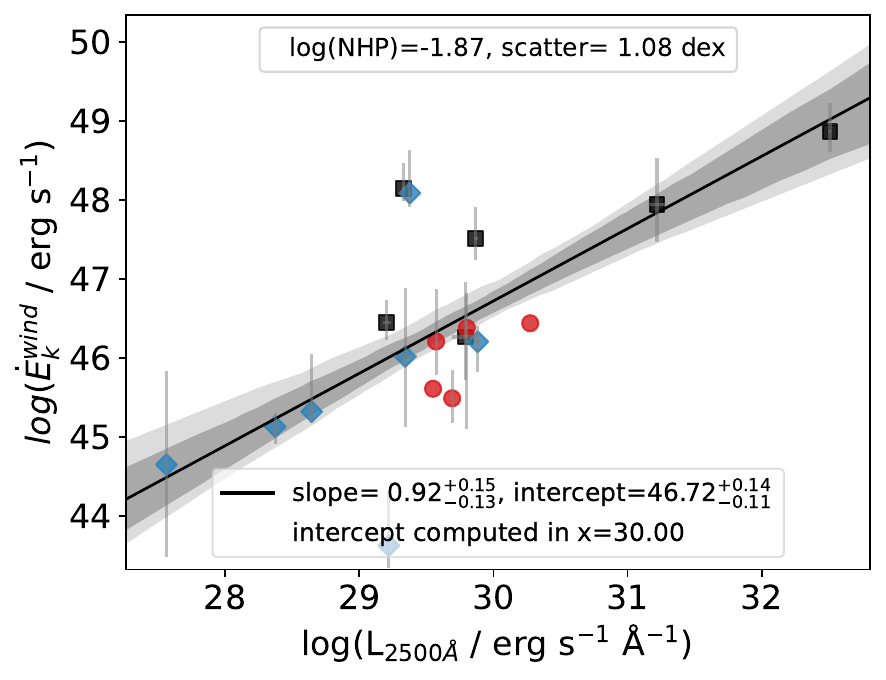}
    \end{subfigure}
    \begin{subfigure}[b]{0.35\textwidth}
        \includegraphics[width=\linewidth]{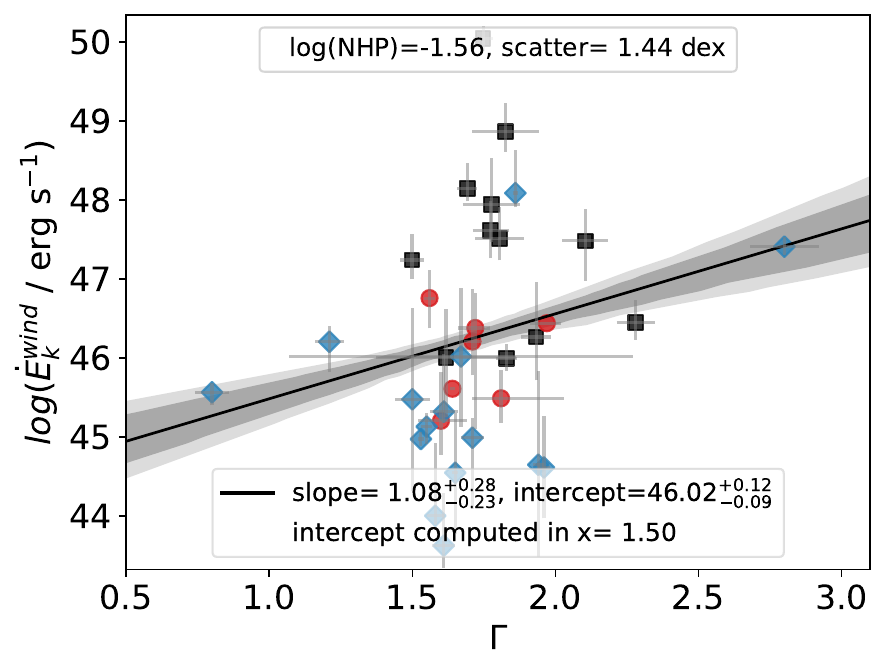}
    \end{subfigure}
    \begin{subfigure}[b]{0.35\textwidth}
        \includegraphics[width=\linewidth]{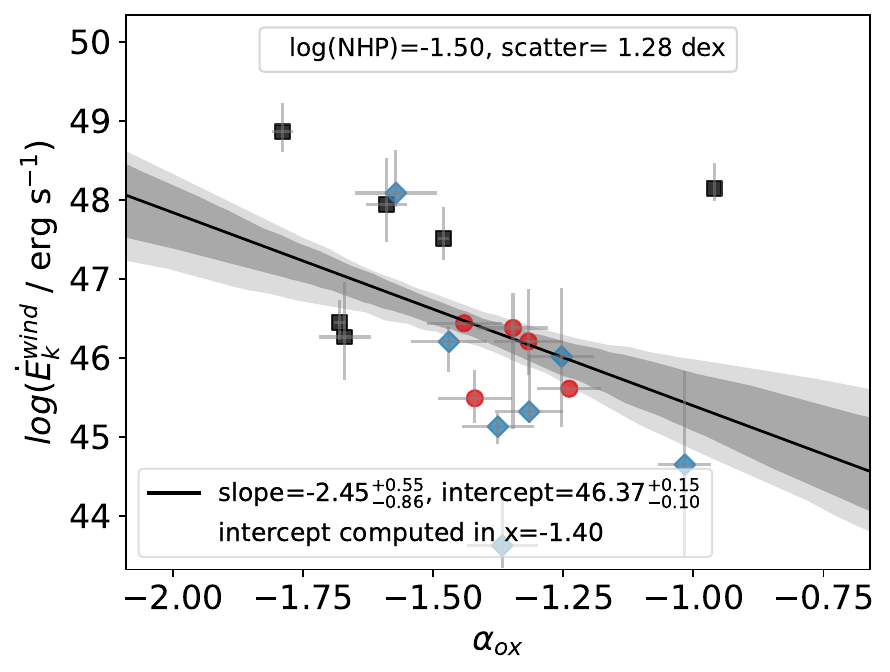}
    \end{subfigure}\\
    \begin{subfigure}[b]{0.25\textwidth}
        \includegraphics[width=\linewidth]{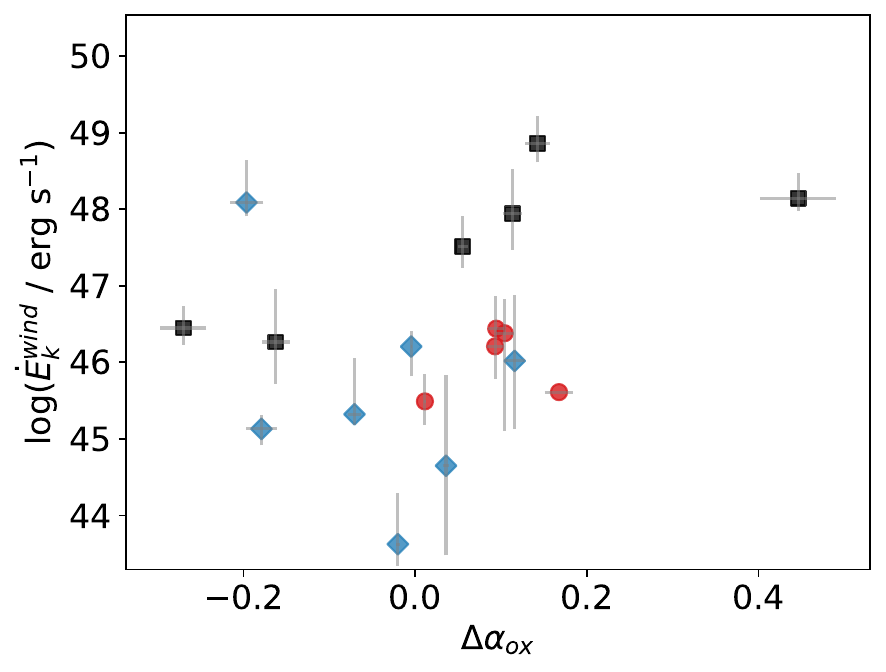}
    \end{subfigure}
        \begin{subfigure}[b]{0.25\textwidth}
        \includegraphics[width=\linewidth]{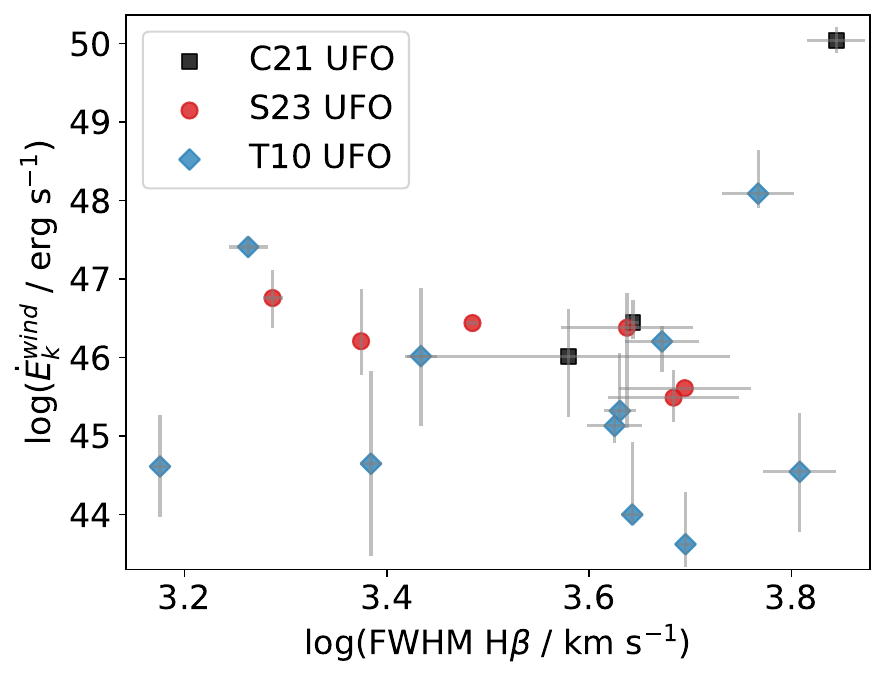}
    \end{subfigure}
    \begin{subfigure}[b]{0.25\textwidth}
        \includegraphics[width=\linewidth]{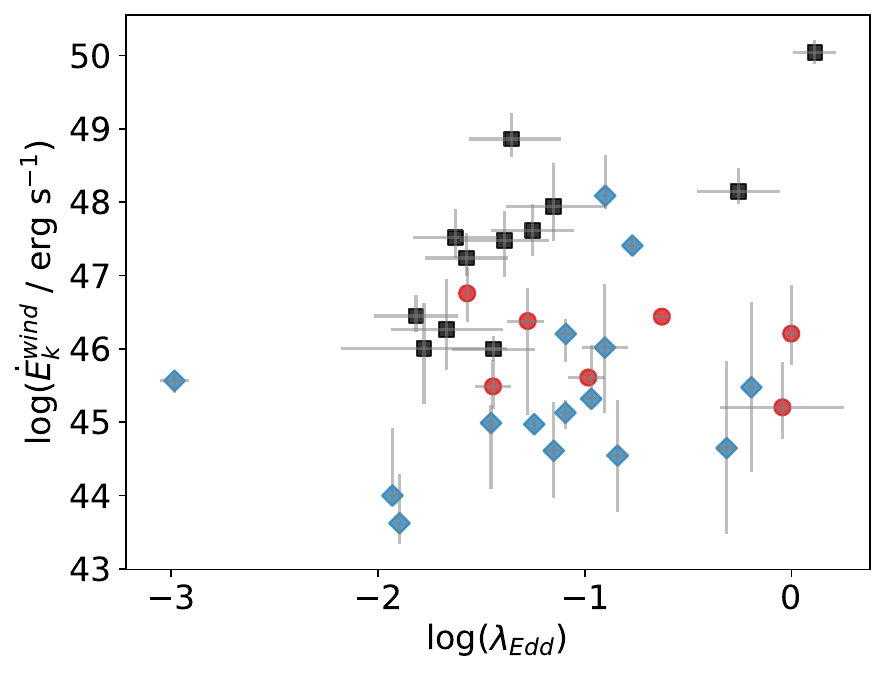}
    \end{subfigure}\\

\caption[]{{UFO kinetic energy versus AGN parameters. Significant and nonsignificant correlations for the S23, T10, and C21 samples. The best-fitting linear correlations, applied exclusively to statistically significant correlations, are presented by the solid black lines and the dark and light gray shadowed areas indicate the 68\% and 90\% confidence bands, respectively. In the legend, we report the best-fit coefficients, $\log\mathrm{NHP}$, and the intrinsic scatters for the correlations.}}
\label{fig:ef12}
\end{figure*}

\begin{figure*}
\centering
    \begin{subfigure}[b]{0.35\textwidth}
        \includegraphics[width=\linewidth]{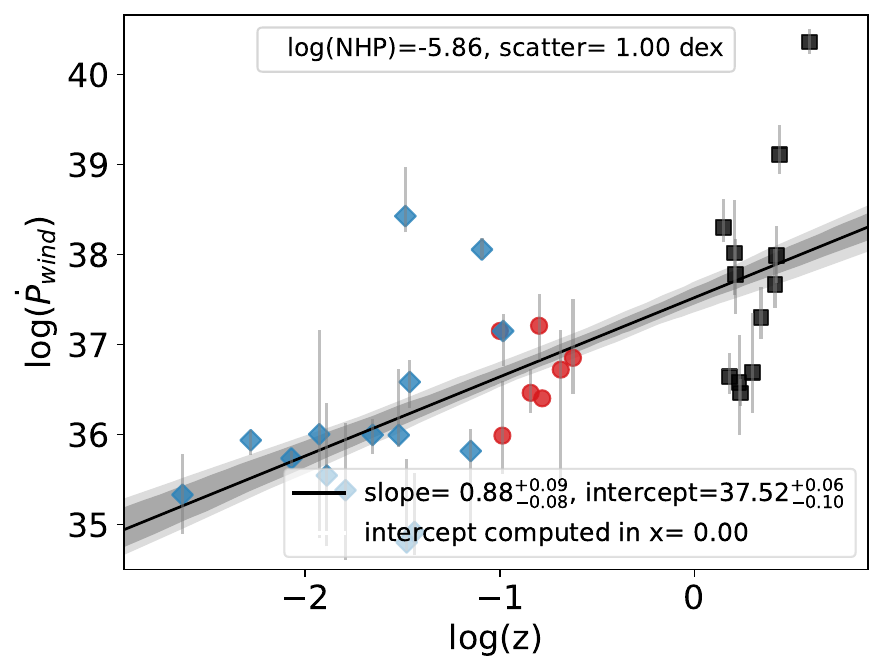}
    \end{subfigure}
    \begin{subfigure}[b]{0.35\textwidth}
        \includegraphics[width=\linewidth]{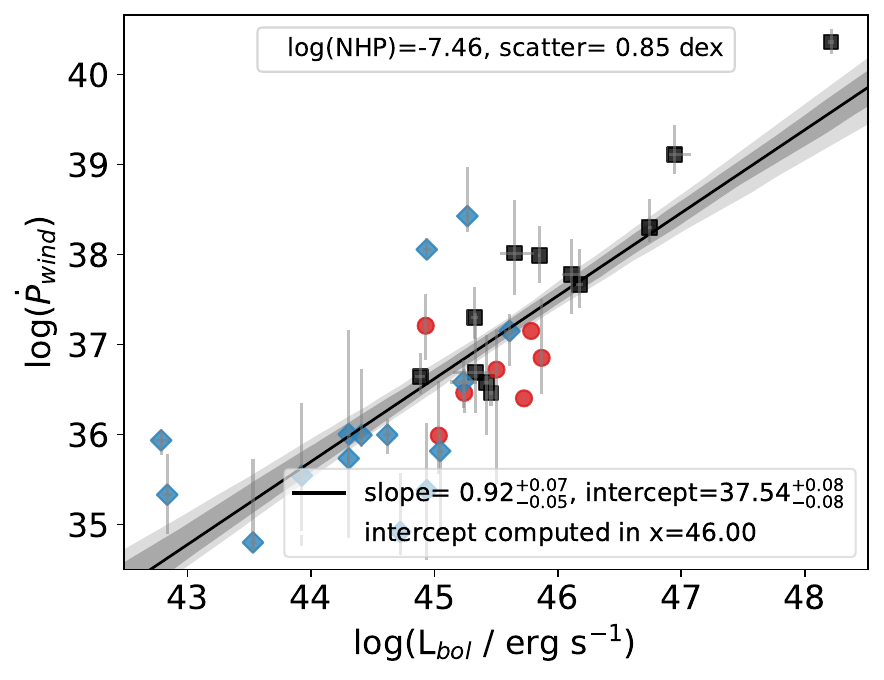}
    \end{subfigure}
    \begin{subfigure}[b]{0.35\textwidth}
        \includegraphics[width=\linewidth]{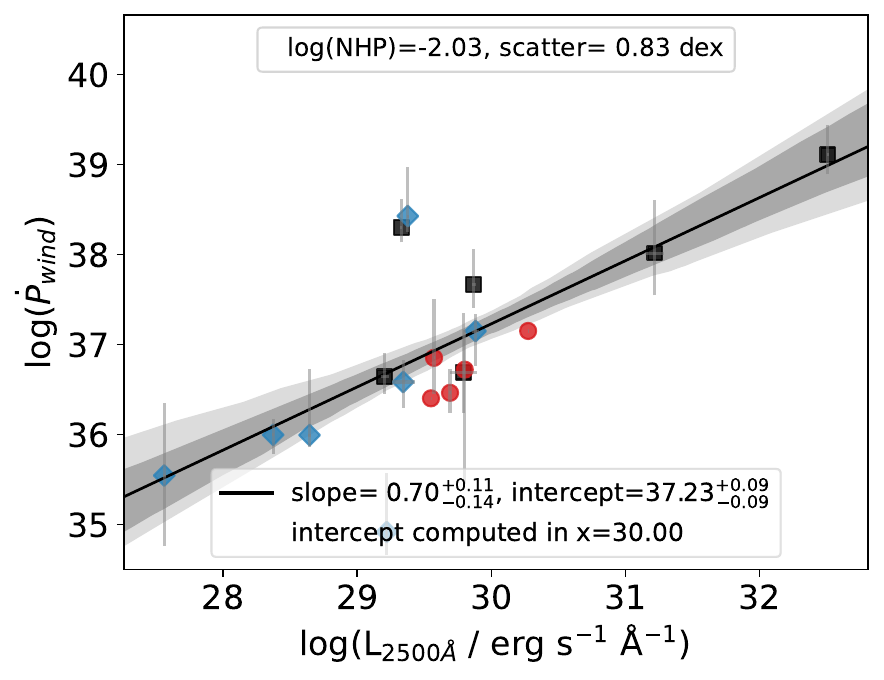}
    \end{subfigure}
    \begin{subfigure}[b]{0.35\textwidth}
        \includegraphics[width=\linewidth]{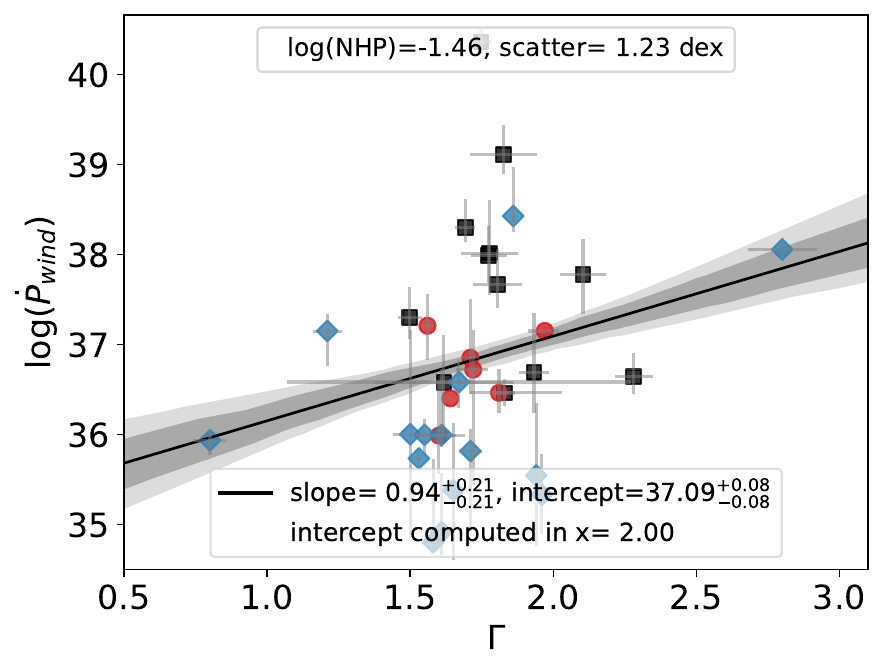}
    \end{subfigure}
    \begin{subfigure}[b]{0.35\textwidth}
        \includegraphics[width=\linewidth]{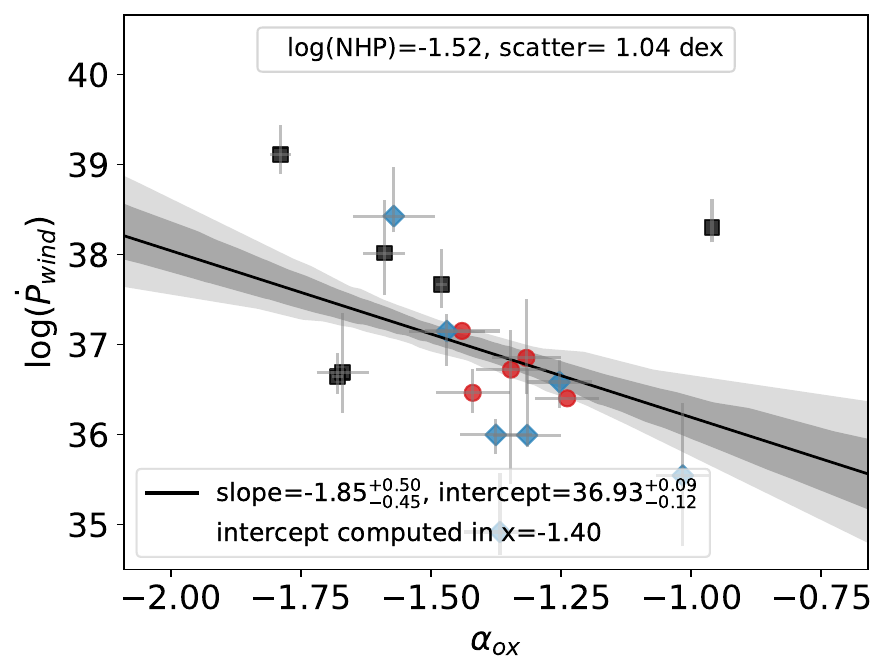}
    \end{subfigure}\\
    \begin{subfigure}[b]{0.25\textwidth}
        \includegraphics[width=\linewidth]{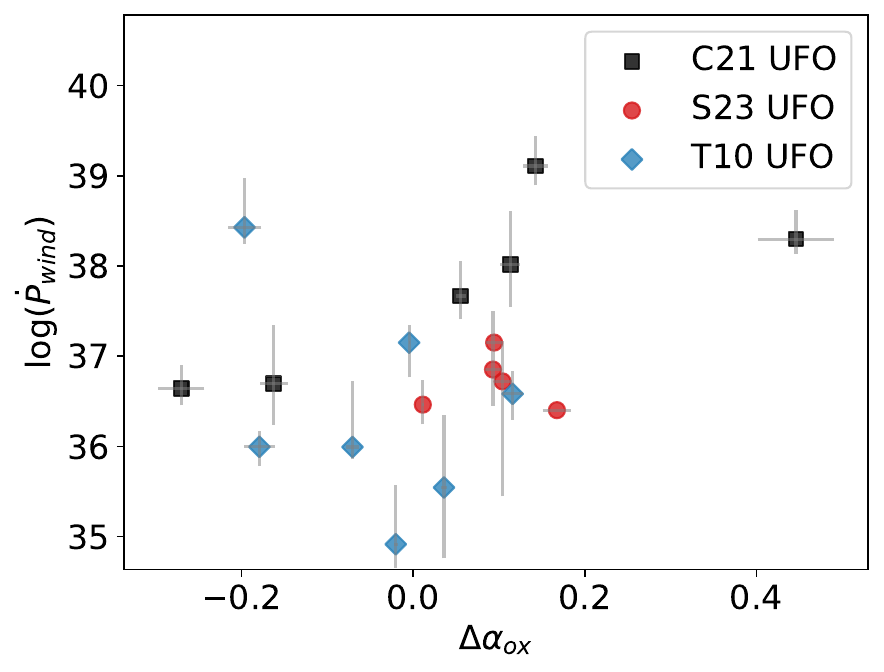}
    \end{subfigure}
        \begin{subfigure}[b]{0.25\textwidth}
        \includegraphics[width=\linewidth]{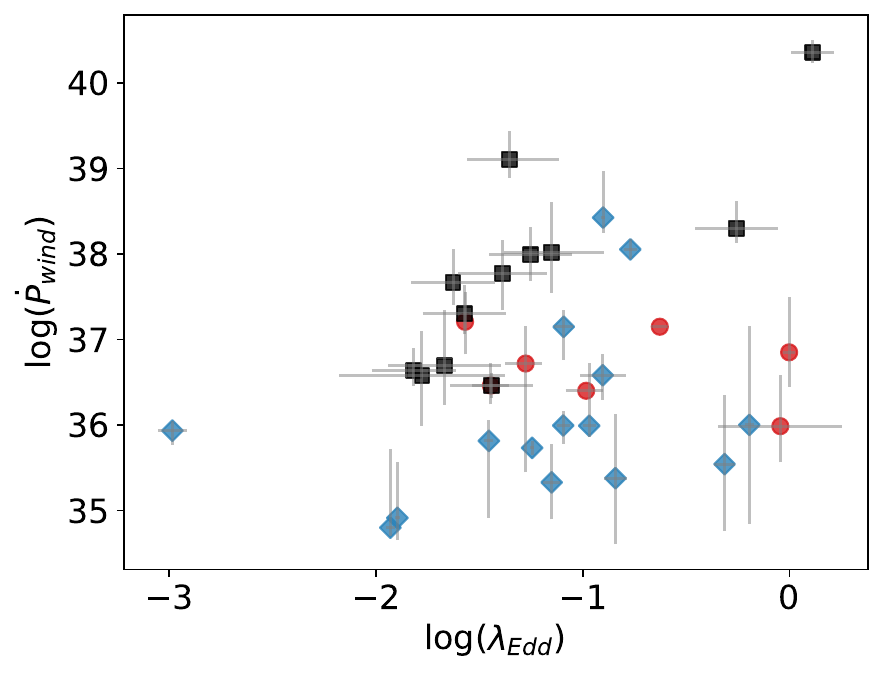}
    \end{subfigure}
        \begin{subfigure}[b]{0.25\textwidth}
        \includegraphics[width=\linewidth]{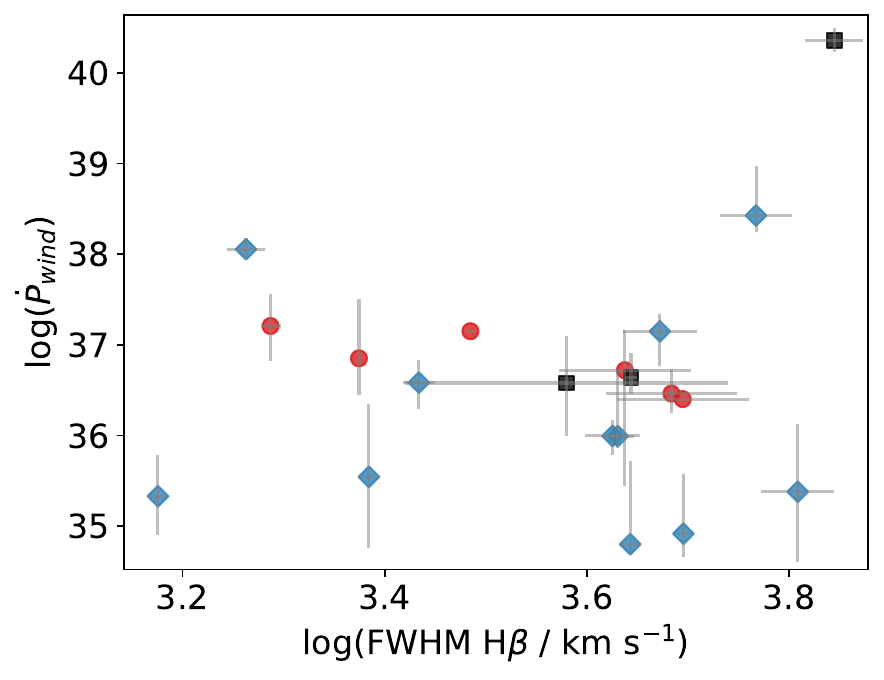}
    \end{subfigure}\\
\caption[]{{UFO momentum rate versus AGN parameters. Significant and nonsignificant correlations for the S23, T10, and C21 samples. The best-fitting linear correlations, applied exclusively to statistically significant correlations, are presented by the solid black lines and the dark and light gray shadowed areas indicate the 68\% and 90\% confidence bands, respectively. In the legend, we report the best-fit coefficients, $\log\mathrm{NHP}$, and the intrinsic scatters for the correlations.}}
\label{fig:ef13}
\end{figure*}

\begin{figure*}
    \centering
    \begin{subfigure}[b]{0.41\textwidth}
        \includegraphics[width=\linewidth]{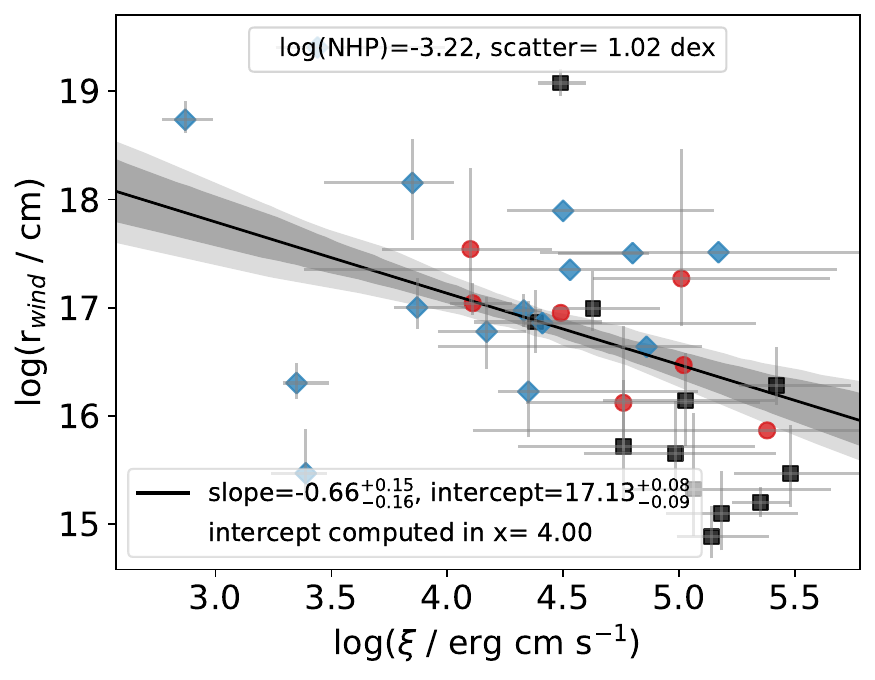}
    \end{subfigure}
    \begin{subfigure}[b]{0.4\textwidth}
        \includegraphics[width=\linewidth]{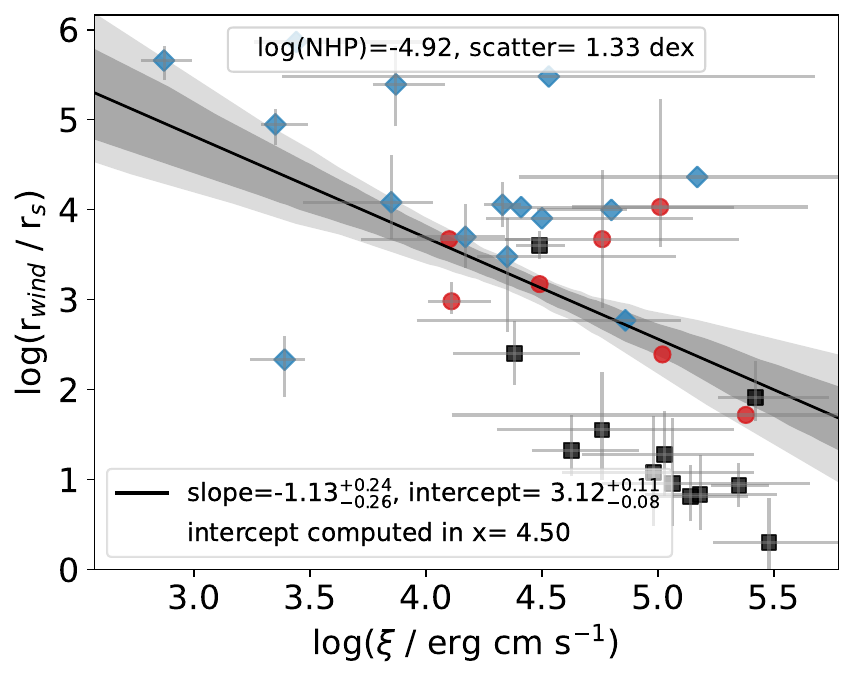}
    \end{subfigure} 
    \begin{subfigure}[b]{0.4\textwidth}
        \includegraphics[width=\linewidth]{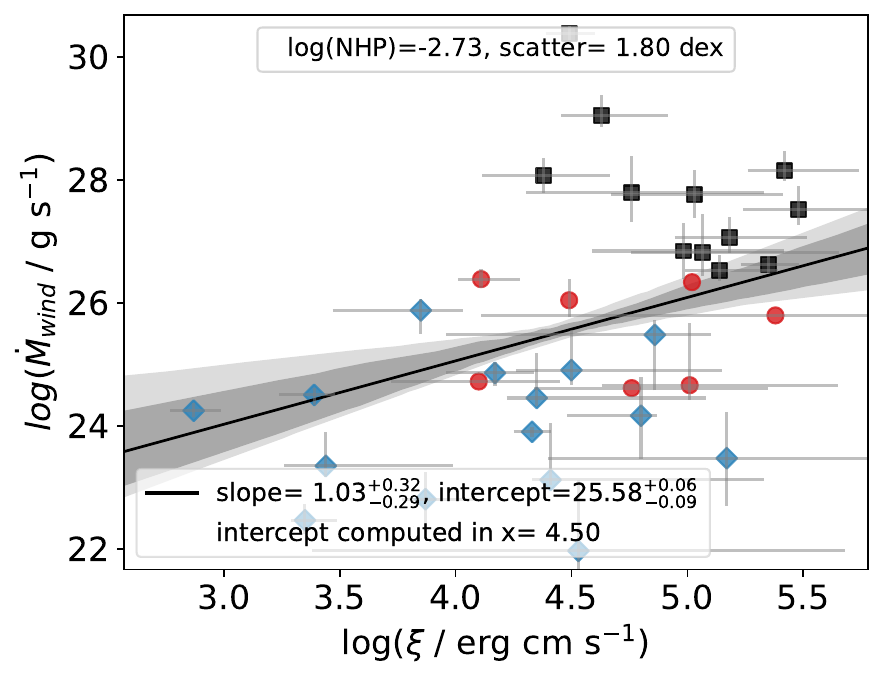}
    \end{subfigure}
    \begin{subfigure}[b]{0.25\textwidth}
        \includegraphics[width=\linewidth]{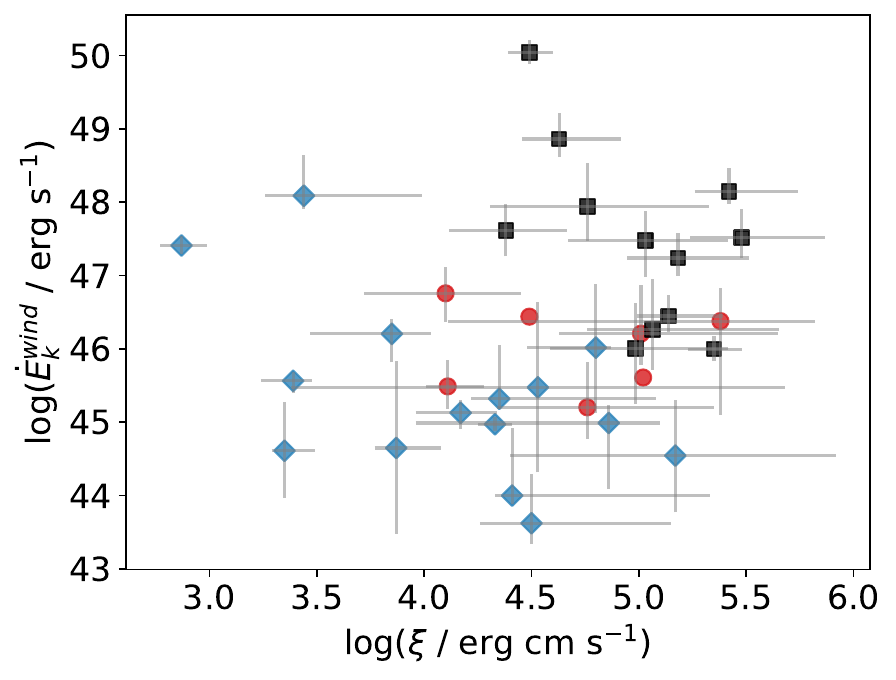}
    \end{subfigure}
    \begin{subfigure}[b]{0.25\textwidth}
        \includegraphics[width=\linewidth]{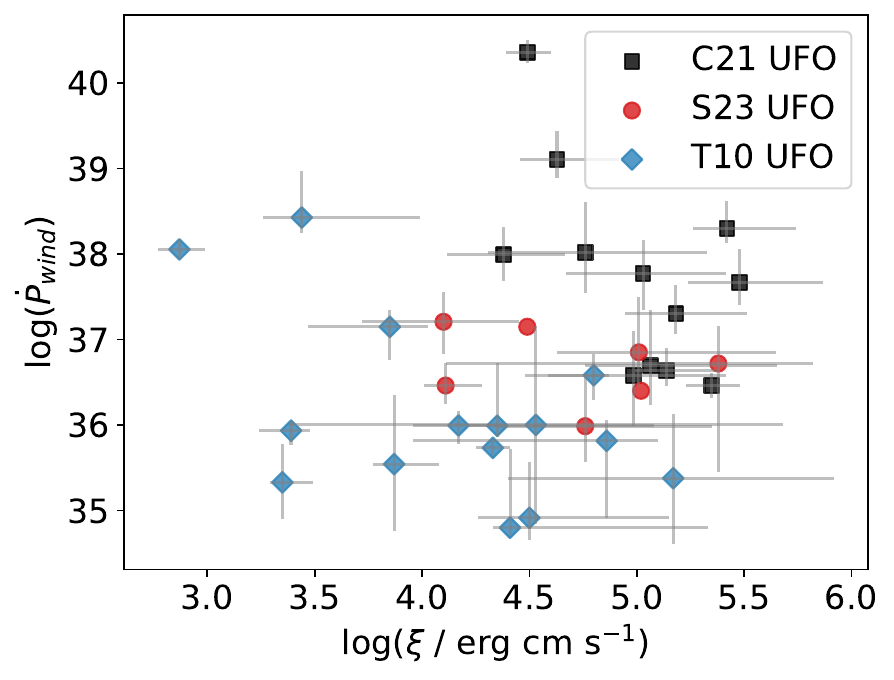}
    \end{subfigure}\\
    \caption[]{{UFO ionization parameter versus derived parameters. Significant and nonsignificant correlations for the S23, T10, and C21 samples. The best-fitting linear correlations, applied exclusively to statistically significant correlations, are presented by the solid black lines and the  dark and light gray shadowed areas indicate the 68\% and 90\% confidence bands, respectively. In the legend, we report the best-fit coefficients, $\log\mathrm{NHP}$, and the intrinsic scatters for the correlations.}}
    \label{fig:ef14}
\end{figure*}

\begin{figure*}
    \centering
    \begin{subfigure}[b]{0.4\textwidth}
        \includegraphics[width=\linewidth]{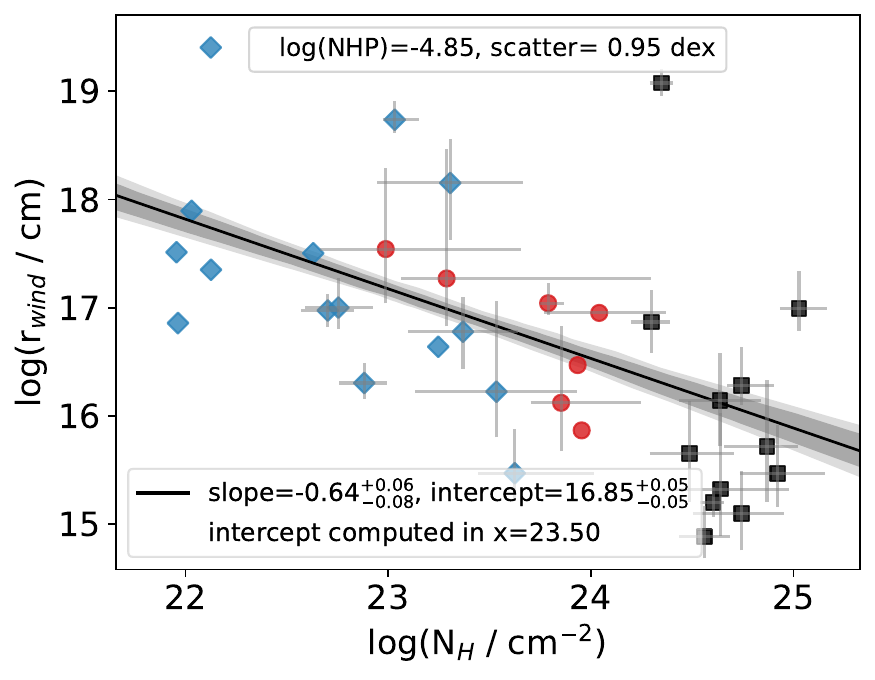}
    \end{subfigure}
    \begin{subfigure}[b]{0.4\textwidth}
        \includegraphics[width=\linewidth]{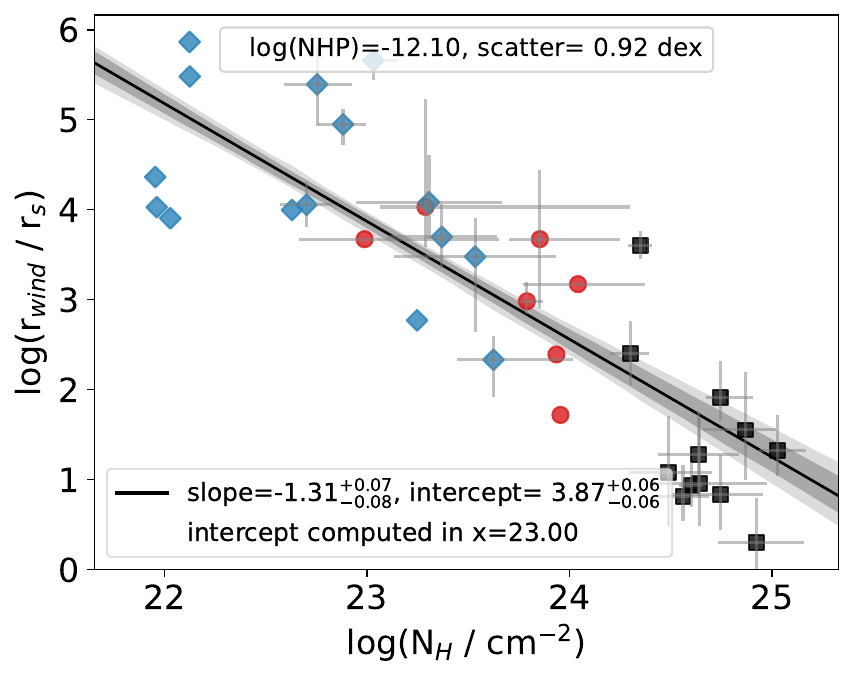}
    \end{subfigure}
    \begin{subfigure}[b]{0.4\textwidth}
        \includegraphics[width=\linewidth]{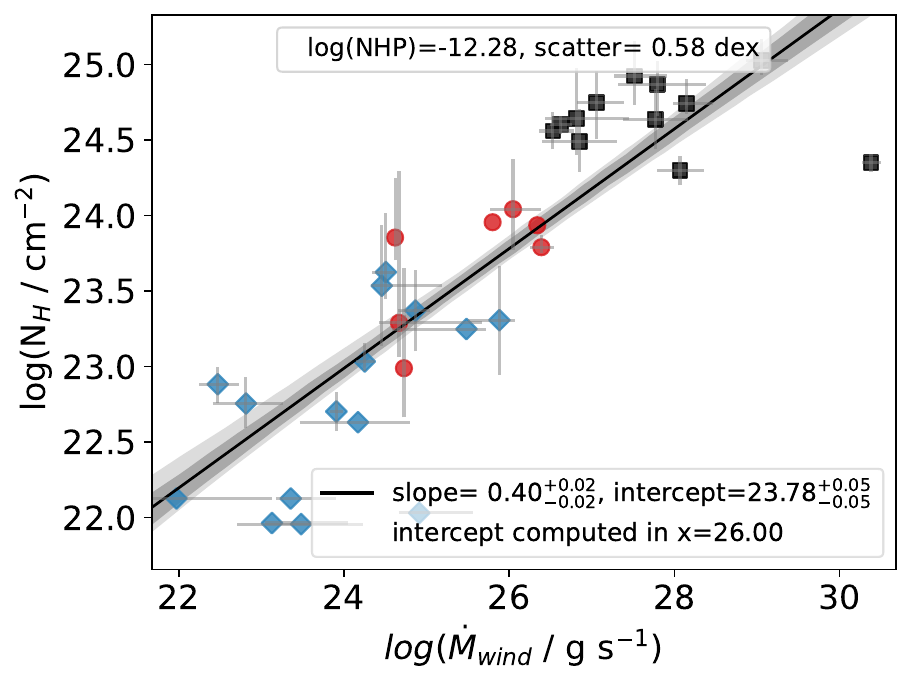}
    \end{subfigure}
    \begin{subfigure}[b]{0.4\textwidth}
        \includegraphics[width=\linewidth]{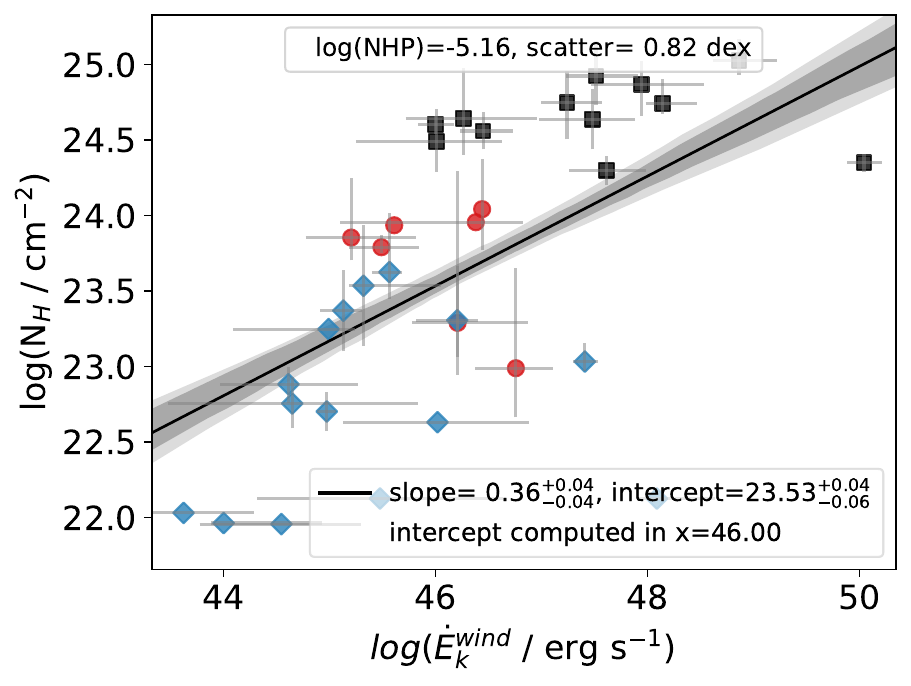}
    \end{subfigure}
    \begin{subfigure}[b]{0.4\textwidth}
        \includegraphics[width=\linewidth]{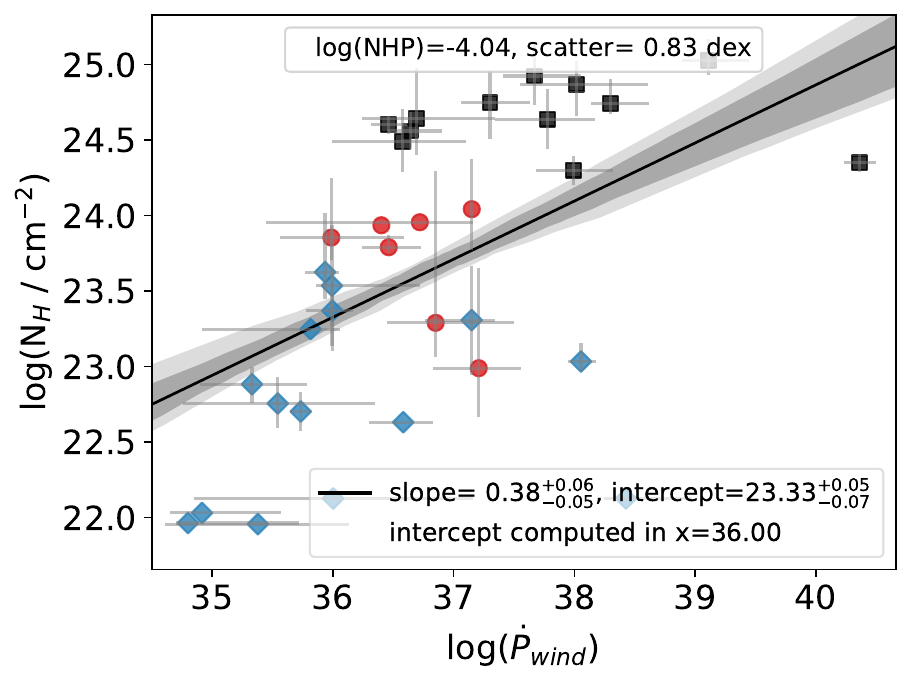}
    \end{subfigure}\\
    \caption[]{{UFO column density versus derived parameters. Significant and nonsignificant correlations for the S23, T10, and C21 samples. The best-fitting linear correlations, applied exclusively to statistically significant correlations, are presented by the solid black lines and the dark and light gray shadowed areas indicate the 68\% and 90\% confidence bands, respectively. In the legend, we report the best-fit coefficients, $\log\mathrm{NHP}$, and the intrinsic scatters for the correlations.}}
    \label{fig:ef15}
\end{figure*}

\begin{figure*}
    \centering
    \begin{subfigure}[b]{0.4\textwidth}
        \includegraphics[width=\linewidth]{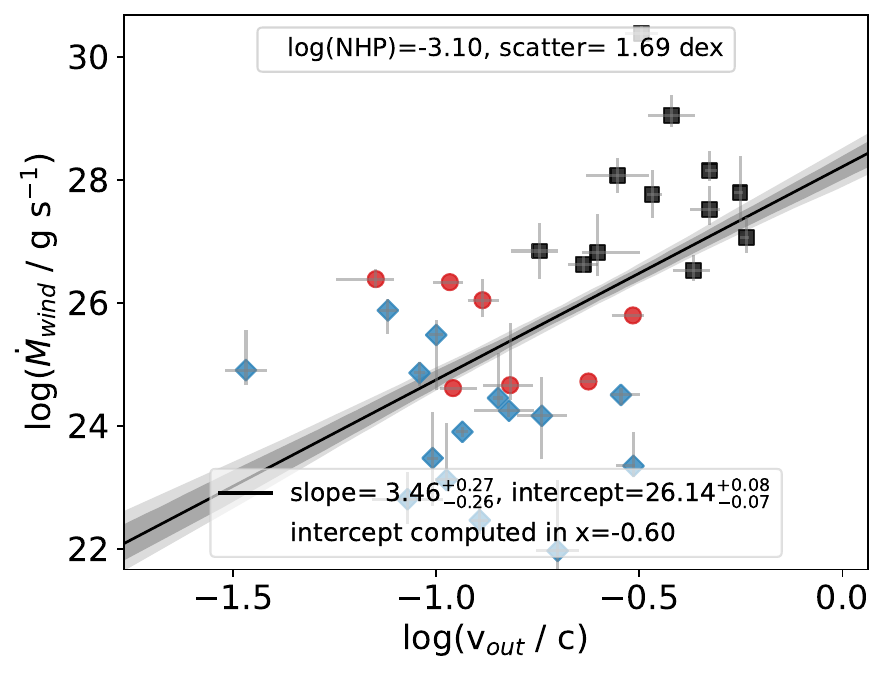}
    \end{subfigure}
    \begin{subfigure}[b]{0.4\textwidth}
        \includegraphics[width=\linewidth]{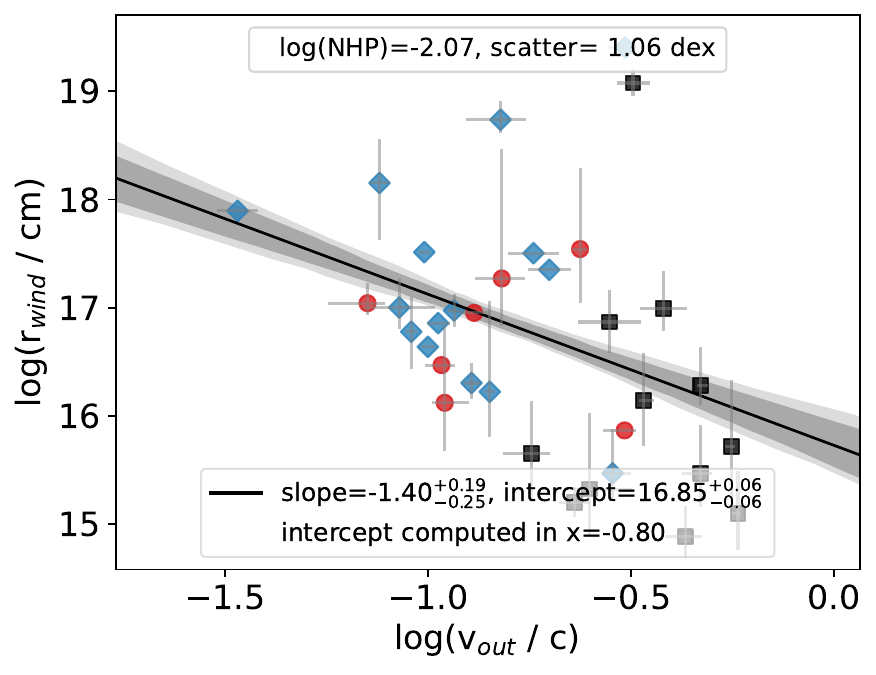}
    \end{subfigure}
    \begin{subfigure}[b]{0.4\textwidth}
        \includegraphics[width=\linewidth]{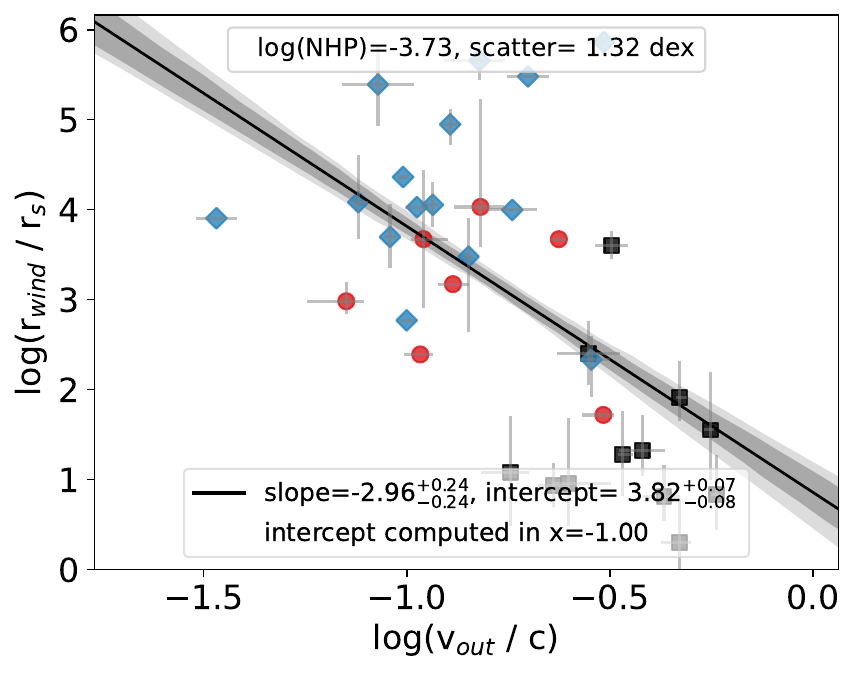}
    \end{subfigure}
    \begin{subfigure}[b]{0.4\textwidth}
        \includegraphics[width=\linewidth]{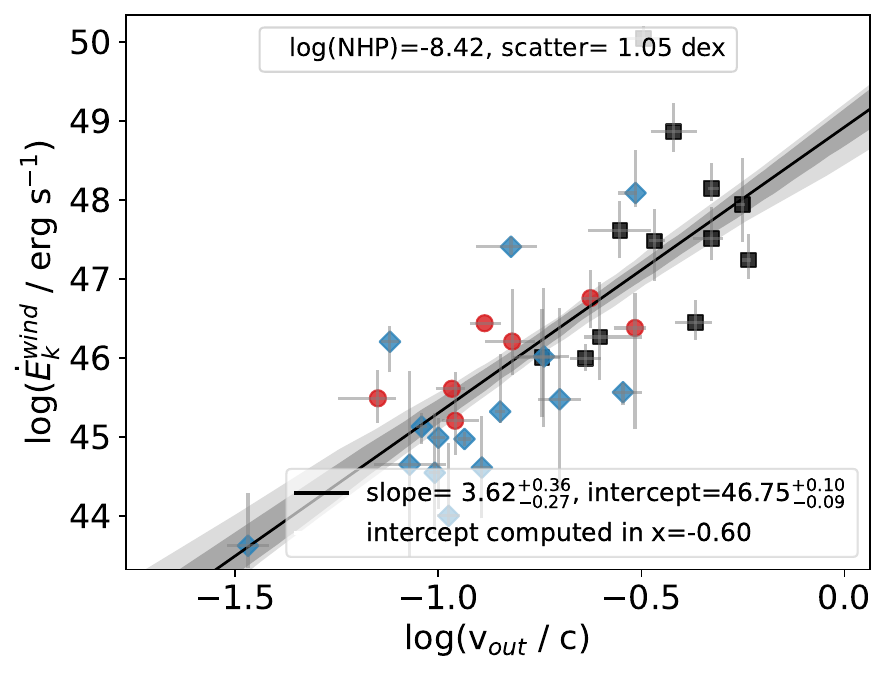}
    \end{subfigure}
    \begin{subfigure}[b]{0.4\textwidth}
        \includegraphics[width=\linewidth]{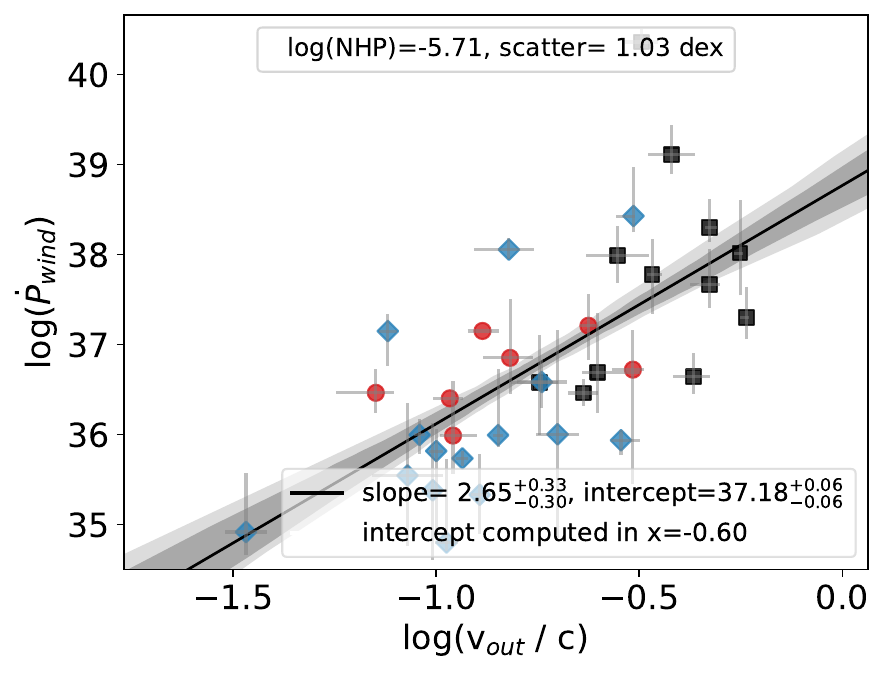}
    \end{subfigure}\\
    \caption[]{{UFO outflow velocity versus derived parameters. Significant and nonsignificant correlations for the S23, T10, and C21 samples. The best-fitting linear correlations, applied exclusively to statistically significant correlations, are presented by the solid black lines and the dark and light gray shadowed areas indicate the 68\% and 90\% confidence bands, respectively. In the legend, we report the best-fit coefficients, $\log\mathrm{NHP}$, and the intrinsic scatters for the correlations.}}
    \label{fig:ef16}
\end{figure*}

\begin{figure*}
    \centering
    \begin{subfigure}[b]{0.4\textwidth}
        \includegraphics[width=\linewidth]{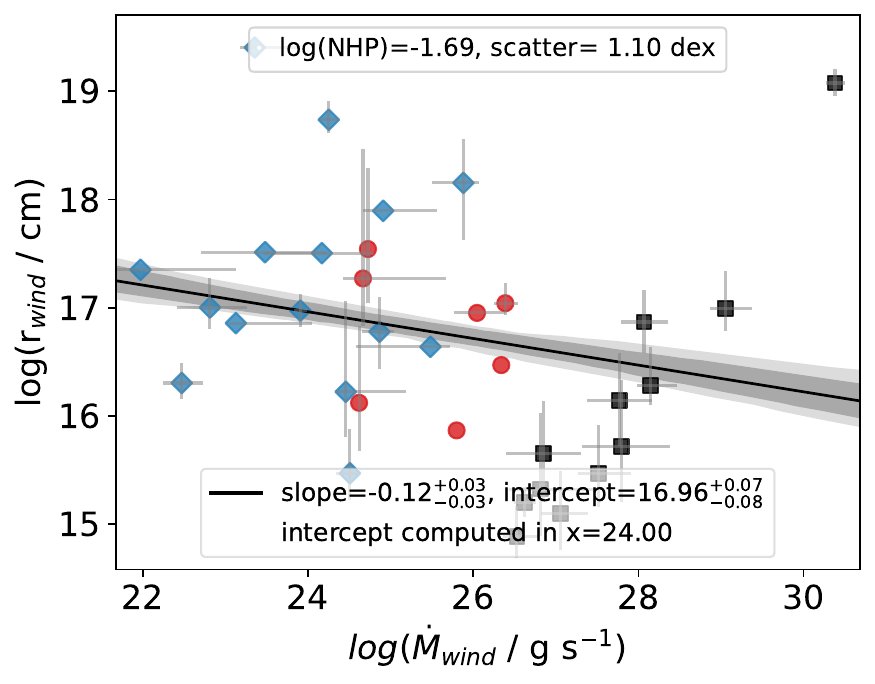}
    \end{subfigure}
    \begin{subfigure}[b]{0.4\textwidth}
        \includegraphics[width=\linewidth]{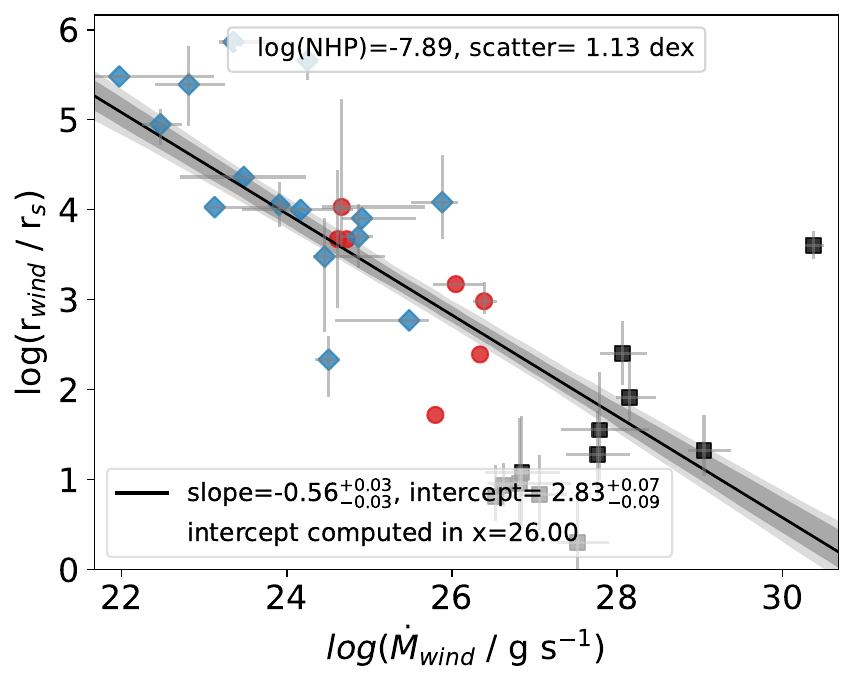}
    \end{subfigure}
    \begin{subfigure}[b]{0.4\textwidth}
        \includegraphics[width=\linewidth]{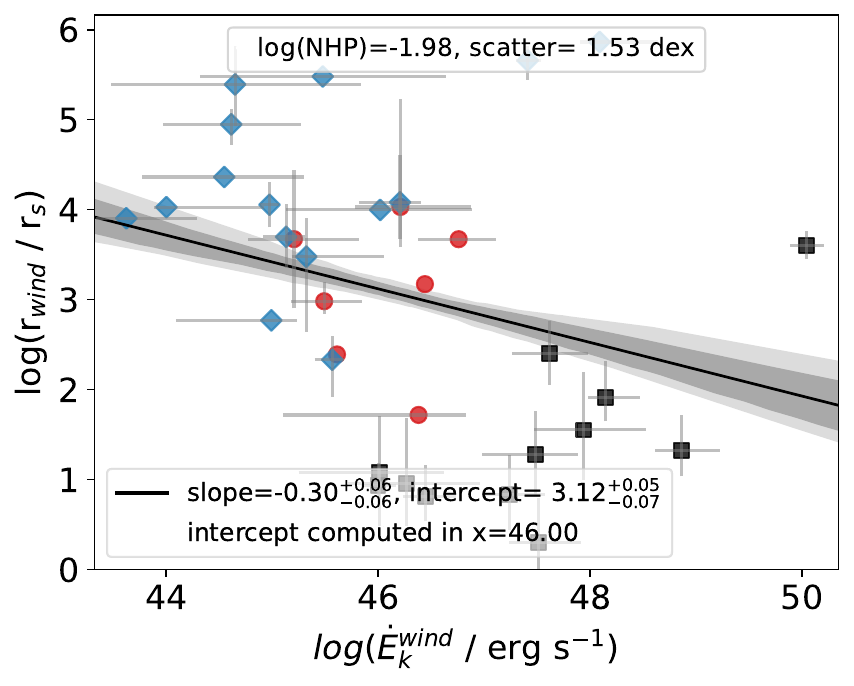}
    \end{subfigure}
        \begin{subfigure}[b]{0.4\textwidth}
        \includegraphics[width=\linewidth]{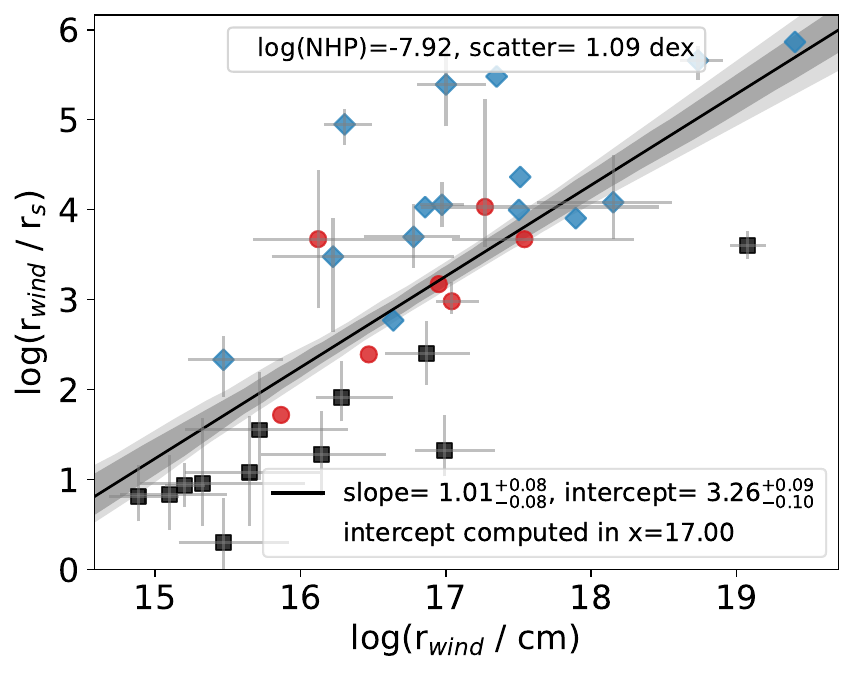}
    \end{subfigure}\\
    \begin{subfigure}[b]{0.25\textwidth}
        \includegraphics[width=\linewidth]{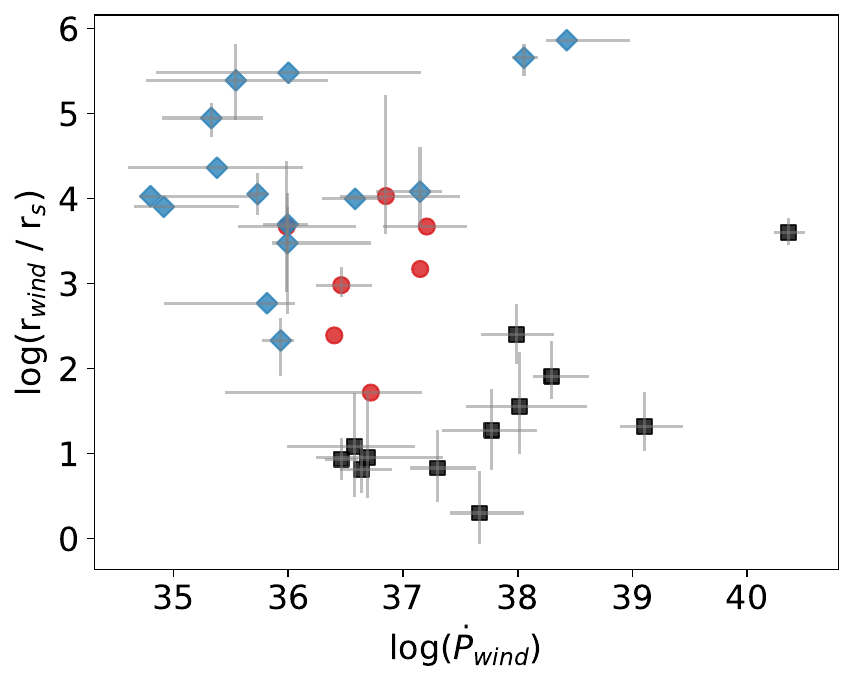}
    \end{subfigure}
    \begin{subfigure}[b]{0.25\textwidth}
        \includegraphics[width=\linewidth]{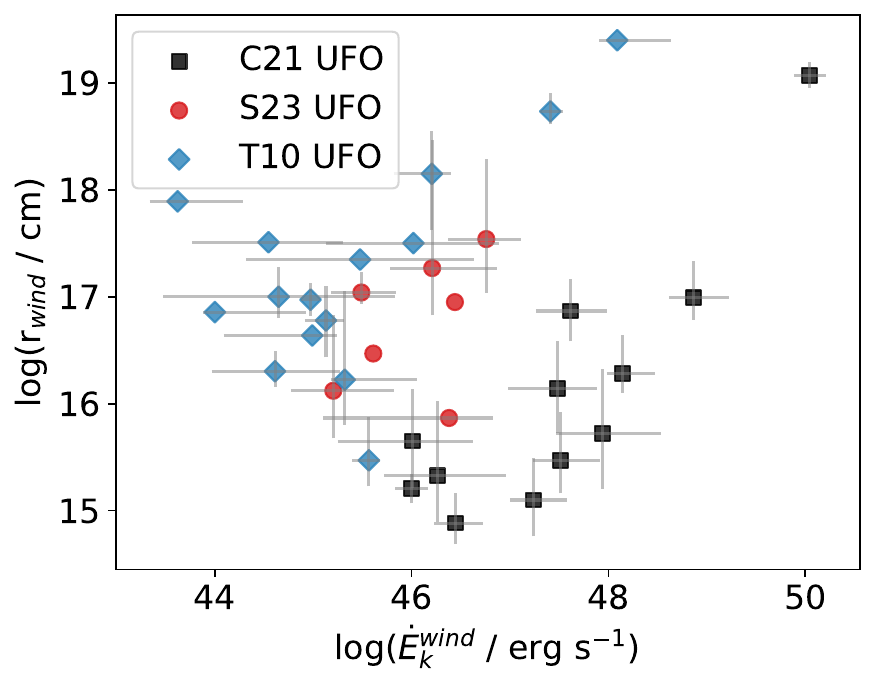}
    \end{subfigure}
    \begin{subfigure}[b]{0.25\textwidth}
        \includegraphics[width=\linewidth]{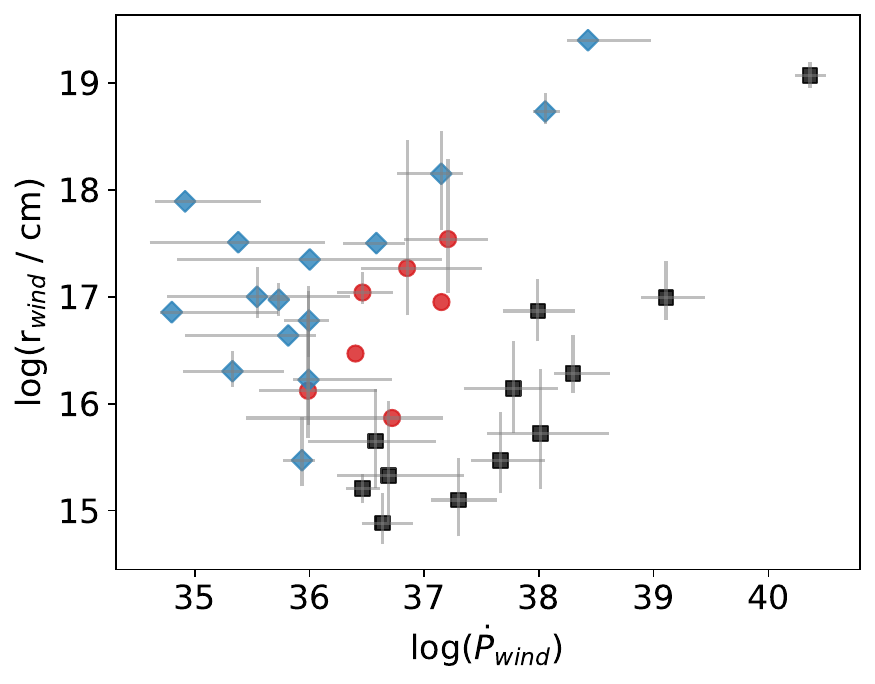}
    \end{subfigure} 
    \caption[]{{UFO radii versus energetics: significant and non correlations for the S23, T10, and C21 samples. The best-fitting linear correlations, applied exclusively to statistically significant correlations, are presented by the solid black lines and the dark and light gray shadowed areas indicate the 68\% and 90\% confidence bands, respectively. In the legend, we report the best-fit coefficients, $\log\mathrm{NHP}$, and the intrinsic scatters for the correlations.}}
    \label{fig:ef17}
\end{figure*}

\begin{figure*}
    \centering
    \begin{subfigure}[b]{0.4\textwidth}
        \includegraphics[width=\linewidth]{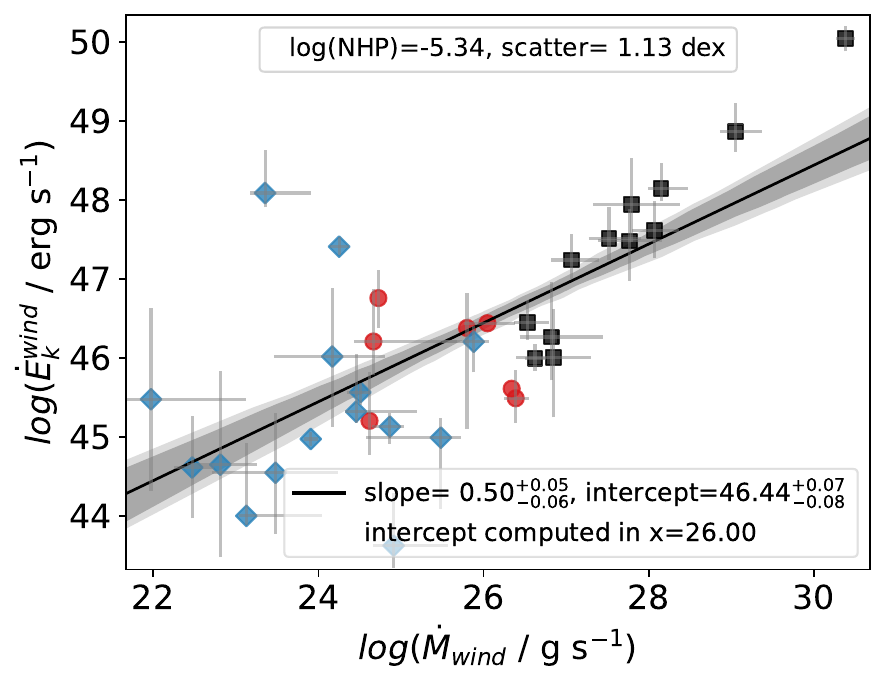}
    \end{subfigure}
    \begin{subfigure}[b]{0.4\textwidth}
        \includegraphics[width=\linewidth]{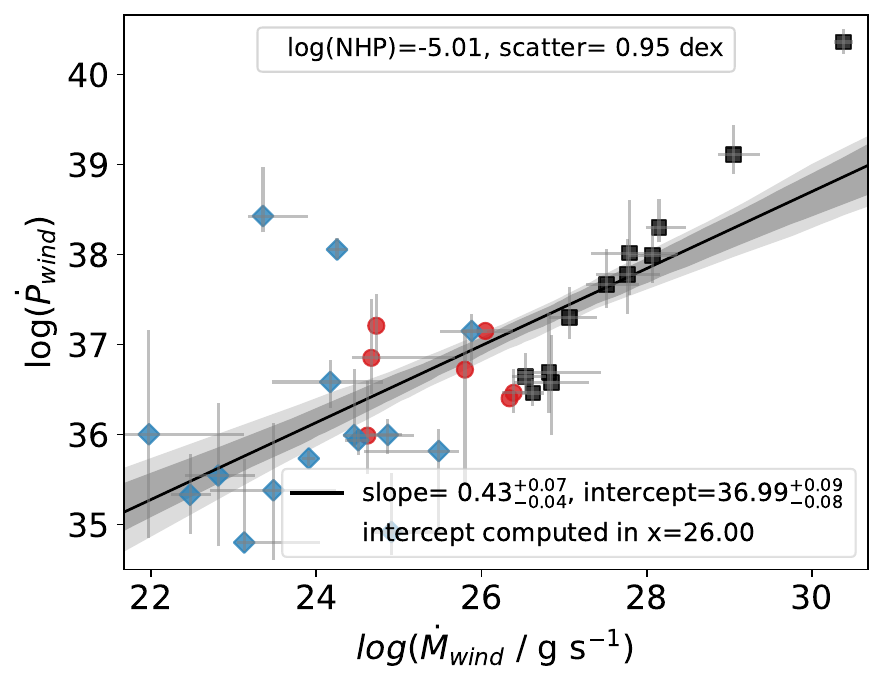}
    \end{subfigure}\\
    \begin{subfigure}[b]{0.4\textwidth}
        \includegraphics[width=\linewidth]{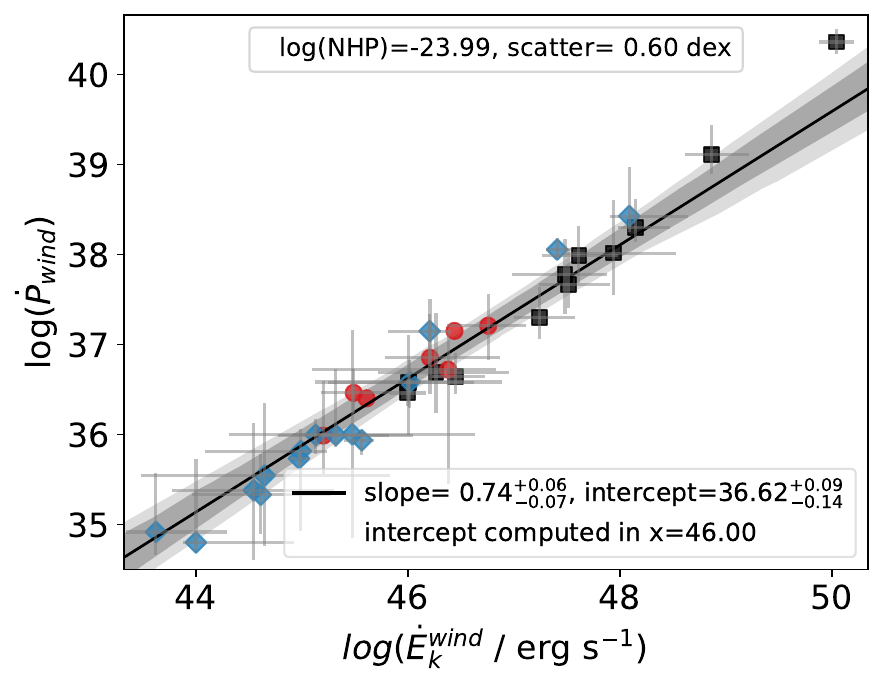}
    \end{subfigure}\\
    \caption[]{{UFO energetics: significant and non correlations for the S23, T10, and C21 samples. The best-fitting linear correlations, applied exclusively to statistically significant correlations, are presented by the solid black lines and the dark and light gray shadowed areas indicate the 68\% and 90\% confidence bands, respectively. In the legend, we report the best-fit coefficients, $\log\mathrm{NHP}$, and the intrinsic scatters for the correlations.}}
    \label{fig:ef18}
\end{figure*}

We first report the correlations (significant and non) obtained between AGN parameters (Figs.~\ref{fig:ef1}-\ref{fig:ef5}), then between UFO characteristics and AGN parameters (Figs.~\ref{fig:ef6}-\ref{fig:ef13}) and finally, only between UFO parameters (Figs.~\ref{fig:ef14}-\ref{fig:ef18}). In particular:

\begin{itemize}
    \item Fig.~\ref{fig:ef1}: SMBH mass versus AGN (observed and derived) parameters;
    \item Fig.~\ref{fig:ef2}: redshift versus AGN parameters;
    \item Fig.~\ref{fig:ef3}: X-ray and UV luminosity versus AGN parameters
    \item Fig.~\ref{fig:ef4}: bolometric luminosity versus AGN parameters;
    \item Fig.~\ref{fig:ef5}: SED parameters (i.e., $\Gamma$, FWHM H$\beta$, $\alpha_{\rm{ox}}$, and $\Delta\alpha_{\rm{ox}}$) versus AGN parameters;
    \item Fig.~\ref{fig:ef6}: ionization parameter of the wind versus AGN parameters;
    \item Fig.~\ref{fig:ef7}: ionized column density of the wind versus AGN parameters;
    \item Fig.~\ref{fig:ef8}: outflow velocity of the wind versus AGN parameters;
    \item Fig.~\ref{fig:ef9}: wind launching radius versus AGN parameters;
    \item Fig.~\ref{fig:ef10}: wind launching radius normalized for the Schwarzschild radius versus AGN parameters;
    \item Fig.~\ref{fig:ef11}: mass outflow rate of the wind versus AGN parameters;
    \item Fig.~\ref{fig:ef12}: kinetic energy of the wind versus AGN parameters;
    \item Fig.~\ref{fig:ef13}: momentum rate of the wind versus AGN parameters;
    \item Fig.~\ref{fig:ef14}: ionization parameter of the wind versus UFO (observed and energetics) parameters;
    \item Fig.~\ref{fig:ef15}: ionized column density of the wind versus UFO parameters;
    \item Fig.~\ref{fig:ef16}: outflow velocity of the wind versus UFO parameters;
    \item Fig.~\ref{fig:ef17}: launching radii versus energetics of the wind;
    \item Fig.~\ref{fig:ef18}: energetics of the wind;
    \item Fig.~\ref{fig:ef19}: significant correlation after the addition of the Fe-K sub-sample (see Sect.~\ref{t10_sample}).
\end{itemize}

\end{appendix}

\end{document}